\documentclass{article} %[12pt,a4paper]{article}
\usepackage{arxiv}

\usepackage[utf8]{inputenc} % allow utf-8 input
\usepackage[T1]{fontenc}    % use 8-bit T1 fonts
\usepackage{lmodern} %{lmodern} %{mlmodern} %{gfsbodoni} 
\usepackage{hyperref}       % hyperlinks
\hypersetup{
    colorlinks=true,
    linkcolor=blue, %blue, \color{ForestGreen} Cerulean
    citecolor=blue,
    pdfpagemode=FullScreen,
    }
\usepackage{url}            % simple URL typesetting
\usepackage{booktabs}       % professional-quality tables
\usepackage{amsfonts}       % blackboard math symbols
\usepackage{nicefrac}       % compact symbols for 1/2, etc.
\usepackage{microtype}      % microtypography
\usepackage{lipsum}		% Can be removed after putting your text content
\usepackage{graphicx}
\usepackage[sort&compress]{natbib} %, style=numeric-comp
\usepackage{doi}
\usepackage{amsthm}
\usepackage{amsmath}
\usepackage{amssymb}
\usepackage{makecell}
\usepackage{cases}
\usepackage{physics}
\usepackage{nicematrix}
\usepackage{blkarray}
\usepackage{comment}
\usepackage{changepage}
\usepackage{tabularx}
\usepackage{multirow}
\usepackage[titletoc]{appendix}

\usepackage{changepage}

\usepackage{enumitem}

\newtheoremstyle{exampstyle}
  {8pt} % Space above %\topsep
  {\topsep} % Space below
  {\itshape} % Body font \itseries
  {} % Indent amount
  {\bfseries} % Theorem head font
  {.} % Punctuation after theorem head
  {.5em} % Space after theorem head
  {} % Theorem head spec (can be left empty, meaning `normal')

\theoremstyle{exampstyle} 
\newtheorem{Property}{Property}

\theoremstyle{exampstyle}
\newtheorem{Theorem}{Theorem}
\newtheorem{Definition}{Definition}
\newtheorem{Corollary}{Corollary}
\theoremstyle{example} \newtheorem{Example}{Example}

\setlist[itemize]{itemsep=0.1pt} %{noitemsep}
\setlist[itemize]{parskip=0.0pt}
\setlist[itemize]{parsep=0.0pt}

\newcommand{\data}{\boldsymbol{\hat{\alpha}}}
\newcommand{\dataprime}{\boldsymbol{\hat{\alpha}}^{\prime}}
\newcommand{\spin}{\boldsymbol{s}}
\newcommand{\state}{\boldsymbol{\alpha}}
\newcommand{\vecmu}{\boldsymbol{\mu}}
\newcommand{\vecnu}{\boldsymbol{\nu}}

\newcommand{\veczero}{\boldsymbol{0}}

\newcommand{\vecg}{\boldsymbol{g}}
\newcommand{\vecr}{\boldsymbol{r}}
\newcommand{\vecell}{\boldsymbol{\ell}}
\newcommand{\vecghat}{\boldsymbol{\hat{g}}}
\newcommand{\basis}{\boldsymbol{b}}
\newcommand{\Basis}{\boldsymbol{B}}
\newcommand{\M}{\mathcal{M}}
\newcommand{\T}{\mathcal{T}}
\newcommand{\Mtilde}{\tilde{\M}} % {\overset{\sim}{\M}} }
\newcommand{\Field}{(\mathbb{Z}/q\mathbb{Z})}
\newcommand{\Ftwo}{(\mathbb{Z}/2\mathbb{Z})}
\newcommand{\FIM}{\mathcal{I}} %{\mathbf{J}}
\newcommand{\I}{i}
\newcommand{\C}{\mathbb{C}}
\newcommand{\R}{\mathbb{R}}
\newcommand{\Tmat}{\boldsymbol{T}}
\newcommand{\Mmat}{\boldsymbol{M}}
\newcommand{\Imat}{\boldsymbol{I}}
\title{Modeling 
Discrete Data\\ 
with High-Order Vector Potts Models}
\author{ 
    Aaron De~Clercq\\
	Helmholtz Institute Erlangen-Nürnberg\\ for Renewable Energy\\
	Forschungszentrum J\"ulich GmbH, 
    Germany\\
	\texttt{a.de.clercq@fz-juelich.de} \\
	\And
    Merijn Moody\\
	Institute for Theoretical Physics\\
    Korteweg-de Vries Institute for Mathematics\\
	University of Amsterdam, %Amsterdam, 
    the Netherlands\\
	\texttt{m.moody@uva.nl} \\
	\AND
    Cl\'elia de~Mulatier\\
	Institute for Theoretical Physics\\
    Informatics Institute\\
	University of Amsterdam,
    the Netherlands\\
	\texttt{c.m.c.demulatier@uva.nl} %\\
}

\renewcommand{\shorttitle}{\textit{arXiv} Template}

\hypersetup{
pdftitle={A template for the arxiv style},
pdfsubject={q-bio.NC, q-bio.QM},
pdfauthor={Aaron De~Clercq, Merijn Moody, Cl\'elia de~Mulatier},
pdfkeywords={statistical inference, discrete data, Potts model, %Spin models
high-order interactions, gauge transformation, cluster expansion, information theory, complexity},
}

\begin{document}
\maketitle

\begin{abstract}
    %\vspace{-0.5cm}
	Modeling high-dimensional data is challenging, yet essential to understanding many complex systems. 
    Maximum entropy models such as Ising and Potts models have been used extensively to capture pairwise interactions from correlation patterns in data, allowing to infer graphical representations of complex systems from observations (e.g., from protein sequences, or neural population activity). 
    Recently, there has been growing interest in modeling higher-order correlation patterns involving simultaneously three or more variables. While progress has been made in binary data with high-order Ising models, we extend this framework to the more general case of discrete data.\\[2mm]
    \indent We introduce $q$-state spin models, a complete family of maximum entropy models that generalize the vector Potts model to include long-range and arbitrary high-order interactions. 
    Even in the pairwise case, compared to the standard vector Potts model, our models allow for more diverse interaction types. 
    We discuss their statistical interpretation through examples and relate these models to discrete Fourier analysis. 
    Using a loop expansion of the partition function, we show that the statistical properties of $q$-state spin models %these models 
    are fully captured by the algebraic structure of their interactions.
    %a set of linear dependencies between the interactions of the model (namely, the loops of the models). 
    We define gauge transformations under which this structure, and therefore the partition function, remains invariant.
    Models equivalent under gauge transformations can thus be seen as different representations of the same abstract statistical model, despite potentially having interactions of different orders, extending results from the binary case.
    For practical application of this framework to data analysis, we focus on a subset of models known, in the binary case, as Minimally Complex Models, generalizing them to discrete data.
    We obtain a closed-form expression for the marginal likelihood of these models, enabling fast model selection. We illustrate their use with simple real-world examples.
\end{abstract}

% keywords can be removed

\newpage
\tableofcontents
\newpage
%%%%%%%%%%%%%%%%%%%%%%%%%%%%%%%%%%%%%%%%%%
\section{Introduction}
\label{sec:intro}

The analysis of high-dimensional data has become a central challenge in modern science, driven by the explosion of high-throughput experiments. % [in fields ranging from biology to neuroscience and social dynamics]. 
Whether analyzing the firing patterns of neural populations, the sequence variations in protein families, or gene regulatory networks, the primary goal is often the same: % there is a common problem
to infer the microscopic laws or structural constraints governing the system 
%\textcolor{red}{from experimental data.} 
based on macroscopic observations obtained from experimental data.
From a statistical modeling standpoint, this problem is frequently addressed using the principle of maximum entropy~\cite{jaynes1957information}, which enables %leads to
%allows %\textcolor{red}{us} to construct 
the construction of the least biased probability distribution consistent with measured observables.
Over the last three decades, this approach has been successful at reconstructing graphical models %representations/
of microscopic interactions in complex systems from discrete data, %using maximum entropy models
specifically pairwise Ising and Potts models. 
%\textcolor{blue}{various complex systems have been successfully modeled using graphical representations reconstructed using this principle}
More generally, physicists have reformulated the task of extracting information from large, high-dimensional data as an inverse statistical problem~\cite{Nguyen2017}. This research has found broad application across the biological sciences, such as the reconstruction of neural functional connectivity from population activity~\cite{Schneidman2006, Cocco2018}, the inference of gene regulatory networks from sequencing data~\cite{Lezon2006}, and the determination of three-dimensional protein structures from sequence variations~\cite{Weigt2009, Morcos2011, Ekeberg2013, Barton2016, jacquin2016benchmarking, rizzato2020inference}.

Many complex systems however have high-order interactions between their components, which are simultaneous interactions between three or more components, and can %\textcolor{red}{[then]} 
exhibit behaviors that are not well explained by pairwise models. %~\cite{Yu2011}.{Remove this citation?}
In data, this can appear for instance as high-order correlation patterns between the states of the components (i.e., correlations between three or more variables) that cannot be explained by the pairwise correlation structure of the data.
This general problem has led to a recent surge of interest for models with high-order interactions in many fields of complex systems~\cite{Battiston2020, Battiston2021}.
From a data modeling perspective, finding which high-order interactions are most relevant to explain the data (if any) is particularly challenging due to the combinatorial explosion of the number of possible interactions.

%%%%%%%%%%%%%%%%%%%%
%\textcolor{blue}{In data, high-order interactions can create high-order correlation patterns [between the states of the components] (i.e., correlations between three or more variables) that cannot be explained by the pairwise correlation structure of the data. From a data modeling perspective, finding which high-order interactions are most relevant to explain the data (if any) is particularly challenging due to the combinatorial explosion of the number of possible interactions.}
%%%%%%%%%%%%%%%%%%%%

%\paragraph{\bf Spin models with high-order interactions (Binary Data).}
Recent progress has been made in addressing this challenge for \textbf{binary data}. To model high-order statistical patterns, References~\cite{mastromatteo2013typical, beretta2018stochastic} proposed a generalization of Ising-like maximum entropy models to include interactions of arbitrarily high order, which the authors generally call ``spin models'' (also known in the literature as ``high-order Ising models'' or ``generalized Ising models''~\cite{wegner1971duality}).
Consider a binary dataset composed of multiple observations of the state $\spin = (s_1,\,\cdots,\,s_n)$ of $n$ binary variables $s_i$ taking values %\footnote{\textcolor{red}{We can always map any values of binary variables to the two values $\pm 1$ without loss in generality.}} 
in $\{-1,+1\}$ (called ``spin'' variables). 
A spin model is a parametric family of probability distributions of the form~\cite{beretta2018stochastic}:
\begin{equation}\label{eq:graph_model}
p(\spin\,|\,\vecg,\mathcal{M})
    =\frac{1}{Z_{\M}(\vecg)}\,\exp \left({\;\sum_{\vecmu\in \M} 
    g_{\vecmu}\,\phi^{\vecmu}(\spin)}\right)\,,
\end{equation}
characterized by the choice of a set $\M=\{\vecmu^{(1)},\,\ldots, \vecmu^{(K)} \}$ of $K$ interactions of arbitrary order between the spins. Each interaction is represented by an $n$-dimensional binary vector $\vecmu$ that indicates which spins are interacting, 
%(the $i$-th entree is `$1$' if the variable $s_i$ is interacting and `$0$' otherwise).
and the functions $\phi^{\vecmu}(\spin)$ define how the variables interact, %the [(functional) form of the] energy of the interaction
%\textcolor{red}{taking here} 
taking %which \textcolor{red}{take[s]} % [here]
the form of product operators:
\begin{equation}\label{eq:spin-operators:q=2}
    \phi^{\vecmu}(\spin)=\prod_{i=1}^{n} {s_i}^{\mu_i}\,,
    \qquad \forall \vecmu\in\Ftwo^n\,.
\end{equation}
For example, in a five-spin system, a three-body interaction between $s_1$, $s_3$ and~$s_4$ is represented by the binary vector $\vecmu=(1,0,1,1,0)$ and corresponds to the operator $\phi^{\vecmu}(\spin) = s_1s_3s_4$.
Ising-like (pairwise) models only have operators of the form $\phi^{\vecmu}(\spin)=s_i$ and $\phi^{\vecmu}(\spin)=s_is_j$ between pairs of spins.
The vector $\vecg=(g_{\vecmu})_{\vecmu\in\M}$ is a vector of real parameters, % $g_{\vecmu}$, 
and the normalization factor $Z_{\M}(\vecg)$, known in physics as the partition function of the model, is given by:
\begin{align}\label{eq:Z:Ising:def}
    Z_{\M}(\vecg) 
    = \sum_{\spin} \,\exp\left({\;\sum_{\vecmu\in\M}g_{\vecmu}\,\phi^{\vecmu}(\spin)}\right)\,.
\end{align}
%
%The coefficients $g_{\vecmu}$ are real parameters that can be interpreted as modulating the strength of each interaction. The normalization factor $Z_{\M}(\vecg)$ is known in physics as the partition function of the model; it can be written as:
%\textcolor{red}{takes the form:} 
%Ising-like (pairwise) models are a subset of these models that contain only operators of the form $\phi^{\vecmu}(\spin)=s_i$ and $\phi^{\vecmu}(\spin)=s_is_j$ between pairs of spins.
%
The spin models~\eqref{eq:graph_model} are maximum entropy models constrained to produce chosen expectation values of the operators in~$\M$. 
%The average of the operators in $\M$ corresponds to (high-order) moments of the multivariate systems, which are considered as the relevant macroscopic observables for the model.
%
These values correspond to (high-order) moments of the multivariate systems that are considered as the relevant macroscopic observables by the model.
%\textcolor{red}{These relevant (macroscopic) observables correspond to (high-order) moments of the multivariate systems.}
%\textcolor{purple}{These (high-order) moments of the multivariate systems are the macroscopic observables that are considered as relevant for the model.}
%%%%%%%%%%%%%%%%%%%%%%
%The spin models~\eqref{eq:graph_model} are maximum entropy models, and the average[s] of the operators in $\M$ are (high-order) moments of the multivariate systems that the model is constrained to reproduce. They are the relevant macroscopic observables for the model.
%%%%%%%%%%%%%%%%%%%%%%
The coefficients~$g_{\vecmu}$ are Lagrange multipliers, which are usually fitted so that the model average $\langle \phi^{\vecmu} \rangle = \sum_{\spin} p(\spin\,|\,\vecg, \M) \,\phi^{\vecmu}(\spin)$ %of the operators in $\M$ match their
of each operator in $\M$ matches its empirical average in the data. 
%These spin models define a complete family of models able to capture any possible patterns of binary data
The family of spin models is thus able to capture any possible patterns of binary data (i.e., % ,/via/through
any choice of the multivariate %high-order
moments). % of the multivariate system). 
In particular, the complete model with all $(2^n-1)$ interactions can fit all high-order correlation patterns.
%\textcolor{red}{Spin models can thus }
In this context, identifying which patterns of the data are relevant is equivalent to %finding which minimal %least complex
%subset of interactions best models the data.
finding 
%\textcolor{blue}{the smallest subset of interactions able to capture the data patterns.} 
%[least complex] 
the subset of interactions that best models the data,
%This is a particularly challenging, because the number of possible models grows super-exponentially with $n$, as $2^{(2^n-1)}$.}
%This is challenging due to the super-exponential number of \textcolor{red}{possible models/combinations}.
%which is [made] challenging [by] the super-exponential number of possible combinations.
which is challenging due to the super-exponential number of possible choices.

%%%%%%%%%%%%%%%%%%%%%%%%%%%%%%%%%%%%%
%\textcolor{red}{This model is a maximum entropy model constrained to reproduced the expectation values of the operators selected in $\M$, and the coefficients~$g_{\vecmu}$ are then Lagrange multipliers that are fitted so that the expectation values $\langle \phi^{\vecmu} \rangle = \sum_{\spin} p(\spin\,|\,\vecg, \M) \,\phi^{\vecmu}(\spin)$ match the expectation values of $\phi^{\vecmu}(\spin)$ over the dataset.}
%Ising-like (pairwise) models are a subset of these models that contain only operators of the form $\phi^{\vecmu}(\spin)=s_is_j$ between pairs of spins.
%The normalization factor $Z_{\M}(\vecg)$ is the partition function:
%\begin{align}\label{eq:Z:Ising:def}
%    Z_{\M}(\vecg) 
%    = \sum_{\spin} \,\exp\left({\;\sum_{\vecmu\in\M}g_{\vecmu}\,\phi^{\vecmu}(\spin)}\right)\,.
%\end{align}
%These spin models define a complete family of models able to capture any possible patterns of binary data. In particular, the complete model with all $2^n-1$ interactions can fit all high-order correlation patterns.
%%%%%%%%%%%%%%%%%%%%%%%%%%%%%%%%%%%%%

\begin{figure}[h]
    \centering
    \includegraphics[width=1.0\textwidth]{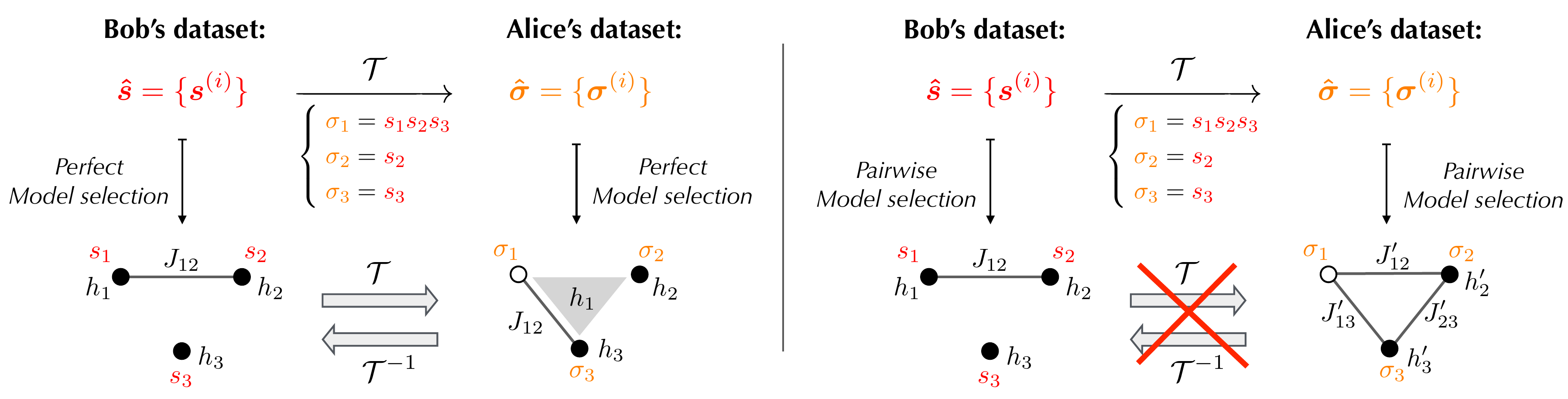} 
\caption{\textbf{Example illustrating equivariant model selection} for binary %discrete 
datasets $\boldsymbol{\hat{s}} = \{\spin^{(i)}\}_{i=1}^N$, where each datapoint $\spin^{(i)}$ is a state of a three-spin system $\spin = (s_1,s_2,s_3) \in \{-1,+1\}^3$. %$s^{(i)} \in \mathbb{F}_q^n$. 
A model selection procedure is equivariant if, for a basis transformation of the dataset $\mathcal{T} \colon \spin \to \boldsymbol{\sigma}$, the models obtained from the model selection procedure on $\boldsymbol{\hat{s}}$ (Bob's data) and $\boldsymbol{\hat{\sigma}}$ (Alice's data) are related by the same transformation $\mathcal{T}$.
The left panel shows an example of an equivariant model selection procedure, for which the spin model found %in Alice's \textcolor{red}{data representation} %representation of the data}
%corresponds to the transformation of Bob's model 
%
in Alice's data representation corresponds to the model found in Bob's representation rewritten in Alice's basis,
%\textcolor{blue}{by Alice corresponds to Bob's model $\M_{\rm Bob}$ rewritten in Alice's basis}, 
i.e. $\M_{\rm Alice} = \T[\M_{\rm Bob}]$.
In the right panel, Alice and Bob have the same data as in the left panel, but they can only select among pairwise models. In this case, Bob will find the same model as in the left panel because his model is already pairwise, but Alice's model cannot be the same because her model has a three body interaction. Instead, Alice will find a pairwise model that likely has similarities with her left-panel model (such as the projection of the three-body interaction onto pairwise interactions), but which cannot be consistent with Bob's findings anymore. This gives an example where it is clear that % in which the 
pairwise model selection is not equivariant.
%
%\textcolor{orange}{This gives an example in which the pairwise approach wouldn't be able to find the consistent solution for Bob and Alice.}
%Equivariance is only possible if we allow for all possible high-order interactions in spin models as $\mathcal{T}$ generally transforms pairwise models into models with high-order interactions.
%Add $\boldsymbol{\sigma}=\basis(\spin)$.
}
\label{fig:Bob_and_Alice}
\end{figure}

%\textcolor{red}{An interesting property which has important consequences for modeling is that this family of models is invariant under the changes of (basis) representations of the variables introduced by~\cite{beretta2018stochastic} as \textbf{gauge transformations}.}
%
An interesting property of this family of models, with %that has/with
important implications %consequences %repercussions %consequences %can reveal important 
for modeling, is that it is closed %invariant
under %a group of
changes of %\textcolor{red}{(basis)} 
representation of the spin variables, %which Ref.~\cite{beretta2018stochastic} called \textbf{gauge transformations}.
called \textbf{gauge transformations} by~\cite{beretta2018stochastic}.
%introduced in~\cite{beretta2018stochastic} as \textbf{gauge transformations}.
%%%%%%%%%%%
%Formally, given a set of $n$ independent operators $\basis(\spin) = (\phi^{\vecmu_1}(\spin), \dots, \phi^{\vecmu_n}(\spin))$ we can define a gauge transformation as the bijective map $\spin' = \basis(\spin)$.
%%%%%%%%%%%
A gauge transformation is a bijective map of the form %$(\sigma_1,\cdots,\sigma_n)=(\phi^{\vecmu_1}(\spin), \dots, \phi^{\vecmu_n}(\spin))=\basis(\spin)$ 
$\boldsymbol{\sigma}=\basis(\spin)$, which defines a %\textcolor{red}{
new basis of spin variables %} % a new spin basis
$\boldsymbol{\sigma}=(\sigma_1,\cdots,\sigma_n)$ in terms of the original variables $\spin$ using a set of %$n$ 
independent operators $\basis(\spin) = (\phi^{\vecmu_1}(\spin), \dots, \phi^{\vecmu_n}(\spin))$ (see example in Fig.~\ref{fig:Bob_and_Alice}). %Under such transformation, a spin operator $\phi^{\vecmu}(\spin)$ is transformed by re-expressing it in terms of the new basis variables $\boldsymbol{\sigma}$ and remains a spin operator. Consequently, the gauge transformation of a spin model is also a spin model.
The probability distribution~\eqref{eq:graph_model} of any spin model can be rewritten in the new basis $\boldsymbol{\sigma}$ by re-expressing each spin operator in terms of the new variables. %\textcolor{red}{, which yields a new spin model}. 
In the new basis, the operators still have the form of spin operators, 
%a spin operator still has the form of a spin operator, 
and the model probability distribution thus remains that of a spin model\footnote{
%Formally, given a set of $n$ independent operators $\basis(\spin) = (\phi^{\vecmu_1}(\spin), \dots, \phi^{\vecmu_n}(\spin))$ we can define a gauge transformation as the bijective map $\spin' = \basis(\spin)$.
Under a gauge transformation, any spin operator $\phi^{\vecmu}(\spin)$ can be transformed by re-expressing it in terms of the new basis variables $\spin' = \basis(\spin)$. Using the inverse transformation $\spin = \basis^{-1}(\spin')$, the transformed operator is given by:
\begin{equation}
    \mathcal{T}_{\basis}[\phi^{\vecmu}](\spin') \doteq \phi^{\vecmu}(\basis^{-1}(\spin'))\,.
\end{equation}
This yields a transformed model $\mathcal{T}_b[\mathcal{M}]$, which is obtained by transforming each operator within $\mathcal{M}$:
\begin{equation}
    \mathcal{T}_{\basis}[\mathcal{M}] \doteq \{\mathcal{T}_{\basis}[\phi^{\vecmu}] \mid \phi^{\vecmu} \in \mathcal{M}\}\,.
\end{equation}
Consequently, the probability of finding the system in state $\spin$ under model $\mathcal{M}$ is identical to the probability of finding the system in the transformed state $\spin'$ under the transformed model $\mathcal{T}_{\basis}[\mathcal{M}]$:
\begin{equation}
    p(\spin\,|\,\vecg,\mathcal{M}) = p(\spin'\,|\,\vecg',\mathcal{T}_{\basis}[\mathcal{M}])\,,
\end{equation}
where the parameter vector $\vecg'$ is a permutation of the original parameters $\vecg$, ensuring that the transformed operator in $\mathcal{T}_{\basis}[\mathcal{M}]$ is coupled to the same parameter as the original operator in $\mathcal{M}$.
}~\cite{beretta2018stochastic}.
We denote $\M'=\mathcal{T}_{\basis}[\M]$ the spin model resulting from the gauge transformation $\boldsymbol{\sigma}=\basis(\spin)$ of a model $\M$.
%After re-expressing it in terms of the new basis variables, a spin operator remains in the form of a spin operator.
Crucially, models related by a gauge transformation can have interactions of different orders. In particular, standard pairwise models can map to models containing high-order interactions. %~\cite{beretta2018stochastic}. 
Moreover, models related by gauge transformations have the same partition function~$Z_{\M}(\vecg)$~\cite{beretta2018stochastic},
%Ref.~\cite{beretta2018stochastic} also showed that models related by gauge transformations have the same partition function. %, and can therefore simply be seen as different representation of the same abstract statistical model.
%\textcolor{red}{Models equivalent under gauge transformation can simply be seen as different representations of the same abstract statistical model.}
and can thus be understood as %seen as 
different representations of the same abstract statistical model.

This gives rise to the concept of \textbf{equivariant model selection}, as illustrated in Figure~\ref{fig:Bob_and_Alice}. Suppose we have a binary dataset~$\boldsymbol{\hat{s}}$ written in terms of the variables~$\spin$ %(for example Bob's dataset) 
(Bob's dataset) and our model selection procedure yields an optimal model~$\M$. % (e.g., Bob's dataset and model).
%
%If we apply a basis transformation to the dataset by rewriting it in terms of new basis variables $\boldsymbol{\sigma}=\basis(\spin)$, we obtain a binary dataset $\boldsymbol{\hat{\sigma}}$ (Alice dataset). 
%If we apply a basis transformation $\basis$ to the dataset to obtain the dataset $\boldsymbol{\hat{s}}'$ and perform the exact same model selection procedure, it yields a new model $\M'$. 
If we rewrite the dataset in terms of the new basis variables $\boldsymbol{\sigma}=\basis(\spin)$ %(for example Alice's dataset)
using a gauge transformation and perform the exact same model selection procedure to the new dataset~$\boldsymbol{\hat{\sigma}}$ (Alice's dataset), it will yield a model~$\M'$ that is a priori different from~$\M$.
%it produces/yields a new model~$\M'$. \textcolor{red}{it will give a model~$\M'$, which should a priori corresponds to the transformation of the model~$\M$.}
%Performing the exact same model selection procedure then yields the new model $\M'$.
%
One can expect an unbiased model selection procedure among all spin models to yield models that are consistent across different representations of the data. In this example, this would mean that $\M'$ is related to $\M$ by the same gauge transformation, $\M' = \mathcal{T}_{\basis}[\mathcal{M}]$.
%\textcolor{blue}{If the model selection procedure is able to select among all spin models, then we expect $\M'$ to be related to $\M$ by the same gauge transformation, $\M' = \mathcal{T}_{\basis}[\mathcal{M}]$. A priori, one can expect an unbiased model selection procedure to yield models that are consistent across different representations of the data.}
Such a model selection procedure is called equivariant.
%\textcolor{purple}{The model selection procedure is considered equivariant if $\M' = \mathcal{T}_{\basis}[\mathcal{M}]$.}
Because the family of pairwise models is not closed under gauge transformations, equivariant model selection is fundamentally impossible unless we encompass high-order interactions within our model selection framework. This underscores the %critical 
%\textcolor{red}{need for/
importance of developing a comprehensive high-order framework for statistical inference.

%a pairwise model selection may not find consistent answers (Fig.~\ref{} gives an example in which the pairwise approach wouldn't be able to find the consistent solution for Bob and Alice).

%%%%%%%%%%%%%%%%%%%%%%%%
%\paragraph{\bf Generalization to Discrete Data.}
%%%%%%%%%%%%%%%%%%%%%%%%
While this framework is well-established for binary variables, a comprehensive framework for \textbf{discrete data} 
%(where variables take values in $\{1, \dots, q\}$) 
(where variables can take $q$ different values with $q\geq 2$) remains under-explored. In this paper, we extend this framework to the more general case of discrete data. 
Taking inspiration from the binary case, 
%\textcolor{blue}{[we introduce planar spins taking $q$ values evenly distributed around the unit circle and]} 
we define a complete family of maximum entropy models able to reproduce %capable of reproducing 
arbitrary multivariate moments of the discrete data.
This family of models, which we call $q$-state spin models, generalizes the vector Potts model to include arbitrary high-order interactions, and reduces to the models in~\eqref{eq:graph_model} when $q=2$.
%%%%%%%%%%%%%%%%%%%%%%%%%%%
%We introduce a complete family of maximum entropy models that generalize the vector Potts model to include arbitrary high-order interactions. We take an approach similar to the binary case to define a complete family of probabilistic models for discrete data. 
%%%%%%%%%%%%%%%%%%%%%%%%%%%
The complete $q$-state spin models, with all $2^{q^n-1}$ parameters, can model any patterns of discrete data. More specifically, we can relate this model to a Fourier decomposition of the log-probability.
%expressing %decomposing/writing 
%the log-probability as a Fourier expansion. 

%We show that this discrete class of statistical models also has gauge transformation invariance allowing for equivariant model selection.
Similarly to the binary case, we identified changes of representation of the %spin system
discrete system (gauge transformations) that transform a $q$-state spin model into another $q$-state spin model, with interactions that can be of different orders. 
Using a loop expansion of the partition function, we show that these models have the same partition function, and thus correspond to different representations of the same abstract statistical model. This defines equivalent classes of $q$-state spin models, which, in particular, allows for the development of equivariant model selection. %enables equivariant model selection.
% [Finally], 
We introduce a linear algebra representation of spin models, which provides a comprehensive mathematical framework for working with spin models. 
In particular, this framework formalizes %encompass
the definition of gauge transformations, the %algebraic
characterization of the loop structure and of equivalence classes, and the definition of symmetries in data and models.
%%%%%%%%%%%%%%%%%
%In particular, it allows for a formal definition of the gauge transformations and model symmetries, and gives an algebraic characterization of the loop structure and a better characterization of the invariants of the model classes.
%%%%%%%%%%%%%%%%%

%Using a loop expansion of the partition function, we show that the statistical properties of $q$-state spin models are fully captured by the algebraic structure of their interactions (invariant of the class).

%%%%%%%%%%%%%%%%%%%%%%%%%%%%%%%%%%%%%%%
%\paragraph{\bf Model selection.}
%%%%%%%%%%%%%%%%%%%%%%%%%%%%%%%%%%%%%%%
While the complete model can be used to infer all probability distributions, it is generally over-parameterized for modeling systems with limited data. The challenge of finding a good model for data lies in selecting the most concise description that captures the relevant statistics without overfitting. Ideally, one would compare all possible models to find the one that best describes a dataset~$\boldsymbol{\hat{s}} = \{\spin^{(i)}\}_{i=1}^N$, i.e., the one achieving an optimal balance between goodness-of-fit and simplicity. %\textcolor{red}{[while minimizing some complexity penalty]}.
%\textcolor{red}{Assuming no specific prior knowledge,} 
Assuming no prior preference between models, we aim to find the model with the largest {\em evidence} (or {\em marginal likelihood})~\cite{kruschke2014doing}:
\begin{equation}\label{eq:evidence}
P(\boldsymbol{\hat{s}}\,|\,\M)
    =
    \int_{\mathbb{R}^M}\!d\vecg \,\prod_{i=1}^N \,p\big(\spin^{(i)}|\,\vecg,\mathcal{M}\big)\;P_0(\vecg\,|\,\M)\,,
\end{equation}
where $P_0(\vecg\,|\,\mathcal{M})$ is a prior distribution over the model parameters. %This approach accounts for model complexity, \textcolor{red}{penalizing more %[flexible] 
%models that can fit more complex patterns}.
%\textcolor{red}{penalizing flexible models that can fit arbitrary patterns}. 
Assuming Jeffreys' prior, in the limit of a large number $N$ of datapoints, maximizing the evidence is also equivalent %to the Minimum Description Length (MDL) principle~\cite{rissanen1986stochastic, rissanen1996fisher, grunwald2007minimum}:
to finding the model minimizing the description length of the data~\cite{rissanen1986stochastic, rissanen1996fisher, grunwald2007minimum, balasubramanian1997statistical, myung2000counting}:
\begin{align}\label{eq:MDL}
    L(\boldsymbol{\hat{s}}\,|\,\M) = \max_{\vecg}\log P(\boldsymbol{\hat{s}}\,|\,\vecg,\,\M) - \frac{K}{2}\,\log\left(\frac{N}{2\pi}\right)-c_{\M}+O\left(\frac{1}{N}\right)\,.
\end{align}
%Here, 
The first term measures goodness-of-fit of the model, while the subsequent terms penalize % \textcolor{red}{[for]} 
model complexity, based on the number~$K$ of parameters and the geometric complexity $c_{\M}$~\cite{myung2000counting, beretta2018stochastic}.
Calculating these quantities for all possible models is computationally intractable. To remedy this, we generalize the definition %\textcolor{red}{concept/notion/definition} 
of minimally complex models (MCMs), introduced for binary data in Ref.~\cite{demulatier2024MCM}, to the discrete case.
MCMs are a sub-family of the $q$-state spin models that correspond to factorizations of the model probability distribution in some spin basis. %, allows us to find an optimal 
%By finding an optimal factorization of the probability distribution that matches the data structure, MCMs enable an effective dimensionality reduction, ensuring that the selected model captures only the statistically relevant patterns.
A remarkable advantage of MCMs is that the model evidence~\eqref{eq:evidence} has a closed-form expression, 
%Moreover, MCMs allow a closed-form expression of the evidence~\eqref{eq:evidence}, 
making Bayesian model selection among MCMs feasible. %By selecting the best MCM based on maximizing the evidence
The MCM with the largest evidence then identifies an optimal factorization of the empirical probability distribution that matches the data structure.

The paper is organized as follows. In Section~\ref{sec:def:HO:vectorPottsModel}, we define high-order vector Potts models, as maximum entropy models constrained to reproduce high-order moments of a discrete spin system. 
In Section~\ref{sec:2a_LoopExpansion}, we obtain a loop expansion of the partition function, which connects the statistical properties of the model to its algebraic structure (the loops of the model), extending results from the binary case to general $q$.
%This connects the statistical properties of the model to its algebraic structure, extending results from the binary case to general $q$. 
%
In Section~\ref{Sec:GT:MatrixRepresentation}, we introduce a linear algebra representation of spin models, which gives a formal framework for working with equivalence classes of $q$-state spin models. 
%%%%%%%%%%%%%%%%%%%%%
%In particular, this formalizes the definition of gauge transformations, the characterization of the loop structure, and the definition of symmetries in data and models.
%%%%%%%%%%%%%%%%%%%%%
Finally, in Section~\ref{sec:3_MCM}, we focus on the subset of Minimally Complex Models and, in Section~\ref{Sec:Applications}, we illustrate their use for model selection in real-world discrete data.

%%%%%%%%%%%%%%%%%%%%%%%%%%%%%%%%%%%%%%%%%%
\section{Complete family of maximum entropy models for discrete data}
%Generalization of the planar Potts model to high-order interactions
\label{sec:def:HO:vectorPottsModel}

In this section, we define a complete family of parametric models for discrete data. Inspired by the high-order generalization of the Ising model in Eq.~\eqref{eq:graph_model}~\cite{mastromatteo2013typical, beretta2018stochastic}, we define maximum entropy models able to reproduce any (high-order) moments of a discrete planar spin system, introducing a high-order generalization of the planar Potts model~\citep{Potts1952} for statistical inference of discrete data.
%\textcolor{blue}{generalizing the planar Potts model~\citep{Potts1952} with %by including
%high-order interactions for the purpose of statistical inference of discrete data.}
The resulting %obtained
family of probability distributions %we obtain 
extends %generalizes 
the %family of 
maximum entropy models in Eq.~\eqref{eq:graph_model} to any discrete values of $q\geq 2$.
In Sec.~\ref{Sec1.3:FT}, we draw attention to the fact that these %the chosen family of models corresponds 
models correspond to a spatial discrete Fourier decomposition of the empirical log-probability (or energy function), which has interesting implications regarding the use of these models. Finally in Sec.~\ref{sec:Potts:interpretation}, we give some insight into the way these models encode patterns of discrete data.

%\noindent {\it Justification:} We would like to define a complete family of parametric models for discrete data. Inspired by the high-order Ising models, we introduce this generalization to discrete data. Idea of finding/defining the most general maximum entropy distribution for a system of $q$-state variables

\subsection{Definitions -- $q$-state spin models with interactions of arbitrary order} % for discrete data}
\label{Sec:1.1}
%(High-order) Maximum entropy models for discrete spin variables} 
%{Complete families of (high-order interaction) models for discrete data.}
%Discrete variables, Spin operators, maximum entropy model for discrete data.} 
%High-order vector/planar Potts model: A complete family of maximum entropy model for discrete data.} %multi-variate discrete systems/data}

We are interested in studying systems of $n$ random variables that can take $q$ different values  (where $q$ is a positive integer).
One can map these $q$ values to integers between $0$ and $(q-1)$ without losing in generality.
We consider discrete variables in this format and define planar spin variables by mapping these $q$ values to evenly spaced positions on the unit circle. 
The choice of this initial mapping must be taken into account when interpreting the interactions of the model, which will be discussed in Sec.~\ref{sec:Potts:interpretation}.

\begin{Definition} 
{\bf Discrete variables: $q$-state spins and colors.}
Consider $n$ discrete random variables $(\alpha_1,\, \cdots,\, \alpha_n)$ taking integer values in $[0,\,q-1]$:
\begin{align}\label{def:alpha_j}
    {\rm for}\;{\rm all}\;j\in\{1, \cdots, n\}, \qquad
    \alpha_j\in \Field\,. %\Z/q\Z\,.
\end{align}
This system can be mapped to the system of $n$ complex variables $(s_1,\, \cdots, s_n)$ taking values in the set of the $q$-th roots of unity, using: %defined by
\begin{align}\label{def:s_j}
    {\rm for}\;{\rm all}\;j\in\{1, \cdots, n\}, \qquad\qquad 
    s_j= \exp\left({\displaystyle\frac{2i\pi}{q}\alpha_j}\right)%\,.
    =z_q^{\;\alpha_j}\,,
    %\qquad {\rm where}\;\;\; z_q=\exp(\frac{2\pi}{q}\,i)}\,,
\end{align}
where\; $z_q=\exp(\frac{2\pi}{q}\,i)$ is the first $q$-th root of unity.
We refer to the variables $s_j$ as {\bf $q$-state spin variables} and to the variables $\alpha_j$ as their {\bf colors}.
\end{Definition}

\begin{figure}[h]
    \centering
    \includegraphics[width=0.92\textwidth]{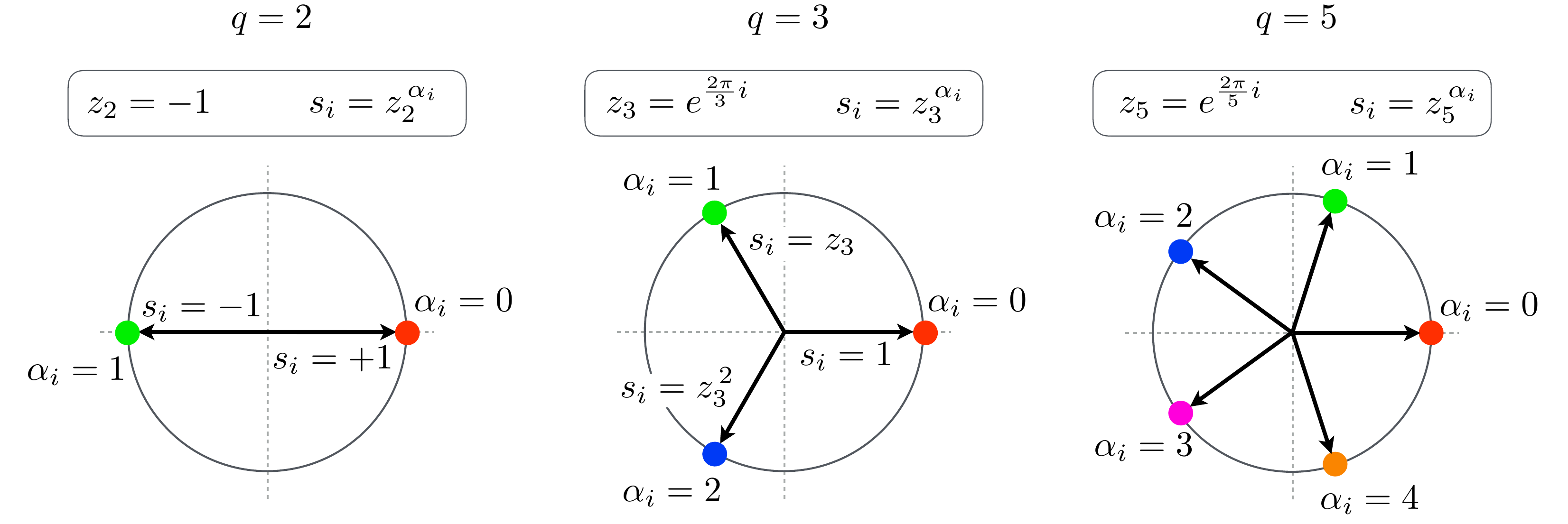} 
    \caption{{\bf Representation for $q$-state spin variables in the complex plane} for different values of~$q$. For a given value of $q$, the color variable $\alpha_i$ can take any integer value modulo $q$, $\alpha_i\in\Field^n$. 
    The corresponding $q$-state spin variable then takes the complex value $s_i=z_q^{\,\alpha_i}$, where $z_q$ is the first $q$-th root of unity. The spin variable can thus be represented in the complex plane by a random unit vector which can point in any of the $q$ different directions corresponding to the $q$-th roots of unity. Observe that $z_q^{q-\alpha_i}$ and $z_q^{\,\alpha_i}$ are complex conjugate spin values, $z_q^{q-\alpha_i}=(z_q^{\,\alpha_i})^*$, which correspond to mirrored directions of $s_i$ with respect to the $x$-axis.
    }
\label{fig:SpinVariables}
\end{figure} 

Equation~\eqref{def:s_j} defines a bijection between the two sets of random variables. In the complex plane, the spin variable~$s_j$ can be seen as a random unit vector whose direction is given by the color variable~$\alpha_j$ (see Fig.~\ref{fig:SpinVariables}), similarly to the planar spins introduced in the very first version of the Potts model~\citep{Potts1952}.
We denote the state of the system by the vector $\state=(\alpha_1,\, \cdots, \alpha_n)\in\Field^n$ when expressed in terms of the variables~$\alpha_j$ and by the complex vector $\spin = (s_1,\, \cdots, s_n)$ when expressed in terms of the variables~$s_j$. In the following, we will use interchangeably the notation with dependency in $\spin$ or in $\state$.
For $q=2$, the variables $s_j$ can take the values $+1$ (for $\alpha_j=0$) or $-1$ (for $\alpha_j=1$), just like in the Ising model and its high-order generalization~\eqref{eq:graph_model}.
One reason for introducing the spin variables~$s_j$ in the form of~\eqref{def:s_j} is that they satisfy the property $s_j^{\,q}=1$. %, \textcolor{red}{which is useful for defining a complete set of high-order moments from which we will construct a complete family of maximum entropy model.}
%
%%%%%%%%%%%%%
%This property is useful for defining gauge transformations in the binary case~\cite{demulatier2024MCM} (which we aim to generalize here) and, as we will see below, to define a complete set of high-order moments from which the maximum entropy model is constructed.
%%%%%%%%%%%%%%%
In the binary case, this property is useful for defining gauge transformations and to obtain a loop expansion of the partition function~\cite{beretta2018stochastic}, both of which we aim to generalize here.
Below, this property is also %immediately 
useful for defining a complete set of high-order moments from which a complete family of maximum entropy model can constructed.

\subsubsection{High-order moments, spin operators, and complete models}
%{Arbitrary-order spin operators and complete models}
\label{subsec:def:CompleteModel}

We aim to define a complete family of parametric models, able to capture any pattern of discrete data. This can be done by defining maximum entropy models constrained to reproduce any choice of moments of the observed system (including high-order moments).
In the following, we construct a family of maximum entropy models able to reproduce arbitrary-order moments of the spin variables $s_j$, as opposed to moments of the colors $\alpha_j$.

\begin{Definition} {\bf Moments of an $n$-spin system.} 
For the system of $n$ $q$-state spin variables $s_j$, there are at most $(q^n-1)$ possible distinct moments, each corresponding to a product of the spins raised to some %a
power between $0$ and $(q-1)$: 
\begin{align}\label{eq:def:multivariate-moments}
    \bigg\langle\, \prod_{j=1}^{n}\, s_j^{\;\mu_j} \,\bigg\rangle\,, 
    \qquad \textrm{for all integer values of }\;\mu_j\in\Field\,, %\{0,\,\cdots,\,q-1\}
\end{align}
except $(\mu_1,\,\cdots,\,\mu_n)=\veczero\,$ for which the product is simply equal to $1$. 
The angle brackets $\langle\,\cdot\,\rangle$ denotes the ensemble average. 
\end{Definition}
Because the spin variables defined in Eq.~\eqref{def:s_j} verify%
\footnote{For all $\mu_j\geq q\,$, $\;s_j^{\mu_j}=\exp\left({\frac{2i\pi}{q}\alpha_j\mu_j}\right) =\exp\left({\frac{2i\pi}{q}\alpha_j(\mu_j\bmod{q})}\right) =s_j^{\mu_j\bmod{q}}\,$.}
that $s_j^{\mu_j}=s_j^{\mu_j\bmod{q}}$,
all the moments corresponding to values of $\mu_j\geq q$ are equal to lower order moments of the form of Eq.~\eqref{eq:def:multivariate-moments}, and therefore there are at most $(q^n-1)$ possible distinct moments. 
%%%%%%%%%%%%% TO KEEP: %%%%%%%%%% %%%%%%%%%% %%%%%%%%%% %%%%%%%%%%
%\footnote{If $q$ is not prime, other moments can also be identical.
%\textcolor{red}{For instance, for $q=4$, $s_j^2 = \exp(\frac{2i\pi}{4} 2\alpha_j) =(-1)^{\alpha_j}$ independently of the values of $\alpha_j$. This means for instance that $s_1^3s_2^2 = s_1^3$ for all $\spin$ values, therefore leading to equal values of the corresponding moments. ISSUE HERE.}}
%%%%%%%%%% %%%%%%%%%% %%%%%%%%%% %%%%%%%%%% %%%%%%%%%% %%%%%%%%%%
These moments define a complete set of patterns of discrete data, i.e., %they describe all possible statistical patterns of discrete data.
%\textcolor{red}{i.e., they fully describe [all possible] the statistical structure of discrete data.}
they can fully %can
describe any possible statistical structure of discrete data.
We introduce the $n$-dimensional vector $\vecmu=(\mu_1,\,\cdots,\,\mu_n)\in\Field^n$, 
such that each value of $\vecmu$ uniquely identifies one of the moments in Eq.~\eqref{eq:def:multivariate-moments} (except for $\vecmu=\veczero$).
The maximum entropy model~\cite{jaynes1957information, tikochinsky1984alternative} constrained to reproduce all these moments %this complete set of moments
then takes the form: % of the parametric family of probability distributions:
\begin{align}\label{eq:def:CompleteM}
    p(\spin\,|\,\vecg)
    =\frac{1}{Z(\vecg)}\,\exp \Bigg({\;\sum_{\vecmu\in [\Field^n]^*} 
    g_{\vecmu}\,\phi^{\vecmu}(\spin)}\Bigg)\,,
\end{align}
where the sum is over all $\vecmu\in\Field^n$ except $\vecmu=\veczero$ (indicated by the star), $Z(\vecg)$ is a normalization factor corresponding to the partition function of the model, 
the $g_{\vecmu}$'s are $(q^n-1)$ complex parameters corresponding to Lagrange multipliers ($\vecg$ is the vector of all the parameters $g_{\vecmu}$), and the $\phi^{\vecmu}(\spin)$'s are the spin operators associated to each of the moments in Eq.~\eqref{eq:def:multivariate-moments} defined as follow.
\begin{Definition}\label{def:spin-operators}
    {\bf (Complex) Spin operators of arbitrary order.} 
    For all states $\spin$ of the $n$-spin system and all vectors $\vecmu=(\mu_1, \cdots, \mu_n)\in \Field^n$, we define the spin operator $\phi^{\vecmu}(\spin)$ as the monomial: 
    \begin{align}\label{eq:def:monomials:s}
        \phi^{\vecmu}(\spin)= \prod_{j=1}^{n} s_j^{\;\mu_j}\,.
    \end{align}
    For binary systems ($q=2$), the exponents $\mu_j$'s can only take the values $0$ or $1$, and we recover the Ising spin operators defined in Eq.~\eqref{eq:spin-operators:q=2}%in the context of binary data
    ~\cite{mastromatteo2013typical, beretta2018stochastic}. 
    %,[in particular $\phi^{\vecmu}(\spin)\in\{\pm 1\} \;\forall \spin$]. 
Using Eq.~\eqref{def:s_j}, the spin operators can also be written in term of the color variables~$\state=(\alpha_1, \cdots, \alpha_n)\in \Field^n$:
\begin{align}\label{eq:def:monomials:alpha}
    \phi^{\vecmu}(\state)
    =  \exp\left(\sum_{j=1}^n{\displaystyle\frac{2i\pi}{q}\alpha_j \mu_j}\right)
    = \exp\left({\displaystyle\frac{2i\pi}{q}\state\cdot\vecmu}\right) %\,,
    =z_q^{\;\vecmu\cdot\state}\,,
    %\qquad {\rm where}\;\;\; z_q=\exp(\frac{2\pi}{q}\,i)}\,,
\end{align}
where %$\state=(\alpha_1, \cdots, \alpha_n)\in \Field^n$ and 
$\state\cdot\vecmu$ is the scalar product of $\state$ and $\vecmu$ in $\Field^n$.
Note that for $q=2$, $\,\phi^{\vecmu}(\state)=(-1)^{\state\cdot\vecmu}\in\{\pm 1\}$. 
\end{Definition}

The ensemble averages of the spin operators in Eq.~\eqref{eq:def:monomials:s} define %correspond to 
all the moments of the spin system%in Eq.
~\eqref{eq:def:multivariate-moments} (except for the operator $\phi^{\veczero}(\spin)=1$).
Thus, at best fit of the maximum entropy model~\eqref{eq:def:CompleteM}, the values taken by these moments in the model are precisely equal to their values in the data, because these are the patterns the model was constrained to reproduce~\cite{jaynes1957information, tikochinsky1984alternative}.

One caveat of the model in Eq.~\eqref{eq:def:CompleteM} is that the term in the exponential can take complex values (as both $g_{\vecmu}$ and $\phi^{\vecmu}(\spin)$ can be complex), and therefore the state probabilities $p(\spin\,|\,\vecg)$ can be complex.
Observe however that for each complex operator $\phi^{\vecmu}(\spin)$, there is another operator $\phi^{-\vecmu}(\spin)=[\phi^{\vecmu}(\spin)]^*$ that is equal to its complex conjugate\footnote{$\phi^{-\vecmu}(\spin)=[\phi^{\vecmu}(\spin)]^*$ because $s_j^{-\mu_j}=\exp\left(-{\frac{2i\pi}{q}\alpha_j\mu_j}\right)=[\exp({\frac{2i\pi}{q}\alpha_j\mu_j})]^* =\big[s_j^{\,\mu_j}\big]^*$\;
for all $\mu_j\in\Field$\,.} in the model~\eqref{eq:def:CompleteM}. 
These operators correspond to the vectors $\vecmu\in\Field^n$ such that $-\vecmu\neq\vecmu$, where $-\vecmu$ denotes\footnote{The vector $-\vecmu$ is defined in $\Field^n$ by applying a ``minus'' sign to all the entries of $\vecmu$ and taking them modulo~$q$.} the vector $-\vecmu=(q-\mu_1,\cdots,$ $q-\mu_n)\in\Field^n$ (for $-\vecmu=\vecmu$, the operators %$\phi^{\vecmu}(\spin)$ 
already take real values $\phi^{\vecmu}(\spin)=\pm 1$ ).
In other words, the two operators $\phi^{\vecmu}$ and $\phi^{-\vecmu}$ impose similar constraints to the model.
%%%%%%%%%%%%%%%%%%
%Observe that, for the vectors $\vecmu\in\Field^n$ such that $-\vecmu\neq\vecmu$, where $-\vecmu$ denotes\footnote{The vector $-\vecmu$ is defined in $\Field^n$ by applying the ``minus'' sign to all the entries of $\vecmu$ and taking them modulo $q$.} the vector $-\vecmu=(q-\mu_1,\cdots,$ $q-\mu_n)\in\Field^n$, the spin operator $\phi^{-\vecmu}(\spin)=[\phi^{\vecmu}(\spin)]^*$ is equal to the complex conjugate\footnote{$\phi^{-\vecmu}(\spin)=[\phi^{\vecmu}(\spin)]^*$ because $s_j^{-\mu_j}=\exp\left(-{\frac{2i\pi}{q}\alpha_j\mu_j}\right)=[\exp({\frac{2i\pi}{q}\alpha_j\mu_j})]^* =\big[s_j^{\,\mu_j}\big]^*$\; for all $\mu_j\in\Field$\,.} of the operator $\phi^{\vecmu}(\spin)$; in other words, the two operators $\phi^{\vecmu}$ and $\phi^{-\vecmu}$ impose similar constraints to the model.
%%%%%%%%%%%%%%%%%%%%%%%%%%%%%%%%%
For the state probabilities $p(\spin\,|\,\vecg)$ %(and their corresponding energies in the exponential) 
to be real, we can then impose %the additional constraints 
that:
\begin{align}\label{eq:g:constraint}
    g_{-\vecmu}=g_{\vecmu}^{\,*}\,,
    \qquad 
    \textrm{for all}\;\,\vecmu=(\mu_1,\,\cdots,\,\mu_n)\in\Field^n\,,
\end{align}
where $g_{\vecmu}^*$ is the complex conjugate of $g_{\vecmu}$.
Consequently, each pair of conjugate operators $\phi^{\vecmu}$ and $\phi^{-\vecmu}$ is only parameterized by two real parameters, $a_{\vecmu}$ and $b_{\vecmu}$ (instead of four), chosen below such that $g_{\vecmu}=(a_{\vecmu}-i b_{\vecmu})/2=g_{-\vecmu}^*$ (see App.~\ref{App:H:CtoR}). 
They constrain the model on reproducing respectively the real and the imaginary parts of $\langle\phi^{\vecmu}(\spin)\rangle$. 
For values of $\vecmu$ such that $-\vecmu=\vecmu$, the operator $\phi^{\vecmu}(\spin)\in\{\pm 1\}$ is real and Eq.~\eqref{eq:g:constraint} implies that $g_{\vecmu}=g_{\vecmu}^*$ is also real. 
Equation~\eqref{eq:def:CompleteM} can thus be re-written under the form of a maximum entropy model constrained to reproduce the real and the imaginary part of all the moments~\eqref{eq:def:multivariate-moments} of a spin system:
\begin{align}\label{eq:def:CompleteM:real}
    p(\spin\,|\,\vecg)
    =\frac{1}{Z(\vecg)}\,\exp \Bigg[\;\sum_{\substack{\vecmu\in [\Field^n]^* \\ \vecmu \leq -\vecmu}}
    \left(a_{\vecmu}\,\varphi^{\vecmu}(\spin)
    +b_{\vecmu}\,\psi^{\vecmu}(\spin)\right) 
    \Bigg]\,,
\end{align}
where the operators $\varphi^{\vecmu}(\spin)={\rm Re}[\phi^{\vecmu}(\spin)]$ and $\psi^{\vecmu}(\spin)={\rm Im}[\phi^{\vecmu}(\spin)]$ are respectively the real and imaginary parts of $\phi^{\vecmu}(\spin)$ (see Definition~\ref{def:spin-operators} -- note that $\psi^{\vecmu}(\spin)=0$ for all $\vecmu$ such that $-\vecmu=\vecmu$). 
For all $\vecmu$ such that %For 
$-\vecmu\neq\vecmu$, contributions from conjugate operators have been grouped together as $a_{\vecmu}=g_{\vecmu}+g_{-\vecmu}$ and $b_{\vecmu}=i(g_{\vecmu}-g_{-\vecmu})$. 
For $\vecmu$ such that $-\vecmu=\vecmu$, %$b_{\vecmu}=0$ and 
we use the convention $a_{\vecmu}=g_{\vecmu}=g_{\vecmu}^*$.
The constraint $\vecmu \leq - \vecmu$ ensures that the sum includes each conjugate pair $(\vecmu, -\vecmu)$ only once%
\footnote{The ordering between operators $\vecmu\in\Field^n$ is based on the ordering of the integers represented by $\vecmu$ in base~$q$.}. %the integers for which $\vecmu$ is the representation in based $q$.
Here, the vector $\vecg$ denotes the vector of %all the 
real parameters $a_{\vecmu}$ and $b_{\vecmu}$.
In Eq.~\eqref{eq:def:CompleteM:real}, it is clear that the state probabilities $p(\spin\,|\,\vecg)$ are now real. 

\begin{Definition}\label{def:spin-operators}
    {\bf (Real) Spin operators of arbitrary order.} 
    For all state $\spin$ of the $n$-spin system
    and all vector $\vecmu=(\mu_1, \cdots, \mu_n)\in \Field^n$, we define the real 
    spin operators $\varphi^{\vecmu}(\spin)$ and $\psi^{\vecmu}(\spin)$ as respectively the real and the imaginary parts of $\phi^{\vecmu}(\spin)$:
\begin{align}\label{eq:def:monomials:spin:real}
    \varphi^{\vecmu}(\spin)= {\rm Re}[\phi^{\vecmu}(\spin)] =\frac{\phi^{\vecmu}(\spin)+\phi^{-\vecmu}(\spin)}{2}
    \qquad {\rm and}\qquad
    \psi^{\vecmu}(\spin)= {\rm Im}[\phi^{\vecmu}(\spin)]=\frac{\phi^{\vecmu}(\spin)-\phi^{-\vecmu}(\spin)}{2i}\,.
\end{align}
    For all $\vecmu$ such that $-\vecmu=\vecmu$, % (this includes all the operators when $q=2$),
    $\varphi^{\vecmu}(\spin)=\phi^{\vecmu}(\spin)\in\{\pm 1\}$ and $\psi^{\vecmu}(\spin)=0$.
    %For $q=2$, $\phi^{\vecmu}(\spin)\in\{\pm 1\}$, and $\varphi^{\vecmu}(\spin) = \phi^{\vecmu}(\spin)$ and $\psi^{\vecmu}(\spin)=0$.
%
These operators can also be written in term of the discrete variables~$\state$ as:
\begin{align}\label{eq:def:monomials:alpha:real}
    \varphi^{\vecmu}(\state)= \cos\left({\displaystyle\frac{2\pi}{q}\state\cdot\vecmu}\right) 
    \qquad {\rm and}\qquad
    \psi^{\vecmu}(\state)= \sin\left({\displaystyle\frac{2\pi}{q}\state\cdot\vecmu}\right)\,,
\end{align}
where $\state=(\alpha_1, \cdots, \alpha_n)\in \Field^n$
and $\state\cdot\vecmu$ is the scalar product of $\state$ and $\vecmu$ in $\Field^n$.
\end{Definition}

{\bf Number of real parameters.} 
%%%%%%%%%%%%%%% %%%%%%%%%%%%%%% %%%%%%%%%%%%%%% %%%%%%%%%%%%%%%
Requiring $p(\spin\,|\,\vecg)$ to be real reduces the effective number of real parameters in the model of Eq.~\eqref{eq:def:CompleteM}. % and \eqref{eq:def:CompleteM:real}. 
%%%%%%%%%%%%%%% %%%%%%%%%%%%%%% %%%%%%%%%%%%%%% %%%%%%%%%%%%%%%
There are two real parameters, $a_{\vecmu}$ and $b_{\vecmu}$, for each pair of conjugate operators and only one real parameter~$a_{\vecmu}$ for each real operator (for which $-\vecmu=\vecmu$).
Therefore, the model has only $(q^n-1)$ real parameters in total (equal to the number of non-zero vector $\vecmu\in\Field^n$).
This is consistent with the number of parameters one would need to model the probability distribution by its $(q^n-1)$ independent state probabilities (where the ``$-1$'' comes from normalization). 
Because the model~\eqref{eq:def:CompleteM} (resp.~\eqref{eq:def:CompleteM:real}) provides an alternative complete description of the discrete system,  
there exits a bijection between the $(q^n-1)$ state probabilities and the $(q^n-1)$ real parameters of the model. This means that for any distribution of the state probabilities\footnote{assuming that all the probabilities are strictly positive}~$p(\spin)$, %assuming that all the states are realizable/accessible,
there is a unique value of the vector of model parameters~$\vecg$ that creates this distribution,
%This will be shown explicitly in Sec.~\ref{Sec1.3:FT}
%This inversion is given in
which will be given in Eq.~\eqref{eq:pdf:FT} (resp.~\eqref{eq:pdf:FT:a_b}).

\subsubsection{Families of maximum entropy models for discrete data}
Equation~\eqref{eq:def:CompleteM} with the constraints~\eqref{eq:g:constraint} gives the most general parametric model able to reproduce all possible patterns of discrete data, expressed under the form of the multivariate moments~\eqref{eq:def:multivariate-moments}. 
In the context of statistical modeling, besides being able to reproduce the data patterns, one also aims to 
distinguish patterns that are significant from those that are just noise (for example, in order to extract structural information about the modeled system, or for the model to better generalize).
In other words, one aims to reduce the set of parameters of the model to only those that captures relevant patterns of the data.
Identifying these relevant parameters can be addressed by comparing (sub-)models of the complete model, in which different sets of parameters have been exactly fixed to zero (before fitting), and selecting the best among them based on model selection criteria, as discussed in the introduction.
In this context, the sub-models of~\eqref{eq:def:CompleteM}, for which some of the complex parameters $g_{\vecmu}$'s have been set exactly to zero, form a complete family of maximum entropy models constrained on reproducing arbitrary subsets of moments of the spin variables from the data.
Each of these sub-models is identified by the choice of a subset $\M=\{\vecmu_1,\cdots,\vecmu_K\}\subseteq[\Field^n]^*$ of operators for which the parameters $g_{\vecmu}$ are not set to zero and which identify the fitted moments. We use this notation to define the family of $q$-state spin models.
%in the definition of

\begin{Definition}\label{eq:def:q:spin-model}
    {\bf ($q$-state spin models) Complete family of maximum entropy models %for (discrete) spin systems 
    constrained on reproducing arbitrary sets of spin moments.}
    Consider a system of $n$ $q$-state spin variables~\eqref{def:s_j} $\spin = (s_1,\,\cdots,\,s_n)$.
    A $q$-state spin model is a parametric probability distribution of the form:
%\begin{align}
\begin{numcases}{}
    &$p(\spin\,|\,\vecg,\mathcal{M})
        =\displaystyle
        \frac{1}{Z_{\M}(\vecg)}\,\exp \left({\;\sum_{\vecmu\in \M} 
    g_{\vecmu}\,\phi^{\vecmu}(\spin)}\right)$\,,\label{eq:q-spin_models}\\[4pt]
    &{\rm with}$\qquad\forall \vecmu\in\M\,,\qquad -\vecmu\in\M$
    \;\;\;\;\,{\rm and}\,\;\;\;
    $g_{-\vecmu}=g_{\vecmu}^*$\,,\label{eq:q-spin_models:conditions}
\end{numcases}
where the set $\M\subseteq[\Field^n]^*$ identifies the empirical moments $\langle\phi^{\vecmu}(\spin)\rangle$ that the model is constrained to reproduce when fitting the parameters $g_{\vecmu}$'s. 
The spin operators $\phi^{\vecmu}(\spin)$ are given by Eq.~\eqref{eq:def:monomials:s} and can be of arbitrary order.  
For the probabilities $p(\spin\,|\,\vecg,\,\M)$ to be real, complex operators must be included by conjugate pairs in $\M$
and the corresponding parameters must satisfy Eq.~\eqref{eq:q-spin_models:conditions}.
The normalization factor $Z_{\M}(\vecg)$ is the partition function of the model:
\begin{align}\label{eq:Z:def}
    Z_{\M}(\vecg) 
    = \displaystyle
    \sum_{\spin} \;\exp\left(\,\sum_{\vecmu\in\M}g_{\vecmu}\,\phi^{\vecmu}(\spin)\right)\,,
\end{align}
where $\vecg$ is a vector containing all the parameters $g_{\vecmu}$'s and the first sum is over the $q^n$ states of $\spin$. The spin models in Eq.~\eqref{eq:q-spin_models} can also be expressed in terms of the discrete variables~$\state$, by simply replacing the dependencies in $\spin$ by dependencies in $\state$.\\ 
For binary systems ($q=2$), all the operators take real (binary) values $\phi^{\vecmu}(\spin)\in\{\pm 1\}$, the parameters $g_{\vecmu}$ are real, 
and one recovers the family of high-order Ising models %spin models
defined in Eq.~\eqref{eq:graph_model}~\cite{mastromatteo2013typical, beretta2018stochastic}. 
\end{Definition}
\noindent Note that, if we relax the constraint that operators must be included in the model by conjugate pairs with conjugate parameters satisfying~\eqref{eq:q-spin_models:conditions}, then the energy function in the exponential can take complex values. This implies complex state probabilities and a complex partition function. Although this may not be physical for classical systems, it could still be worthy of study. 
For instance, extending the partition function to the complex plane has been commonly used to study phase transitions in finite size classical many-body systems~\cite{yang1952statistical}, including on complex networks~\cite{krasnytska2015violation, krasnytska2016partition}.

%\textcolor{red}{In the spin model~\eqref{eq:q-spin_models}, fitting a given complex parameter $g^{\vecmu}$ ensures that the model reproduces both the real and the imaginary parts of the corresponding moment $\langle\phi^{\vecmu}\rangle$.}
%\textcolor{purple}{In the spin model~\eqref{eq:q-spin_models}, by fitting the complex parameter $g^{\vecmu}$, both the real and the imaginary parts of the corresponding moment $\langle\phi^{\vecmu}\rangle$ are reproduced by the model.}
In the spin model~\eqref{eq:q-spin_models}, upon fitting the complex parameter $g^{\vecmu}$, the model reproduces both the real and the imaginary parts of the corresponding moment $\langle\phi^{\vecmu}\rangle$.
More generally, one could choose to set to zero any of the $(q^n-1)$ real parameters $a_{\vecmu}$ and $b_{\vecmu}$ in Eq.~\eqref{eq:def:CompleteM:real}.
This is different from the previous setting, because it allows the modeler to choose to reproduce only the real or the imaginary part of a given moment (as opposed to both of them simultaneously).
This approach then corresponds to defining the following family of maximum entropy models, constrained to reproduce the real and/or the imaginary part of a selected set of moments of the spin variables:
%\textcolor{blue}{This corresponds to fitting separately the imaginary and real parts of the discrete moments.}
\begin{Definition}\label{def:spinmodel:cos-sin}
    {\bf Complete family of maximum entropy models for discrete data.}
    Consider a system of $n$ $q$-state discrete variables~\eqref{def:alpha_j} $\state = (\alpha_1,\,\cdots,\,\alpha_n)$. The following exponential family of probability distributions defines a complete family of parametric models for discrete data:
    %
%\begin{adjustwidth}{-\extralength}{0cm}
\begin{equation}\label{eq:q-spin_models:real}
    p(\state\,|\,\vecg,\mathcal{M})
        =\displaystyle
        \frac{1}{Z_{\M}(\vecg)}\,
        \exp\Bigg[        
    %\displaystyle\sum_{\vecmu\in \M_0}  
    %\frac{a_{\vecmu}}{2}\,\cos\left({\displaystyle\frac{2\pi}{q}\state\cdot\vecmu}\right)+
    \displaystyle\sum_{\vecmu\in \M_{0}\cup\M_{Re}} 
    a_{\vecmu}\,\cos\left({\displaystyle\frac{2\pi}{q}\state\cdot\vecmu}\right)
    +\displaystyle\sum_{\vecmu\in \M_{Im}} 
    b_{\vecmu}\,\sin\left({\displaystyle\frac{2\pi}{q}\state\cdot\vecmu}\right)
    \Bigg]
\end{equation}
%\end{adjustwidth}
where $a_{\vecmu}$ and $b_{\vecmu}$ are real parameters. %\textcolor{red}{For $\vecmu=-\vecmu$,} 
$\M_0$ identifies an arbitrary set of real moments (for $\vecmu=-\vecmu$), $\langle\phi^{\vecmu}(\state)\rangle=\langle\varphi^{\vecmu}(\state)\rangle$, that the model is constrained to reproduce, and %, for $\vecmu\neq -\vecmu$, 
$\M_{Re}$ and $\M_{Im}$ identify two arbitrary sets of complex moments (for $\vecmu\neq -\vecmu$) for which the model is constrained to reproduce respectively the real part $\langle\varphi^{\vecmu}(\state)\rangle$ and the imaginary part $\langle\psi^{\vecmu}(\state)\rangle$.
These sets are chosen such that:
\begin{equation}\label{test}
    \begin{cases}
        \;\M_0&\subseteq\{\,\vecmu\in[\Field^n]^* \mid \vecmu=-\vecmu\,\}\,,\\[1pt]
        \;\M_{Re}&\subseteq\{\,\vecmu\in[\Field^n]^* \mid \vecmu<-\vecmu\,\}\,,\\[1pt]
        \;\M_{Im}&\subseteq\{\,\vecmu\in[\Field^n]^* \mid \vecmu<-\vecmu\,\}\,,
    \end{cases}
\end{equation}
where the ordering between vectors $\vecmu\in\Field^n$ is based on the ordering of the integers represented by $\vecmu$ in base $q$.
The condition that $\vecmu<-\vecmu$ is used so that $\M_{Re}$ and $\M_{Im}$ only contains one vector $\vecmu$ for each conjugate pair ($\forall\vecmu\in\M_{Re}\,,\;-\vecmu\notin\M_{Re}$). % and the same for $\M_{Im}$,
%which is why there's no factor one half in front of the sums in Eq.~\eqref{eq:q-spin_models:real} compared to Eq.~\eqref{eq:def:CompleteM:real}.

\noindent {\bf Special cases:}
\begin{itemize}
    \item[--] For odd values of $q$  , $\M_0=\emptyset$.
    \item[--] The sets $\M_{Re}$ and $\M_{Im}$ can overlap. When these sets are equal, the model parametric distribution in Eq.~\eqref{eq:q-spin_models:real} is identical to the one in Eq.~\eqref{eq:q-spin_models} with $\M=\M_0\cup\M_{Re}\cup(-\M_{Re})$, and the parameters $a_{\vecmu}$ and $b_{\vecmu}$ are related to the $g_{\vecmu}$'s by $g_{\vecmu}=(a_{\vecmu}-i b_{\vecmu})/2=g_{-\vecmu}^*$ for all $\vecmu\in\M_{Re}$ and $g_{\vecmu}=a_{\vecmu}$ for all $\vecmu\in\M_0$. %\textcolor{purple}{Actually here it is $g_{\vecmu}=a_{\vecmu}$, because there's no factor one half in front of the sum (contrary to Eq.~(13)).}
    \item[--] For the binary case ($q=2$), $\M_{Re}=\M_{Im}=\emptyset$ and, for all $\vecmu\in\M_0$, $\varphi^{\vecmu}(\state)=\phi^{\vecmu}(\state)=(-1)^{\state\cdot\vecmu}$. We recover the family of binary spin models defined in Eq.~\eqref{eq:graph_model}.
\end{itemize}
\end{Definition}

The families of models in Eq.~\eqref{eq:q-spin_models} and Eq.~\eqref{eq:q-spin_models:real} both generalize the family of binary spin models in Eq.~\eqref{eq:graph_model} to values of $q\geq 2$. 
%\textcolor{red}{specify: for both complex and real Hamiltonian.}

%%%%%%%%%%%%%%%%%%%%%%%%%%%%%%%%%% %%%%%%%%%%%%%%%%%%%%%%%%%%%%%%%%%% 
%\begin{Definition}
    %{\bf Complete family of maximum entropy models for discrete data.}
    %Consider a system of $n$ $q$-state discrete variables~\eqref{def:alpha_j} $\state = (\alpha_1,\,\cdots,\,\alpha_n)$. The following exponential family of probability distribution defines a complete family of parametric models for discrete data:
%\begin{equation}\label{eq:q-spin_models:alpha:cos-sin}
%    \begin{cases}
%    &p(\state\,|\,\vecg,\mathcal{M})
%        =\displaystyle \frac{1}{Z_{\M}(\vecg)}\,
%        e^{-H(\state\,|\,\vecg,\,\mathcal{M})}\,,\\[10pt]
%    &{\rm with}\;\;
%    H(\state\,|\,\vecg,\,\mathcal{M})=-\displaystyle\sum_{(\vecmu, -\vecmu)\in \M} 
%    \left[\,
%    a_{\vecmu}\,\cos\left({\displaystyle\frac{2\pi}{q}\state\cdot\vecmu}\right) +
%    b_{\vecmu}\,\sin\left({\displaystyle\frac{2\pi}{q}\state\cdot\vecmu}\right)
%    \right]\,,\\
%    &{\rm with}\qquad\forall \vecmu\in\M\;{\rm such\;that}\;-\vecmu=\vecmu,\qquad b_{\vecmu}=0\,,
%    \end{cases}
%\end{equation}
%where the sum is over pairs of conjugate operators in $\M$, and where $a_{\vecmu}$ and $b_{\vecmu}$ are real parameters that can be separately set to zero. These parameters are related to the $g_{\vecmu}$'s in Eq.~\eqref{eq:q-spin_models} by $g_{\vecmu}=(a_{\vecmu}-i b_{\vecmu})/2$ when $-\vecmu\neq\vecmu$, and by $a_{\vecmu}=g_{\vecmu}$ and $b_{\vecmu}=0$ when $-\vecmu=\vecmu$.
%\end{Definition}
%%%%%%%%%%%%%%%%%%%%%%%%%%%%%%%%%% %%%%%%%%%%%%%%%%%%%%%%%%%%%%%%%%%%

\subsubsection{Interactions of arbitrary order between the variables} 

In the maximum entropy models~\eqref{eq:q-spin_models} and \eqref{eq:q-spin_models:real}, the term in the exponential, which we denote $-H(\spin)$, can be interpreted as (minus) the energy of the spin system, resulting from interactions between the spins.
%In the maximum entropy models~\eqref{eq:q-spin_models} and \eqref{eq:q-spin_models:real}, the term in the exponential, which we denote $-H(\spin)$, can be interpreted/seen as minus the energy of the state[s] of the spin system. This energy can interpreted as resulting from the interactions between the spins.
More precisely, the $q$-state spin models in Eq.~\eqref{eq:q-spin_models} have the same form as the binary models in Eq.~\eqref{eq:graph_model}, but generalized to $q\geq 2$.
By analogy to the binary case, one can interpret the discrete vectors~$\vecmu$ as encoding interactions (of arbitrary order) between the spins $s_j$ for which $\mu_j\neq 0$, 
%[which are associated] 
with an interaction energy $H^{\vecmu}(\spin)=-g_{\vecmu}\,\phi^{\vecmu}(\spin)=-g_{\vecmu}\,\prod_{j=1}^n s_{j}^{\mu_j}$ that depends only on the state of these variables.
%\textcolor{blue}{A vector $\vecmu$ is associated with a term in the exponential of the form $H^{\vecmu}(\spin)=-g_{\vecmu}\,\phi^{\vecmu}(\spin)=-g_{\vecmu}\,\prod_{j=1}^n s_{j}^{\mu_j}$, which depends only on the state of the variables~$s_j$ for which $\mu_j\neq 0$ and which can be interpreted as an energy of interaction between these spins.}
%
However, unlike the binary case, the contribution of each spin to the interaction energy is now weighted differently depending on the value of the $\mu_j$'s, and therefore there can be different types of interactions involving the same set of spins.

%%%%% Examples: %%%%%%%%%
For example, for a binary system~\eqref{eq:graph_model} with $n=4$ spins, 
the vector $\vecmu=(1,1,0, 1)$ identifies the spin operator $\phi^{\vecmu}(\spin)=s_1s_2s_4$ and is interpreted as encoding a 3-body interaction between the spins $s_1$, $s_2$, and $s_4$. 
In the ternary case $q=3$, the vector $\vecmu=(2,1,0, 1)$ identifies the spin operator $\phi^{\vecmu}(\spin)=s_1^2 s_2s_4$ and can similarly be interpreted as encoding a 3-body interaction between the same spins, but the spin $s_1$ is now weighted twice in the interaction.
The vector $\vecmu=(1,1,0, 1)$ also encodes a 3-body interaction between the same spins, but 
%it is a different interaction (as it has a different interaction energy).
it has a different interaction energy.
Indeed, different values of $\mu_j$ weight the colors $\alpha_j$ differently in the operator $\phi^{\vecmu}(\state)=\exp\big((2i\pi/q)\sum_{j=1}^n \mu_j\,\alpha_j\big)$, thus resulting in different contributions to the energy landscape.
In Sec.~\ref{sec:Potts:interpretation}, we discuss the interpretation of %how to interpret 
different types of interaction involving the same spins.

Similarly, in Eq.~\eqref{eq:q-spin_models:real}, each vector $\vecmu$ can be interpreted as encoding an interaction, whose energy $H^{\vecmu}(\state)=-a_{\vecmu}\varphi^{\vecmu}(\state)-b_{\vecmu}\psi^{\vecmu}(\state)=-a_{\vecmu}\cos((2\pi/q)\,\vecmu\cdot\state)-b_{\vecmu}\sin((2\pi/q)\,\vecmu\cdot\state)$ only depends on the values taken by the spins $s_j$ for which $\mu_j\neq 0$.
More generally:
\begin{Definition}
    {\bf Arbitrary-order interactions.} For a system of $n$ random variables $\state=(\alpha_1,\,\cdots,\,\alpha_n)$, 
    one can identify an interaction of arbitrary order by:
    \begin{itemize}
        \item[--] an $n$-dimensional vector $\vecmu\in\mathbb{N}^n$, whose non-zero entries identify which variables interact,
        \item[--] and an interaction energy $H^{\vecmu}(\state)$ %$\textcolor{red}{g_{\vecmu}}f^{\vecmu}(\state)$
        that defines how the variables interact and whose functional form depends only on the variables $\alpha_j$ for which $\mu_j\neq0$. 
        %%%%%%%%
        %As functional form $f^{\vecmu}(\spin)$ that depends solely on the variables identified by $\vecmu$ and with a multiplicative/power factor that depends on the value of $\mu_i$.
        %%%%%%%%
        For example, for the models discussed in this paper, $H^{\vecmu}(\state)$ depends on the scalar product $\vecmu\cdot\state = \sum_j \mu_j\,\alpha_j$.
    \end{itemize} 
%and associate it with an energy term $g_{\vecmu}f^{\vecmu}(\state)$ with a functional form that depends only on the variables identified by $\vecmu$ and 
\end{Definition}
%
%\textcolor{purple}{$H^{\vecmu}(\state)=-g_{\vecmu}\,\phi^{\vecmu}(\state)-g_{\vecmu}^*\,\phi^{-\vecmu}(\state)$ (for $\vecmu\neq-\vecmu$)}

\noindent In the cases considered here, the interaction energies, $H^{\vecmu}(\state)=g_{\vecmu}\phi^{\vecmu}(\state)$ or $H^{\vecmu}(\state)=a_{\vecmu}\varphi^{\vecmu}(\state) + b_{\vecmu}\psi^{\vecmu}(\state)$, are %proportional to 
non-linear functions of a linear combination $\vecmu\cdot\state= \sum_j \mu_j\,\alpha_j$ of the colors of the spins involved in the interaction encoded by~$\vecmu$.

\begin{Definition}\label{def:interaction_order}
    {\bf Order of an interaction.}
    The order of an interaction corresponds to the number of variables involved in the interaction. For an interaction identified by a vector $\vecmu$, it is given by the numbers of non-zero entries in $\vecmu$, which is $|\vecmu|=\sum_{j=1}^n (1-\delta_{0,\mu_j})$.
    Note that this is not the same as the order of the operator identified by $\vecmu$ (or of the corresponding moment), which is $\sum_{i=1}^n \mu_i$. 
\end{Definition}

%%%%%%%%%%%%%%%%%%%%%%%%%%%%%%%%%%%%%%%%% DO NOT ERASE:
%%%%%%%%%%%%%%%%%%%%%%%%%%%%%%%%%%%%%%%%%
% DEFINITION MODELS:
% as Hamitonian expanded over all possible multisets of vertices.
%%%%%%%%%%%%%%%%%%%%%%%%%%%%%%%%%%%%%%%%%
%In general $H^{\vecmu}=g_{\vecmu}f^{\vecmu}(\state)$, where $g_{\vecmu}$ is a complex parameter and $f_{\vecmu}(\state)$ is a complex function.
%%%%%%%%%%%%%%%%%%%%%%%%%%%%%%%%%%%%%%%%%
% If besides, the function $H^{\vecmu}$ is real, then it [the interaction energy] should take the form: $H^{\vecmu}=g_{\vecmu}f^{\vecmu}(\state) + g_{\vecmu}^{*} f^{-\vecmu}(\state)$.
%%%%%%%%%%%%%%%%%%%%%%%%%%%%%%%%%%%%%%%%%
%or $g_{\vecmu}f^{\vecmu}(\state) + g_{\vecmu}^*f^{-\vecmu}(\state)$ if the functions are real.
%%%%%%%%%%%%%%%%%%%%%%%%%%%%%%%%%%%%%%%%%
%%%%%%%%%%%%%%%%%%%%%%%%%%%%%%%%%%%%%%%%%

\subsection{Comparison with other models for discrete systems}
\subsubsection{Examples of pairwise discrete models}

\begin{Example}\label{ex:vector_Potts}
    {\bf The planar Potts model~\citep{Potts1952}} (also known as vector Potts model or clock model).
    In this model, spins take $q$ discrete values uniformly distributed around the unit circle, as in Eq.~\eqref{def:s_j}, and are placed on a lattice or, more generally, on the nodes of a network. 
    Spins connected by a link in the network interact and the Hamiltonian of the system is given by:
    \begin{align}\label{eq:PlanarPotts}
        H(\state)=-\displaystyle\sum_{\langle i, j\rangle} 
    J_{ij}\,\cos\left(\displaystyle\frac{2\pi}{q}(\alpha_i-\alpha_j)\right)\,,
    \end{align}
    where $\langle i,j\rangle$ denotes the pairs of nodes $i,j$ that are connected by a link in the network. 
    In the original model, there is only one parameter, $J_{ij}=J$ for all pairs $\langle i,j\rangle$. Here, we adopted a more general notation, where the coupling parameters $J_{ij}$ can take different values. %\\
    %\textcolor{orange}{{\bf $Z_q$-model.} Add information about the $Z_q$ model, which is maybe closer to our model.}
\end{Example}

The models we introduced in~\eqref{eq:q-spin_models} and~\eqref{eq:q-spin_models:real} can be seen as high-order generalizations of the planar Potts model.
Indeed, it is straightforward to check that the Hamiltonian in Eq.~\eqref{eq:PlanarPotts} corresponds to a model of the form of Eq.~\eqref{eq:q-spin_models:real}, in which $b_{\vecmu}=0$ for all $\vecmu$ and all the interactions are pairwise of the type: 
for each pair of nodes $(i,j)$ connected by a link, there is one interaction vector $\vecmu$ such that%
\footnote{Which of the two nodes is `$i$` or `$j$` doesn't matter. One could choose to take the conjugate operators instead in $\M_0\cup\M_{Re}$, i.e. the operator~$\vecmu$ with $\mu_i=(q-1)$ and $\mu_j=1$. This would give the exact same model.}
\begin{align}
    \mu_i=1\,,\qquad \qquad
    \mu_j=(q-1)\,,\qquad {\rm and }\qquad
    \mu_k=0\;\qquad \forall k\neq i,j\,.
\end{align}
The corresponding operator $\varphi^{\vecmu}(\state)$ in~\eqref{eq:q-spin_models:real} is then given by:
\begin{align}
    \cos\left(\frac{2\pi}{q}\state\cdot\vecmu\right) 
    = \cos\left(\frac{2\pi}{q} \left(\alpha_i+(q-1)\alpha_j\right)\right)
    = \cos\left(\frac{2\pi}{q} (\alpha_i-\alpha_j)\right)\,.
\end{align}

Note that the contribution of each pairwise interaction in Eq.~\eqref{eq:PlanarPotts} can also be written as a scalar product between the vectors representing the two spins in the complex plane:
\begin{align}
        H(\spin)=-\displaystyle\sum_{\langle i,j\rangle} 
    J_{ij}\;\vec{s}_i\cdot\vec{s}_j\,,
\end{align}
where $\vec{s}_{j}=(\cos(\frac{2\pi\alpha_j}{q}), \sin(\frac{2\pi\alpha_j}{q}))$.
Similarly, a maximum entropy model constrained to reproduce the imaginary parts of the same moments would take the form: 
\begin{align}\label{eq:vector-Potts:im}
    H(\state)
        = - \sum_{\langle i,j\rangle} 
    J_{ij}\, \sin\left(\displaystyle\frac{2\pi}{q}(\alpha_i-\alpha_j)\right)
        = -\Bigg(\displaystyle\sum_{\langle i,j\rangle} 
    J_{ij}\;\vec{s}_i\times\vec{s}_j\Bigg)\cdot\vec{e}_z\,.
\end{align}
where $\times$ denotes the vector product between the spin vectors $\vec{s}_j=(\cos(\frac{2\pi\alpha_j}{q}), \sin(\frac{2\pi\alpha_j}{q}), 0)$ representing the planar spins in a 3-dimensional space, and $\vec{e}_z=(0,0,1)$. This produces a model with antisymmetric interactions in the values taken by the pair $\alpha_i$ and $\alpha_j$. 

Although the planar Potts model wasn't introduced in the context of statistical inference, one may wonder if it could be used to model the pairwise structure of discrete data. 
In Eq.~\eqref{eq:PlanarPotts}, each pairwise interaction is parameterized by 1~parameter. If we were to also fit the corresponding imaginary part, we would have 2~parameters. 
However, to fully parameterize all the pairwise moments based on a given pair of spins, one needs a total of $(q-1)^2$ real parameters\footnote{The number of pairwise operators on a given pair of spins is equal to the number of possible values of the pair $(\mu_i, \mu_j)\in\{1,\cdots,(q-1)\}^2$ ($\mu_i$ and $\mu_j$ must be both non-zero for $\vecmu$ to encode an interaction between $\alpha_i$ and $\alpha_j$).
Besides, there are two independent real parameters for each pair of conjugate operators, and thus one can count the number of independent parameters as being equal to the number of operators (see Sec.~\ref{subsec:def:CompleteModel}).}. 
For example, for $q=3$ and $n=2$ spins, one needs 4 real parameters: 2 to parameterize the conjugate operators $(1,2)$ and $(2,1)$, and 2 for the conjugate operators $(1,1)$ and $(2,2)$ (see Sec.~\ref{sec:Potts:interpretation} for a discussion on the meaning of these different interactions).
These latter are not accounted for in Eq.~\eqref{eq:PlanarPotts}, but are needed to fully parameterize statistical patterns of pairs of spins.
In the context of statistical inference, a common parametric model that can account for all first and second order patterns of discrete data is the following.

\begin{Example}
    {\bf Generalized pairwise Potts model for statistical inference.} A generalization of the Potts model commonly used in statistical inference~\cite{Nguyen2017} takes the form:
    \begin{align}\label{Eq:Ex:Potts:stat_inference}
        H(\state) 
            = \sum _{<i,j>}J_{ij}(\alpha_i,\alpha_j)+\sum_{i} h_i(\alpha_i)\,,
    \end{align}
    where the coupling parameters $J_{ij}$ and field parameters $h_i$ can take different values for each state of the nodes $\alpha_i$ and $\alpha_j$,
    i.e., $J_{ij}(\alpha_i,\alpha_j)$ is the energy of the interaction between nodes $i$ and $j$ when they take the value $(\alpha_i,\alpha_j)$. Note that $J_{ij}(\alpha,\alpha^{\prime})$ is not necessarily equal to $J_{ij}(\alpha^{\prime},\alpha)$. 
\end{Example}
The form of this model is different from the one in Eq.~\eqref{eq:q-spin_models} and~\eqref{eq:q-spin_models:real} reduced to all first and second order interactions.
Although each of these models is able to fit all the pairwise patterns of discrete data, the model~\eqref{Eq:Ex:Potts:stat_inference} has more parameters than the number of pairwise constraints in the system. 
In total, there are $q$ real parameters for $h_i(\alpha_i)$ (one for each value of $\alpha_i$) and $q^2$ parameters for each pairwise interactions (one for each value of $(\alpha_i,\alpha_j)$), while there are only $(q-1)$ moments based on a given spin and $(q-1)^2$ moments on a given pair of spins. 
This overparameterization results in the parameters of Eq.~\eqref{Eq:Ex:Potts:stat_inference} being defined up to a gauge~\cite{Ekeberg2013, Barton2016, rizzato2020inference}, which is not the case for Eq.~\eqref{eq:q-spin_models} and~\eqref{eq:q-spin_models:real}.
The parameters of~\eqref{Eq:Ex:Potts:stat_inference} are related to the parameters of~\eqref{eq:q-spin_models} and \eqref{eq:q-spin_models:real} by:
\begin{align}
    \begin{cases}
    & h_{j}(\alpha_j) = \displaystyle
    \sum_{\mu_j=1}^{q-1} 
    g_{\mu_j} 
    \exp\left(\frac{2i\pi}{q} \mu_j\alpha_j\right)\,,
    \qquad \forall \alpha_j\\
    & J_{ij}(\alpha_i, \alpha_j) = \displaystyle 
    \sum_{\mu_i=1}^{q-1} \sum_{\mu_j=1}^{q-1} 
    g_{\mu_i\mu_j} 
    \exp\left(\frac{2i\pi}{q} (\mu_i\alpha_i+\mu_j\alpha_j)\right)\,, 
    \qquad \forall (\alpha_i,\alpha_j)
    \end{cases}
\end{align}
%where $g_{\mu_j}$ (resp. $g_{\mu_i \mu_j}$) denotes the parameter $g_{\vecmu}$ for which $\vecmu$ verifies $\mu_k=0$ for all $k\neq j$ (resp. $k\neq i,j$) and $\mu_k=\mu_j$ for $k=j$ (resp. $k=i,j$). 
%\textcolor{red}{denotes the vector, for which the $j$-th entree is $\mu_j$ and all the other entrees are $0$.}
where $g_{\mu_j}$ (resp.~$g_{\mu_i \mu_j}$) denotes the parameter $g_{\vecmu}$ for which the $j$-th (resp.~$i$-th and $j$-th) element of $\vecmu$ is $\mu_j$ (resp.~$\mu_i$ and $\mu_j$) and all the other elements are $0$.
We recall that $g_{-\vecmu}=g_{\vecmu}^*=(a_{\vecmu}+i b_{\vecmu})/2$.
On the left-hand-side, there are $q+q^2$ parameters, 
while on the right-hand-side, there are $(q-1)+(q-1)^2$ independent real parameters $a_{\vecmu}$ and $b_{\vecmu}$.

For example, for a system with $q=3$ and $n=2$ variables $\state=(\alpha_1,\alpha_2)$, the one-body terms are parametrized as:
\begin{align}
    h_j(\alpha_j) 
        &= g_{(10)}z_3^{\alpha_j}+  g_{(20)}z_3^{2\alpha_j}\,,\qquad {\rm with}\; g_{20}=g_{10}^*\nonumber\\
        &=a_{(10)} \cos\left(\frac{2\pi}{3}\alpha_j\right)+
        b_{(10)} \sin\left(\frac{2\pi}{3}\alpha_j\right)\,,\nonumber
\end{align}
where there are three parameters $h_j$ on the left-hand side and only two real parameters on the right-hand side. The pairwise interaction is parametrized as:
\begin{align}
    J_{12}(\alpha_1,\alpha_2) 
        &=g_{(11)}z_3^{\alpha_1+\alpha_2}+
        g_{(22)}z_3^{2\alpha_1+2\alpha_2}+
        g_{(12)}z_3^{\alpha_1+2\alpha_2}+
        g_{(21)}z_3^{2\alpha_1+\alpha_2}\,,\;{\rm with}\; g_{22}=g_{11}^*\,,\;g_{21}=g_{12}^*\nonumber\\
        %&\qquad\qquad\qquad  {\rm with}\; g_{22}=g_{11}^*\;{\rm and}\;g_{21}=g_{12}^*\nonumber\\
        &=a_{(11)} \cos\Big(\frac{2\pi}{3}(\alpha_1+\alpha_2)\Big)+
        b_{(11)} \sin\Big(\frac{2\pi}{3}(\alpha_1+\alpha_2)\Big)\nonumber\\
        &\qquad +a_{(12)} \cos\Big(\frac{2\pi}{3}(\alpha_1-\alpha_2)\Big)+
        b_{(12)} \sin\Big(\frac{2\pi}{3}(\alpha_1-\alpha_2)\Big)\,,
        \nonumber
\end{align}
where there are nine parameters $J_{ij}$ on the left-hand side and only four real parameters on the right-hand side

\begin{Example}{\bf Continuous limit and XY model.}
In the limit $q\to \infty$, the planar Potts model~\eqref{eq:PlanarPotts} becomes the XY model (after adding one-body contributions):
\begin{align}
    H(\spin) = 
    -\displaystyle\sum_{\langle i, j\rangle} 
    J_{ij}\,\cos\left(\theta_i-\theta_j\right)
    -\displaystyle\sum_{j} h_j\,\cos\left(\theta_j\right)
    \,,
\end{align}
%in which we added one-body contributions, and
where the variable $\theta_j\in[0,\,2\pi]$ corresponds to the angle that the planar spin $\vec{s}_j$ forms with the $x$-axis. 
Before taking the limit $q\to \infty$, $\theta_j$ is related to the variable  $\alpha_j\in\Field^n$ by $\theta_j=\frac{2\pi}{q}\alpha_j$.
\end{Example}
Similarly, defining the angle $\theta_j=\frac{2\pi}{q}\alpha_j$ for each value of $\alpha_j\in\Field^n$ and taking the limit $q\to \infty$, the model in Eq.~\eqref{eq:q-spin_models:real} can be seen as a generalization of the XY model to higher-order interactions, where now the vectors $\vecmu\in\mathbb{N}^n$. 
Note that after taking $q\to \infty$, $\theta_j$~takes real values in $[0,2\pi]$, but $\alpha_j$ remains an integer, $\alpha_j \in\mathbb{N}$.

\subsubsection{Examples of high-order discrete models}
%{Comparison with the standard Potts model and its high-order generalization}
%{\bf Other types of high-order interaction Potts models.}

\begin{Example}{\bf High-order generalization of the standard Potts model.}
Ref.~\cite{grimmett1994potts} introduced a high-order generalization of the standard Potts~\cite{wu1982potts} 
model of the form:
\begin{align}\label{eq:PottsModel:standard}
    H(\state\,|\,E) = 
    -\displaystyle\sum_{e\in E} 
    J_{e}\,\delta_e(\state)
    \,,
    \qquad
    {\rm where}\qquad
    \delta_e(\state) = 
        \begin{cases}
        \;1 & {\rm if}\;\; \alpha_i=\alpha_j\;\;{\rm for\; all}\;\;i,j\in e\\
        \;0 & {\rm otherwise}
        \end{cases}
\end{align}
where is $E$ set of hyperedges $e$ between the variables.  $\delta_e(\state)=1$ only if all the spins connected by~$e$ have the same value.
When all hyperedges are pairwise, one recovers the standard Potts model. 
%The standard Potts model corresponds to the case where all the hyperedges are pairwise.
\end{Example}

Although this model wasn't introduced in the context of statistical inference, one may wonder if it could be used for this purpose. Similarly to our own case, each operator~$\delta_e(\state)$ defines a (high-order) interaction between the spins connected by the hyperedge~$e$.
%%%%%
%However, this model encodes information differently. 
%%%%%
However, the interaction energy $-J_e\delta_e(\state)$ can only take two values (instead of $q$ in our case) 
%, $-J_e$ when all the variables connected by $e$ have the same states and $0$ for all the other states, 
and is unable to distinguish states %of the variables $\alpha_i$ connected by $e$.}
for which the variables $\alpha_i$ connected by~$e$ 
%don't all take the same values.
take non-equal values. 
In particular, 
%it doesn't allow to weigh differently \textcolor{red}{the contribution[s]} of a given spin to an interaction.
it doesn't allow to weigh differently the contributions of different spins to the same interaction.
This is an issue for statistical inference, because one aims %where one would like to be able 
to model how similar or different the values taken by the random variables are. %\textcolor{blue}{(not just if they are all equal).} % if they are equal or not.
Besides, each hyperedge~$e$ is only associated to one interaction (compared to $(q-1)^{|e|}$ %interactions 
in our case).
%, which doesn't allow to weigh differently the contribution of a given spin to an interaction.}
This means that there are at most %is a total of 
$(2^n-1)$ different $\delta_e$-functions, % in total, % (one for each hyperedge). 
which is not sufficient to form a basis of $\mathbb{R}[\Field^n]$ for $q>2$,
and therefore a full model of the form of Eq.~\eqref{eq:PottsModel:standard} wouldn't be able to capture all %possible 
patterns of discrete data for $q > 2$.

For $q=2$, this model could a priori model all possible patterns of binary data, if the set of $\delta_e$-functions forms a basis of $\mathbb{R}[(\mathbb{Z}/2\mathbb{Z})^n]$ (this remains to be checked) and after adding terms of order one (self-interactions). If this were the case, this basis would differ from the Ising spin basis studied here, % and would thus %extract/
decomposing multi-variate information in a different manner. Note that models of the form~\eqref{eq:PottsModel:standard} %with only one-body and pairwise interactions,
with only pairwise interactions $e=(ij)$, can be mapped to pairwise Ising models using $\delta_{ij}(\state)=(1+\phi^{(ij)}(\state))/2$, where $(ij)$ denotes the vector $\vecmu$ in which $\mu_i=\mu_j=1$ and all the other elements are zero.

\begin{Example}{\bf Other high-order models for discrete systems.} %data.}
%%%% Blume-Capel:
%%% https://gitpages.physik.uni-wuerzburg.de/marqov/webmarqov/post/2020-05-15-blume-capel/
%%%%%%%%%%%%%%%
%Other models have been used for modeling high-order moments of discrete systems in condensed matter and statistical physics (such as %the Blume-Capel model or 
%the Blume-Emery-Griffiths~\cite{blume1971ising} model for three state systems). However, so far they haven't been used directly for statistical modeling~\cite{waldorp2026blume}.
Ref.~\cite{waldorp2026blume} recently introduced another model for the statistical inference of three-state systems inspired by the Blume-Capel model~\cite{blume1966theory, capel1966possibility} from condensed matter.
This model corresponds to a maximum entropy model constrained on reproducing multivariate moments of the discrete variables~$\state$, instead of moments of the spin variables~$\spin$. %not on reproducing moments of the spins, but moments of the variables. 
So far, this model only includes interactions up to pairwise, but can easily be generalized by constraining the model on higher-order moments as well.
Another approach in the context of statistical modeling is the one by Ref.~\cite{Hindriks2024Higher}, which defines models constrained to reproduce high-order cumulants from multivariate data. %high-order correlation patterns of multi-variate data.}
\end{Example}

\subsection{Properties of $q$-state spin models and discrete Fourier analysis}
\label{Sec1.3:FT}
%%%%%%%%% %%%%%%%%%  CHECK THESE LINKS:   %%%%%%%%% %%%%%%%%% 
% https://dsp.stackexchange.com/questions/1406/real-discrete-fourier-transform
% Paper: discrete Hartley transform: https://ieeexplore.ieee.org/document/1164687
% Paper: Real-valued fast Fourier transform algorithms: https://ieeexplore.ieee.org/document/1165220
% Book on signal processing: http://www.dspguide.com/ch8/2.htm
%%%%%%%%% %%%%%%%%%  %%%%%%%%% %%%%%%%%%  %%%%%%%%% %%%%%%%%%  

In this section, we show that the set of $q^n$ complex spin operators~\eqref{eq:def:monomials:s} forms an orthonormal basis of the vector space of complex functions over $\Field^n$, and thus that the model~\eqref{eq:def:CompleteM} corresponds to a particular choice of basis decomposition of the multivariate log-probability, more specifically to a (spatial) Fourier decomposition. 
Similarly, Eq.~\eqref{eq:def:CompleteM:real} corresponds to a decomposition of the log-probability over the (spatial) trigonometric Fourier basis.
%Future works could study other forms of basis decompositions~\cite{Peerbooms2025, Hindriks2024Higher}.

The properties discussed below extend results known in the binary case~\cite{beretta2018stochastic} to the more general discrete case $q\geq 2$. They are direct consequences of the operators being Fourier basis functions.
%Recognizing the expression of the log-probability in Eq.~\eqref{eq:def:CompleteM} and \eqref{eq:def:CompleteM:real} as Fourier decomposition
The advantage of mapping Eq.~\eqref{eq:def:CompleteM} and \eqref{eq:def:CompleteM:real} to Fourier series is that it %allows to use/
enables the use of Fourier analysis tools for statistical inference
%%%%%%%%%%%%%%%%%%%
%The advantage of recognizing Eq.~\eqref{eq:def:CompleteM} and \eqref{eq:def:CompleteM:real} as expansions over a Fourier basis is that one may be able to use tools developed in Fourier analysis, 
%%%%%%%%%%%%%%%%%%%%%%%
%the possibility to use \textcolor{red}{Fast Fourier transform} algorithms to compute the Fourier coefficients
(e.g., efficient algorithms to compute the Fourier coefficients can be used to compute the model parameters~\cite{kato2010haplotype, buchman2012sparse, mosseri2015ising, Peerbooms2026}).
%\textcolor{orange}{It can also help us better understand/study how properties of the system in the probability space translate to properties of the model in the parameter space (Fourier space). [For example, the impact of symmetries in probability space (such as certain states occurring with the same probability) on %the degeneracy of 
%the parameters of the model~\cite{Gresele2017}.]} %(such as a state occurring with the same probability) 

\subsubsection{Spin models as Fourier decompositions of the log-probability}
%{Spatial Fourier bases / Fourier decompositions of the log-probability} 
%Spatial Fourier decompositions
%Complex Fourier decomposition
%Fourier decomposition of the log-probability
%\subsubsection{Group of monomials.} 
\begin{Property}
    {\bf Group $\Omega$ of complex spin operators.} The set $\Omega$ of all the $q^n$ spin monomials~\eqref{eq:def:monomials:s} forms a finite multiplicative group with identity element $\phi^{\veczero}(\spin) = 1$. It is the cyclic group of order~$q$ %$C_q^{\,n}$ %multiplicative 
 generated by the spin variables $s_1, s_2, \cdots, s_n$:
\begin{align}
    \Omega \doteq \{\phi^{\vecmu}(\spin) \mid \vecmu\in  \Field^n\} \,= \;\langle s_1, s_2, \cdots, s_n\rangle_{gen}\,. % = Z_q^{\,n}
\end{align}
%where the variables $s_j$ are the $q$-th roots of unity~\eqref{def:s_j}.
where, for all $j$, the variable $s_j$ takes values in the $q$-th roots of unity~\eqref{def:s_j}.
%Each element of $\Omega$ is at most of order $q$, as:
%\begin{align}
%    [\phi^{\vecmu}(\spin)]^q = \exp\left({\displaystyle 2i\pi\state\cdot\vecmu}\right) = 1\,,
%    \qquad\qquad %{\rm for}\;{\rm all}\;
%    \forall\;\vecmu \in \Field^n
%    \,.
%\end{align}
\end{Property}
\begin{proof} %Indeed, 
From Definition~\ref{def:spin-operators}, it is straightforward that, for all $(\vecmu, \vecnu)\in[\Field^n]^2$, 
\begin{align}\label{eq:monomials:group_structure}
    %{\rm for all}\; (\vecmu, \vecnu)\in[\Field^n]^2, 
    \phi^{\vecmu}(\spin)\,\phi^{\vecnu}(\spin) = \phi^{\vecmu+\vecnu}(\spin)\in\Omega\,,
\end{align}
is also a monomial that belongs to $\Omega$, where the sum $\vecmu+\vecnu$ is performed on $\Field^n$ (i.e. with modulo $q$). % from the definition Eq.~\eqref{eq:def:monomials}.
The inverse of an operator is $[\phi^{\vecmu}(\spin)]^{-1}=\phi^{-\vecmu}(\spin)=[\phi^{\vecmu}(\spin)]^*$. In the binary case, 
%the inverse of an operator is the operator itself.
each operator is its own inverse.
Each element of $\Omega$ is at most of order $q$, as:
\begin{align}
    [\phi^{\vecmu}(\spin)]^q = \exp\left({\displaystyle 2i\pi\,\state\cdot\vecmu}\right) = 1\,,
    \qquad\qquad %{\rm for}\;{\rm all}\;
    \forall\;\vecmu \in \Field^n
    \,.
\end{align}
\end{proof}

\begin{Property}
    \label{ppty:sum:phi}
    {\bf Sum %Equi-repartition} 
    of spin operator values.} %Using the property Eq.~(\ref{eq:ppty:root_unity}), all the monomials verify that: %the properties that:
    The monomials $\phi^{\vecmu}(\spin)\in\Omega$ verify that:
\begin{align}\label{eq:ppty:monomials}
    \sum_{\spin} \phi^{\vecmu}(\spin)
        = \sum_{\state\in\Field^n} \phi^{\vecmu}(\state)
        = q^n\,\delta_{\vecmu, \veczero}\,,
    \qquad\qquad %{\rm for}\;{\rm all}\;
    \forall\;\vecmu \in \Field^n\,,
\end{align}
where the two sums are respectively over all possible $q^n$ states of $\spin$ and $\state$, 
and where the delta function~$\delta_{\vecmu, \veczero}$ is equal to $1$ if $\vecmu=\veczero$ and to $0$ otherwise. \\
Replacing the expression of $\phi^{\vecmu}(\state)$ in Eq.~\eqref{eq:ppty:monomials}, and identifying the real and the imaginary parts on the two sides of the equality, one obtains a similar property for $\varphi^{\vecmu}$ and $\psi^{\vecmu}$:
%$\forall\vecmu \in\Field^n$,
\begin{align}\label{eq:ppty:monomials:cos-sin}
    \forall\vecmu \in\Field^n\,,
        \;\;
    \sum_{\state\in\Field^n} \cos\left(\frac{2\pi}{q}\state\cdot\vecmu\right)= q^n\,\delta_{\vecmu, \veczero}
        \qquad{\rm and}\;\;
    \sum_{\state\in\Field^n} \sin\left(\frac{2\pi}{q}\state\cdot\vecmu\right)=0\,.
\end{align}
\end{Property}
\noindent Equation~\eqref{eq:ppty:monomials} %This
%comes from the fact that [...]
%caries from the fact that [...]
is a direct consequence %follows directly from
of the fact that the variables $s_j$ are roots of unity, which %thus %verify that
satisfy\footnote{For all $\mu_j\in\Field$,\;
%Indeed,
%\begin{align}
    $\sum_{s_j} s_j^{\mu_j} = \sum_{\alpha_j = 0}^{q-1} e^{\frac{2i\pi}{q}\alpha_j\,\mu_j} = \frac{1-\Big(e^{\frac{2i\pi}{q}\mu_j}\Big)^q}{1-e^{\frac{2i\pi}{q}\mu_j}} = 0\,.$
%\nonumber
%\end{align}
} 
$\sum_{s_j} s_j^{\mu}=0$ for all strictly positive integer $\mu$.
%This is a direct consequence of the following property of roots of unity $\sum_{s_j} s_j^{\mu}=0$ for all positive integer $\mu$.
It reflects %\textcolor{red}{the fact} 
that, 
similarly to $s_j$, 
%similarly to the roots of unity, 
the values taken by $\phi^{\vecmu}(\spin)$ as one varies $\spin$ are evenly distributed over 
%\textcolor{red}{the direct product of $|\vecmu|$ unit circles (where $|\vecmu|$ denotes the number of non-zero entries in $\vecmu$).}
the $q$-th roots of unity\footnote{Or evenly distributed over the $q'$-th roots of unity, if $q$ is non-prime and $q'<q$ is the smallest integer satisfying $q'\vecmu=\veczero$ (see Sec.~\ref{sec1:interaction_interpretation:symmetry}).}. % (see Sec.~\ref{sec1:interaction_interpretation:symmetry}). 

\begin{Property}
    {\bf Orthogonality of %between 
    complex spin operators.} We denote by $\C[\Field^n]$ the vector space of functions from $\Field^n$ to $\C$ endowed %provided 
    with the following %Hermitian
    inner product:\\
    for two complex functions $f$ and $h$ in $\C[\Field^n]$,
    %$f(\state)$ and $h(\state)$,
    %and define the following inner product on $\C[\Field^n]$: 
    %define the inner product on $\C[\Field^n]$, between two complex functions $f(\state)$ and $h(\state)$, by:
\begin{align}\label{eq:def:HermitianInnerProd}
    \langle \,f\,|\, h\,\rangle
        &= \frac{1}{q^n}\sum_{\state\in\Field^n} [f(\state)]^*\,h(\state)\,,
\end{align}
where $[f(\state)]^*$ %represents 
is the complex conjugate of $f(\state)$.
Applying this scalar product to any pair of %two 
operators in $\Omega=\{\phi^{\vecmu}(\state) \mid \vecmu\in  \Field^n\}$, one obtains that they are orthonormal (see App.~\ref{app:FourierBasis:C}): % vectors of $\C[\Field^n]$:
\begin{align}\label{eq:linearly_indep_monimials}
    \;\hspace{-2.8cm}
    {\rm for}\;{\rm all}\;
    (\phi^{\vecmu}, \phi^{\vecnu})
    %(\phi^{\vecmu}(\state), \phi^{\vecnu}(\state))
    \in\Omega^2\,,\qquad \qquad
    \langle \,\phi^{\vecmu}\,|\, \phi^{\vecnu}\,\rangle
        &= \delta_{\vecmu, \vecnu}\,.        
\end{align}
Thus, the set $\Omega$ of all spin operators form an orthonormal set of vectors of $\C[\Field^n]$.
%forms a set of linearly independent vectors of $\C[\Field^n]$. 
%
%%%%%%%%%%
%The set $\Omega=\{\phi^{\vecmu}(\state)\;\slash \; \vecmu\in  \Field^n\}$ of all spin operators form an orthonormal set of vectors of $\C[\Field^n]$.\\
%Indeed, applying this scalar product to any pair of monomials in $\Omega$, one obtains that they are orthonormal:
%%%%%%%%%%
\end{Property}

%\subsubsection{Discrete Fourier transform of log-probability.}
Besides, there are exactly $q^n$ such independent vectors, which is also the dimension of the vector space $\C[\Field^n]$. 
%Besides, $\Omega$ contains exactly $q^n$ monomials.
The set $\Omega$ of monomials thus forms a basis of $\C[\Field^n]$:

\begin{Theorem}
    {\bf Complex spin model~\eqref{eq:def:CompleteM} as Fourier decomposition of the log-probability.}
    %{\bf Fourier Basis of complex functions over $\Field^n$. } 
    %%%%%% 
    The set $\Omega=\{\phi^{\vecmu}(\state) \mid \vecmu\in  \Field^n\}$ of all spin operators (including $\phi^{\veczero}(\state)=1$) forms an orthonormal basis of the vector space $\C[\Field^n]$ of complex functions over $\Field^n$.
    %Therefore, 
    As a consequence, any complex function % over $\Field^n$
    of the variables $\state\in\Field^n$ can be decomposed over this basis. In particular, the complete spin model in Eq.~\eqref{eq:def:CompleteM} corresponds to an expansion of the log-probability %(similarly, of the Hamiltonian)
    %(i.e., of the energy function)
    over this basis: \;\;for all $\state\in\Field^n$,
    \begin{align}\label{eq:expansion:Fourier_basis}
    \log P(\state)\, = \sum_{\vecmu\in\Field^n} \, g_{\vecmu}\, \phi^{\vecmu}(\state)\;
        = \sum_{\vecmu\in\Field^n} \, g_{\vecmu} \,
        \exp\left({\displaystyle\frac{2i\pi}{q}\,\state\cdot\vecmu}\right)\,,
    \end{align}
    where the $g_{\vecmu}$'s are  $q^n$ complex coefficients. The coefficient $g_{\veczero}$ is fixed 
    by the normalization of the probability distribution $P(\state)$ (recalling that $\phi^{\veczero}(\state)=1$):
    %by requiring that the probability distribution $P(\state)$ is normalized (using that $\phi^{\veczero}(\state)=1$):
    %From the normalization, $c_0=-\log \sum = -\log Z$.
    %Requiring that the probability distribution $P(\state)$ is normalized also fix the value of $g_{\veczero}$:
    \begin{align}
        g_{\veczero}
            =-\log \sum_{\state\in\Field^n} e^{\,
            %\sum_{\vecmu\in[\Field^n]^*} 
            \sum_{\vecmu\neq\veczero}
            g_{\vecmu}\phi^{\vecmu}(\state)}
            %=-\log \sum_{\state\in\Field^n} \exp\bigg(\sum_{\vecmu\in[\Field^n]^*}g_{\vecmu}\phi^{\vecmu}(\state)\bigg)
            =-\log Z(\vecg)\,,
    \end{align}
    where we identified the partition function $Z(\vecg)$.
    One can easily recognize that Eq.~\eqref{eq:expansion:Fourier_basis} is a (spatial) discrete Fourier decomposition of the log-probability%
    \footnote{This is also valid for the (high-order) Ising model $q=2$, for which $s_i=\exp(\frac{2i\pi\alpha_i}{q})=(-1)^{\alpha_i}$.}%
    , and that the set $\Omega$ of monomials is the discrete Fourier basis of complex functions on $\Field^n$.
\end{Theorem}
\noindent The high-order Ising model~\eqref{eq:graph_model} is a particular case of~\eqref{eq:expansion:Fourier_basis} for $q=2$, and can thus also be understood as a Fourier decomposition of the log-probability. % a decomposition of the log-probability over a Fourier basis. %In particular, the generalization of the Ising model in Eq.~\eqref{eq:graph_model}

%%%%%%%%% %%%%%%%%% %%%%%%%%% %%%%%%%%% 
%Any sub-model of the form~\eqref{eq:q-spin_models} can then be seen as a complete model with some harmonics $g_{\vecmu}$ exactly set to zero, while preserving the constrained that $g_{-\vecmu}=g_{\vecmu}^*$ (i.e., that inverse operators). It is associated to a subset of interactions $\tilde{\M}\subseteq\Omega\setminus\{1\}$ that always contain an element $\phi^{\vecmu}(\state)$ and its inverse $\phi^{-\vecmu}(\state)$. %In the more general case here the inverse of an operator $\phi^{\vecmu}(\state)$ is the operator $\phi^{-\vecmu}(\state)$.
%%%%%%%%% %%%%%%%%% %%%%%%%%% %%%%%%%%% 

Similarly, one can %easily 
show that the complete model in Eq.~\eqref{eq:def:CompleteM:real} corresponds to an expansion of the log-probability over the trigonometric Fourier basis of $\mathbb{R}[\Field^n]$.

\begin{Theorem}
    {\bf Real spin model~\eqref{eq:def:CompleteM:real} as trigonometric Fourier decomposition of $\log P(\state)$.} % the log-probability.}
    %{\bf Fourier Basis of real functions over $\Field^n$. } 
    The set%
    \footnote{In the definition of $\Omega_R$, the ordering between vectors $\vecmu\in\Field^n$ is defined by the ordering of the integers represented by $\vecmu$ in base $q$.  %for which $\vecmu$ is the representation in base $q$.
    The ordering is used to % prevent identical or parallel vectors in the $\Omega_R$. 
    ensure %make sure 
    that $\Omega_R$ doesn't contain identical 
    ($\varphi^{-\vecmu}(\state)=\varphi^{\vecmu}(\state)$ for $-\vecmu\neq\vecmu$) or parallel elements
    ($\psi^{-\vecmu}(\state)=-\psi^{\vecmu}(\state)$ for $-\vecmu\neq\vecmu$).}
    %\textcolor{red}{doesn't contain twice the same basis element, because, for $\vecmu\neq-\vecmu$, $\varphi^{-\vecmu}(\state)=\varphi^{\vecmu}(\state)$ and $\psi^{-\vecmu}(\state)=-\psi^{\vecmu}(\state)$.}}
    $\Omega_R$ of real spin operators~\eqref{eq:def:monomials:alpha:real}
    $\varphi^{\vecmu}(\state)=\cos(\frac{2 \pi}{q}\state\cdot\vecmu)$ and  $\psi^{\vecmu}(\state)=\sin(\frac{2 \pi}{q}\state\cdot\vecmu)$,
\begin{align}\label{eq:Omega:real}
    \Omega_R
    &=\{\varphi^{\vecmu}(\state) \mid \vecmu\in  \Field^n \;{\rm and}\; \vecmu \leq -\vecmu\}\cup \{\psi^{\vecmu}(\state) \mid \vecmu\in  \Field^n \;{\rm and}\; \vecmu < -\vecmu\}\,,
\end{align}
    forms an orthogonal basis of the vector space $\mathbb{R}[\Field^n]$ of real functions over $\Field^n$ (see proof in App.~\ref{app:FourierBasis:R}), 
    which can be recognized as the trigonometric discrete Fourier basis.
    %is a set of $q^n$ orthogonal vectors of the vector space of real functions over $\Field^n$ (see App.~\ref{app:FourierBasis:R}). It is thus an orthogonal basis of $\mathbb{R}[\Field^n]$, which can be recognized as the trigonometric discrete Fourier basis.
    The complete model~\eqref{eq:def:CompleteM:real} corresponds to an expansion of the log-probability over this basis, where the coefficient of the operator $\varphi^{\veczero}(\state)=1$ is fixed by normalization to $a_{\veczero}=-\log Z(\vecg)$.
    %\textcolor{orange}{Add Eq of Fourier decomposition here?}
\end{Theorem}

%%%%%%%%% %%%%%%%%% %%%%%%%%% %%%%%%%%% %%%%%%%%% %%%%%%%%% %%%%%%%%% %%%%%%%%%
%Similarly, this model~\eqref{eq:def:CompleteM:real} can be recognized as a real (spatial) discrete Fourier decomposition of the log-probability, in which the coefficient of the operator $\varphi^{\veczero}(\state)=1$ was obtained by normalization, $a_{\veczero}=-\log Z(\vecg)$.
%In particular:
%%%%%%%%% %%%%%%%%% %%%%%%%%% %%%%%%%%% %%%%%%%%% %%%%%%%%% %%%%%%%%% %%%%%%%%%

More generally, the spin models~\eqref{eq:q-spin_models} and~\eqref{eq:q-spin_models:real} correspond to an expansion of the log-probability over a subset of Fourier basis elements, i.e. in which some harmonics have been (artificially) set exactly to zero.
%In the models of Eq.~\eqref{eq:q-spin_models:real}, some harmonics have been (artificially) set exactly to zero.

Finally, exchanging the roles of $\vecmu$ and $\state$ in Eq.~\eqref{eq:linearly_indep_monimials} leads to another property of the complex spin operators,
%called completeness property in the binary case
called completeness of $\Omega$ in the binary case~\cite{beretta2018stochastic, mastromatteo2013typical}:
\begin{Property}
    {\bf Completeness property of $\Omega$.} For all states $(\state, \state')\in[\Field^n]^2$,
    \begin{align}
        \frac{1}{q^n} \sum_{\vecmu\in\Field^n} \phi^{\vecmu}(\state)\,\phi^{-\vecmu}(\state')
        =\delta_{\state, \state'}\,.
    \end{align}
    This relation can also be written under the form of the scalar product:
    \begin{align}\label{ppty:op:complete}
        \langle \phi_{\state'} \mid \phi_{\state}\rangle
        = \frac{1}{q^n} \sum_{\vecmu\in\Field^n} \, \phi_{\state}(\vecmu) \,
        %\phi_{-\state'}(\vecmu)
        [\phi_{\state'}(\vecmu)]^*
        =\delta_{\state, \state'}\,,
    \end{align}
    where, for each state $\state\in\Field^n$, we introduce the function $\phi_{\state}(\vecmu)\doteq\phi^{\vecmu}(\state)=\exp\left({\frac{2i\pi}{q}\,\state\cdot\vecmu}\right)$ for all $\vecmu\in\Field^n$.
\end{Property}

\subsubsection{Inversion formulas for the model parameters}
%{Spatial Fourier transform and \textcolor{red}{[model parameters as]} Fourier coefficients.}
The complete models in Eq.~\eqref{eq:def:CompleteM} and \eqref{eq:def:CompleteM:real} correspond respectively to the complex and real Fourier series of $\log P(\state)$, and thus the Fourier coefficients $g_{\vecmu}$, $a_{\vecmu}$, and $b_{\vecmu}$ can be obtained by Fourier transform of $\log P(\state)$.
Equations~\eqref{eq:pdf:FT} and~\eqref{eq:pdf:FT:a_b} below generalize to $q\geq 2$ the inversion formulas known for binary spin models~\cite{mastromatteo2013typical, beretta2018stochastic} (for $q=2$, $g_{\vecmu}=a_{\vecmu}$ and $b_{\vecmu}=0$). %for all $\vecmu$).
 
\begin{Property}
{\bf The coefficients $g_{\vecmu}$} 
in Eq.~\eqref{eq:def:CompleteM} can be obtained by Fourier Transform of the log-probability function~$\log P(\state)$
%of the complex Fourier decomposition of $\log P(\state)$ in Eq.~\eqref{eq:def:CompleteM} can be obtained by Fourier Transform
(see App.~\ref{app:Fouriercoeff}): \;\;for all $\vecmu\in[\Field^n]^*$,
\begin{align}\label{eq:pdf:FT:1}
    g_{\vecmu} = \mathcal{F}\,[\log P](\vecmu) \,
        &= \,\frac{1}{q^n}  \sum_{\state\in\Field^n} \, \log P(\state) \,
        \exp\left({\displaystyle-\frac{2i\pi}{q}\,\state\cdot\vecmu}\right)\,,
\end{align}
which can also be written under the form of the inner product~\eqref{eq:def:HermitianInnerProd}:
\begin{align}\label{eq:pdf:FT}
    g_{\vecmu} &= \,\frac{1}{q^n} \sum_{\state\in\Field^n} \, \log P(\state) \,
        \left[\phi^{\vecmu}(\state)\right]^*
        =\left\langle \phi^{\vecmu}\,|\,\log P\right\rangle
        \,.
\end{align}
\end{Property}

%%%%%%%%% %%%%%%%%% %%%%%%%%% %%%%%%%%% %%%%%%%%% %%%%%%%%% 
%\footnote{Note the less usual notation with the coefficient $1$ in front of the sum for the inverse Fourier transform and the coefficient $1/q^n$ for the Fourier transform; compare to a coefficient $1/\sqrt{q^n}$ for both transforms in the usual case.}
%%%%%%%%% %%%%%%%%% %%%%%%%%% %%%%%%%%% %%%%%%%%% %%%%%%%%% 

%%%%%%%%% %%%%%%%%% %%%%%%%%% %%%%%%%%% %%%%%%%%% %%%%%%%%% 
%There is a bijection between the $q^n$ values of $g_{\vecmu}$ and the $q^n$ probability $P(\state)$. Equation~\eqref{eq:pdf:FT} generalizes to system of $q$-state variables the inversion relations already known for the complete model in the context of binary data~\cite{mastromatteo2013typical}. 
%%%%%%%%% %%%%%%%%% %%%%%%%%% %%%%%%%%% %%%%%%%%% %%%%%%%%% 
%Similarly:
\begin{Property}
    {\bf The coefficients $a_{\vecmu}$ and $b_{\vecmu}$} in Eq.~\eqref{eq:def:CompleteM:real} can be obtained by real Fourier Transform of the log-probability function $\log P(\state)$
    (see App.~\ref{app:Fouriercoeff}): \;\;for all $\vecmu\in[\Field^n]^*$,
    \begin{align}\label{eq:pdf:FT:a_b}
    \begin{cases}
    \;\;\;{\rm Re}[g_{\vecmu}] = \displaystyle\frac{1}{q^n}\sum_{\state\in\Field^n} \log P(\state) \cos\left(\frac{2\pi}{q}\,\state\cdot\vecmu\right)
    =\left\langle \varphi^{\vecmu}\,|\,\log P\right\rangle 
    =
    \begin{cases}
            \, a_{\vecmu} &{\rm for\;} \vecmu=-\vecmu\\[2pt]
            \, a_{\vecmu}/2 & {\rm else} %{\rm for\;} \vecnu\neq-\vecnu\\
    \end{cases}
    \\[15pt] 
    -{\rm Im}[g_{\vecmu}] = \displaystyle\frac{1}{q^n}\sum_{\state\in\Field^n} \log P(\state) \sin\left(\frac{2\pi}{q}\,\state\cdot\vecmu\right)
    =\left\langle \psi^{\vecmu}\,|\,\log P\right\rangle
    =
    \begin{cases}
            \, 0 &{\rm for\;} \vecmu=-\vecmu\\[2pt]
            \, b_{\vecmu}/2 &{\rm else} %{\rm for\;} \vecnu\neq-\vecmu\\
    \end{cases}
    \end{cases}
\end{align}
\end{Property}

%%%%%%%%%%%%%%%%%%%%%%%%%%
%There is a bijection between the $q^n$ values of $g_{\vecmu}$ and the $q^n$ state probabilities $P(\state)$ in Eq.~\eqref{eq:def:CompleteM}, and between the $q^n$ parameters $a_{\vecmu}$ and $b_{\vecmu}$ and the $q^n$ state probabilities $P(\state)$ in Eq.~\eqref{eq:def:CompleteM:real}.
%Equations~\eqref{eq:pdf:FT} and~\eqref{eq:pdf:FT:a_b} generalize to system of $q$-state variables the inversion relations already known for the complete %high-order Ising 
%model in the context of binary data~\cite{mastromatteo2013typical, beretta2018stochastic} (we recall that in the binary case $g_{\vecmu}=a_{\vecmu}$ and $b_{\vecmu}=0$ for all $\vecmu$).
%This also shows that the formula known for the binary case is simply a Fourier transform of the data with $q=2$.
%%%%%%%%%%%%%%%%%%%%%%%%%%
%As for binary spin models~\cite{mastromatteo2013typical}, there is a bijection between the $q^n$ probability parameters $p_{\vecmu}$ and the $q^n$ real parameters $a_{\mu}$ and $b_{\mu}$.

%Reciprocally, the coefficient $C(\vecmu)$ are then obtained by taking the {\it Fourier Transform} of $\log P(\state)$. 

Reciprocally, %Similarly, 
choosing a set $\M$ of interactions and their coefficients $\vecg=(g_{\vecmu}\in\mathbb{C}\mid \forall \vecmu\in\M)$, and defining the discrete function $G(\vecmu) = \sum_{\vecnu\in\M}g_{\vecnu}\delta_{\vecmu, \vecnu}$ for all $\vecmu\in\Field^n$, the %log-
probability distribution of the spin model~\eqref{eq:def:CompleteM} can be obtained by inverse Fourier transform: 
%Considering the complex discrete function $G(\vecmu) = \sum_{\vecnu}g_{\vecnu}\delta_{\vecmu, \vecnu}$ for all $\vecmu\in\Field^n$. The function $\log P(\state)$ is the {\it Inverse Fourier Transform} of $G(\vecmu)$: for all $\state\in\Field^n$,
%
\begin{align}\label{eq:pdf:IFT}
    \log P(\state\mid \M,\,\vecg) 
        &= \mathcal{F}^{-1}[G](\state) \,
        = \sum_{\vecmu\in\Field^n} \, G(\vecmu) \,
        \exp\left({\displaystyle\frac{2i\pi}{q}\,\state\cdot\vecmu}\right)
        %&= \sum_{\vecmu\in\Field^n} \, G(\vecmu) \, |\,\phi^{\vecmu}(\state)\rangle\,.\\
        = q^n \,\langle G^* \mid \phi_{\state}\rangle\,.
\end{align}
where we introduced the functions $\phi_{\state}(\vecmu)=\phi^{\vecmu}(\state)$ for all $\vecmu$ and $\state$ in $\Field^n$ (see App.~\ref{app:Fouriercoeff}).  %which represents the values of all the operators for a given state $\state$.
%\textcolor{blue}{(such that the $q^n$-dimensional vector $\mid \phi_{\state}\rangle$ contains the values of all the operators for a given state $\state$).}

For systems with up to about $25$ spins, this approach can be used to perform direct sampling from any chosen high-order spin model, or to compute exactly the coefficient $g^{\vecmu}$ from the empirical probability distribution. %for reasonably small systems %(for small numbers of variables). 
For instance, in the binary case, Ref.~\cite{AgrimiCaputo2026}
%\footnote{in a few second for systems of the order of $25$ spins on a laptop} 
uses %this approach
the Walsh–Hadamard transform to perform exact sampling from high-order spin models with $n=20$ spins in order to create benchmark data. % with chosen [community-like] patterns
Ref.~\cite{kato2010haplotype, buchman2012sparse, mosseri2015ising, Peerbooms2026} also use the Walsh–Hadamard transform to compute efficiently the parameters of specific pairwise %models with specific geometries 
and high-order Ising models.
%\textcolor{orange}{In Sec.~\ref{section:Ex:gen_fit}, we give an example for $q=3$ and $n=20$ spins, where we use this approach to both sample data and recover the model parameters %, using this procedure 
%in different sampling regimes.}

%%%%  EMAIL: Ilya Schurov  %%%%   %%%%   %%%%  
%Ref.~\cite{} also uses the Walsh–Hadamard transform to compute efficiently the $g_{\vecmu}$-parameters, and obtained an approximate value for the parametric complexity of quantum spin models from the number of parameters needed to capture the sign function.
%%%%

%%%%%%%%%%%%%%%% %%%%%%%%%%%%%%%% %%%%%%%%%%%%%%%% %%%%%%%%%%%%%%%%
%In the binary case, one can use the Walsh–Hadamard transform to computing efficiently all the $g^{\vecmu}$-parameters\footnote{in a few second for systems of the order of $25$ spins on a laptop}~\cite{} or to sample exactly from the maximum entropy model distribution at chosen values of the parameters~\cite{}.
%%%%%%%%%%%%%%%% %%%%%%%%%%%%%%%% %%%%%%%%%%%%%%%% %%%%%%%%%%%%%%%%

%\textcolor{orange}{\subsubsection{[Discussion: choices of basis decomposition]}}

Using $q$-state spin models for data corresponds to decomposing the model log-probability over a specific choice of basis functions (the Fourier basis functions).
%%%%%%%%%%%%%
%\textcolor{purple}{means fitting the model log-probability to/with a specific set of basis functions (the Fourier basis functions).}
%\textcolor{blue}{Using $q$-state spin models for data corresponds to [making] a specific choice of decomposition of the log-probability over basis functions.} 
%%%%%%%%%%%%%
%(or, in the physical sense/more physically to a specific choice of family of many-body interaction energies).}
One could choose to use other forms of basis functions, each of them corresponding to a different way of decomposing the information of the multivariate system into basis patterns~\cite{amari2002information}. %\textcolor{orange}{[Mention exponential families?]

\subsection{Interpretation of $q$-state spin interactions}
%Interpretation of the interactions
\label{sec:Potts:interpretation}

\begin{figure}[h]
    \includegraphics[width=\textwidth]{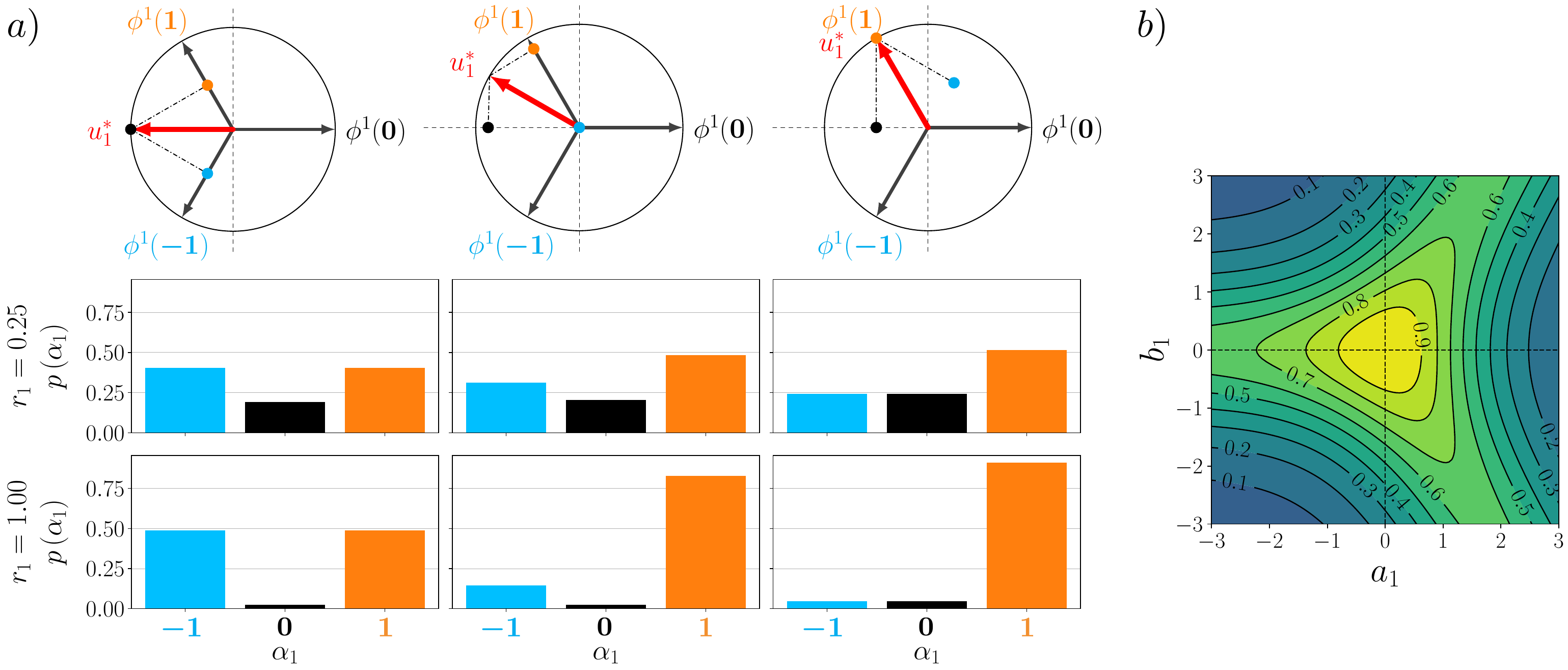} 
\caption{{\bf Example for $q=3$.}
    One-spin system $\state=(\alpha_1)$ modeled by a single interaction $\vecmu=(1)$, corresponding to the operator $\phi^1(\state) = \exp\big({\frac{2i\pi}{3}\alpha_1}\big)$ with parameter $g_1$ (and its c.c.).
    {\bf a) Probability distribution of the model} for different directions and strengths of the parameter $\vec{g}_1^* = r_1 \vec{u}_1^*$. % \textcolor{red}{parametrizing the interaction}. %associated with the interaction $\phi^1$.
    Each column correspond to a fix direction of $\vec{g}_1^*$, indicated by the vector $\vec{u}_1^*$ in the complex plane representation (see first row).
    %%%%%%%%%%%%%%%%%%%%%
    The energy of a state~$\state$ is given by the scalar product~\eqref{eq:energy_projection} between %the vector representations of $g_1^*$ (direction indicated by the red arrow) and $\phi^1(\state)$ (black arrows) in the complex plane. 
    $\vec{g}_1^*$ (direction $\vec{u}_1^*$ indicated by red arrow) and $\vec{\phi}^1(\state)$ (black arrows) in the complex plane.
    The dots indicate the projection of the vector $\vec{u}_1^*$ on each possible direction of $\vec{\phi}^{1}(\state)$.
    The state~$\state$ for which $\vec{\phi}^{1}(\state)$ is best aligned with $\vec{g}_{1}^*$ has the highest probability, and states become less likely as $\vec{\phi}^{1}(\state)$ points further away from $\vec{g}_{1}^*$. % [along the unit circle].}
    In the left column, $\vec{g}_{1}^*$ points halfway between the values of $\phi^1(\alpha_1)$ for $\alpha_1=1$ and $\alpha_1=-1$,
    and thus these two states have equal probability (recall that $\alpha=-1=2$ in $\mathbb{Z}/3\mathbb{Z}$). 
%%%%%%
    %In the middle column, it is closer to $\alpha_1=1$, which then has the highest probability.
%%%%%%
    In the right column, $\vec{g}_{1}^*$ is exactly aligned with %state 
    $\alpha_1=1$, which then has the highest probability, the two other states having equal probability.
    %Each row corresponds to a different value for the parameter strength $r=||g_1^*||$, increasing from top to bottom.
    The probability distributions are shown for two values of the parameter strength, $r_1=0.25$ and $r_1=1$.
    %
%%%%%%
    %For increased strength, the distribution becomes more extreme, \textcolor{red}{the difference between probabilities in the distribution are more marked/sharper/stronger}, leading to a decrease in the entropy [of the distribution].
    %\textcolor{red}{the differences between the probabilities of the states in the distribution become stronger (see figure b)).}
%%%%%%
    % the difference between the probabilities of the states become more significant, and thus the stronger is the pattern associated to that interaction (i.e., the lower is the entropy of the operator). 
    %The larger amplitude of the parameter, the stronger the pattern.
    %
    {\bf b) Entropy of the model} for varying values of the parameters $a_1=2\,{\rm Re}[g^{*}_1] = 2\,r_1\cos\big(\theta_{u_1^*}\big)$ and $b_1=2\,{\rm Im}[g^{*}_1] =2\,r_1\sin\big(\theta_{u^*_1}\big)$, where $\theta_{u^*_1}$ is the angle formed by $\vec{u}^*_1$ along the unit circle.
    %At fixed value of the strength $r_1$, the entropy is maximal when $\vec{u}_{1}^*$ points exactly between two consecutive possible directions of $\vec{\phi}^{\,1}(\state)$, and minimal when its direction coincides with one of the $\vec{\phi}^{1}(\state)$ (positions of the roots of unity). 
    At fixed value of the strength $r_1$, the entropy is minimal when the direction of $\vec{u}_{1}^*$ coincides with one of the three possible directions of $\vec{\phi}^{1}(\state)$ %(which are the cube roots of unity) 
    (%cube 
    roots of unity) and maximal when it points exactly between two consecutive directions of $\vec{\phi}^{\,1}(\state)$.
    %\textcolor{orange}{(indicated by the red dashed lines)}.
    }
\label{fig:prob_distr_q3_n1}
\end{figure} 

%%%%%%%%%%%%%%%%%
In this section, we give some insight into the way $q$-state spin models %interactions
encode patterns of data. We discuss how spin operators shape the energy landscape of models, and what distinguish %differentiate 
different interactions based %/acting 
on the same set of variables (e.g., for $q=5$, there are two types of one-body interaction -- see Fig.~\ref{fig:prob_distr_q5_n1}). %$\phi^{(1)}(\spin)=s_1$ and $\phi^{(2)}(\spin)=s_1^{\,2}$).
%and what is the meaning of multiple interactions based on the same set of variables.
%\textcolor{blue}{We discuss properties of spin models resulting from the use of a given spin operator.}

%%%%%%%%%%%%%%% %%%%%%%%%%%%%%% %%%%%%%%%%%%%%% %%%%%%%%%%%%%%% %%%%%%%%%%%%%%%
%Physics interpretation?
%Each monomial (interaction) implement a torsion between a chosen set of spins, which will tend to align spins in different angles (different colors of the spins). Monomials are seen as a way to align different colors of the spins?

In Eq.~\eqref{eq:q-spin_models}, the energy of the spin system (Hamiltonian) for a given model $\M\subseteq \Field^n$ with $K=|\M|$ operators can be represented as a scalar product in $\mathbb{C}^K$: %the $K$-dimensional complex space.
%%%%%%%%%%%%%%%% QUESTION: 
%%%%%% Should we uniformize notations with~\eqref{eq:pdf:IFT}?
%Can be seen as the same scalar product as in Eq.~\eqref{eq:pdf:IFT}, but in which some $g^{\vecmu}$ are null.}
%%%%%%%%%%%%%%%%
\begin{align}
    H(\state\,|\,\M,\,\vecg) 
        = -\sum_{\vecmu\in\M} g_{\vecmu} \,\phi^{\vecmu}(\state)
        %= -\langle \,\vecg^*\,|\,\phi_{\state}\,\rangle_{\M}\,,
        = - \,\vecg^* \cdot \boldsymbol{\phi}(\state)\,,
\end{align}
where $\vecg^*$ is a $K$-dimensional complex vector containing the values of the $K$ parameters~$g_{\vecmu}^*$ (conjugate of $g_{\vecmu}$), and $\boldsymbol{\phi}(\state)$ is a $K$-dimensional complex vector containing the values of the $K$ operators $\phi^{\vecmu}(\state)$ evaluated on the state $\state$ (both vectors are indexed by $\vecmu$).
The energy of the system is minimal if these two vectors are aligned and maximal if they are anti-aligned.
However, this is difficult to visualize or interpret in the $K$-dimensional space.
Besides, for the Hamiltonian $H(\state\,|\,\M,\,\vecg)$ to be real, we must take into account that $g_{-\vecmu}=g_{\vecmu}^*$ for all $\vecmu\neq-\vecmu$.
In the following, we focus on the contribution of pairs of conjugate operators $H^{\vecmu}(\state)=-g_{\vecmu}\,\phi^{\vecmu}(\state)-g_{\vecmu}^*\,\phi^{-\vecmu}(\state)$ (for $\vecmu\neq-\vecmu$) to the Hamiltonian and give a geometrical representation and interpretation of the effect %impact 
of a given interaction on the energy of the states, %a state~$\state$, 
and therefore on the probability distribution.

\subsubsection{Geometrical representation of the interaction energy}

\begin{Definition} %{\bf Representations in the complex plane:}
{\bf Representation of $\phi^{\vecmu}(\state)$ and $g_{\vecmu}$ as vectors in the complex plane:}
Similarly to the planar spin variables, the states of an operator $\phi^{\vecmu}(\state)=z_q^{\state\cdot\vecmu}$ and the complex parameters $g^*_{\vecmu}$ %$g^*_{\vecmu}=(a_{\vecmu}+i b_{\vecmu})/2$
can be represented by vectors in the complex plane with coordinates:
\begin{align}
\begin{cases}
&\vec{\phi^{\vecmu}}(\state) = \left(\cos\left(\frac{2\pi}{q}\vecmu\cdot\state\right),\, \sin\left(\frac{2\pi}{q}\vecmu\cdot\state\right)\right)
=\left(\varphi^{\vecmu}(\state),\, \psi^{\vecmu}(\state)\right)\,\\[3mm]
&\vec{g^*_{\vecmu}} = \left({\rm Re}[g_{\vecmu}^*],\, {\rm Im}[g_{\vecmu}^*]\right) =
\begin{cases}
\left(\frac{a_{\vecmu}}{2},\,\frac{b_{\vecmu}}{2}\right)
& {\rm for}\;\; -\vecmu\neq\vecmu \\[2mm]
\left(a_{\vecmu},\,0\right)
 & {\rm for}\;\; -\vecmu=\vecmu
 \end{cases}
\end{cases}
\end{align}
where $a_{\vecmu}$ and $b_{\vecmu}$ are %real parameters. 
the real parameters from Eq.~\eqref{eq:def:CompleteM:real}.
Let us also denote $\vec{g^*_{\vecmu}} = r_{\vecmu}\,\vec{u^*_{\vecmu}}$\,, where $r_{\vecmu}=\lVert g_{\vecmu}\rVert$ is the amplitude of the parameter %$g_{\vecmu}$ 
and $\vec{u}_{\vecmu}^*$ is its direction.
\end{Definition}

\begin{Property}
{\bf Interaction energy.} Using this vector representation, the energy $H^{\vecmu}(\state)$ %associated to
of an interaction~$\vecmu$ is given by: %can be represented as
\begin{align}\label{eq:energy_projection}
H^{\vecmu}(\state)
= - [a_{\vecmu} \varphi^{\vecmu}(\state) + b_{\vecmu} \psi^{\vecmu}(\state)]
=
\begin{cases}
-2\;\vec{g_{\vecmu}^*}\cdot\vec{\phi^{\vecmu}}(\state) & {\rm for}\;\; -\vecmu\neq\vecmu \\
 -\;\vec{g_{\vecmu}^*}\cdot\vec{\phi^{\vecmu}}(\state) & {\rm for}\;\; -\vecmu=\vecmu
\end{cases}
\end{align}
which contains the contributions of both %the two
%conjugate operators.
$\phi^{\vecmu}(\state)$ and its complex conjugate (c.c.) if $-\vecmu\neq\vecmu$, or just the single real operator if $-\vecmu=\vecmu$.
The total Hamiltonian is then %$H=\sum_{\vecmu\in\M:\\\vecmu\leq-\vecmu} H^{\vecmu}$,
\begin{align}\label{eq:energy_projection:total-H}
H(\state)=\sum_{\vecmu\in\M:\vecmu\leq-\vecmu} H^{\vecmu}(\state)\,,
\end{align}
where the contribution of $H^{\vecmu}$ can be visualized by the projection of $\vec{g}_{\vecmu}^*$ onto the vector $\vec{\phi^{\vecmu}}(\state)$.
In the following, we analyze $H^{\vecmu}$ for values of $\vecmu$ such that $\vecmu\leq-\vecmu$ (the conjugate term being identical $H^{-\vecmu}=H^{\vecmu}$).
\end{Property}

%%%%%%% {\bf Message 1:}\\ %%%%%%%  %%%%%%%  %%%%%%%  %%%%%%%  %%%%%%% 
%The probability of the states of the system depends how well the operator evaluated at that state align with the parameter.\\
%on the align between the operator and the parameter (see Fig. 1): link between the alignment and the probability: $p = <g | \phi>$\\
%
%This is because the energy associated to the interaction is given by $-(g \phi + g^* \phi^*) = -<g | \phi>$, so the better align they are, the lower the energy, and therefore the more likely the state.\\
%%%%%%%  %%%%%%%  %%%%%%%  %%%%%%%  %%%%%%%  %%%%%%%  %%%%%%%  %%%%%%% 

\begin{figure}[h]
    \centering
    \includegraphics[width=\textwidth]{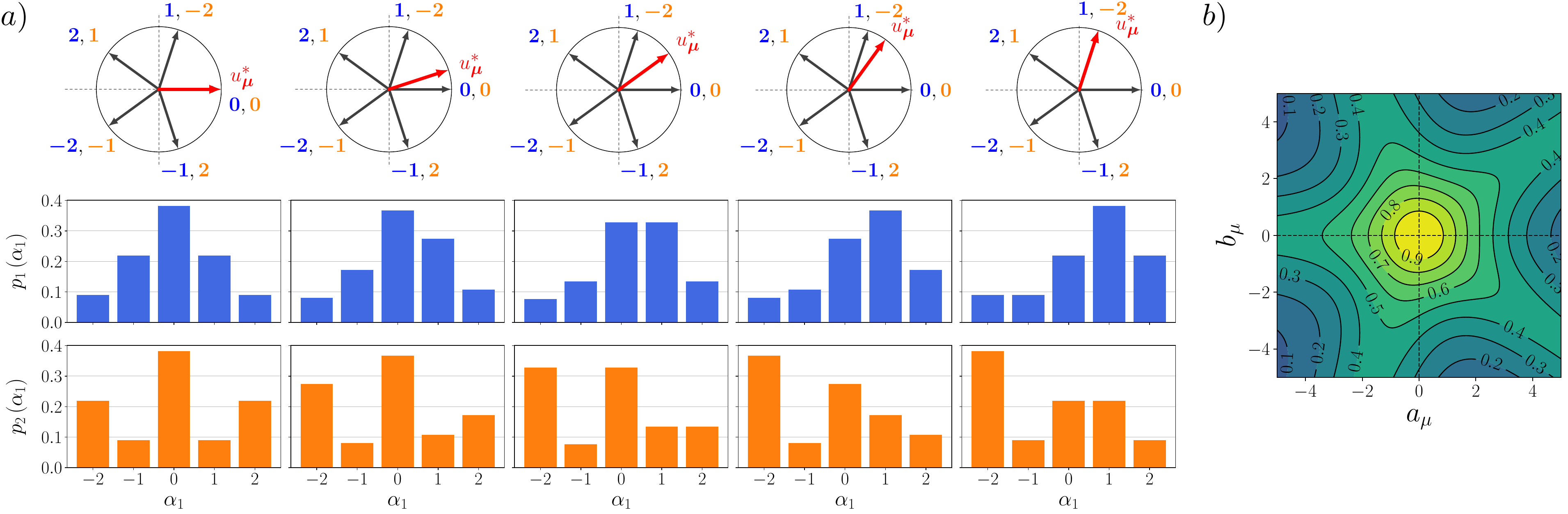} 
\caption{
    {\bf Example for $q=5$ and two types of 1-body operator.}
    %One-spin system $\state=(\alpha_1)$ with two models of the form $p_{\vecmu}(\state)= \exp\big(\vec{g}_{\vecmu}^*\cdot\vec{\phi}^{\vecmu}(\state)\big)/Z_{\vecmu}$ with a  single interaction $\vecmu=(\mu_i)$ corresponding to the operator $\phi^{\vecmu}(\state) = \exp\big({\frac{2i\pi}{5}\mu_i\alpha_1}\big)$ (and its c.c.).
    %
    One-spin system $\state=(\alpha_1)$ modeled by two examples of a single-spin interaction: an interaction $\vecmu=(1)$ corresponding to the operator $\phi^1(\state) = \exp\big({\frac{2i\pi}{5}\alpha_1}\big)=z_5^{\,\,\alpha_1}$ (and its c.c.) with probability distribution denoted $p_1(\state)$; and an interaction $\vecmu=(2)$ corresponding to the operator $\phi^2(\state) = \exp\big({\frac{2i\pi}{5}\,2\alpha_1}\big)=z_5^{\,\,2\alpha_1}$ (and its c.c.) with probability distribution denoted $p_2(\state)$.
    %$\vecmu=(1)$ for the model $p_1(\state)= \exp\big(\vec{g}_1^*\cdot\vec{\phi}^{1}(\state)\big)/Z_1$ with $\phi^1(\state)=$; and $\vecmu=(2)$ for the model $p_2(\state) = \exp(\vec{g}_2^*\cdot\vec{\phi}^{2}(\state))$.\\
    {\bf a)} Probability distributions for the models $p_1(\state)$ (in blue) and $p_2(\state)$ (in orange) for different directions of the parameter $\vec{g}_{\vecmu}^* = r_{\vecmu} \vec{u}_{\vecmu}^*$ at fixed strength $r_{\vecmu}=0.4$; the respective directions of $\vec{u}_{\vecmu}^*$ are indicated by a red arrow in the circles above the distributions.
%    The parameter strength used to make the distributions is 0.4.
% 
    We indicated around the unit circle, at the five positions that can be taken by $\vec{\phi}^{\vecmu}(\alpha_1)$ (black arrows located at the fifth roots of unity), the values of the state of $\alpha_1$ that yields this direction, in blue for $\phi^1$ and in orange for $\phi^2$.
    %\textcolor{blue}{We have represented in color the value of the state $\alpha_1$ that gives the corresponding location of $\phi^{\vecmu}(\alpha_1)$ on the unit circle, in blue for $\vecmu=(1)$ and in orange for $\vecmu=(2)$ (matching the color of the corresponding probability distributions).}
    %On the left, $\vec{u}_i^*$ (red arrow) is aligned with $\phi^1(0)$ and $\phi^2(0)$, and on the right it is aligned with $\phi^1(1)$ and $\phi^2(-2)$. 
    On the left, $\vec{u}_{\vecmu}^*$ %(red arrow) 
    is aligned with $\phi^1(0)$ and $\phi^2(0)$, which implies that $\alpha_1=0$ is the state of highest probability in both $p_1$ and $p_2$; the probability of the other states depends on how far they are from $\vec{u}^*_{\vecmu}$. On the right $\vec{u}_{\vecmu}^*$ is aligned with $\phi^1(1)$ and $\phi^2(-2)$.
    From left to right, the angle of $\vec{u}_{\vecmu}^*$ is progressively increased by $\pi/10$.
    {\bf b)} 
    %Entropy of the model for varying values of the parameters $a_{\vecmu}=2\,r_{\vecmu}\cos\big(\theta_{u_{\vecmu}^*}\big)$ and $b_{\vecmu}=2\,r_{\vecmu}\sin\big(\theta_{u^*_{\vecmu}}\big)$, where $\theta_{u^*_{\vecmu}}$ is the angle formed by $\vec{u}^*_{\vecmu}$ along the unit circle. 
    Entropy of the model for varying values of the parameters $a_{\vecmu}=2\,{\rm Re}[g_{\vecmu}^*]$ and $b_{\vecmu}=2\,{\rm Im}[g_{\vecmu}^*]$.
    The two models $p_1(\state)$ and $p_2(\state)$ have exactly the same entropy plot. 
    This can be appreciated in the probability distributions of panel~a), where, for each value of the parameter, the probability of the states are simply shuffled between $p_1$ and $p_2$. 
    %\textcolor{orange}{We denote by $u$ the complex number corresponding to the vector $\vec{u}$ in the complex plane.}
    %\textcolor{orange}{Observe the change in periodicity of the distributions between $p_1$ and $p_2$ (frequency is doubled from $p_1$ to $p_2$), reflecting the fact that they correspond to different Fourier modes.}
    }
\label{fig:prob_distr_q5_n1}
\end{figure}

%%%%%%%%%%%%%%%%%%%%%% %%%%%%%%%%%%%%%%%%%%%% %%%%%%%%%%%%%%%%%%%%%%
%Each interaction vector $\vecmu\in\Field^n$ defines a linear combination $\vecmu\cdot\state$ of the colors of the spins involved in the interaction. The direction of the spin monomial $\vec{\phi}^{\vecmu}(\state)$ in the complex plane results from this linear combination~\eqref{eq:def:monomials:alpha:real}: 
%\begin{align}\label{eq:def:monomials:alpha:bis}
%    \phi^{\vecmu}(\state) =\exp(\frac{2\pi i}{q}\,\vecmu\cdot\state)
%    =z_q^{\;\vecmu\cdot\state}\,,\qquad {\rm where}\qquad z_q=\exp(\frac{2\pi}{q}\,i)
%\end{align}
%is the first $q$-th root of unity, and thus rotates as one changes the state of the variables $\state$ (see black arrows in Fig.~\ref{fig:prob_distr_q3_n1}.a and~\ref{fig:prob_distr_q5_n1}.a).
%%%%%%%%%%%%%%%%%%%%%% %%%%%%%%%%%%%%%%%%%%%% %%%%%%%%%%%%%%%%%%%%%%
Let's discuss on the contribution $H^{\vecmu}(\state)$ of a given interaction to the energy of the system.
For a given interaction~$\vecmu$, the direction of the operator $\vec{\phi}^{\vecmu}(\state)$ in the complex plane rotates as one changes the state of the variables $\state$  (see black arrows in Fig.~\ref{fig:prob_distr_q3_n1}.a and~\ref{fig:prob_distr_q5_n1}.a).
The contribution of an interaction~$\vecmu$ to the energy of the state~$\state$ then depends on how well the direction of the operator $\vec{\phi^{\vecmu}}(\state)$ aligns with the direction of its parameter~$\vec{g}_{\vecmu}^*$, as expressed by Eq.~\eqref{eq:energy_projection}: the better the alignment, the lower the interaction energy, and therefore the higher the (relative) probability $P(\state)$ of the state (see Fig.~\ref{fig:prob_distr_q3_n1}.a and~\ref{fig:prob_distr_q5_n1}.a, and interactive widget in Supplementary Material). 
Moreover, increasing the strength~$r_{\vecmu}$ of the parameter amplifies the differences between the state probabilities, making the pattern associated with the interaction more significant and in particular leading to a decrease of the entropy of the operator (see Fig.~\ref{fig:prob_distr_q3_n1} and~\ref{fig:prob_distr_q5_n1}.b).
%%%%%%%%%%% %%%%%%%%%% %%%%%%%%%% %%%%%%%%%% 
%\textcolor{blue}{In order to minimize its energy for a given set of parameters, the system will favor states for which the monomials are mostly aligned with their respective \textcolor{red}{parameter[s]} $g_{\vecmu}^*$, while weighting the contributions of different monomials based on the relative strengths of the parameters.}
%%%%%%%%%%% %%%%%%%%%% %%%%%%%%%% %%%%%%%%%% 
Overall, in order to minimize its energy, % for a given model, 
the system will favor states for which the operators $\vec{\phi}^{\vecmu}(\state)$ of the model are best aligned with the direction~$\vec{u}_{\vecmu}^*$ of their respective (conjugate) parameter, 
while %prioritizing/
weighting the alignments of the operators
%\textcolor{red}{while weighting the contributions of different operators} 
based on the relative strengths~$r_{\vecmu}$ of their parameters.
Below we discuss examples of different types of interactions between the same variables. 
%\textcolor{blue}{We recall that this doesn't happen in the binary case.} 
Note that this effect is specific to non-binary discrete models, as in the binary case there is only one type of spin operator over each subset of spins.

\subsubsection{Different types of interactions between the same variables and examples}
\label{Sec:HOI:interpretation}
%{\bf Message: different operators order the states differently around the unit circle, which impacts which states are close to each other in terms of their probability.}\\

Each interaction vector $\vecmu\in\Field^n$ defines a linear combination $\vecmu\cdot\state=\sum_i \mu_i\alpha_i$ of the colors $\alpha_i$ of the spins involved in the interaction (for which $\mu_i\neq 0$).
The direction of the spin monomial $\vec{\phi}^{\vecmu}(\state)$ in the complex plane results from this linear combination, as $\phi^{\vecmu}(\state)=z_q^{\vecmu\cdot\state}$, where $z_q$ is the first $q$-th root of unity.
For example in Fig.~\ref{fig:prob_distr_q5_n1}.a, the direction of the operator $\phi^1(\alpha_1)=z_5^{\,\,\alpha_1}$ rotates by $2\pi/5$ each time $\alpha_1$ is incremented by $1$, while the direction of $\phi^2(\alpha_1)=z_5^{\,\,2\alpha_1}$ rotates by $4\pi/5$.  
As a result, the operators $\phi^1$ and $\phi^2$ order the states differently around the unit circle.
%%%%%%%%%%%%%%%%%%%%%%%% %%%%%%%%%%%%%%%%%%%%%%%% %%%%%%%%%%%%%%%%%%%%%%%% 
%Because the probability of a state (under a single interaction) depends on how far $\vec{\phi}^{\vecmu}(\state)$ is from $\vec{g}_{\vecmu}^*$ around the unit circle~\eqref{eq:energy_projection}, states that are closer to each other around the unit circle are also closer to each other in terms of their probabilities. Different orderings of the states around the unit circle by different interactions thus constraint the probability distribution resulting from/associated with a single interaction, and therefore the energy landscape created by that interaction.
%%%%%%%%%%%%%%%%%%%%%%%%
%Because the probability of a state (under a single interaction) depends on how far $\vec{\phi}^{\vecmu}(\state)$ is from $\vec{g}_{\vecmu}^*$ around the unit circle~\eqref{eq:energy_projection}, the ordering of the state by the interaction imposes constraints on the probability distribution resulting from the single interaction, and therefore the energy landscape \textcolor{red}{created by/associated with} that interaction. In particular, it constrains the probability distribution of models with a single interaction, as illustrated in Fig.~\ref{fig:prob_distr_q5_n1}.a.
%%%%%%%%%%%%%%%%%%%%%%%% 
Because $H^{\vecmu}(\state)$ increases when the direction of $\vec{\phi}^{\vecmu}(\state)$ is further away from $\vec{g}_{\vecmu}^*$% on the unit circle
~\eqref{eq:energy_projection}, this ordering thus constrains the energy landscape $H^{\vecmu}(\state)$ created by the interaction (e.g., states that are directly on each side of the direction of $\vec{g}_{\vecmu}^*$ have the lowest energy). %highest probability).

In particular, we can observe the impact of this ordering on the shape of the probability distribution associated with a single interaction.
For example in Fig.~\ref{fig:prob_distr_q5_n1}.a, the shapes of $p_1(\alpha_1)$ and $p_2(\alpha_1)$ reflect the respective ordering of the states of $\alpha_1$ by $\phi^1$ and $\phi^2$ around the unit circle. 
%%%%%%%%%%%%%%%%%%%%%%% 
%Once you choose which state(s) has the highest probability, you automatically impose an ordering in the probability of the other states.
For each model, 
choosing the states with highest probability (by aligning $\vec{g}_{\vecmu}$ with the direction of the operator $\vec{\phi}^{\vecmu}(\alpha_1)$ for these states) automatically imposes %creates
an order on the %\textcolor{red}{[values of the]} 
probabilities of the other states, given by the %\textcolor{red}{states'} %their
ordering of the states around the unit circle.
It is not possible for instance
to recreate the orange probability distribution in the bottom left of Fig.~\ref{fig:prob_distr_q5_n1}.a with the model $p_1(\alpha_1)$, no matter the value of the parameters. %by tuning the parameters of the distribution $p_1$. 
Indeed in $p_1(\alpha_1)$, for $\alpha_1=0$ to have the largest probability, $\vec{g}^*_{1}$ must be closely aligned with $\vec{\phi}^1(0)$, which then imposes 
%the probability of $\alpha_1=\pm 2$ to be smaller than the probability of $\alpha_1=\pm 1$,
that $\alpha_1=\pm 2$ has a smaller probability than $\alpha_1=\pm 1$,
due to the state ordering created by $\phi^{1}$.
%the ordering of the states created %imposed 
%by $\phi^{1}$ around the circle.

\begin{figure}[h]
%\isPreprints{\centering}{} % Only used for preprints
    \centering
    \includegraphics[width=\textwidth]{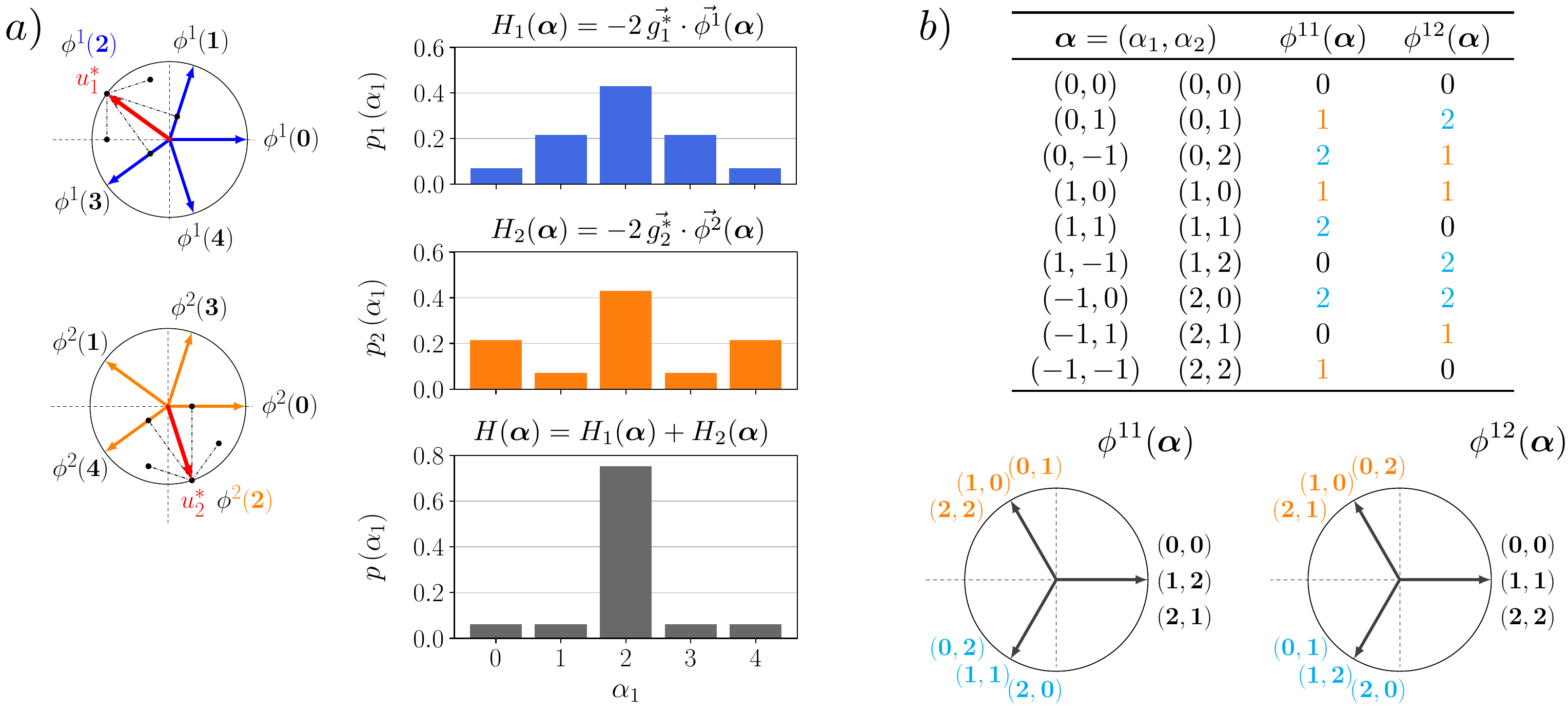} 
\caption{
    {\bf a) Example of a model for $q=5$ with two types of one-body interactions.}
    %One-spin system $\state=(\alpha_1)$ modeled by the same single-spin operators as in Fig.~\ref{fig:prob_distr_q5_n1}: $\phi^1(\state)=z_5^{\alpha_1}$ and $\phi^2(\state)=z_5^{2\alpha_1}$.
    One-spin system $\state=(\alpha_1)$ modeled by the operators $\phi^1(\state)=z_5^{\alpha_1}$ and $\phi^2(\state)=z_5^{2\alpha_1}$ (same %operators
    as in Fig.~\ref{fig:prob_distr_q5_n1}).
    At the top is the probability distribution $p_1(\alpha_1)$ of the model with operator $\phi^1$ (and its c.c.) parameterized by $g_1^* = r_1 u_1^*$ with $u_1^*=z_5^{\,2}$ (see red arrow in the unit circle).
    In the middle is the distribution $p_2(\alpha_1)$ %of a model with only $\phi^2$
    with operator $\phi^2$ (and c.c.) parameterized by $g_2^* = r_2 u_2^*$ with $u_2^*=z_5^{\,4}$. 
    In both cases, the parameter 
    %points in the same direction as
    is aligned with the direction of 
    $\phi^{\vecmu}(\alpha_1=2)$ and has a strength $r_{\vecmu}=0.5$.
    At the bottom is the distribution $p(\alpha_1)$ of the model containing both operators (and their c.c.) with the same parameters $g_1^*$ and $g_2^*$ as above, %. This model gives 
    giving 
    the largest probability to the state $\alpha_1=2$ and equal probability to the other states.
    {\bf b)~Examples of two types of pairwise interactions %in a system 
    for $q=3$.} The table reports the values of $\vecmu\cdot\state$ for two pairwise operators: $\phi^{11}(\spin)=s_1s_2$ for $\vecmu=(1,1)$ and $\phi^{12}(\spin)=s_1s_2^2$ for $\vecmu=(1,2)$. 
    %In the first and second columns, 
    In the two first columns, the values of $\alpha_i$ are reported using two notations: %respectively, 
    $\alpha_i\in\{-1,0,1\}$  and $\alpha_i\in\{0,1,2\}$ (using that $\alpha=2=-1\,{\rm mod}\,3$).
    %in $\mathbb{Z}/3\mathbb{Z}$).
    Because $\phi^{\vecmu}(\state)= z_q^{\,\vecmu\cdot\state}$, where $z_q$ is the first $q$-th root of unity, 
    %We recall that $\phi^{\vecmu}(\state)= z_q^{\,\vecmu\cdot\state}$, where $z_q$ is the first $q$-th root of unity, and thus 
    the value of $\vecmu\cdot\state$ gives the position of the operators $\phi^{\vecmu}(\state)$ around the unit circle, as shown below the table. %Multiple states corresponds to the same position.
    States are uniformly distributed over the three $3$-rd roots of unity.
%\label{tab:PairwiseInteractions}
}
    \label{fig:different_interactions:phi1+2:PM}
\end{figure}

\begin{Property}{\bf Effect of different operators %  interactions
on the energy landscape of a model.}
    Different operators yield different orderings of the states around the unit circle, imposing different constraints on the energy landscape $H^{\vecmu}(\state)$ created by an interaction~$\vecmu$.
    These different orderings are what distinguishes different types of interactions %spin operators 
    acting on the same sets of variables.
    Examples of such interactions are given in Fig.~\ref{fig:prob_distr_q5_n1}.a for one-body interactions and in Fig.~\ref{fig:different_interactions:phi1+2:PM}.b for pairwise interactions.
\end{Property}
\noindent Another way to think about this is from the point of view of the Fourier decomposition. %For instance, observe 
The differences between the distributions $p_1$ and $p_2$ comes from the change in periodicity of the distributions % between $p_1$ and $p_2$
(frequency is doubled from $p_1$ to $p_2$), which reflects the fact that they correspond to different Fourier modes.

%\textcolor{purple}{Different orderings of the states around the unit circle by different interactions thus imposes different constraints on the probability distribution associated with a single interaction, and on the energy landscape created by that interaction.}

We discuss below examples in which certain types of first- or second-order interactions may be more appropriate to model the system.

\begin{Example}{\bf First-order interactions when modeling nominal or ordinal categorical data.} 
For modeling many systems, it is common to include one-body interactions in the model. However, for $q\geq 5$, there are more than one type of first-order interactions, and one may wonder which to include. 
We saw that each of these interactions encodes %implies
an ordering on the probability of the states %of~$\alpha_i$
of the variable~$\alpha_i$ affected by the interaction
(see e.g. $\phi^1$ and $\phi^2$ for $q=5$ in Fig~\ref{fig:prob_distr_q5_n1}.a). 
But in general, variables in the studied systems may or may not have a natural ordering. 

{\bf For ordinal data,} such as answers to questionnaires of the type $\{$``{\it Strongly agree}'', ``{\it Agree}'', ``{\it Neutral}'', ``{\it Disagree}'', ``{\it Strongly disagree}''$\}$, there is a natural order between the values taken by the variables (see example of personality test in Sec.~\ref{Sec:MCM:Big5}). 
In such cases, it is natural to reflect this order, as well as possible opposite valence of the variables, %symmetries, % [of the variables],
in the choice of the integer value~$\alpha_i$ assigned to each category. 
It is then possible that certain single-variable patterns in the data are best modeled by a subset of one-body %single-variable 
interactions instead of all (see for instance the patterns in Fig.~\ref{fig:prob_distr_q5_n1}.a).
%
% To reflect opposite valence between opposite answers, it seems natural to assign opposite values of $\alpha_i$ to opposite answers.
In the example above, a natural choice reflecting the order and opposite valence between the possible answers
%In the example above, a natural choice 
is to assign %take 
values of $\alpha_i$ ranging from $-2$ for ``{\it Strongly disagree}'' to $2$ for ``{\it Strongly agree}'', passing by $\alpha_i=0$ for ``{\it Neutral}''.
In this case, choosing an informed assignment of the discrete states over other assignments can help with the interpretation of the uncovered interactions.
%\textcolor{orange}{Give the example of people being highly polarised or not.}
%In some cases, choosing one mapping over the other may lead to easier ways to interpret the uncovered interactions.}
%\textcolor{red}{One could also assign values from $0$ for ``{\it Strongly disagree}'' to $4$ for ``{\it Strongly agree}''. Shuffling these mappings can change the interactions uncovered in the ``best'' model, but should not change the interpretation of the final result.} %\textcolor{blue}{[but not the interpretation]}
%This can change the [interactions of the] uncovered ``best'' model, but not the interpretation of the result. 
%%%%%%
%but not the interpreted result.
%%%%%%
%(it should be possible to map one version to the other). %(one would have to re-mapped the variables). 
%\textcolor{blue}{doesn't change the results, but can make interpretation easier depending on the situation).}
%\textcolor{red}{can make it simpler to interpret the results depending on the situation).}
%%%%%%%%%%%%%
%In such cases, there may be situations where single-variable patterns in the system are \textcolor{red}{best/more naturally} modeled by a subset of single-variable interactions (see for instance patterns of the types of the distributions in the two first columns of Fig.~\ref{fig:prob_distr_q5_n1}.a).
%%%%%%%%%%%%%

{\bf For nominal data} (e.g., amino-acids in protein sequencing), ordering the categories by mapping them to integer values $\alpha_i\in\Field^n$ has no specific meaning and the mapping choice is often arbitrary.
In this case, the constraints on single-variable patterns imposed by using a subset of single-spin operators (e.g., %either 
$\phi^{1}$ or $\phi^2$ in Fig.~\ref{fig:prob_distr_q5_n1}.a) are a priori arbitrary, and one can expect to include all one-body operators in the model, from which any one-body pattern can be created. 
In Fig.~\ref{fig:different_interactions:phi1+2:PM}.a for example, the two operators $\phi^1$ and $\phi^2$ for $q=5$ are combined to give the largest probability to a specific value of $\alpha_i$ and equal probability to the other states, a pattern that can't be obtained with a single operator.
\end{Example}

%\textcolor{red}{Use the word ``Category'' instead of values taken by the variables?}
%{\bf Explains Figure 4-5}, Categorical/non-ordered variables: as a consequence of this effect. For the type of data where the labels are categorical, i.e. for which the order / ``distance'' between categories doesn’t matter, then you would expect to have to use all operators to remove the ordering (along the circle) between the categories. 

\begin{Example}{\bf Encoding symmetric or antisymmetric pairwise patterns.} \label{Ex:Pairwise_interactions}
For $q=3$ for instance, pairwise interactions can be encoded by two different types of spin operators (and their c.c.): 
$\phi^{(11)}(\spin)=s_1s_2$ (with c.c. $\phi^{(22)}(\spin)=s_1^2s_2^2$) and $\phi^{(12)}(\spin)=s_1^{\,}s_2^2$ (with c.c. $\phi^{(21)}(\spin)=s_1^{\,2}s_2^{\,}$).
Fig.~\ref{fig:different_interactions:phi1+2:PM}.b shows how these two operators place the states of a two-spin system around the unit circle: states are divided in three groups with three states each. Observe that the interaction encoded by the operator $\phi^{12}$ and a parameter with direction $u^*_{12}=0$ (along the $x$-axis) gives the largest probability to symmetric states\footnote{We recall that the values $\alpha_i=2$ and $\alpha_i=-1$ are equal %corresponds to the same value 
in $\mathbb{Z}/3\mathbb{Z}$.} $(\alpha_1,\alpha_2)\in\{(0,0), (1,1), (-1,-1)\}$, while the interaction encoded by the operator $\phi^{11}$ and a parameter direction $u^*_{11}=0$ gives the largest probability to anti-symmetric states of the type $(\alpha_1,\alpha_2)\in\{(0,0), (-1,+1),(+1,-1)\}$ (imagine for instance a system in which $\alpha_i=0$ is a neutral state, while $\alpha_i=\pm 1$ encode opposite states).

To %understand
interpret these observations from a statistical modeling perspective,  %in the context of statistical modeling/
consider a dataset for two discrete variables $\state=(\alpha_1,\,\alpha_2)$ with $q=3$
%in which the symmetric states $(\alpha_1,\alpha_2)\in\{(0,0), (1,1), (2,2)\}$ are observed for most of the datapoints. 
for which a large fraction of the datapoints are in a symmetric state $(\alpha_1,\alpha_2)\in\{(0,0), (1,1), (2,2)\}$.
For all these symmetric states, the interaction $\vecmu=(1,2)$ verifies that $\vecmu\cdot\state=0\,({\rm mod}\,3)$.
%which means that the operator $\phi^{12}$ takes the value $\phi^{12}(\state)=z_3^{\,0}=1$ for a large fraction of the datapoints. This gives a low entropy to %lowers the entropy of 
%the operator $\phi^{12}$, thus increasing its significance in comparison to the other pairwise operator $\phi^{11}$\footnote{\textcolor{blue}{In the context in which the null hypothesis (operator not included in the model) corresponds to a pattern that has an entropy of $1$ trits.}}.
This means that the operator $\phi^{12}$ takes the value $\phi^{12}(\state)=z_5^{\,0}=1$ for a large fraction of the datapoints, which lowers the entropy of $\phi^{12}$ and increases its significance in comparison to the other pairwise operator $\phi^{11}$. 
Alternatively, if the symmetric pattern is particularly unlikely in the data, this would also reduce the entropy of $\phi^{12}$, thus also increasing the significance of this patterns. Patterns can be significant because they are either more predominant or more rare than what can be expected from chance.
Similarly, $\phi^{11}$ would be more relevant than $\phi^{12}$ in a dataset dominated by the anti-symmetric state mentioned above.
Note that the converse is not necessarily true: a low entropy of $\phi^{12}$ doesn't necessarily imply that the symmetric pattern is dominant in the data, because any of the two other groups of states (orange and blue states in %right-most circle of 
Fig.~\ref{fig:different_interactions:phi1+2:PM}.b) %\textcolor{blue}{[, or a combination of them,]} 
could also be dominant. %\textcolor{blue}{significant} %important/present
However, these patterns can be harder to interpret a priori. %\textcolor{blue}{alternatively the symmetric pattern could also be particularly unlikely in the data}. % [(in this case, cyclic permutations of $(\alpha_1, \alpha_2)$, while $\phi^{11}$ encodes an anti-cyclic permutations between $\alpha_1$ and $\alpha_2$)].

More generally for larger values of $q$, the pairwise interaction $\vecmu=(1,\,q-1)$ groups together, in the direction of $u^*_{\vecmu}=0$, the states $\state=(\alpha_1,\,\alpha_2)$ that satisfy $\vecmu\cdot\state = \alpha_1-\alpha_2=0$, which are the symmetric states %of the form %\textcolor{blue}{[verifying]} % $(\alpha_1-\alpha_2)=0$ in $\Field$. 
$\alpha_1=\alpha_2$ in $\Field$. 
Similarly, the pairwise interaction $\vecmu=(1,\,1)$ groups, in the direction of $u^*_{\vecmu}=0$, the states $\state=(\alpha_1,\,\alpha_2)$ that satisfy $\vecmu\cdot\state= \alpha_1+\alpha_2=0$, which are the anti-symmetric states of the form $\alpha_1=-\alpha_2$ in $\Field$. 
These types of interactions are thus a good choice to model significant symmetric or anti-symmetric pairwise patterns. % in data.
%\textcolor{red}{Reciprocally, low entropy of these operators may indicate the presence of symmetric or antisymmetric states (which should then be confirmed in the data).}
In particular, \textbf {the pairwise interactions used in the planar Potts model~\citep{Potts1952} (see Example~\ref{ex:vector_Potts}) are of the symmetric form} with $u^*_{\vecmu}$ aligned with the $x$-axis (as $b_{\vecmu}=0$); this model thus favors or disfavors symmetric pairwise patterns depending on the sign of the parameter $J_{ij}$.

As an example, the most prominent voting patterns in the US Supreme Court dataset analyzed in Sec.~\ref{Sec:Ex:SCOTUS} are symmetric votes between pairs of justices, which explains why the best basis uncovered in that case (after artificially increasing $q$) has mostly interactions of the type $\vecmu=(1,\,q-1)$ (and not $\vecmu=(1,\,1)$ as one could naively expect).

\end{Example}

\subsubsection{Entropy of spin operators, symmetry and invariance}\label{sec1:interaction_interpretation:symmetry}

A final observation is that each interaction~$\vecmu$ divides the space of states into equal-size groups of states that have %with 
the same interaction energy $H^{\vecmu}(\state)$.
%(always be pointing in one of the directions given by the $q$ roots of unity)
%
Indeed, by definition~\eqref{eq:def:monomials:alpha}, an operator $\phi^{\vecmu}(\state)$ only takes values in the $q$ roots of unity, corresponding to $q$ locations around the unit circle. %which corresponds to $q$ evenly distributed locations around the unit circle. 
%For a given interaction~$\vecmu$, 
The states~$\state$ for which the dot product $\vecmu\cdot\state$ takes the same value correspond to the same location %[on the unit circle] 
(see example in Fig.~\ref{fig:different_interactions:phi1+2:PM}.b) %for two types of pairwise interactions with $q=3$) 
and thus have the same interaction energy $H^{\vecmu}(\state)$. % (independently of the value of the parameter).
%The number of states at each (accessible) location is the same.
Besides, there are the same number of states at each these %(accessible) 
locations\footnote{
which is also why an operator verifies Property~\ref{ppty:sum:phi} on the sum of spin operators.}%
%\textcolor{red}{(as an operator verifies Property~\ref{ppty:sum:phi})}, 
, and therefore a single spin operator divides the state space into %$q$
equal-sized groups of states of equal interaction energy (see Fig.~\ref{fig:different_interactions:phi1+2:PM}.b).
%For $q=3$, Fig.~\ref{fig:different_interactions:phi1+2:PM}.b lists the values of $\vecmu\cdot\state$ for two types of pairwise interactions.

A consequence is that %As a consequence, %
a model with a single interaction~$\vecmu$ divides the state space into equal-sized groups of states of equal probability, thus encoding certain symmetries in the model probability distribution. 
%As a consequence, a model with a single interaction gives the same probability to all these states, %the states that are in the same location on the circle.
%encoding a certain symmetry in the data.
If $q$ is prime, or if $\vecmu$ verifies for all integer $q'\in[0,q-1]$ that $\,q'\vecmu\neq\veczero$ in $\Field^n$, %$c\in\Field$, 
then there are $q$ such groups (otherwise there are fewer groups%
%if these conditions are not satisfied
\footnote{
%If $q$ is not prime and this condition is not satisfied, then 
In this case, %the operator values of 
the states are equally distributed over a subset of roots, corresponding to the $q'$-th roots of unity, 
%of a smaller $q'=q/c$ where $c$ is the smallest integer verifying $\,c\vecmu\neq\veczero$ in $\Field^n$.} 
%of $q'$, 
where $q'$ is the smallest integer %divisor of $q$
verifying that $\,q'\vecmu=\veczero$ in $\Field^n$ ($q'$ is a divisor of $q$).
}).
Moreover, %for each independent interaction added to the model%
each time an independent interaction\footnote{A set of operators $\{\phi^{\vecmu}\}$ is said independent if the corresponding set of interaction vectors $\{\vecmu\}$ is linearly independent in $\Field^n$ (see Sec.~\ref{Sec3:subsec:ClusterExpansion} on independent models). There at most $n$ independent operators.} 
is added to the model, each group is divided further into $q$ equal-size groups of states of equal probabilities. Here the independence between interactions is defined by the linear independence between the corresponding vectors $\vecmu$ in $\Field^n$ (see Sec.~\ref{Sec3:subsec:ClusterExpansion} on independent models), and %there are at most $n$ independent operators.
at most $n$ independent operators can be added to a model.
%Moreover, each additional independent interaction, where the independence between interactions is defined by the linear independence between the corresponding vectors $\vecmu$ in $\Field^n$, further divides each group into $q$ groups of states of equal probabilities \textcolor{red}{(see Sec.~\ref{} about independence between operators)}.
%
For example, for $q=2$, a model with a single interaction~$\vecmu$ %system with a single spin operator
maps each state $\state$ to the location $x=1$ %$(1,0)$ 
(for $\vecmu\cdot\state=0$)
or  $x=-1$ %$(-1,0)$ 
(for $\vecmu\cdot\state=1$)
on the unit circle depending on the parity %of the sum 
of the variables~$\state_{\mid\vecmu}$ involved in the interaction, %(i.e., based on the parity of $\vecmu\cdot\state$), 
thus dividing the states into two groups of equal probability. Adding another independent interaction $\vecmu'$ will further divide each group in two depending on the parity of the variables~$\state_{\mid\vecmu'}$ involved in $\vecmu'$.
%this corresponds to placing the states $\alpha$ at the location $(1,0)$ or $(-1,0)$ on the unit circle depending on the parity of the variables involved in the interactions (i.e. on the parity of $\vecmu\cdot\state$). For larger $q$, it is less trivial. 
%For $q=3$, Fig.~\ref{fig:different_interactions:phi1+2:PM}.b lists the values of $\vecmu\cdot\state$ for two types of pairwise interactions.

The equal distribution of the states around the unit circle by an operator $\phi^{\vecmu}$ implies a permutation in the energy $H^{\vecmu}$ of the states as one rotates the parameter $\vec{g}_{\vecmu}^*$ by $2\pi/q$ at fixed strength $r_{\vecmu}$ (as the amplitudes of the projections of $g_{\vecmu}^*$ on the $\phi^{\vecmu}(\state)$'s remain unchanged, just shuffled). In particular, for a model with a single interaction~$\vecmu$, this means that the amplitudes of the probability distribution over the states are simply shuffled when the parameter $\vec{g}_{\vecmu}^*$ is rotated by $2\pi/q$ at fixed strength.
For example, rotating the vector $\vec{g}_{\vecmu}^*$, % (at fixed strength $r_{\vecmu}$), 
from being aligned with one of the $q$-th roots of unity (i.e., aligned with the operator value of a group of states) to being aligned with another root (i.e., with the operator value of another group of states) results in shuffling the amplitude of the probability distribution over the states,
as shown in the first and last columns of Fig.~\ref{fig:prob_distr_q3_n1}.a and~\ref{fig:prob_distr_q5_n1}.a.
%(see for example the distributions in the first and last columns of Fig.~\ref{fig:prob_distr_q3_n1}.a and~\ref{fig:prob_distr_q5_n1}.a).
%
This effect %is still present when $g_{\vecmu}^*$ is not aligned with a root of unity 
is reflected in the invariance of the entropy of a model with a single interaction~$\vecmu$ when the parameter $\vec{g}_{\vecmu}^*$ is rotated by $2\pi/q$, thus giving a periodic structure to the entropy of the model as $\vec{g}_{\vecmu}^*$ is continuously rotated at fixed strength 
(see Fig.~\ref{fig:prob_distr_q3_n1}.b and~\ref{fig:prob_distr_q5_n1}.b).
%(see Fig.~\ref{fig:prob_distr_q3_n1}.b,~\ref{fig:prob_distr_q5_n1}.b, and Fig.~\ref{app:fig:Entropy} for larger values of $q$).
%
At fixed value of the strength $r_{\vecmu}$, the entropy is maximal when $\vec{g}_{\vecmu}^*$ points exactly between two consecutive directions of $\vec{\phi}^{\vecmu}$, and minimal when its direction coincides with the direction of one of the vectors $\vec{\phi}^{\vecmu}$. 

More generally, the entropy landscape of an operator $\phi^{\vecmu}(\state)$ under a spin model with the single interaction~$\vecmu$ as one varies the parameter $g_{\vecmu}$ is the same for all operators at a given value of $q$ (see Fig.~\ref{app:fig:Entropy} for entropy plot at different values of $q$). This is valid under the condition that $q$ is prime, or that $\vecmu$ verifies for all integer $q'\in[0,q-1]$ that $\,q'\vecmu\neq\veczero$ in $\Field^n$. Otherwise, the entropy plot will be that of the smallest value $q'$ such that $\,q'\vecmu=\veczero$.
Indeed, for a $q$-state spin model with a single interaction~$\vecmu$ satisfying the condition above, the energy of the states can only take $q$ different values,
$H^{\vecmu}(\state)=-2\,\vec{g_{\vecmu}^*}\cdot\vec{z}_q^{\,j}$ for $j\in\Field$, where $\vec{z}_q^{\,j}$ %is the direction of the root $z_q^j$ in the complex plane. 
is the vector in the complex plane corresponding to the root $z_q^j$.
The entropy of the model is then (see Appendix~\ref{app:entropy:single-mu}):
\begin{align}
    S[P_{\mathcal{M}}] = (n-1)\log q + S[\phi^{\vecmu}]\,, % - \sum_{j=0}^{q-1} p_j\log p_j\,.
\end{align}
which corresponds to the entropy of $(n-1)$ purely random $q$-state variables, plus the entropy of the operator $\phi^{\vecmu}$ under the model. This later is given by:
\begin{align}
    S[\phi^{\vecmu}] = -\sum_{j=0}^{q-1} p_j\log p_j\,,\qquad {\rm where}\;\; p_j=\frac{e^{\,2\,\vec{g^*_{\vecmu}}\cdot \vec{z}_q^{\,j}}}{Z_q}\,\,
    {\rm and}\;\;
    Z_q = \sum_{j=0}^{q-1} e^{\,2\,\vec{g^*_{\vecmu}}\cdot \vec{z}_q^{\,j}} \,.
\end{align}
%where $\vec{z}_q^{\,j}$ is the vector in the complex plane corresponding to the $j$-th roots of unity.
%And the entropy of the model is given by:
%\begin{align}
%    S[P_{\mathcal{M}}] = (n-1)\log q - \sum_{j=0}^{q-1} p_j\log p_j\,.
%\end{align}
%
At fixed $q$, the entropy of an operator in a model with a single interaction~$\vecmu$ is the same function of the parameters $g_{\vecmu}^*$ for all $\vecmu$ (if $\vecmu$ %is not divisible) (see App.~\ref{} and Fig.~\ref{}). 
verifies that $\,c\vecmu\neq\veczero$ for all integer $c\in[0,q-1]$).
%\textcolor{red}{if $\vecmu$ is linearly independent.}
We will see in Sec.~\ref{Sec:GT:MatrixRepresentation} that this is because all the $q$-state spin models with a single operator $\vecmu$ (verifying that $\forall c\in\Field$,  $\,c\vecmu\neq\veczero$) % for all integer $c\in[0,q-1]$) 
are equivalent.

\section{Loop and cluster expansion of the partition function} 
%and Equivalence classes of spin models} 
\label{sec:2a_LoopExpansion}

%%%% Possible general introduction: %%%%
%Ref.~\cite{beretta2018stochastic} shows that the partition function~\eqref{eq:Z:Ising:def} of the high-order Ising model can be written as a sum over a set of ``loops'' of the model. Models with the same loop structure then have the same partition function up to permutation of their parameters and thus encode the same statistical structure (because they have the same Fisher Information Matrix). From this result, Ref.~\cite{beretta2018stochastic} introduces the notion of equivalent classes of spin models, by defining gauge transformations. Here we generalize these results to the more general case of discrete data.
%%%%%%%%%%%%%%%%%%%%%%%%%%%%%%%%%%%%%%%%%%%

In the binary case, Ref.~\cite{beretta2018stochastic} showed that multiple spin models can have the same partition function (up to permutation of their parameters) and therefore describe the same statistical structure.
This was obtained using a cluster expansion of the partition function, expressing it as a sum over the loops of the model: models with the same loop structure have the same partition function.
Introducing spin basis transformations, Ref.~\cite{beretta2018stochastic} showed that these models are different representations of the same abstract model seen in different bases. 
These basis transformations, called gauge transformations in~\cite{beretta2018stochastic}, divide the space of models into equivalence classes of models describing the same statistical structure: models of the same class have the same partition function, and thus the same Fisher information matrix and model complexity.
They are characterized by the same number of operators, the same rank, and the same loop structure.
Here, we extend these results to the more general case of $q$-state spin models with $q\geq 2$, starting with the generalization of the loop expansion of the partition function in this section, and followed by the notion of gauge transformations and equivalent models in Sec.~\ref{Sec:GT:MatrixRepresentation}.

\subsection{Loop expansion of the (complex) partition function}
For $q=2$, the partition function~\eqref{eq:Z:Ising:def} of the high-order Ising model can be written as a sum over its loops $\mathcal{L}(\M)$~\cite{beretta2018stochastic}:
\begin{align}\label{eq:Z:Ising}
    {\rm for}\;\;q=2,\;\;\;
    &Z_{\M}(\vecg) 
    = 2^n \bigg(\prod_{\vecmu\in\M} \cosh(g_{\vecmu})\bigg)\sum_{\ell\in\mathcal{L}(\M)}  \prod_{\vecmu\in\ell} \tanh(g_{\vecmu})\,,\\
    \label{eq:Z:Ising:def:loop}
    &{\rm where} \;\; \mathcal{L}(\M)
    = \Big\{\ell\subseteq \M\;\Big|\; \sum_{\vecmu\in\ell} \vecmu = \veczero\Big\}\,.
\end{align}
In Eq.~\eqref{eq:Z:Ising:def:loop}, the sum over $\vecmu$ is taken in $(\mathbb{Z}/2\mathbb{Z})^n$ (i.e., with a modulo 2) and $\mathcal{L}(\M)$ contains the empty set $\ell=\emptyset$, for which the product over the elements of $\ell$ in Eq.~\eqref{eq:Z:Ising} is equal to $1$. Note that Ref.~\cite{beretta2018stochastic} defines a loop as a subset of operators $\phi^{\vecmu}(\state)$ of the model whose product is equal to 1 for all states~$\state$ of the system; this is equivalent to the sum of the corresponding binary representation $\vecmu$ being equal to $\veczero$, as written in Eq.~\eqref{eq:Z:Ising:def:loop}. 
Equation~\eqref{eq:Z:Ising} implies that all the models with the same loop structure also have the same partition function (up to permutations of the parameters), and thus they have the same statistical properties (in particular, they have the same Fisher Information Matrix).

Similarly, for $q\geq 2$, %to the binary case, 
we found that the partition function of the high-order discrete spin models in Eq.~\eqref{eq:q-spin_models} can also be written as a sum over the ``loops'' of the model.

\begin{Property}{\bf Loop expansion of the partition function of $q$-state spin models~\eqref{eq:q-spin_models}.}
\label{ppty:Z:loop-expansion}
Consider a $q$-state spin model of the form~\eqref{eq:q-spin_models} defined by $\M\subseteq (\mathbb{Z}/q\mathbb{Z})^n$. We denote by $K=|\M|$ the number of operators in the model and arbitrarily index the operators from $1$ to $K$: $\M = \{\vecmu_1,\,\cdots,\,\vecmu_K\}$.
Using the generalized Euler formula~\eqref{eq:Euler:general} for $e^{\,g_{\vecmu}\phi^{\vecmu}(\state)}$~\cite{muldoon1996beyond}, the partition function~\eqref{eq:Z:def} of $\M$ can be written under the form (see proof in App.~\ref{app:Z:loops:V1}): 
\begin{align}\label{eq:Z:loop-expansion:V2}
    Z_{\M}(\vecg) =q^n\,\sum_{\vecell\in\mathcal{L}_q(\mathcal{M})}
    \prod_{k=1}^K c_{\ell_k}(g_{\vecmu_k})\,,
    \qquad {where}\;\; 
    c_\ell(x) =\sum_{k\geq 0}\frac{x^{kq+\ell}}{(kq+\ell)!} %\,,
\end{align}
are the generalized Euler coefficients~\cite{muldoon1996beyond} (see App.~\ref{app:EulerFormula}), and
\begin{align}\label{eq:loop:V2}
    \mathcal{L}_q(\M) = \bigg\{\vecell\in (\mathbb{Z}/q\mathbb{Z})^K\;\Big|\; \sum_{k=1}^K \ell_k\,\vecmu_k = \veczero\bigg\}%\;,
\end{align}
is the set of loops of the model $\M$, in which $\ell_k$ denotes the $k$-th coefficient of $\vecell$. The set $\mathcal{L}_q(\M)$ always contains the null vector $\vecell=\veczero$, for which the product over $k$ in Eq.~\eqref{eq:Z:loop-expansion:V2} is equal to $\prod_{k=1}^K c_0(g_{\vecmu_k})$.
\end{Property}
\noindent In Equation~\eqref{eq:loop:V2}, we used a more general notation for %the loops of 
$\mathcal{L}_q(\M)$ compared to Eq.~\eqref{eq:Z:Ising:def:loop}: a loop is denoted as a vector $\vecell$ in $(\mathbb{Z}/q\mathbb{Z})^K$ and $\ell_k$ denotes the $k$-th coefficient of $\vecell$. 
For $q=2$, the coefficients~$\ell_k$ only take the values $0$ or $1$ and the loop $\vecell$ is equivalently identified by
%the vector $\vecell$ thus identifies a 
the subset of elements of $\M$ for which $\ell_k=1$, as defined in Eq.~\eqref{eq:Z:Ising:def:loop}.
Moreover, for $q=2$, the Euler coefficients in Eq.~\eqref{eq:Z:loop-expansion:V2} are $c_0(x)=\cosh(x)$ and $c_1(x)=\sinh(x)$, which reduces Eq.~\eqref{eq:Z:loop-expansion:V2} to the expansion in Eq.~\eqref{eq:Z:Ising}. % known in the binary case.
Equation~\eqref{eq:Z:loop-expansion:V2} thus generalizes the loop expansion of the partition function to all $q$-state spin models for $q\geq 2$.

The loops of a spin model, defined in Eq.~\eqref{eq:loop:V2}, are a set of linear dependencies between the operators of the model that result from the choice of the model architecture (i.e., which operators are chosen in the model).
%%%%%%%%%%%%%%%%%%%
%The set of linear dependencies between the operators of a spin model, encoded in the loop structure in Eq.~\eqref{eq:loop:V2}, constrains/captures the statistical properties of the model, as shown by the loop expansion of the partition function in Eq.~\eqref{eq:Z:loop-expansion:V2}.
%%%%%%%%%%%%%%%%%%%
%They capture the statistical constraints of the probabilistic model.
%%%%%%%%%%%%%%%%%%%
%They capture the constraints satisfied by the statistical structure of the probabilistic model. 
%capture the constraints of the statistical structures [that can be] modeled by the model, thus defining the statistical properties of the parametric model.
%imposes constraints to the statistical structure of the parametric probabilistic model, thus defining its statistical properties.
%%%%%%%%%%%%%%%%%%%
These dependencies capture %\textcolor{blue}{constrain/capture/encapsulate} 
the statistical properties of the parametric model, as shown by the loop expansion of the partition function in Eq.~\eqref{eq:Z:loop-expansion:V2}.
The properties of the set of loops $\mathcal{L}_q(\M)$ of a spin model are further discussed in Sec.~\ref{sec:loop-structure}. Figure~\ref{fig:GT_and_Loops} gives an example of the set of loops of a $q$-state spin model for $q=3$. Note that the figure doesn't show trivial loops resulting from the structure of complex conjugate operators (see next section).

%%%%%%%%%%%%%%%%%%%%%%%%%%%%%%%%%%%%%%%%%%%%%%%%%%%%%%%
%\textcolor{blue}{Note: (discussion section) In Eq.~\eqref{eq:expansion:Fourier_basis}, there are no requirements about the fact that $P(\state)$ needs to be real. Some of the discussion in the following part of this paper, also do not require that $P(\state)$ to be real (for instance the loop expansion of the partition in Eq.~\ref{}). This may be interesting to explore in the context of quantum spin systems.}

\subsection{Loop expansion of the real partition function over non-trivial loops (accounting for c.c.~operators)} %(due to c.c.~operators)}

The expansion of the partition function in Eq.~\eqref{eq:Z:loop-expansion:V2} is obtained for any $q$-state spin model, % of the form~\eqref{eq:q-spin_models}, which can be real or complex valued.
real or complex valued. 
In general, operators are not necessarily included by conjugate pairs in the model and the Hamiltonian $H(\state)$ and the partition function in Eq.~\eqref{eq:Z:loop-expansion:V2} can take complex values.
In order to interpret a $q$-state spin model as a probabilistic model, the Hamiltonian of the model must be real. %This \textcolor{red}{[can be used to]} reduces the sets of loops \textcolor{red}{needed} in the loop expansion, as described below.
This introduces ``trivial'' loops that can be removed from the expansion.

For the Hamiltonian to be real, operators should be included by conjugate pairs $(\vecmu,-\vecmu)$ in the model and their parameters should satisfy the constraint $g_{-\vecmu}=g_{\vecmu}^*$ (see Definition~\ref{eq:def:q:spin-model}).
For example, for $q=3$, a model containing the operator~$\phi^{\vecmu}(\spin)=s_1$ must also have its complex conjugate (c.c.) operator\footnote{We recall that the conjugate of an operator $\phi^{\vecmu}(\spin)$ is the operator $\phi^{-\vecmu}(\spin)$, where $-\vecmu\in(\Field)^n$ is the vector with elements of the form $(q-\mu_i)$ (see Sec.~\ref{sec:def:HO:vectorPottsModel}). 
For example, for a one-spin system $\spin=(s_1)$ with $q=3$, the operator $\phi^{\vecmu_1}(\spin) = s_1$ is identified by the (one-dimensional) vector $\vecmu_1=(1)$, and the operator $\phi^{\vecmu_2}(\spin) = s_1^2$ by the vector $\vecmu_2=(2)$. The latter can be re-written as
$\vecmu_2 = (q-1) = -(1)=-\vecmu_1$ over $\mathbb{Z}/3\mathbb{Z}$, which means that the operators $\phi^{\vecmu_1}(\spin)$ and $\phi^{\vecmu_2}(\spin)$ are conjugate operators (see Sec.~\ref{sec:def:HO:vectorPottsModel}).
}%
~$\phi^{-\vecmu}(\spin)=s_1^{\,2}$.
By definition, conjugate operators, $\vecmu_1$ and $\vecmu_2$, verify that $\vecmu_1+\vecmu_2=\veczero$ in $\Field^n$, and as a consequence they always form a loop in Eq.~\eqref{eq:loop:V2}.
However, this is a ``trivial'' loop, in the sense that conjugate operators are included together by construction for the Hamiltonian of the model to be real.
%%%%%%%%%%%%%%%
%[and thus don't encode any additional non-trivial statistical dependencies in the model.] 
This loop thus encodes a ``trivial'' statistical dependency of the model.

A model can have many such trivial loops %due to c.c.~operators 
contributing to its partition function~\eqref{eq:Z:loop-expansion:V2}. A model with a single pair of conjugate operators has already multiple such contributions. For example, for $q=3$, a model with a single pair of conjugate operators $\M=\{\vecmu_1, \vecmu_2\}$ with $\vecmu_2=-\vecmu_1$ (e.g., $\Mtilde=\{s_1, s_1^2\}$) has two trivial loops %in Eq.~\eqref{eq:loop:V2},
resulting from the conjugate relation: $\vecmu_1 + \vecmu_2=0$ and $2\vecmu_1 + 2\vecmu_2=0$. Although these equations encode the same dependence relation, they correspond to two different loops %vectors
$\vecell$ in Eq.~\eqref{eq:loop:V2} and therefore to two different terms in the expansion of the partition function. 
The set of loops of $\M$ is then $\mathcal{L}_q(\M)=\{(0,0), (1,1), (2,2)\}$, and the partition function~\eqref{eq:Z:loop-expansion:V2} then takes the form:
\begin{align}\label{Ex:Zind:q=3:n=1}
    Z_{\M}(\vecg)
    &= 3^n\,
    \Big[
    c_0(g_{\vecmu_1})\,c_0(g_{\vecmu_1}^*)
    +    c_1(g_{\vecmu_1})\,c_1(g_{\vecmu_1}^*)
    +    c_2(g_{\vecmu_1})\,c_2(g_{\vecmu_1}^*)
    \Big]\,,
\end{align}
where the explicit expressions of the functions $c_0$, $c_1$ and $c_2$ for $q=3$ are given in Eq.~\eqref{eq:Euler:coeff:q=3}, and where we used that $g_{\vecmu_2}=g_{\vecmu_1}^*$ because $\vecmu_2=-\vecmu_1$.

The loops~\eqref{eq:loop:V2} of a spin model capture the statistical constraints of the probabilistic model imposed by the choice of the model architecture (i.e., of the set of interactions).
Because the constraints imposed by complex-conjugate operators are there by construction, we can compute explicitly their contribution to the partition function and remove them from the set of loops. The latter will thus better capture non-trivial constraints of the model. In the following, we reformulate the partition function with this approach, separating trivial c.c.~dependencies of the model from the non-trivial ones. 
%\textcolor{red}{[We remove trivial loops due to c.c.~operators from the loop structure by computing explicitly their contribution to the partition function.]}

\begin{Definition}\label{def:M:M0-M1}
To group complex conjugate terms together, we organize the operators of $\M$ into three disjoint sets, $\M_0$, $\M_1$ and $\M^*_1$, such that $\M=\M_0\cup\M_1\cup\M_1^*$ and
\begin{align}
    \M_0 &= \{\vecmu\in\M \mid \vecmu=-\vecmu\}\,,\label{eq:M0:def}\\
    \M_1 \cup \M^*_1 &= \{\vecmu\in\M \mid \vecmu\neq-\vecmu\}\,, 
    \qquad{\rm with }\;\;
    \forall \vecmu\in \M_1,\;-\vecmu\in\M^*_1\;.
    \label{eq:M1:def}
\end{align}
As all three sets are disjoint, if $\vecmu\in\M_1$, then $-\vecmu\notin\M_1$. 
For all odd values of $q$, $\M_0=\emptyset$.
The sets $\M_1$ and $\M^*_1$ are not uniquely defined, but any choice that respects Eq.~\eqref{eq:M1:def} is fine. For example, one way to construct $\M_1$ and $\M_1^*$ is, for each element $\vecmu\in \M_1\cup\M_1^*$, place $\vecmu$ in $\M_1$ if $\vecmu < -\vecmu$\footnote{As for Definition~\ref{def:spinmodel:cos-sin}, the ordering %between $\vecmu$ and $-\vecmu$
is based on the order of the integer representations of $\vecmu$ and $-\vecmu$ in base $q$, i.e., $\vecmu<-\vecmu$, if its first entry (most significant qits) $\mu_i$ such that $\mu_i\neq -\mu_i$ is in the interval $[0,\frac{q}{2}[$.},
and place it in $\M_1^*$ otherwise. 
%if its first entree $\vecmu_1$ $\mu_1$ is in the interval $[0,\frac{q}{2}[$ and place it in $\M_1^*$ otherwise.
Another way is to successively remove conjugate pairs from the set $\{\vecmu\in\M \mid \vecmu\neq-\vecmu\}$ and to place one element of the pair in $\M_1$ and the other in $\M_1^*$.
\end{Definition}

This organization gives the following expression for the partition function:
\begin{align}\label{eq:Z:def:2}
    Z_{\M}(\vecg) 
    = \sum_{\state} 
    \left(\prod_{\vecmu\in\M_0} e^{\;g_{\vecmu}\phi^{\vecmu}(\state)}\right)
    \left(\prod_{\vecmu\in\M_1} e^{\;g_{\vecmu}\phi^{\vecmu}(\state)+g_{\vecmu}^*\phi^{-\vecmu}(\state)}\right)\,,
\end{align}
in which we grouped together c.c.~contributions $(\vecmu, -\vecmu)$ for the operators in $\M_1$ (operators $\vecmu\in\M_0$ are their own c.c.~by definition~\eqref{eq:M0:def}).
Using the generalized Euler formula~\eqref{eq:Euler:general}, the contribution coming from complex conjugate terms can be re-written as a single sum over powers of $\phi^{\vecmu}$ 
(see App.~\ref{app:Euler:M1}):
\begin{align}
    \label{eq:Euler:general:M1}
    &\forall \vecmu\in\M_1, \qquad
    e^{\;g_{\vecmu}\,\phi^{\vecmu}(\state)+g_{\vecmu}^*\,\phi^{-\vecmu}(\state)}
    =\sum_{r=0}^{q-1} B_{r}(g_{\vecmu}) \,\left[\phi^{\vecmu}(\state)\right]^r\,,\\
    \label{eq:Euler:general:M1:B}
    &{\rm where}\;\;\;
        \forall r\in\{0,\,\cdots,\,q-1\}\,,\;\;\;
        B_{r}(x) = \sum_{m=0}^{q-1} c_m(x)\,c_{[(m-r)\bmod{q}]}(x^*)\,,
\end{align}
where the coefficients $c_m(x)$ are given in Eq.~\eqref{eq:Z:loop-expansion:V2}.
Besides, by definition~\eqref{eq:M0:def}, all the elements $\vecmu\in\M_0$ verify that $2 \vecmu=\veczero$ in $\Field^n$ and, as a consequence, that $\left[\phi^{\vecmu}(\state)\right]^2=1$ for all states $\state\in\Field^n$. 
This means that for $\vecmu\in\M_0$, one can use the original Euler formula for $\exp(g_{\vecmu}\phi^{\vecmu}(\state))$ (see App.~\ref{app:Euler:M0}): 
\begin{align}
    \label{eq:Euler:general:M0:maintext}
    &\forall \vecmu\in\M_0, \qquad
        e^{g_{\vecmu}\phi^{\vecmu}(\state)}
        = A_{0}(g_{\vecmu}) + A_{1}(g_{\vecmu})\; \phi^{\vecmu}(\state)\,,\\
    \label{eq:Euler:general:M0:A:maintext}
    &{\rm where}\;\;\; 
        A_{0}(x) = \sum_{k=0}^{\frac{q}{2}-1} c_{2k} (x) = \cosh(x)
        \;\;\; {\rm and}\;\;\; 
        A_{1}(x) = \sum_{k=0}^{\frac{q}{2}-1} c_{2k+1} (x) = \sinh(x)\,.
\end{align}
As a sanity check, if $q=2$, then $\M_1=\M_1^*=\emptyset$, and Eq.~\eqref{eq:Z:def:2} combined with Eq.~\eqref{eq:Euler:general:M0:maintext} returns the original definition of the partition function of the high-order Ising model~\eqref{eq:Z:Ising:def}.
Finally, substituting these formulas in Eq.~\eqref{eq:Z:def:2} and summing over the states~$\state$, we obtain the following loop expansion of the partition function (see proof in App.~\ref{app:ClusterExpansion}), where the set of loops does not contain any trivial contributions from c.c.~operators. 
%This loop structure thus better captures non-trivial dependencies imposed by the model architecture.

\begin{Property}\label{ppty:LoopExpansion:R-Model}
{\bf (for real-valued $q$-state spin models) Loop expansion of the partition function without trivial loops from c.c.~operators.} Consider a spin model $\M=\M_0\cup\M_1\cup\M_1^*$.
Let us denote $K_0 = |\M_0|$ and $K_1 = |\M_1|$ the number of operators respectively in $\M_0$ and in~$\M_1$, and arbitrarily index the operators $\M_0 = \{\vecmu_{0, 1}, \cdots,\,\vecmu_{0, K_0}\}$ and $\M_1 = \{\vecmu_{1, 1}, \cdots,\,\vecmu_{1, K_1}\}$. 
The partition function can be written under the form (see App.~\ref{app:ClusterExpansion}):
\begin{align}\label{eq:Z:loop-expansion:V3}
Z_{\M}(\vecg) 
    =q^n\,
    \sum_{(\vecell_0, \vecell_1)\in\mathcal{L}^{cc}_q(\mathcal{M}_0,\mathcal{M}_1)}
    \prod_{i=1}^{K_0} A_{\ell_{0,i}}(g_{\vecmu_{0,i}}) 
    \prod_{j=1}^{K_1} B_{\ell_{1,j}}(g_{\vecmu_{1,j}})\,,
\end{align}
where the coefficients $A_{0}(x)$, $A_{1}(x)$, and $B_{r}(x)$ are given by Eq.~\eqref{eq:Euler:general:M0:A:maintext} and Eq.~\eqref{eq:Euler:general:M1:B}, and where we defined the reduced set of loops of a model as:
\begin{align}\label{eq:loop:V3}
    \mathcal{L}^{cc}_{q}(\M_0,\M_1) = \bigg\{(\vecell_0,\vecell_1)\in (\mathbb{Z}/2\mathbb{Z})^{K_0}\times(\mathbb{Z}/q\mathbb{Z})^{K_1}\;\Big|\; \sum_{i=1}^{K_0} \ell_{0,i}\,\vecmu_{0,i} + \sum_{j=1}^{K_1} \ell_{1,j}\,\vecmu_{1,j} = \veczero\bigg\}\;.
\end{align}
A loop is now defined by two vectors, $\vecell_0\in(\mathbb{Z}/2\mathbb{Z})^{K_0}$ weighting the elements of $\M_0$ and $\vecell_1\in(\mathbb{Z}/q\mathbb{Z})^{K_1}$ weighting the elements of $\M_1$. The coefficients $\ell_{0,i}$ and $\ell_{1,i}$ denote respectively the $i$-th elements of $\vecell_{0}$ and $\vecell_{1}$.
Note that the entries of $\vecell_0$ are taken in $\{0,1\}$, but they are thought of as elements of $\Field$.
The sum over elements of $\M_0$ and $\M_1$ is thus well defined as a sum in $\Field^n$:
%This allows to define the sum on the right-hand-side: 
the sum over the elements $\vecmu_{0,i}$ is implicitly taken modulo $2$ because $2\vecmu_{0,i}=\veczero$, while the sum over $\vecmu_{1,i}$ is modulo $q$. The resulting two vectors should sum to zero in $\Field^n$.
\end{Property}
\noindent We added the superscript $cc$ to $\mathcal{L}_q^{cc}$ to indicate that this set of loops is obtained after removing the contributions from c.c. operators. 
One can easily check that there are indeed no such contributions left in $\mathcal{L}_q^{cc}$,  as none of the operators $\vecmu_1\in\M_1$ and $\vecmu_0\in\M_0$ are c.c.~to each other. 
The reduced loop structure~$\mathcal{L}_q^{cc}$ thus better captures non-trivial dependencies imposed by the model architecture.
Figure~\ref{fig:GT_and_Loops} gives an example of the set of loops $\mathcal{L}_q^{cc}$ of a $q$-state spin model for $q=3$ after removing trivial loops from c.c.~operators.
Note that besides removing c.c.~contributions, the approach also removes contributions from combinations of these trivial loops with other loops. % For instance [move text from sec.3 to here].}

\subsection{Independent models and cluster expansion of the partition function}\label{Sec3:subsec:ClusterExpansion}
In the binary case, Ref.~\cite{beretta2018stochastic, demulatier2024MCM} define {\em independent models}, as models whose set of operators $\M_{ind}$ is an independent set, or equivalently whose set of loops only contains the empty loop, $\mathcal{L}(\M_{ind}) = \{\emptyset\}$. % (which is rewritten as $\mathcal{L}_2(\M_{ind}) = \{\veczero\}$ in the more general form of Eq.~\eqref{eq:loop:V2}).
%The cluster expansion of the partition function usually consists of expanding %the partition function 
%around the partition function of a %the 
%non-interaction system (which is, in this case, %for spin models
%an independent model). 
%which, here, would 
%\textcolor{red}{usually} %normally
%consists of expanding the partition function around the case of the non-interaction system, which 
%around its non-interacting part, i.e. around the partition function of the 
The cluster expansion of the partition function usually consists of expanding the partition function around the approximation of a ``non-interacting'' system, 
which, in the context of spin models, corresponds to the partition function of the model computed as if it was an independent model. %\textcolor{red}{[(defined above)]}.
%i.e., \textcolor{red}{in this case,} around the partition function of the model computed as if it was an independent model.
For binary systems ($q=2$), this is the case of the expressions of the partition function in Eq.~\eqref{eq:Z:Ising} and~\eqref{eq:Z:loop-expansion:V2}, % and~\eqref{eq:Z:Mind:q2},
%for which having no loop between operators return:  
%\textcolor{red}{This means $\mathcal{L}(\M_0) = \{\emptyset\}$ in the form of Eq.~\eqref{eq:loop:V1}, or $\mathcal{L}(\M_0) = \{\veczero\}$ in the form of Eq.~\eqref{eq:loop:V2}, and one recovers the known expression for the partition function of an independent model with $q=2$:}
for which having no loop between operators %directly 
yields the known expression of the partition function of an independent model $\M_{ind}$~\cite{beretta2018stochastic}:
\begin{align}\label{eq:Z:Mind:q2}
    Z_{\M_{ind}}(\vecg) 
    =2^n\,
    \prod_{\vecmu\in\M_{ind}} \cosh(g_{\vecmu})
    \,,\qquad {\rm for}\;\; q=2
    \,.
\end{align}
The cluster expansion can then be made explicit by rewriting Eq.~\eqref{eq:Z:Ising} under the form: 
\begin{align}\label{eq:Z:Ising:clusterExp}
    {\rm for}\;\;q=2,\qquad
    &Z_{\M}(\vecg) 
    = Z^{ind}_{\M}(\vecg) \left[1+\sum_{\ell\in\mathcal{L}(\M)\backslash \emptyset}  \;\prod_{\vecmu\in\ell} \tanh(g_{\vecmu})\right]\,,
\end{align}
where $Z^{ind}_{\M}(\vecg)$ denotes the partition function of $\M$ computed as if $\M$ were an independent model, given by Eq.~\eqref{eq:Z:Mind:q2}, and where the sum over the loops now excludes the empty loop.
%
%Note that the fact that the operators are independent, doesn't mean that they the 
Note that the fact that a model is independent doesn't mean that the spins don't interact. An independent model can have interactions of any order between the spins. % (not necessarily just one-body interactions).
Instead, {\it independence} means that the interactions don't constrain each other (no loops), so in some sense there is no ``interaction'' between the operators. %, as there are no loops.} %, that they are not related to each other.}

Similarly, one can factorize the term corresponding to the empty loop in the expansion of the partition function for $q>2$, both in  Eq.~\eqref{eq:Z:loop-expansion:V2} and Eq.~\eqref{eq:Z:loop-expansion:V3}. However, this term doesn't necessarily correspond to the partition function one would get for an independent model. %if the model were an independent model.
In the following, we give the definition of independent models for complex- and real-valued $q$-state spin models, 
which we then use to write the partition function in the form of a cluster expansion.
%and write the partition function under a form in which the empty loop represents the contribution of the model as if it were independent.
%\textcolor{red}{In the following, we aim to write the partition function under a form, where the empty loop represents the contribution of the model as if it were independent.}

\subsubsection{General expression}
%\subsubsection{Independent models (strong independence)}
%\label{sec:IM:strong-indep}
The definition of independent models~\cite{beretta2018stochastic, demulatier2024MCM} can be easily extended to $q\geq 2$. %, \textcolor{red}{(with some subtleties when $q$ is not prime, see [...])}.
To do so, we first introduce the notion of independence between spin operators, from which follows a natural definition of independent models.

\begin{Definition}\label{def:independent:operators}
{\bf Independent set of operators.} Consider a system of $n$ $q$-state spin variables. 
A set of operators $\{\phi^{\vecmu_1},\,\dots,\phi^{\vecmu_K}\}$ is said independent if and only if the set of vectors $\mathcal{S}=\{\vecmu_1,\,\dots,\vecmu_K\}$ is linearly independent in $\Field^n$:
\begin{align}\label{eq:condition:independence}
    \forall (r_k)%_{k=\{1..K\}}
    \in\Field^K, \;\;\;\;\sum_{k=1}^K r_k\,\vecmu_k=\veczero\;({\rm mod}\;q)
    \qquad\Rightarrow \qquad
    r_k=0 \;\;\;\;\forall\, k\,.
\end{align}
This is equivalent to the set of loops of $\mathcal{S}$ containing only the null vector: $\mathcal{L}_q(\mathcal{S})=\{\veczero\}$.
\end{Definition}

\begin{Definition}\label{def:independent:model:real}
{\bf Independent model.} %\textcolor{red}{[for complex-valued $q$-state model]}.} 
A $q$-state spin model is said {\it independent} if the set $\M$ of operators of the model is an independent set, % of operators.
or equivalently, if the set of loops of the model only contains the null vector (empty loop) $\mathcal{L}_q(\M)=\{\veczero\}$.
\end{Definition}

\begin{Property}{\bf Partition function of an independent model:} 
Consider an independent model~$\M_{ind}$ for any $q\geq 2$. Taking the term corresponding to the empty loop in Eq.~\eqref{eq:Z:loop-expansion:V2}, the partition function of $\M_{ind}$ takes the form:
\begin{align}\label{eq:Z:Mind:q}
{\rm for}\;q\geq 2\,,
%\;\;\;\;\;
\qquad
Z_{\M_{ind}}(\vecg) 
    =q^n\,
    \prod_{\vecmu\in\M_{ind}} c_0(g_{\vecmu})\,,
    %\;\;\;\;\;
    \qquad {where}\;\; 
    c_{0}(x) =\sum_{k\geq 0}\frac{x^{kq}}{(kq)!}\,.
\end{align}
%where $c_0(x)=\sum_{k\geq 0} x^{kq}/(kq)!\;$. 
This equation generalizes Eq.~\eqref{eq:Z:Mind:q2} to larger values of $q$, as $c_0(x)=\cosh(x)$ for $q=2$.
\end{Property}
\noindent The partition function~\eqref{eq:Z:loop-expansion:V2} can then be rewritten under the following form.
\begin{Property}
{\bf Cluster expansion of the partition of $q$-state spin models:}
Consider a $q$-state spin model $\M$ with $K$ operators arbitrarily indexed as $\M = \{\vecmu_{1}, \cdots,\,\vecmu_{K}\}$. For all $q\geq2$, the partition function of $\M$ can be written under the cluster expansion form:
\begin{align}\label{eq:Z:cluster-expansion:complex}
Z_{\M}(\vecg) 
    =Z^{ind}_{\M}(\vecg) \,\left[\,1 +
    \sum_{\vecell\in\mathcal{L}_q(\mathcal{M})\backslash \veczero}\;
    \prod_{k=1}^{K} \frac{c_{\ell_{k}}}{c_0}(g_{\vecmu_{k}})\,\right]\,,
\end{align}
where the coefficients $c_{\ell}(x)$ are given in Eq.~\eqref{eq:Z:loop-expansion:V2}, and where $Z^{ind}_{\M}(\vecg)$ is the partition function of $\M$ computed as if $\M$ was an independent model, which is given by Eq.~\eqref{eq:Z:Mind:q}.\\
For $q=2$, this equation reduces to the known Eq.~\eqref{eq:Z:Ising:clusterExp}, using that $c_0(x)=\cosh(x)$ and $c_1(x)=\sinh(x)$. 
\end{Property}

\subsubsection{Real-valued spin models}
For real-valued spin models, we recall that operators must be included by conjugate pairs in the model for the Hamiltonian to be real. It is clear that two conjugate operators $\vecmu_1$ and $\vecmu_2$ are not independent from each other, as they satisfy $\vecmu_1=-\vecmu_2$. %As seen previously, we can obtain a more refined expression of the loop expansion of the partition function 
However, we previously obtained a more explicit form of %\textcolor{red}{more refined/closer expression for/of} %form 
the loop expansion of the partition function~\eqref{eq:Z:loop-expansion:V3} by %\textcolor{red}{[explicitly]} 
computing analytically the contributions of complex conjugate operators (thus removing them from the loop structure).
To obtain a similar form for %use the same approach for 
the cluster expansion, we extend the definition of independent models to real-valued $q$-state spin models, %by basing the %notion
%\textcolor{red}{criterion/requirement of independence} only on one element for each conjugate pair.
by requiring independence only for one element for each conjugate pair.
%%%%%%
%For real-valued spin models, we can obtain a more refined expression of the partition function by computing explicitly the contributions from complex conjugate operators (as done previously for the loop expansion). We recall that for the Hamiltonian to be real, operators must be included by conjugate pairs in the model. It is clear that two conjugate operators $\vecmu_1$ and $\vecmu_2$ are not independent from each other, as they satisfy $\vecmu_1=-\vecmu_2$. However, we can extend the notion of independent models to real-valued $q$-state spin models, by basing the definition only on one element for each conjugate pair.
%%%%%%%%%%%%%%%%%%%%%%%%%%%

\begin{Definition}%[Independent model]
    \label{def:strongindependent:model}
    {\bf Real-valued independent model.} % (for real-valued spin models).} 
%    {\bf Independent model (strong independence).} 
    Using the decomposition of Definition~\ref{def:M:M0-M1}, a real-valued spin model $\M=\M_0\cup\M_1\cup\M_1^*$ is said {\it independent} if $\M_0\cup\M_1$ is an independent set of operators.
\end{Definition}
\noindent Although the decomposition of $\M=\M_0\cup\M_1\cup\M_1^*\,$ is not unique, this definition doesn't depend on the precise assignment of the conjugate operators to $\M_1$ and $\M_1^*$, because the elements of $\M_1$ and $\M_1^*$ are related by just a sign change.
The absence of dependencies among the operators in $\M_0\cup\M_1$ of an independent model means that the model doesn't contain any (non-empty) loop, except for %the empty loop and 
the trivial loops due to c.c.~operators in $\M_1\cup\M_1^*$. Reciprocally, a real-valued model that doesn't contain any loop, except due to c.c.~operators, % except for these exceptions,
verifies the definition of independence.

\begin{Corollary}
    \textbf{A real-valued independent model} %(for real-valued spin models)
    is a model whose set of loops in Eq.~\eqref{eq:loop:V2} only contains: 1) the empty loop, and 2) trivial loops due to c.c.~operators. 
    Equivalently, it is a model whose reduced set of loops in Eq.~\eqref{eq:loop:V3} solely contains the empty loop, $\mathcal{L}^{cc}_q=\{(\veczero,\,\veczero)\}$.
\end{Corollary}
\noindent The partition function expansion in Eq.~\eqref{eq:Z:loop-expansion:V3} thus already separates the contributions obtained as if the real-valued model were independent, for which $(\vecell_0,\vecell_1)=(\veczero,\,\veczero)$, 
from the contributions resulting from non-trivial loop constraints between the operators, for which $(\vecell_0,\vecell_1)\neq(\veczero,\,\veczero)$.
%%%%%%%%%%
%The partition function of a (strongly) independent model is thus given by \eqref{eq:Z:loop-expansion:V3} in which the set of loops only contains the empty loop $\mathcal{L}^{cc}_q=\{(\veczero,\,\veczero)\}$.
%%%%%%%%%%

Note that we kept $\M_0$ in definition~\ref{def:strongindependent:model}, because it allows the definition to be also valid for spin models with $q=2$ (which are already real-valued without having to introduce the structure of c.c.~operators). Indeed in the binary case, $\M_1=\emptyset$ and the definition and corollary above are consistent with the previous definition of independent models~\cite{beretta2018stochastic, demulatier2024MCM}. 
%In this case, $\M_1=\emptyset$, and one recovers the partition function of independent models~\eqref{eq:Z:Mind:q2} and the cluster expansion~\eqref{eq:Z:Ising:clusterExp}.
%
For $q>2$, %consider a model $\M=\M_0\cup\M_1\cup\M_1^*\,$. 
any operator $\vecmu\in\M_0$ verifies that $2\vecmu=\veczero$ (by definition of $\M_0$) and therefore doesn't satisfy the condition of independence in Eq.~\eqref{eq:condition:independence}.
This means that, for $q>2$, independent models have $\M_0=\emptyset$. The term corresponding to the empty loop ($\vecell_1=\veczero$) in the partition function~\eqref{eq:Z:loop-expansion:V3} then gives the partition function of a real-valued independent model for $q>2$. 
\begin{Property}{\bf Partition function of a real-valued independent model for $q>2$:} 
Consider a real-valued independent model $\M_{Rind}$ for $q>2$. 
%Using the decomposition of Definition~\ref{def:M:M0-M1}, the model can be decomposed under the form $\M=\M_1\cup\M_1^*$
Using the decomposition of Definition~\ref{def:M:M0-M1}, $\M_{Rind}=\M_1\cup\M_1^*$ and the partition function of $\M_{Rind}$ takes the form
(term $\vecell_1=\veczero$ in Eq.~\eqref{eq:Z:loop-expansion:V3}):
% (given by the empty-loop term in Eq.~\eqref{eq:Z:loop-expansion:V3}):
\begin{align}\label{eq:Z:loop-expansion:V3:ind:q-prime}
{\rm for}\;q>2\,, %\;q\;{\rm prime,}
%\;\;\;\;\;
\qquad
Z_{\M_{Rind}}(\vecg)
    &=q^n\,
    \prod_{\vecmu_1\in\M_1} \,
    \left[\, \sum_{m=0}^{q-1} \left\lVert c_{m}(g_{\vecmu_{1}}) \right\rVert^2 
    \,\right]\,,
    \;\;\;\;\;
    %\;\;\;\;{\rm for}\;q>2\,, %\;q\;{\rm prime,}
\end{align}
where the $c_m(x)$ are complex coefficients given in Eq.~\eqref{eq:Z:loop-expansion:V2} and $\lVert c_{m}(x) \rVert = \sqrt{c_{m}(x)\,c_{m}(x^*)}$ denotes the norm of $c_{m}(x)$ (because $c_{m}(x^*)=[c_{m}(x)]^*$).
\end{Property}
\noindent To obtain Eq.~\eqref{eq:Z:loop-expansion:V3:ind:q-prime} from Eq.~\eqref{eq:Z:loop-expansion:V3}, we used that $B_0(x)=\sum_{m=0}^{q-1} c_m(x)c_m(x^*)$ from Eq.~\eqref{eq:Euler:general:M1:B}.
For example, for $q=3$,
\begin{align}\nonumber
    %{\rm for}\;q=3\,,\;\;\;
Z_{\M_{Rind}}(\vecg) 
    =3^n\,
    \prod_{\vecmu_1\in\M_1} \Big[
    c_0(g_{\vecmu_1})\,c_0(g_{\vecmu_1}^*)
    +    c_1(g_{\vecmu_1})\,c_1(g_{\vecmu_1}^*)
    +    c_2(g_{\vecmu_1})\,c_2(g_{\vecmu_1}^*)
    \Big]\,,
\end{align}
which gives back Eq.~\eqref{Ex:Zind:q=3:n=1}, previously obtained for a model with a single conjugate pair.

For $q>2$, for models for which $\M_0=\emptyset$ (this is always the case if $q$ is prime), the partition function~\eqref{eq:Z:loop-expansion:V3} can then be written under the following cluster expansion form.
\begin{Property}
%\textbf{Cluster expansion of the partition around the independent model for a model $\M$ with no self-dependent operators with $q>2$:}
{\bf Cluster expansion of the partition of a real-valued model $\M=\M_1\cup\M_1^*$ (for $q>2$):}
Consider a real-valued spin model $\M=\M_1\cup\M_1^*$, for which $\M_0=\emptyset$ in the decomposition of Definition~\ref{def:M:M0-M1}.
%Consider a spin model $\M$ that has no self-dependent operators (which implies in particular that $\M_0=\emptyset$). Using the decomposition of Definition~\ref{def:M:M0-M1}, the model can be written under the form $\M=\M_1\cup\M_1^*$. 
We denote by $K_1 = |\M_1|$ the number of operators in~$\M_1$ and arbitrarily index the operators, $\M_1 = \{\vecmu_{1, 1}, \cdots,\,\vecmu_{1, K_1}\}$. The partition function of $\M$ can then be written under the cluster expansion form:
\begin{align}\label{eq:Z:cluster-expansion}
%\hspace{-5mm}
%\textbf{\textrm{for\;$q>2$}}\,,\;\;\;\;\;\;
Z_{\M}(\vecg) 
    =Z^{Rind}_{\M}(\vecg) \,\left[\,1 +
    \sum_{\vecell_1\in\mathcal{L}^{cc}_q(\mathcal{M}_1)\backslash \veczero}\;
    \prod_{j=1}^{K_1} \frac{B_{\ell_{1,j}}}{B_0}(g_{\vecmu_{1,j}})\,\right]\,,
\end{align}
where the coefficients $B_{r}(x)$ are given by Eq.~\eqref{eq:Euler:general:M1:B}, and where $Z^{Rind}_{\M}(\vecg)$ is the partition function of $\M$ computed as if it were a real-valued independent model, which is given by Eq.~\eqref{eq:Z:loop-expansion:V3:ind:q-prime}.
%Here, 
For clarity, we simplified the notation $(\vecell_0, \vecell_1)\in\mathcal{L}^{cc}_q(\mathcal{M}_0,\mathcal{M}_1)\backslash (\veczero,\veczero)$ into $ \vecell_1\in\mathcal{L}^{cc}_q(\mathcal{M}_1)\backslash \veczero$, because %in this case 
$\M_0=\emptyset$.\\
This cluster expansion is {\bf always valid when $q>2$ is prime}. %, because in this case there are no self-dependent operators.
\end{Property}

%%%%%%%%%%%%%%%%%%%%%%%%%%%%%%%%%%%%%%%%
%\begin{Property} {\bf Cluster expansion of the partition of a real-valued model $\M$, for $q>2$:}
%Consider a spin model $\M$ that has no self-dependent operators (which implies in particular that $\M_0=\emptyset$). Using the decomposition of Definition~\ref{def:M:M0-M1}, the model can be written under the form $\M=\M_1\cup\M_1^*$. We denote $K_1 = |\M_1|$ the number of operators in~$\M_1$ and arbitrarily index the operators, $\M_1 = \{\vecmu_{1, 1}, \cdots,\,\vecmu_{1, K_1}\}$. The partition function of $\M$ can then be written under the cluster expansion form:
%\begin{align}\label{eq:Z:cluster-expansion}
%Z_{\M}(\vecg) 
    %=Z^{Rind}_{\M}(\vecg) \,\left[\,1 + \sum_{\vecell_1\in\mathcal{L}^{cc}_q(\mathcal{M}_1)\backslash \veczero}\; \prod_{j=1}^{K_1} \frac{B_{\ell_{1,j}}}{B_0}(g_{\vecmu_{1,j}})\,\right]\,,
%\end{align}
%where the coefficients $B_{r}(x)$ are given by Eq.~\eqref{eq:Euler:general:M1:B}, and where $Z^{Rind}_{\M}(\vecg)$ is the partition function of $\M$ computed as if $\M$ was a real-valued independent model, which is given by Eq.~\eqref{eq:Z:loop-expansion:V3:ind:q-prime}. For clarity, we simplified the notation $(\vecell_0, \vecell_1)\in\mathcal{L}^{cc}_q(\mathcal{M}_0,\mathcal{M}_1)\backslash (\veczero,\veczero)$ into $ \vecell_1\in\mathcal{L}^{cc}_q(\mathcal{M}_1)\backslash \veczero$, as in this case $\M_0=\emptyset$.\\
%This cluster expansion is {\bf always valid when $q$ is prime}, because in this case there are no self-dependent operators.
%\end{Property}
%%%%%%%%%%%%%%%%%%%%%%%%%%%%%%%%%%%%%%%%

More generally, for real-valued models with $\M_0\neq\emptyset$, we can factorize the term corresponding to the empty loop $(\vecell_0,\vecell_1)=(\veczero,\veczero)$ in Eq.~\eqref{eq:Z:loop-expansion:V3}, which gives the following general expression of the cluster expansion:
%%%%%%%%%%%%%
%\begin{align}\label{eq:Z:cluster-expansion:general}
%Z_{\M}(\vecg) 
%    =q^n\,\prod_{i=1}^{K_0} A_0(g_{\vecmu_{0,i}}) \prod_{j=1}^{K_1} B_0(g_{\vecmu_{1,j}})
%    \,\left[\,1 +
%    \sum_{(\vecell_0, \vecell_1)\in\mathcal{L}^{cc}_q(\mathcal{M}_0,\mathcal{M}_1)\backslash (\veczero,\veczero)}\;
%    \prod_{i=1}^{K_0} \frac{A_{\ell_{0,i}}}{A_0}(g_{\vecmu_{0,i}})
%    \prod_{j=1}^{K_1} \frac{B_{\ell_{1,j}}}{B_0}(g_{\vecmu_{1,j}})\,\right]\,,
%\end{align}
%%%%%%%%%%%%%
\begin{align}\label{eq:Z:cluster-expansion:general}
Z_{\M}(\vecg) 
    =\frac{Z^{ind}_{\M_0}(\vecg_0)}{2^n}\,Z^{Rind}_{\M_1\cup\M_1^*}(\vecg_1) \,\left[\,1 +
    \sum_{(\vecell_0, \vecell_1)\in\mathcal{L}^{cc}_q(\mathcal{M}_0,\mathcal{M}_1)\backslash (\veczero,\veczero)}\;
    \prod_{i=1}^{K_0} \frac{A_{\ell_{0,i}}}{A_0}(g_{\vecmu_{0,i}})
    \prod_{j=1}^{K_1} \frac{B_{\ell_{1,j}}}{B_0}(g_{\vecmu_{1,j}})\,\right]\,,
\end{align}
where we can now recognize the factorized term as the partition function of $\M$ computed as if: 1) the elements in $\M_1\cup\M_1^*$ were to form a real-valued independent model $Z^{Rind}_{\M_1\cup\M_1^*}(\vecg_1)=q^n \prod_{j=1}^{K_1} B_0(g_{\vecmu_{1,j}})$ as in Eq.~\eqref{eq:Z:loop-expansion:V3:ind:q-prime}, and 2) the elements in $\M_0$ were to form an binary independent model $Z^{ind}_{\M_0}(\vecg_0)=2^n \prod_{i=1}^{K_0} A_0(g_{\vecmu_{0,i}})$ as in Eq.~\eqref{eq:Z:Mind:q2} (with $A_0(x)=\cosh(x)$).
%The vector $\vecg_0$ (resp.~$\vecg_1$) denotes to the subset of parameters of $\vecg$ associated with the operators of the sub-model $\M_0$ (resp.~$\M_1\cup\M_1^*$).

\subsubsection{Additional considerations when $q$ is not prime}
\label{sec:cluster:q-non-prime}
%{Self-loops and weak independence for $q$ non-prime}

%%%%%%%%%%%%%%%%%%%%%%%%%%%%%%%
%When $q$ is not prime, some operators~$\vecmu$ other than those in $\M_0$ can create ``{\it self-loops}'' of the form $c\vecmu=0$, where $c$ is an integer strictly smaller than $q$. Just like for the operators in $\M_0$, these operators cannot contribute to an independent model, because they don't satisfy Eq.~\eqref{eq:condition:independence}. We call them {\it self-dependent} operators.
%For example, consider a model for $q=6$ with a single pair of c.c.~operators $\M=\{\vecmu,-\vecmu\}$ that verifies the self-dependence relation $3\vecmu=0$ (e.g., the one-body operator $\vecmu=(2)$ for a system with a single spin variable). This model has two loops in Eq.~\eqref{eq:loop:V3}, the empty loop $\vecell_1=(0)$ and the self-loop $\vecell_1=(3)$, which yields the partition function~\eqref{eq:Z:loop-expansion:V3}:
%\begin{align}\label{eq:Ex:Zwind:q6}
%Z_{\M}(\vecg) 
%    =6\,[B_{0}(g_{\vecmu}) + B_{3}(g_{\vecmu})]\,.
%\end{align}
%For such models, one can get a more explicit form of the partition function by computing the contribution of the self-loops to the partition function (similarly to what was done in Eq.~\eqref{eq:Z:loop-expansion:V3} for the operators of $\M_0$). The terms corresponding to the non-trivial loops would then capture the correction to the partition function due to the non-trivial dependencies (no self-loops) between the different operators.
%%%%%%%%%%%%%%%%%%%%%%%%%%%%%%%

When $q$ is not prime, some operators~$\vecmu$ other than those in $\M_0$ can create ``{\it self-loops}'' of the form $\eta\,\vecmu=0$, where $\eta$ is an integer strictly smaller than $q$. Just like for the operators in $\M_0$, these operators cannot contribute to an independent model, because they don't satisfy Eq.~\eqref{eq:condition:independence}. We call them {\it self-dependent} operators.
Because of these operators, the notion of independent models may seem ambiguous.
%For example, a model for $q=6$ with a single pair of c.c.~operators $\M=\{\vecmu,-\vecmu\}$ verifying the self-dependence relation $3\vecmu=0$ is not an independent model (because of the self-loop).
For example, consider a model for $q=6$ with a single pair of c.c.~operators $\M=\{\vecmu,-\vecmu\}$ that verifies the self-dependence relation $3\vecmu=0$ (e.g., the one-body operator $\vecmu=(2)$ for a system with a single spin variable). This model has two loops in Eq.~\eqref{eq:loop:V3}, the empty loop $\vecell_1=(0)$ and the self-loop $\vecell_1=(3)$, which yields the partition function~\eqref{eq:Z:loop-expansion:V3}:
\begin{align}\label{eq:Ex:Zwind:q6}
Z_{\M}(\vecg) 
    =6^n\,[B_{0}(g_{\vecmu}) + B_{3}(g_{\vecmu})]\,.
\end{align}
This model is not independent (because of the self-loop). However, from a modeling perspective, a model with a single pair of c.c.~operator could be seen as ``independent'', in the sense that: 1) it models (at least partially) one dimension of the system, and 2) it doesn't encode %\textcolor{red}{encode/create} 
constraints between different operators. %\textcolor{red}{with other operators} 
%(no loop with other operators).
%
For $q$ non-prime, we extend the notion of independent models to also include models with self-dependent operators that do not create non-trivial constraints (loops) with the other operators of the model.
We call these models {\it weakly independent}\footnote{
%However, it is as if the number of dimensions spanned by the model is equal to the number of interactions in the model, i.e., as if each interaction in the model corresponds to a different dimension.
%To better define weak independence, we need a notion of dimension. 
The notion of weakly independent model is best defined using the definition of length of a module.
As $\Field^n$ is not a vector space, but a module \cite{robinson2012course} we introduce the length of a module, a generalization of dimension from vector spaces to modules.
%\begin{Definition}[Length of a Module]
The length of a module $M$, denoted $l(M)$, is the number of non-zero submodules in the longest possible chain of distinct submodules:
\[ 0 \subset M_1 \subset M_2 \subset \dots \subset M_{l(M)} = M. \]
%\end{Definition}
For a finite set of vectors $\vecnu_1,\dots,\vecnu_K\in\Field^n$, we denote by $\langle\vecnu_1,\dots,\vecnu_K\rangle$ the submodule of $\Field^n$ generated by these vectors, i.e., the set of all $\Field$-linear combinations $\sum_k r_k\vecnu_k$ with $r_k\in\Field$. Using this notion, a real-valued spin model
$\M=\M_0\cup\M_1\cup\M_1^*$ is weakly independent if and only if
\begin{equation}
    l(\langle \M_0 \cup \M_1\rangle_{gen}) = \sum_{\vecmu\,\in\,\M_0\cup\M_1} l\bigl(\langle\vecmu\rangle\bigr).
\end{equation}}.
\begin{Corollary}
    \textbf{A real-valued weakly independent model (for $q$ non-prime)} is a model whose set of loops in Eq.~\eqref{eq:loop:V2} solely contains the following % [types of] 
    loops or their combinations: 1) the empty loop; 2) trivial loops due to c.c.~operators; 3) %(only for $q$ non-prime) 
    trivial loops due to self-dependent operators
    %(i.e., of the type $c\,\vecmu=\veczero$, where $c<q$).
    %(i.e., resulting from relations of the type $c\,\vecmu=\veczero$, where $c<q$).
    %(i.e., which are consequence of the equation $c\,\vecmu=\veczero$, where $c<q$).
    (i.e., of the type $\eta\,\vecmu=\veczero$ with $\eta<q$). %; or combinations of these loops.
\end{Corollary}
\noindent For models with self-dependent operators, one can get a more explicit form of the partition function by computing the contribution of the self-loops to the partition function (similarly to what was done in Eq.~\eqref{eq:Z:loop-expansion:V3} for the operators of $\M_0$). 
The terms corresponding to the non-trivial loops would then capture the correction to the partition function due to the non-trivial dependencies between the different operators (i.e., not due to self-loops or c.c.~operators).

%%%%%%%%%%%%%%%%%%%%%%%%
%\textcolor{red}{Despite only having a single pair of c.c.~operators, the partition function of this model has more than just the empty loop term in~\eqref{eq:Z:loop-expansion:V3} due to the self-dependency of the operator.}
%%%%%%%%%%%%%%%%%%%%%%%%

%%%%%%%%%%%%%%%%%%%%%%%%
%\textcolor{orange}{For models that have self-dependent operators, it is also interesting to write the partition function in a cluster-expansion-like form, in which the term corresponding to the empty loop represents the partition function of the model as if it were a (weakly) independent model. The terms corresponding to the non-trivial loops would then capture the correction to the partition function due to the non-trivial dependencies between the different operators. To write partition function under this form, we extract the self-loops from the loop structure and compute their contributions explicitly.}
%%%%%%%%%%%%%%%%%%%%%%%%

The contributions of self-loops of the form $2\vecmu=\veczero$, which come from the operators of $\M_0$ (when $q$ is even),
%(for even values of $q$), 
have been already computed explicitly in Eq.~\eqref{eq:Z:loop-expansion:V3}. % in Sec.~\ref{}
This was done by grouping odd and even powers of $\phi^{\vecmu}$ in the Euler formula~\eqref{eq:Euler:general:M0:maintext} for $e^{g_{\vecmu}\phi^{\vecmu}(\state)}$.
%
%%%%%%%%%%%%
%The model can also have self-dependent operators in $\M_1\cup\M_1^*$.
%%%%%%%%%%%%
%Other possible self-dependent operators in the model must be in $\M_1\cup\M_1^*$.
Similarly, consider a self-dependent operator~$\vecmu\in\M_1\cup\M_1^*$ and the smallest integer $\eta$ such that $\eta\vecmu=\veczero$ forms a self-loop ($\eta$ is a factor of $q$).
The contribution of this self-loop can be explicitly computed by grouping together the terms that correspond to 
$\phi^{\vecmu}$ raised to the same power modulo $\eta$ in the generalized Euler formula of Eq.~\eqref{eq:Euler:general:M1} (using the fact that $\eta\vecmu=\veczero$ implies $[\phi^{\vecmu}(\state)]^\eta=1$):  
\begin{align}\label{eq:Euler:general:M1:self-loop}
    &%\textrm{for}\;\vecmu\in\M_1,\; \textrm{s.t.}\; p\vecmu=\veczero \qquad
    e^{\;g_{\vecmu}\,\phi^{\vecmu}(\state)+g_{\vecmu}^*\,\phi^{-\vecmu}(\state)}
    =\sum_{r=0}^{q-1} B_{r}(g_{\vecmu}) \,\left[\phi^{\vecmu}(\state)\right]^r
    =\sum_{r'=0}^{\eta-1} C^{\eta}_{r'}(g_{\vecmu}) \,\left[\phi^{\vecmu}(\state)\right]^{r'}\,.%\\
    %&{\rm where}\;\;\;
    %    \forall r\in\{0,\,\cdots,\,p-1\}\,,\;\;\;
    %    C^p_{r}(x) = \sum_{k=0}^{(q/p)-1} B_{kp+r}(x)\,,
\end{align}
Here the coefficient $C^{\eta}_{r'}(x)$ collects %\textcolor{red}{regroups} 
the coefficients $B_{r}(x)$ that correspond 
%to the same order of $\phi^{\vecmu}$ modulo $p$: $r\,{\rm mod}\,q=r'$, $r'=kp+r$.
%group together powers of $\phi^{\vecmu}$ that correspond to the same order modulo $p$.
to the same power~$r'$ (modulo $\eta$) of $\phi^{\vecmu}(\state)$:
\begin{align}\label{eq:Euler:general:M1:C}
    \forall r'\in\{0,\,\cdots,\,\eta-1\}\,,\;\;\;
        C^{\eta}_{r'}(x) = \sum_{k=0}^{(q/\eta)-1} B_{k\eta+r'}(x)\,,
\end{align}
and $B_{r}(x)$ is given by Eq.~\eqref{eq:Euler:general:M1:B}. We recall that $\eta$ is a factor of $q$, and thus $q/\eta$ is an integer.
This expression is valid also when $\eta=q$, in which case $C_r^{q}(x)=B_r(x)$.
Finally, substituting Eq.~\eqref{eq:Euler:general:M0:maintext} and~\eqref{eq:Euler:general:M1:self-loop} in Eq.~\eqref{eq:Z:def:2} and computing the sum over the states, we obtain the following loop expansion for the partition function:
\begin{Property}{\bf Loop expansion of the partition function of real-valued $q$-state spin models~\eqref{eq:q-spin_models}, 
%from the weakly independent model.} 
without trivial loops from c.c.~operators and without self-loops.}
Consider a spin model $\M=\M_0\cup\M_1\cup\M_1^*$ and
%Let us denote $K_0 = |\M_0|$ and $K_1 = |\M_1|$ the number of operators respectively in $\M_0$ and in~$\M_1$, and 
arbitrarily index the operators $\M_0 = \{\vecmu_{0, 1}, \cdots,\,\vecmu_{0, K_0}\}$ and $\M_1 = \{\vecmu_{1, 1}, \cdots,\,\vecmu_{1, K_1}\}$. 
The partition function can be written under the form:
\begin{align}\label{eq:Z:loop-expansion:V4}
Z_{\M}(\vecg) 
    =q^n\,
    \sum_{(\vecell_0, \vecell_1)\in\mathcal{L}^{cc,s}_q(\mathcal{M}_0,\mathcal{M}_1)}
    \prod_{i=1}^{K_0} A_{\ell_{0,i}}(g_{\vecmu_{0,i}}) 
    \prod_{j=1}^{K_1} C^{\eta_j}_{\ell_{1,j}}(g_{\vecmu_{1,j}})\,,
\end{align}
where %, for each operator $\vecmu_{1,j}\in\M_1$, 
$\eta_j$ denotes the smallest integer such that $\eta_j \vecmu_{1,j}=\veczero$ ($\eta_j$ is equal to $q$ if $\vecmu_{1,j}$ is not self-dependent, and is a factor of $q$ otherwise).
Here, the contributions from self-loops are explicitly computed and the loop structure $\mathcal{L}^{cc,s}_{q}$ doesn't contain self-loops anymore:
%in which we removed the contribution from self-loops:
%in which self-loops were removed from the loop structure $\mathcal{L}$ and their contribution were explicitly computed.
\begin{align}\label{eq:loop:V4}
    \mathcal{L}^{cc,s}_{q}(\M_0,\M_1) = \bigg\{(\vecell_0,\vecell_1)\in N_2^{\,K_0}\bigotimes_{j=1}^{K_1} N_{\eta_j}\;\Big|\; \sum_{i=1}^{K_0} \ell_{0,i}\,\vecmu_{0,i} + \sum_{j=1}^{K_1} \ell_{1,j}\,\vecmu_{1,j} = \veczero\bigg\}\;,
\end{align}
where $N_{\eta}=\{0,\dots,\eta-1\}\subseteq \mathbb{Z}/q\mathbb{Z}$.
The coefficients $A_{0}(x)$, $A_{1}(x)$, and $C_{r}^{\eta}(x)$ are given by Eq.~\eqref{eq:Euler:general:M0:A:maintext} and Eq.~\eqref{eq:Euler:general:M1:C}. We added the superscript $s$ to $\mathcal{L}_q^{cc,s}$ to indicate that this set of loops is obtained after removing %the contributions due to 
self-loops.
\end{Property}
 
Finally, the term corresponding to the empty loop in Eq.~\eqref{eq:Z:loop-expansion:V4} now gives the expression of the partition function for a weakly independent model:
\begin{Property}\textbf{Partition function of a weakly independent model $\M_{wind}=\M_0\cup\M_1\cup\M_1^*$.}
%For $q$ prime, or $q=2p$ with $p$ prime:
\begin{align}\label{eq:Z:loop-expansion:V4:ind}
Z_{\M_{wind}}(\vecg)
    &=q^n\,
    \prod_{\vecmu_0\in\M_0} \cosh(g_{\vecmu_{0}}) \,
    \prod_{\vecmu_1\in\M_1} 
    C_{0}^{\eta_{\vecmu_1}}(g_{\vecmu_1})\,,
    %\left[ \sum_{m=0}^{q-1} \left\lVert c_{m}(g_{\vecmu_{1}}) \right\rVert^2 \right]\,,
\end{align}
where $\eta_{\vecmu}$ denotes the smallest integer such that $\eta_{\vecmu}\,\vecmu=\veczero$ for all operator $\vecmu\in\M_{1}$, % ($\eta_j$ is equal to $q$ if $\vecmu_{1,j}$ is not self-dependent, and is a prime factor of $q$ otherwise).
and $C_{0}^{\eta}$ is given by Eq.~\eqref{eq:Euler:general:M1:C}.
\end{Property}
\noindent For example, for a model $\M=\{\vecmu,-\vecmu\}$ verifying $3\vecmu=\veczero$  with $q=6$, we recover Eq.~\eqref{eq:Ex:Zwind:q6}.
As a sanity check, for an independent model with $q=2$, one has that $\M_1=\emptyset$ and we recover Eq.~\eqref{eq:Z:Mind:q2}. For a real-valued independent model with $q>2$, one has that $\M_0=\emptyset$ and $\eta_{\vecmu}=q$ for all $\vecmu\in\M_1$, and we recover Eq.~\eqref{eq:Z:loop-expansion:V3:ind:q-prime}.
Finally, substituting Eq.~\eqref{eq:Z:loop-expansion:V4:ind} in Eq.~\eqref{eq:Z:loop-expansion:V4}, we obtain: %one obtains the cluster expansion of the partition around the weakly independent model:
\begin{Property}\textbf{Cluster expansion of the partition around the weakly independent model.}
Consider a real-valued spin model $\M=\M_0\cup\M_1\cup\M_1^*$. The cluster expansion of the partition function of $\M$ around the weakly independent model is given by:
%obtained after removing trivial loops (from c.c.~operators and from self-loops) is:
%For $q$ prime,  or $q=2p$ with $p$ prime:
%\qquad{\rm for}\;\,q=2p\;\, {\rm with}\;\, p\;\,{\rm prime}
\begin{align}\label{eq:Z:cluster-expansion:V4}
Z_{\M}(\vecg) 
    =Z^{wind}_{\M}(\vecg) \,\left[\,1 +
    \sum_{(\vecell_0, \vecell_1)\in\mathcal{L}^{cc,s}_q(\mathcal{M}_0,\mathcal{M}_1)\backslash (\veczero,\veczero)}\;
    \prod_{i=1}^{K_0} \frac{A_{\ell_{0,i}}}{A_0}(g_{\vecmu_{0,i}}) 
    \prod_{j=1}^{K_1} \frac{C_{\ell_{1,j}}^{\eta_j}}{C_0^{\eta_j}}(g_{\vecmu_{1,j}})\,\right]\,,
\end{align}
where $Z^{wind}_{\M}(\vecg)$ denotes the partition function of $\M$ computed as if it were % the model $\M$ was 
weakly independent, which is given by Eq.~\eqref{eq:Z:loop-expansion:V4:ind}. The remaining set of loops $\mathcal{L}^{cc,s}_q(\mathcal{M}_0,\mathcal{M}_1)$ doesn't contain any trivial loops (from c.c.~operators or self-loops).
\end{Property}

\subsection{Rank and dimension of a spin model} %Notions of independence; 

The definition~\ref{def:independent:operators} of an independent set of operators allows us to extend the definition of the rank of a spin model, 
originally introduced in the binary case~\cite{beretta2018stochastic, demulatier2024MCM}, to $q\geq 2$:
%, which was introduced in the binary case by~\cite{beretta2018stochastic, demulatier2024MCM}:
\begin{Definition}\label{def:M:rank}
    {\bf Rank of a spin model.} The rank of a spin model $\M$, denoted by $\rank(\M)$, is the cardinality of the largest linearly independent set of operators in~$\M$. 
    %Using the matrix representation of spin models in Definition~\ref{eq:model:matrix-representation}, it corresponds to the rank of the matrix representation of $\M$ in $\Field^n$. 
    By definition, in an $n$-spin system, $\,\operatorname{rank}(\M)\leq n$. 
\end{Definition}
\noindent For example, by definition~\ref{def:independent:model:real}, the rank of an independent model is equal to the number of operators in the model.
%\textcolor{blue}{For instance, for an independent model (definition~\ref{def:independent:model:real}), the rank of the model is equal to the number of operators in the model. 
For a real-valued independent spin model with $q>2$, $\M=\M_1\cup\M_1^*$, the rank of the model is equal to the number of operators in $\M_1$.

%%%%%%%%%%%%%%%%%%%%%%%%%%
%When $q$ is not prime, because of self-dependent operators, the \textcolor{red}{concept/notion} of independent model could be thought of/seen as ambiguous. For example, a model for $q=6$ with a single pair of c.c.~operators $\M=\{\vecmu,-\vecmu\}$ verifying the self-dependence relation $3\vecmu=0$ is not an independent model (because of the self-loop). However, from a modeling perspective, a model with a single interaction could be seen as ``independent'', in the sense that: 1) it models (at least partially) one dimension of the system, and 2) it doesn't encode constraints between different operators. Moreover, the rank of this model is zero, while it is as if the number of dimensions spanned by the model is equal to one. The operator $\vecmu$ lives in one dimension, but doesn't span the whole $(\mathbb{Z}/6\mathbb{Z})^n$.
%%%%%%%%%%%%%%%%%%%%%%%%%%

When $q$ is not prime, the rank of the model is not sufficient to capture %\textcolor{blue}{the span of the model} 
the number of dimensions in which the model is embedded.
%/ the number of dimensions spanned/covered by the model.
Consider for example a model for $q=6$ with a single pair of c.c.~operators $\M=\{\vecmu,-\vecmu\}$ verifying the self-dependence relation $3\vecmu=0$.
The rank of this model is zero, but %whereas %while
the model is embedded in at least %at least in
one dimension. %the smallest number of dimensions in which the model is embedded is one.
%\textcolor{purple}{it is as if the number of dimensions spanned by the model is equal to one.}
This is because the operator $\vecmu$ lives in one dimension, %\textcolor{red}{even though it} 
but doesn't span the whole $(\mathbb{Z}/6\mathbb{Z})$.
To clarify this distinction when $q$ is not prime, we introduce the definition of the (embedding) dimension of a $q$-state spin model.
This definition is inspired by the definition of basis of a spin model, which was introduced in the binary case~\cite{demulatier2024MCM} as a minimal set of independent operators that can generate all the operators of the model.
%This definition is inspired by the definition (in the binary case) of a basis of a spin model~\cite{demulatier2024MCM}, which is a minimal set of independent operators that can generate all the operators of the model.
%Inspired by the definition (in the binary case) of basis of a spin model~\cite{demulatier2024MCM}, as the minimal set of independent operators that can generate all the operators of the model, we give the following definition.
%
\begin{Definition}
    \textbf{Dimension of a spin model.} The dimension of a spin model $\M$, denoted $\dim(\M)$, is the cardinality of the minimal independent set of operators that can generate all the operators of the model~$\M$. By definition, in an $n$-spin system, $\,\dim(\M)\leq n$.
\end{Definition}
\noindent For example, for $q=6$, for the one-spin model $\M=\{\vecmu,-\vecmu\}$ with the one-body operator $\vecmu=(2)$ and its c.c. $-\vecmu=(4)$, the set with the single operator $\{\vecmu_1=(1)\}$ is a minimal independent set %of operators 
that can generate all the operators of $\M$ in $\mathbb{Z}/6\mathbb{Z}$ (the operator $\vecmu_1$ is not self-dependent, as oppose to the operators $(2)$ and $(4)$). The dimension of $\M$ is therefore $\dim(\M)=1$.

For models without self-dependent operators, % \textcolor{red}{(which are any spin model when $q$ is prime)}, 
any maximal independent set of operators in the model can generate all the operators of the model, which means that %for these models $\rank(\M)=\dim(\M)$.
the dimension of the model is equal to its rank. 
This is always the case for models with $q=2$, and in general for any $q$ prime, for which there is therefore no need to introduce a distinction between rank and dimension~\cite{demulatier2024MCM}. 
This is also the case for independent models (complex- or real-valued) at any value of $q$ (prime or not).
%Similarly, for a strongly independent model $\M_{ind}$ for any value of $q$, we have that $\rank(\M_{ind})=\dim(\M_{ind})=|\M_0\cup\M_1|$.
%
%When $q$ is not prime, 
For models with self-dependent operators (which exist only when $q$ is not prime), the dimension of the model can be larger than its rank. In the example above, $\M=\{(2), (4)\}$ for $q=6$, the rank of the model is $\rank(\M)=0$, but its dimension is $\dim(\M)=1$.

\subsection{Discussion: Loops as algebraic constraints defining the statistical properties of the spin model}
%algebraic constraints of the statistical model

The loops of a $q$-state spin model, defined in Eq.~\eqref{eq:loop:V2}, are %\textcolor{red}{are/identify a set of} 
linear dependencies between the operators of the model that result from the choice of the model architecture (i.e., which operators are in the model).
Equation~\eqref{eq:Z:loop-expansion:V2} connects these structural dependencies, captured by the loop structure on the r.h.s., to the statistical properties of the model, captured by the partition function on the l.h.s.
More specifically, the Hessian of the log-partition function corresponds to the Fisher information matrix, a covariance matrix that characterizes how susceptible the model probability distribution is to perturbations around any given value of the parameters.
The structural dependencies encoded by the loops $\mathcal{L}_q(\M)$ thus ultimately shape the Fisher information matrix, making the model more or less susceptible to changes in the parameters, or, from a modeling perspective, more or less %flexible/adaptable/
malleable to fit a wide variety of data patterns. 
%\textcolor{red}{For integrable models,} 
This flexibility of a statistical model can be summarized in a single characteristic called the geometric complexity of the model~\cite{balasubramanian1997statistical, myung2000counting}, corresponding to the term $c_{\M}$ in Eq.~\eqref{eq:MDL} (when the integral is defined). 
%These structural dependencies are related to the statistical properties of the model via the Fisher information matrix. 
In the binary case, Ref.~\cite{beretta2018stochastic} explores the impact of %\textcolor{red}{impact of/connections between} 
the loop structure of a spin model~\eqref{eq:Z:Ising:def:loop} on the geometric complexity.
%explores the dependency of the geometric complexity of a spin model on its loop structure~\eqref{eq:Z:Ising:def:loop}. 

In summary, equation~\eqref{eq:Z:loop-expansion:V2} draws a useful connection between %the connection between the statistical structure of a spin model and its algebraic properties. 
the statistical properties of a spin model and algebraic properties of %/constraints between 
its operators, i.e., %encoded %represented by
the linear dependencies between the operators encoded in %by
the loop structure~\eqref{eq:loop:V2}. Equation~\eqref{eq:Z:loop-expansion:V2} describes how these structural dependencies %, resulting from the choice of the operators of the model, 
constrain the statistical properties of the model.
%in the loops of the model
%\textcolor{purple}{(encoded by the loops of the model)}
% algebraic properties. 
%\textcolor{red}{to algebraic properties of the model (represented by the linear dependencies between the operators of the model~\eqref{}).}
%\textcolor{red}{Equation~\eqref{eq:Z:loop-expansion:V2} thus can be seen as connecting the statistical properties of a spin model to its algebraic properties.}
In the following section, we introduce % propose
an algebraic representation of spin models, which will allow for a better characterization of this connection.

%%%%%%%%%%%%%% %%%%%%%%%%%%%%% %%%%%%%%%%%%%% %%%%%%%%%%%%%%%
%Better connect the notion of loops to the idea of ``algebraic property'': linearly dependent set of operators.
%%%%%%%%%%%%%% %%%%%%%%%%%%%%% %%%%%%%%%%%%%% %%%%%%%%%%%%%%%

%%%%%%%%%%%%%%%%%%%%%%%%%%%%%%%%%%%%%%%%%%
\section{Gauge transformations (GTs) and equivalence classes of $q$-state spin models}
%\label{sec:2b_EquivalenceClasses}
\label{Sec:GT:MatrixRepresentation}
%\subsection{Gauge transformations and equivalence classes of $q$-state spin models} 
\label{Sec:GT:MatrixRepresentation}

%%%%%%%%%%%%%%%%%%%%%%%%%%%%%%%%%%%%%%%%%%
%"Linear algebra" over Z/nZ:
%https://mathoverflow.net/questions/44576/linear-algebra-over-z-nz-reference-please
%%%%%%%%%%%%%%%%%%%%%%%%%%%%%%%%%%%%%%%%%%

In the binary case, Ref.~\cite{beretta2018stochastic} observed that all the models that have the same loop structure~\eqref{eq:Z:Ising:def:loop} have the same partition function~\eqref{eq:Z:Ising} up to permutation of their parameters. 
They introduce bijective transformations of the spin variables (and thus of the state space) that preserve the structure of spin operators, which they call gauge transformations.
These transformations map a spin model onto another spin model (with different operators), while keeping the loop structure invariant, therefore preserving the partition function. %(up to permutation of the parameters). 
The transformed spin models can be seen as different representations of the same abstract model.
%These transformations define equivalence classes of binary spin models.
Let us generalize these ideas to $q$-state spin models.

%\textcolor{blue}{Ref.~\cite{beretta2018stochastic} introduces bijective transformations of the state space that map a spin models onto another spin models that have the same loop structure and thus the same partition function.}

%%%%%%%%% Possible points: %%%%%%%%%%%%% %%%%%%%%%%%%% %%%%%%%%%%%%% 
% - Note that this representation of spin models, transform the problem of model selection with spin models, into a problem that can be expressed using linear algebra.
% - For the part when introducing the basis transformation: outside of the exponential there is a multiplicative group, and inside the exponential there is an additive group.
%%%%%%%%%%%%% %%%%%%%%%%%%% %%%%%%%%%%%%% %%%%%%%%%%%%% %%%%%%%%%%%%% 

%%%%%%%%%%%%% %%%%%%%%%%%%% %%%%%%%%%%%%% %%%%%%%%%%%%% %%%%%%%%%%%%% 
%\textcolor{red}{Here we can also generalise the notion of gauge transformation, as bijective transformations of the state space that preserve the structure of spin operators. We will see that these correspond to automorphisms of the state space $\Field^n$.}
%%%%%%%%%%%%% %%%%%%%%%%%%% %%%%%%%%%%%%% %%%%%%%%%%%%% %%%%%%%%%%%%% 

\subsection{Gauge transformations as changes of spin representation} % -- definition and matrix representation}
\label{sec:GT:def}
Following Ref.~\cite{beretta2018stochastic}, 
we use the notion of independence between spin operators (see Definition~\ref{def:independent:operators}) to generalize the definition of gauge transformations to systems of $q$-state spin variables with $q>2$. 
Ref.~\cite{demulatier2024MCM} also introduced a linear algebra formalism for spin models and gauge transformations, which we generalize to larger values of $q$. Let us start by recalling the definition of independence between spin operators:

\begin{Definition}\label{def:independent:operators:bis}
{\bf Independent set of operators.} Consider a system of $n$ $q$-state spin variables. A set of operators $\{\phi^{\vecmu_1},\,\dots,\phi^{\vecmu_K}\}$ is independent if and only if the set of vectors $\mathcal{S}=\{\vecmu_1,\,\dots,\vecmu_K\}$ is linearly independent in $\Field^n$:
\begin{align}\label{eq:linear-indep:bis}
    \forall (r_k)%_{k=\{1..K\}}
    \in\Field^K, \;\;\;\;\sum_{k=1}^K r_k\,\vecmu_k=0\;({\rm mod}\;q)
    \qquad\Rightarrow \qquad
    r_k=0 \;\;\;\;\forall\, k\,.
\end{align}
This is equivalent to the set of loops of $\mathcal{S}$ containing only the null vector: $\mathcal{L}_q(\mathcal{S})=\{\veczero\}$.
\end{Definition}
\noindent We recall that spin operators are defined in Eq.~\eqref{eq:def:monomials:s} as products of the spin variables identified by the vector $\vecmu$. Equation~\eqref{eq:linear-indep:bis} thus means that no operator $\phi^{\vecmu}$ in the set can be obtained as a product of other operators in the set raised to any power. Moreover, in a system of $n$ spin variables, an independent set of operators can have at most $n$ operators, because there are at most $n$ independent vectors in $\Field^n$.

Note that $\Field^n$ is not a vector space when $q$ is not prime. However, the definition above in terms of linear independence between elements of $\Field^n$ still holds in this case, as well as the other definitions and the linear algebra representation discussed in this section (see the notion of {\em modules} in mathematics~\cite{robinson2012course}).

\begin{Definition}{\bf Gauge transformations (GTs).}\label{def:GT}
    Consider a system of $n$ $q$-state spin variables $\spin=(s_1,\,\dots,\,s_n)$. 
    A gauge transformation $\mathcal{T}$ is a change of spin basis identified by an ordered set of $n$ independent operators $(\phi^{\vecmu_1},\,\dots,\,\phi^{\vecmu_n})$, where the new basis variables $\spin'=(s_1',\,\dots,\,s_n')$ are defined as: 
    \begin{align}\label{def:GT:s_j}
        \mathcal{T}:\spin\rightarrow
        \begin{cases}
            s_1'= \phi^{\vecmu_1}(\spin)\\
            \;\;\cdots\\
            s_n'= \phi^{\vecmu_n}(\spin)
        \end{cases}\,.
    \end{align}
%%%%%% %%%%%% %%%%%% %%%%%% %%%%%% %%%%%%
%\textcolor{blue}{A gauge transformation $\mathcal{T}$ is a change of spin basis, in which the new spin variables $\spin'=(s_1',\,\dots,\,s_n')$ are defined as:}
%    \begin{align}\label{def:GT:s_j}
%        \mathcal{T}:\spin\rightarrow
%        \begin{cases}
%            s_1'= \phi^{\vecmu_1}(\spin)\\
%            \;\;\cdots\\
%            s_n'= \phi^{\vecmu_n}(\spin)
%        \end{cases}\,,
%    \end{align}
%    \textcolor{blue}{where $(\phi^{\vecmu_1},\,\dots,\,\phi^{\vecmu_n})$ is an ordered set of $n$ independent operators identifying the transformation.}
%%%%%% %%%%%% %%%%%% %%%%%% %%%%%% %%%%%%
    The transformation can equivalently be written in terms of the discrete variables $\state=(\alpha_1,\,\dots,\,\alpha_n)\in\Field^n$, where each $\alpha_j$ is the color of the spin $s_j$.
    Under the gauge transformation $\T$ above, the new variables $\state'=(\alpha_1',\,\dots,\,\alpha_n')\in\Field^n$ are defined as:
    \begin{align}\label{def:GT:alpha_j}
    {\rm for}\;{\rm all}\;j\in\{1, \cdots, n\}, \qquad
    \alpha_j'= \state\cdot\vecmu_j\,, %\Z/q\Z\,.
    \end{align}
    where each variable $\alpha_j'$ is the color of the spin $s_j'$ and where the scalar product is taken in $\Field^n$.
    Figure~\ref{fig:GT_and_Loops}.a gives an example of a gauge transformation for a spin model with $q=3$.
\end{Definition}
\noindent Note that self-dependent operators (operators $\vecmu$ for which there exists a scalar $c\in\Field$ such that $c\vecmu=\veczero$) %-- see Sec.~\ref{sec:IM:strong-indep}) 
cannot contribute to a set of independent operators, and therefore, they cannot be used to define a GT.

Using the definition of spin operators in Eq.~\eqref{eq:def:monomials:alpha}, it is straightforward to check that the new variables $s_j'$ defined by the transformation~\eqref{def:GT:s_j} are indeed still $q$-state spin variables~\eqref{def:s_j}:
\begin{align}\label{GT:alpha}
    s_j'=\phi^{\vecmu_j}(\spin) =z_q^{\,\state\cdot\vecmu_j} = z_q^{\,\alpha_j'}\,,
    \qquad{\rm where}\;\;
    \alpha_j'=\state\cdot\vecmu_j
\end{align}
are the corresponding colors. This also proves the expression~\eqref{def:GT:alpha_j} of the GT in terms of the colors. 
Because the set of $n$ vectors $(\vecmu_1,\,\dots,\,\vecmu_n)$ are linearly independent, they form a new basis of $\Field^n$, and Eq.~\eqref{def:GT:alpha_j} simply re-writes the states of $\state$ into this new basis. 
Hence, GTs written for the variables $\state$ correspond to linear basis transformations in $\Field^n$, which can be identified as the automorphisms of $\Field^n$.

\begin{Property}\label{def:GT:matrix}
    {\bf Matrix representation of gauge transformations.} 
    For convenience, the gauge transformation in Eq.~\eqref{def:GT:alpha_j} can be written as the vector-matrix product over $\Field$~\cite{demulatier2024MCM}:
    \begin{align}\label{def:GT:matrix:alpha}
        \state'=\state\,\Tmat\,,
    \end{align}
    where $\state=(\alpha_1,\,\dots,\,\alpha_n)$ and $\state'=(\alpha_1',\,\dots,\,\alpha_n')$ are two row vectors containing respectively the states of the original variables $\alpha_j$ and of the new variables $\alpha_j'$\,, and $\Tmat=(\vecmu_1,\,\dots,\,\vecmu_n)$ is an $n\times n$-matrix whose columns are the linearly independent vectors $\vecmu_j$ defining the gauge transformation.
\end{Property}
\noindent When performing the matrix product over $\Field$ in Eq.~\eqref{def:GT:matrix:alpha}, the elements of the resulting vector are taken modulo $q$.
The $n$ columns of $\Tmat$ being linearly independent in $\Field^n$ means that $\Tmat$ is invertible in $\Field^n$. 
Multiplying Eq.~\eqref{def:GT:matrix:alpha} on both side by the inverse matrix $\Tmat^{-1}$ allows one to recover the original variables from the transformed ones, thus defining the inverse gauge transformation (also confirming that GTs are indeed bijective).
%
%%%%%%%%%%%%%%%%%%
\begin{Property}\label{def:GT:inverse}
    {\bf Inverse gauge transformation.} Gauge transformations (GTs) are bijective. Following a GT, the original variables $\state$ can be recovered from the transformed variables $\state'$ using:
    \begin{align}
        \state=\state'\,\Tmat^{-1}\,,
    \end{align}
    where $\Tmat^{-1}$ is the inverse of the matrix representation $\Tmat$ of the GT.
\end{Property}

In the following, we will often denote by $\basis = (\phi^{\vecmu_1},\,\dots,\,\phi^{\vecmu_n})$ the set of $n$ independent operators used to define the new basis, by $\mathcal{T}_b$ the corresponding gauge transformation, and by $\Tmat_{\basis}$ its matrix representation.
The change of basis will then be denoted by $\spin'=\basis(\spin)$ or, equivalently, by $\state'=\state\,\Tmat_{\basis}$.
Similarly, we will denote by $\T_{\basis}^{-1}$ the inverse GT, which is represented by %defined via
the inverse matrix $\Tmat_{\basis}^{\,-1}$. 
The columns of $\Tmat_{\basis}^{\,-1}$ identify $n$~independent vectors $\vecmu_j'$\,, which correspond to the operators defining the basis of the inverse GT, $\,\basis^{-1} = (\phi^{\vecmu_1'},\,\dots,\,\phi^{\vecmu_n'})$ .

For a system of $n$ $q$-state spin variables, each choice of gauge transformation $\spin'=\basis(\spin)$ can be performed on 
the vector of discrete variables $\state$ by multiplying it with the corresponding invertible $n\times n$-matrix $\Tmat_{\basis}$ in $\Field^n$, as described in Property~\ref{def:GT:matrix}. Reciprocally, the columns of each invertible $n\times n$-matrix $\Tmat$ in $\Field^n$ uniquely identify $n$ linearly independent vectors $\vecmu_i\in\Field^n$, $\Tmat=(\vecmu_1,\,\dots,\,\vecmu_n)$, and therefore uniquely identify a gauge transformation. 
It is then easy to see that GTs, when written in terms of the variables $\state\in\Field^n$, are the automorphisms of $\Field^n$ (i.e., corresponding to all possible invertible $n\times n$-matrices over $\Field$).

\begin{Property}
    Consider a system of $n$ $q$-state spins characterized by the discrete color variables $\state=(\alpha_1,\dots,\alpha_n)\in\Field^n$. The gauge transformations defined for spin systems in Definition~\ref{def:GT}
    are the automorphisms of the state space of the variables $\state$, i.e., they are the automorphisms of the cyclic group $\Field^n$:
    \begin{align}
        \mathcal{G}_q(n)=Aut[\Field^n]\,,
    \end{align}
    where $\mathcal{G}_q(n)$ denotes the set of all gauge transformations.
\end{Property}

\begin{Property}
    {\bf Total number of GTs.} It follows that, for $q$ prime, the number of possible gauge transformations is:
    \begin{align}\label{eq:nb_GT}
    \textrm{for}\;q\;\textrm{prime},\;\qquad
    |\mathcal{G}_q(n)|  
    = \prod_{i=0}^{n-1} (q^n-q^i)\,.
    \end{align}
    For $q$ not prime, let's consider the prime factorization $q = \prod_{k=1}^K p_k^{\,m_k}$, where the $p_k$'s are prime numbers and the $m_k$'s are their multiplicity. The number of automorphisms of $\Field^n$ is given by:
\begin{align}\label{eq:nb_GT:q-non-prime}
    \textrm{for}\;q\;\textrm{non prime},\;\qquad
    |\mathcal{G}_q(n)| 
    = \prod_{k=1}^{K} p_k^{\,(m_k-1)n^2}\,
    \prod_{i=0}^{n-1} (p_{k}^{\,n}-p_{k}^{\,i})\,.
\end{align}
\end{Property}

When $q$ is prime, the number of GTs~\eqref{eq:nb_GT} is straightforward to compute: after choosing the $i$ first independent operators~$\vecmu_i$ of the basis in Eq.~\eqref{def:GT:s_j}, there are exactly $(q^n-q^i)$ possibilities to choose the $(i+1)$-th operator~$\vecmu_{i+1}$ such that it is independent from the $i$ operators already chosen. 
This is because the set of $i$ independent vectors $\vecmu$ already chosen can generate $q^i$ different vectors (and thus, different operators) through linear combinations in $\Field^n$, and $\vecmu_{i+1}$ cannot be chosen among them.

If $q$ is not prime, then for some operators $\vecmu\in\Field^n$ there exists a strictly positive integer $c<q$ such that $c\,\vecmu=0\,({\rm mod}\,q)$\, (for example, for $q=4$ and $n=2$, the vector $\vecmu=(2,2)$ verifies that $2\vecmu=0$). 
These operators don't satisfy the condition of linear independence, and thus cannot be used to construct a set of independent operators.
The total number of GTs in this case is therefore smaller than the number in Eq.~\eqref{eq:nb_GT}. Computing the precise number~\eqref{eq:nb_GT:q-non-prime} uses notions from group theory and is provided in App.~\ref{app:GT:counting}. 
For example, for $n=1$ spin with $q=4$, there are three possible operators, corresponding to the vectors $\vecmu\in\{(1),(2),(3)\}$, as $\vecmu=(2)$ cannot be used to define a new basis because $2\,\vecmu=0\,({\rm mod}\,4)$. We thus find that the number of GTs in this simple case is $|\mathcal{G}_4(1)|=2$, which is also obtained from Eq.~\eqref{eq:nb_GT:q-non-prime}.

Gauge transformations of a spin system correspond to changes of representation of the spin variables. As a result, a GT induces a transformation of any dataset or spin model expressed in these variables. In the following, we discuss how these transformations affect data and spin models.

\subsection{Gauge transformations of data}
\label{sec:GT:data}
Consider a dataset composed of multiple observations (datapoints) of the states of a spin system. 
For instance, this can be a discrete dataset, composed of observations of the discrete color variables $\state\in\Field^n$.
A GT changes the basis in which the spin system is represented and thus induces a transformation of the dataset, which is then observed in a new basis of discrete variables $\state'\in\Field^n$.
A dataset is therefore transformed under a GT by re-writing each datapoint in this new basis.

This operation can be performed efficiently using the vector-matrix product in Eq.~\eqref{def:GT:matrix:alpha}, where $\Tmat$ is the matrix representing the GT, and $\state$ and $\state'$ are the discrete vectors representing respectively the original and the new value of a given datapoint.
Defining the data matrices $\boldsymbol{D}$ and $\boldsymbol{D}'$, whose $i$-th row contains the $i$-th datapoint written respectively in the original and in the new basis, the GT of a dataset can be obtained by the matrix product over $\Field$:
\begin{align}\label{def:GT:matrix:D}
        \boldsymbol{D}'=\boldsymbol{D}\,\Tmat\,.
\end{align}
Because a gauge transformation corresponds to a change of basis, there is no loss of information between the original and the transformed data; the new dataset is just the old one represented in a different basis.

Under a gauge transformation, each state $\state\in\Field^n$ of the discrete system is mapped to a unique new state $\state'\in\Field^n$. A gauge transformation thus draws a bijection in state space, between the $q^n$ states of the original variables and the $q^n$ states of the new variables. As a result, the empirical distribution (i.e., frequency of occurrence of the states) remains unchanged, but the states corresponding to each frequency are now shuffled.

\subsection{Gauge transformations of $q$-state spin models}
\label{Sec:GTSpinModel}

GTs are changes of the basis in which the spin variables are represented. As a result, spin operators, which are functions of the spin variables, are also modified under GTs.
Remarkably, by construction of GTs, the transformation of a spin operator remains a spin operator, and as a result, a spin model is transformed into another spin model, as described in this section.

\begin{Definition}\label{def:GT:Op}
    {\bf Transformed operator.} A spin operator $\phi^{\vecmu}(\spin)$ is transformed under a GT~$\mathcal{T}_{\basis}$ by rewriting $\phi^{\vecmu}(\spin)$ in terms of the new variables $\spin'$, using the inverse GT $\spin= \basis^{-1}(\spin')$:
    \begin{align}\label{def:eq:GT:Op}
        \mathcal{T}_{\basis}[\phi^{\vecmu}](\spin') \doteq \phi^{\vecmu}\big(\basis^{-1}(\spin')\big)\,. 
    \end{align}
    The resulting transformed operator is still a spin operator and is given by:
    \begin{align}\label{def:eq:GT:Op:2}
        \mathcal{T}_{\basis}[\phi^{\vecmu}](\spin') = \phi^{\vecmu'}(\spin')\,,
        \qquad
        {\rm where}
        \qquad
        \vecmu' = \Tmat_{\basis}^{-1}\vecmu
    \end{align}
    is a matrix-vector multiplication over $\Field$, $\Tmat_{\basis}^{-1}$ is the matrix representation of the inverse gauge transformation $\spin'=\basis^{-1}(\spin)$ (see Property~\ref{def:GT:inverse}), and $\vecmu\in\Field^n$ and $\vecmu'\in\Field^n$ are two column vectors representing the operator, respectively in the original and in the new basis.
\end{Definition}

\begin{proof}
    We obtain this result by using the matrix representation of the inverse GT $\state = \state' \Tmat_{\basis}^{-1}$ in the expression of the operator $\phi^{\vecmu}$ written in terms of the colors: 
    \begin{align}
        \phi^{\vecmu}(\state) = z_q^{\state\cdot\vecmu} 
        = z_q^{(\state' \Tmat_{\basis}^{-1})\cdot\vecmu}
        = z_q^{\state'\cdot \vecmu'}
        \doteq \phi^{\vecmu'}(\state')\,,
        \qquad{\rm where}\;\; \vecmu' = \Tmat_{\basis}^{-1}\vecmu\,.
    \end{align}
\end{proof}

Transforming a spin model with a gauge transformation $\mathcal{T}_{\basis}$ consists in re-writing the Hamiltonian (energy landscape) of the model in the new basis $\spin'=\basis(\spin)$, mapping the energy of a state $\spin$ to the new state $\spin'$.
As the transformation of a spin operator under a GT is also a spin operator, the new Hamiltonian remains that of a spin model, and takes the following form. 

\begin{Definition}\label{def:GT_M}
    {\bf Transformed model.} 
    Under a gauge transformation $\mathcal{T}_{\basis}$, the Hamiltonian of a spin model, with operators $\M$ and parameters $\vecg$, is rewritten in the new basis $\spin'=\basis(\spin)$. This gives the Hamiltonian of a new spin model $\M'$ with parameters $\vecg'$ (see App.~\ref{App:GT}):
    %\textcolor{blue}{The energy of a state $\spin$ under a model $\M$ is thus mapped to the energy of the state $\spin'=\basis(\spin)$ of a new spin model $\M'$:}
    \begin{align} \label{eq:GT:Hamiltonian}
        H(\spin\mid \M,\,\vecg) 
        \overset{\spin=\basis^{-1}(\spin')}{=}
        H(\spin'\mid \M',\,\vecg')\,,
        %= H(\basis(\spin)\mid \mathcal{T}_{\basis}[\vecg],\,\mathcal{T}_{\basis}[\M])\,,
    \end{align}
    where %whose 
    the set of operators $\M'=\mathcal{T}_{\basis}[\M]$ is obtained by transforming each operator of $\M$%
    \footnote{%We recall that 
    We use the notation $\tilde{\M}$ to denote the set of spin operators $\phi^{\vecmu}(\spin)$ of the model %, to distinguish it from 
    and the notation $\M$ for the set of their corresponding vector representations $\vecmu\in\Field^n$.
    }:
    \begin{align}
        \tilde{\M}'= \{\mathcal{T}_{\basis}[\phi^{\vecmu}](\spin') \mid \phi^{\vecmu}(\spin)\in\tilde{\M}\}
        \qquad\Leftrightarrow \qquad
        \M'= \{\Tmat_{\basis}^{-1}\vecmu \mid \vecmu\in\M\}\,,
    \end{align}
    and where the %whose 
    vector of parameters $\vecg'=\mathcal{T}_{\basis}[\vecg]$ is a permutation of the original parameters~$\vecg$:
    %of the elements of the vector %of parameters of the original model:
    \begin{align}
        g_{\vecmu'}=g_{\vecmu}\,,\qquad {with}\qquad
        \vecmu'=T_{\basis}^{-1}\vecmu\,.
    \end{align}
    The permutation is such that the transformed operator $\phi^{\vecmu'} = \mathcal{T}_{\basis}[\phi^{\vecmu}]$ in $\M'$ is parameterized by the same parameter as the operator $\phi^{\vecmu}$ in $\M$.
%%%%%%%%%%%
%The gauge transformation $\M'$ of a model $\M\subseteq [\Field^n]^*$ is obtained by transforming each operator of $\M$:
%%%%%%%%%%%    
%Under a GT $\mathcal{T}_{\basis}$, a spin model with operators $\M\subseteq [\Field^n]^*$ is transformed into a new model $\M'=\mathcal{T}_{\basis}[\M]$ with operators/by transforming each operator of $\M$ in the new basis:
%%%%%%%%%%%  
\end{Definition}
\noindent Because GTs are bijections of the state space, the gauge transformation of a spin model maps the energy of a state $\spin$ to another (unique) state $\spin'$, thus drawing a one-to-one mapping in state space that rearranges %shuffles
the energy landscape. In the context of real-valued $q$-state spin models, this yields a permutation of the state probabilities %probability of the states
in the model probability distribution, just like % analogous/similar to
the one described for GTs of data.

\begin{figure}[!h] %[!ht]
\centering
    \centering 
    \includegraphics[width=0.95\textwidth]{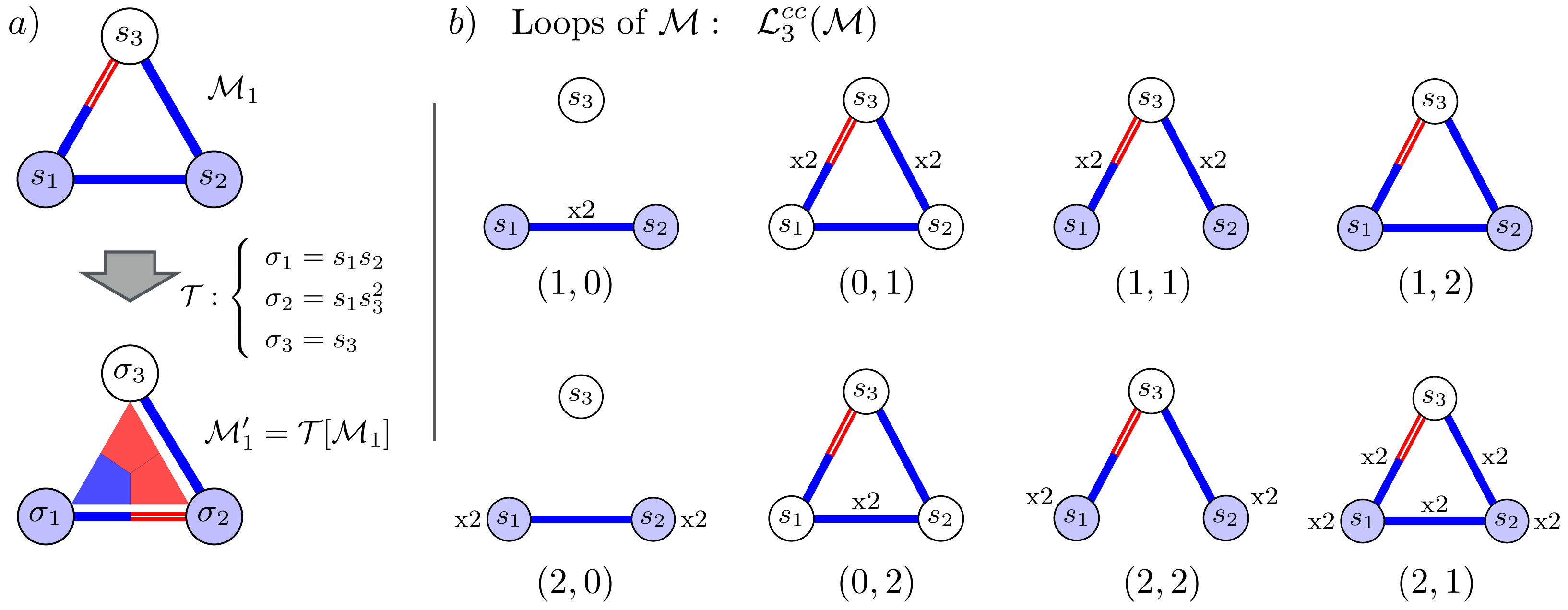} 
\caption{
    \label{fig:GT_and_Loops}
%%%%%%%%%%%%%%%%%%%% GAUGE TRANSFORMATION:
{\bf a) Example of a gauge transformation between spin models for $q=3$.}
    Graphical representation of the 3-spin model
    %\footnote{We use the notation $\tilde{\M}$ for the set of spin operators $\phi^{\vecmu}(\spin)$ of the model and the notation $\M$ for the set of their corresponding vector representations $\vecmu\in\Field^n$.}
    $\Mtilde = \{s_1, s_1^2, s_2, s_2^2, s_1s_2, s_1^2s_2^2, s_1s_3^2, s_1^2s_3, s_2s_3, s_2^2s_3^2\}$ (top) and of the model $\Mtilde^{\prime}=\{\sigma_2\sigma_3,\sigma_2^2\sigma_3^2, \sigma_1\sigma_2^2\sigma_3^2, \sigma_1^2\sigma_2\sigma_3,\sigma_1,\sigma_1^2,\sigma_2,\sigma_2^2,\sigma_1\sigma_2^2, \sigma_1^2\sigma_2\}$ (bottom) obtained from the gauge transformation $\mathcal{T}$ of $\M$. Because these models consist of 5 pairs of conjugate operators, we only represented one operator for each pair, choosing arbitrarily
    $\Mtilde_1=\{s_1, s_2, s_1s_2, s_1s_3^2, s_2s_3\}$ for $\Mtilde$ and $\Mtilde_1^{\prime} = \mathcal{T}[\Mtilde_1]=\{\sigma_2\sigma_3, \sigma_1\sigma_2^2\sigma_3^2,\sigma_1,\sigma_2,\sigma_1\sigma_2^2\}$ for $\Mtilde^{\prime}$.
    Variables are indicated by circles labeled from $1$ to $3$.
    One-body spin operators %(i.e., operators based on a single variable) 
    are represented by colored circles on a single variable, pairwise interactions by lines between two variables, and 3-body interactions by triangles connecting three variables. 
    The colors indicate the power to which a variable is raised within an interaction: blue for~1 and red for~2.
    For example, the operator $s_1s_3^2$ is indicated by a bicolor line, % with 2 colors, 
    blue on the side of $s_1$ and red on the side of $s_3$.
    Observe that the top model is pairwise (i.e., it has only operators up to second order), while the bottom model obtained after GT is not.
    The matrix forms of the two models and of the GT are given in Eq.~\eqref{eq:exFig:GT:M} and~\eqref{eq:exFig:GT:T}.
%
%%%%%%%%%%%%%%%%%%%% LOOP STRUCTURE:
    {\bf b)~Example of the loop structure of a model for $q=3$.} Graphical representation of the loops $\mathcal{L}^{cc}_q$ of the spin model $\M$ in panel a) Top, 
    excluding trivial loops due to c.c.~operators (which are of the type $\vecmu_1+\vecmu_2=0$, and their combinations with other loops).
    Operators are visualized using the same convention as in panel a). In this example, the model can be expressed as $\M=\M_1\cup\M_1^*$, where $\M_1$ is represented in panel a). In this case, the loop $\vecell\in\mathcal{L}^{cc}_q(\M)$ is a vector $\vecell\in\Field^{K_1}$ such that $\sum_{j=1}^{K_1}\ell_j\vecmu_j=\veczero$, where the operators $\vecmu_j\in\M_1$ (see Eq.~\eqref{eq:loop:V3}).
    For each loop $\vecell$, we only visualized the operators $\vecmu_k$ for which the coefficient $\ell_k$ is non-zero, and we indicated with a ``$\times 2$'' the operators for which $\ell_k=2$.
    Every spin model also contains the empty loop $\vecell=\veczero$, which is not represented here. 
    All the loops of the model in $\mathcal{L}^{cc}_q(\M)$ can be obtained from two %independent
    generators (see Sec.~\ref{sec:loop-structure}). 
    Here, we chose the two loops on the top left as generators, represented by the vectors denoted $\vecell_a$ and $\vecell_b$; the other loops are obtained by linear combinations of these two vectors.
    The two-dimensional coordinate vector below each loop diagram indicates which combination of $\vecell_a$ and $\vecell_b$ produced the loop. 
    For instance, the last loop of the first row corresponds to $\vecell_a+2\vecell_b$.
    We organized the diagrams such that  
    each loop in the second row corresponds to the complex conjugate of the loop directly above it (notice that their coordinate vectors are conjugate for $q=3$).
    }
\end{figure}

Figure~\ref{fig:GT_and_Loops}.a gives an example of a spin model for $q=3$ and of a gauge transformation of that model. Observe that the order of the operators is not preserved under GT: the top model has only operators up to order $2$ (i.e., operators connecting at most two variables), whereas the transformed model at the bottom has an operator of order $3$ (and its c.c.), connecting all three variables. In general, the order of an operator is not preserved under GT. In Sec.~\ref{sec:Ppty:EquivalentClasses}, we will discuss properties of spin models that remain invariant under GT.

For convenience, the gauge transformation of a spin model can be written as a matrix product. To do so we first introduce a matrix representation of spin models.
\begin{Definition}\label{def:model:matrix-representation}
    {\bf Matrix representation of spin models.} The matrix representation $\Mmat$ of a spin model $\M=\{\vecmu_1,\,\dots,\,\vecmu_K\}$ is the $n\times K$ matrix over $\Field$, whose $i$-th column is the vector~$\vecmu_i$\,:
    \begin{align}
        \Mmat=(\vecmu_1,\,\dots,\,\vecmu_K)\,.
    \end{align}
\end{Definition}
\noindent Note that the rank of a spin model (see Definition~\ref{def:M:rank}) is equal to the rank of its matrix representation over $\Field$.
%\textcolor{red}{Using the matrix representation of spin models in Definition~\ref{eq:model:matrix-representation}, it corresponds to the rank of the matrix representation of $\M$ in $\Field^n$.}

\begin{Property}\label{eq:model:matrix-representation}
    {\bf Matrix representation of gauge transformed models.} 
    For convenience, the gauge transformation of a spin model
    $\M=\{\vecmu_1,\,\dots,\,\vecmu_K\}$ can be written as a matrix product over $\Field$, by extending Eq.~\eqref{def:eq:GT:Op:2} to multiple operators:
    \begin{align}\label{eq:GT:model:matrix}
        \Mmat'=\Tmat^{-1}_{\basis}\,\Mmat\,,
    \end{align}
    where $\Tmat^{-1}_{\basis}$ is the matrix representation of the inverse gauge transformation $\spin'=\basis^{-1}(\spin)$ (see Definition~\ref{def:GT:inverse}), and where $\Mmat=(\vecmu_1,\,\dots,\,\vecmu_K)$ and $\Mmat'=(\vecmu_1',\,\dots,\,\vecmu_K')$ are two $n\times K$ matrices in $\Field$, whose columns are the vector representations of the operators, respectively in the original model $\M$ and in the new models $\M'$.
    Note that the number $K=|\M|$ of operators in the original model is preserved under gauge transformation.
\end{Property}

The definitions and notations introduced so far for GTs are valid for all $q$-state spin models (with real or complex Hamiltonian). For $q$-state spin models to be probabilistic models, their Hamiltonian must be real, which requires operators to be included by conjugate pairs in the model (see Definition~\ref{eq:def:q:spin-model}). In the following, we discuss further simplifications in this case. 

\subsection{Examples and Gauge transformations of real-valued $q$-state spin models} 
\label{Sec:GT:RealM}

Fig.~\ref{fig:GT_and_Loops}.a gives an example of a gauge transformation for $q=3$ between two spin models with $n=3$ spins. Operators are included by conjugate pairs in these models.
The gauge transformation can be written in the matrix-product form %$\Mmat'=\Tmat^{-1}_{\basis}\,\Mmat$,
of Eq.~\eqref{eq:GT:model:matrix}, where the initial model $\M$ and the new model $\M'$ are respectively represented by the following matrices:
\begin{align}\label{eq:exFig:GT:M}
\begingroup
    \setlength{\arraycolsep}{1.4pt}
    \renewcommand{\arraystretch}{1} 
\Mmat = 
\begin{pNiceArray}{cccccccccc}[first-row,last-col,nullify-dots]
\mu_1 & \mu_2 & \mu_3 & \mu_4 & \mu_5 & \mu_6 & \mu_7 & \mu_8 & \hspace{-0.05mm}\mu_9 & \hspace{-0.8mm}\mu_{10}\hspace{-10mm}&\hspace{-5mm}\\
    1 & \textcolor{gray}{2} & 0 & \textcolor{gray}{0} & 1 & \textcolor{gray}{2} & 1 & \textcolor{gray}{2} & 0 & \textcolor{gray}{0} & s_1\\ 
    0 & \textcolor{gray}{0} & 1 & \textcolor{gray}{2} & 1 & \textcolor{gray}{2} & 0 & \textcolor{gray}{0} & 1 & \textcolor{gray}{2} & s_2\\
    0 & \textcolor{gray}{0} & 0 & \textcolor{gray}{0} & 0 & \textcolor{gray}{0} & 2 & \textcolor{gray}{1} & 1 & \textcolor{gray}{2} & s_3
\end{pNiceArray}
    \;\;\;\;\;\;
\Mmat' = 
\begin{pNiceArray}{cccccccccc}[first-row,last-col,nullify-dots]
\mu_1' & \mu_2' & \mu_3' & \mu_4' & \mu_5' & \mu_6' & \mu_7' & \mu_8' & \hspace{-0.05mm}\mu_9' & \hspace{-0.8mm}\mu_{10}'\hspace{-10mm} &\\
    0 & \textcolor{gray}{0} & 1 & \textcolor{gray}{2} & 1 & \textcolor{gray}{2} & 0 & \textcolor{gray}{0} & 1 & \textcolor{gray}{2} & \sigma_1\\ 
    1 & \textcolor{gray}{2} & 2 & \textcolor{gray}{1} & 0 & \textcolor{gray}{0} & 1 & \textcolor{gray}{2} & 2 & \textcolor{gray}{1} & \sigma_2\\
    1 & \textcolor{gray}{2} & 2 & \textcolor{gray}{1} & 0 & \textcolor{gray}{0} & 0 & \textcolor{gray}{0} & 0 & \textcolor{gray}{0} & \sigma_3
\end{pNiceArray}
\endgroup
\end{align}
and where the gauge transformation~$\mathcal{T}$ and its inverse $\mathcal{T}^{-1}$ are given by the matrices: 
\begin{align}\label{eq:exFig:GT:T}
   \qquad 
    \Tmat = 
    \begin{blockarray}{cccc}
    \sigma_1 & \sigma_2 & \sigma_3 &\\
    \begin{block}{(ccc)c}
    1 & 1 & 0 & s_1\\ 
    1 & 0 & 0 & s_2\\
    0 & 2 & 1 & s_3\\
    \end{block}
    \end{blockarray}
    \qquad\Leftrightarrow\qquad
  \Tmat^{-1}=
  \begin{blockarray}{cccc}
    s_1 & s_2 & s_3 &\\
    \begin{block}{(ccc)c}
    0 & 1 & 0 & \sigma_1\\ 
    1 & 2 & 0 & \sigma_2\\
    1 & 2 & 1 & \sigma_3\\
    \end{block}
  \end{blockarray}
  \;\;.
\end{align}
To help the reader, we added labels to each column and row of the matrices. It is straightforward to check using Eq.~\eqref{eq:GT:model:matrix} that the $j$-th column of the matrix $\Mmat'$ (operator $\vecmu_j'$) is the GT of the $j$-th column of the matrix $\Mmat$ (operator $\vecmu_j$). 
We recall that $\Tmat$ and $\Mmat$ are not acting linearly on the variables $s_j$, but on their exponents $\alpha_j$. 
Here, we labeled rows/columns with $s_j$ instead of~$\alpha_j$ for a more direct connection to Fig.~\ref{fig:GT_and_Loops}.a.
For instance, each column of the matrix $\Tmat$ can be read as encoding the dependency of the new variable $\sigma_j$ in terms of the old variables $\spin=(s_1,s_2,s_3)$ (e.g., the second column corresponds to $\sigma_2=s_1s_3^2$), while the matrix $\Tmat^{-1}$ encodes the reverse (e.g., the second column corresponds to $s_2=\sigma_1\sigma_2^2\sigma_3^2$).
In the matrices $\Mmat$ and $\Mmat'$ above, we highlighted complex conjugate operators with alternating black and gray colors\footnote{%
We recall that operators that have opposite vector representations, $\vecmu$ and $-\vecmu$, in $\Field^n$ are complex conjugate: $\phi^{-\vecmu}(\spin)=[\phi^{\vecmu}(\spin)]^{*}$ (see Definition~\ref{def:spin-operators}).}: 
each column in gray represents the complex conjugate of the previous column in black. For clarity in the models of Fig.~\ref{fig:GT_and_Loops}.a, we represented only one operator for each conjugate pair (the operators in black in the matrices).
Observe that the matrix representation $\Mmat$ of a (high-order) spin model is the incidence matrix of the hypergraph representing the model: each column of the matrix encodes the weighted (hyper)-edge corresponding to one of the operator~$\vecmu$ of the model, with weights given by the $\mu_j$'s.

By default, operators of a real-valued spin model are either their own conjugate or included by conjugate pairs (see Definition~\ref{eq:def:q:spin-model}), as in the examples above. We can thus simplify the representation of a spin model by keeping only one operator for each conjugate pair, both in the graphical representation of the model (as in Fig.~\ref{fig:GT_and_Loops}.a) and in its matrix representation.
For the previous example, this corresponds to replacing the matrix $\Mmat$ and $\Mmat'$ above by the following reduced matrices:
\begin{align}\label{eq:exFig:GT:M1}
\begingroup
    \setlength{\arraycolsep}{1.4pt}
    \renewcommand{\arraystretch}{1} 
\Mmat_1 = 
\begin{pNiceArray}{ccccc}[first-row,last-col,nullify-dots]
\mu_1 & \mu_3 & \mu_5 & \mu_7 & \mu_9&\hspace{-5mm}\\
    1 & 0 & 1 & 1 & 0 & s_1\\ 
    0 & 1 & 1 & 0 & 1 & s_2\\
    0 & 0 & 0 & 2 & 1 & s_3
\end{pNiceArray}
    \qquad{\rm and}\qquad
\Mmat_1' = 
\begin{pNiceArray}{ccccc}[first-row,last-col,nullify-dots]
\mu_1' & \mu_3' & \mu_5' & \mu_7' & \mu_9' &\\
    0 & 1 & 1 & 0 & 1 & \sigma_1\\ 
    1 & 2 & 0 & 1 & 2 & \sigma_2\\
    1 & 2 & 0 & 0 & 0 & \sigma_3
\end{pNiceArray}
    \;\;.
\endgroup
\end{align}
Using the model decomposition $\M=\M_0\cup\M_1\cup\M_1^*$ in Definition~\ref{def:M:M0-M1}, this reduction corresponds to representing only the elements of $\M_0\cup\M_1$, excluding the elements of $\M_1^*$ (in our examples, $\M_0=\emptyset$, which is always the case for $q>2$ and prime).
\begin{Definition}{\bf Reduced representation of real-valued spin models.}
\label{def:reducedrep}
Consider a real-valued $q$-state spin model $\M$. Using the decomposition of Definition~\ref{def:M:M0-M1}, $\M=\M_0\cup\M_1\cup\M_1^*$, we defined the reduced model $\M_r=\M_0\cup\M_1$, such that $\M=\M_r\cup\M_r^*$, where the operators in $\M_r^*$ are the c.c.~of the operators in $\M_r$ (note that the operators of $\M_0$ are both in $\M_r$ and $\M_r^*$).
The reduced matrix representation of $\M$ is the matrix whose column are the operators of $\M_r$. The choice of the reduced representation $\M_r$ for a given model $\M$ is not unique.
\end{Definition}

Besides simplifying the representation of real-valued spin models, the reduced matrix representation can also be used for the %to perform 
gauge transformation of these models, thanks to the following property (see proof in App.~\ref{App:M_r}):
\begin{Property}\label{ppty:GT:cc-op}
    Conjugate operators $(\vecmu,-\vecmu)$ stay conjugate under a GT: $\,\mathcal{T}[-\vecmu]=-\mathcal{T}[\vecmu]$.
\end{Property}
\noindent Therefore, for each conjugate pair $(\vecmu,-\vecmu)$, the GT of the operator non-represented in the reduced model, let's say $-\vecmu$, is the conjugate of the GT of the represented operator~$\vecmu$. 
Manipulations of the model such as gauge transformations can thus be directly performed on the reduced model, and the GT of the full model $\M'=\T[\M]$ can be recovered from the GT of the reduced model $\M_r'=\T[\M_r]$, using $\M'=\M_r'\cup {\M_r'}^*$.
\begin{Property}
    Consider a model $\M=\M_0\cup\M_1\cup\M_1^*$ with reduced representation $\M_r=\M_0\cup\M_1$, and denote $\M'=\T[\M]$ the transformation of $\M$ under a given GT $\T$. Using the reduced representation $\M_r$, the transformed model $\M'$ can be computed as $\M'=\M_r'\cup\M_r'^{\,*}$, where $\M_r'=\T[\M_r]$ is the gauge transformation of the reduced model. 
\end{Property}
\noindent Using the decomposition of Definition~\ref{def:M:M0-M1} for the transformed model, $\M'=\M_0'\cup\M_1'\cup\M_1'^*$, Property~\ref{ppty:GT:cc-op} also implies that $\T[\M_0]=\M_0'$ (i.e., for all $\vecmu_0\in\M_0$, the operator $\vecmu_0'=\T[\vecmu_0]$ is its own conjugate) and $\T[\M_1\cup\M_1']=\M_1'\cup\M_1'^{\,*}$ (see App.~\ref{App:M_r}).

\subsection{Algebraic representation of $q$-state spin models} 
%{Algebraic representation and properties of $q$-state spin models}
%Loop structure of a model: from statistical to algebraic interpretation/representation of spin models}
\label{sec:loop-structure}

As hinted at by the matrix representations introduced above, spin models can be studied from a linear algebra perspective. In this section, we formalize this framework by introducing a linear map representation of spin models.
\begin{Definition}\label{def:linearmap}
    {\bf Linear map representation of spin models.}
    Consider a system of $n$ discrete variables $(\alpha_1, \dots,\alpha_n)\in\Field^n$ and a $q$-state spin model $\M$ for this system, specified by the choice of $K$~operators, $\M=\{\vecmu_1,\dots,\vecmu_K\}\subseteq [\Field^{n}]^*$.
    The model $\M$ can be represented as a linear map $f_{\M}$ from $\Field^K$ %-- the space of operators -- 
    to $\Field^n$ %-- the space of state variables --, 
    which defines each operator of the model in terms of the state variables. 
    We denote $(\boldsymbol{u}_1, \dots,\boldsymbol{u}_K)$ a basis of the domain $\mathcal{U}=\Field^K$, which can be thought of as the model's space; 
    we associate each of these basis elements with one of the $K$~operators of the model~$\M$.
    We denote $(\boldsymbol{v}_1, \dots,\boldsymbol{v}_n)$ a basis of the codomain $\mathcal{V}=\Field^n$, which can be thought of as the space of the spin system, %the space of the system's variables, 
    and associate each of these basis elements with one of the $n$ discrete (color) variables of the system.
    The linear map $f_{\M}$ is defined by specifying the following image for each basis element of $\mathcal{U}$: 
\begin{align}\label{eq:def:f_M}
    \textrm{for all } k\in\{1,\dots,K\}\,,
    \qquad\qquad
    f_{\M}(\boldsymbol{u}_k) 
    %= \sum_{i=1}^n \mu_j^{(i)}\,\alpha_i \,a_i\,, %= \vecmu_j\cdot\state\,,
    = \sum_{j=1}^n \mu_k^{(j)}\,\boldsymbol{v}_j
    = \vecmu_k\,,
    \qquad\qquad
\end{align}
where the coefficient $\mu_k^{(j)}$ corresponds to the $j$-th element of the vector $\vecmu_k\in\Field^n$ defining the $k$-th operator of $\M$. Here the vector $\vecmu_k=(\mu_k^{(1)},\dots,\mu_k^{(n)})$ is considered as an element of the codomain.
\end{Definition}

Following this approach, the matrix representation of the linear map $f_{\M}$ defined in Eq.~\eqref{eq:def:f_M} corresponds precisely to the $n\times K$-matrix $\Mmat$ introduced in Definition~\ref{def:model:matrix-representation} as the matrix representation of the spin model $\M$.
Note that $\Field$ is a vector space when $q$ is prime, it is a module otherwise. The linear map representation is valid in both cases. 
For example, the model $\M$ shown in Fig.~\ref{fig:GT_and_Loops}.a top, with $K=10$ operators over $n=3$ spins, corresponds to the following linear map: %$f_{\M}: \Field^{10}\to\Field^{3}$,
$f_{\M}: (\mathbb{Z}/3\mathbb{Z})^{10}\to(\mathbb{Z}/3\mathbb{Z})^{3}$,
\begin{align}\label{eq:ex:fM}
    \begin{cases}
        \;f_{\M}(\boldsymbol{u}_1)= \vecmu_1=\boldsymbol{v}_1\,;
        \qquad
        &f_{\M}(\boldsymbol{u}_2)= \vecmu_2 = 2\boldsymbol{v}_1\\
        \;f_{\M}(\boldsymbol{u}_3)= \vecmu_3=\boldsymbol{v}_2\,;
        \qquad
        &f_{\M}(\boldsymbol{u}_4)= \vecmu_4 = 2\boldsymbol{v}_2\\
        \;f_{\M}(\boldsymbol{u}_5)= \vecmu_5=\boldsymbol{v}_1+\boldsymbol{v}_2\,;
        \qquad
        &f_{\M}(\boldsymbol{u}_6)= \vecmu_6 = 2\boldsymbol{v}_1+2\boldsymbol{v}_2\\
        \;f_{\M}(\boldsymbol{u}_7)= \vecmu_7=\boldsymbol{v}_1+2\boldsymbol{v}_3\,;\;
        \qquad
        &f_{\M}(\boldsymbol{u}_8)= \vecmu_8 = 2\boldsymbol{v}_1+\boldsymbol{v}_3\\
        %&\qquad\qquad\cdots\\
        \;f_{\M}(\boldsymbol{u}_9)= \vecmu_9=
        \boldsymbol{v}_2 + \boldsymbol{v}_3\,;\qquad\qquad
        &f_{\M}(\boldsymbol{u}_{10})= \vecmu_{10} = 2\boldsymbol{v}_2 + 2\boldsymbol{v}_3
    \end{cases}\,,
\end{align}
whose matrix representation is the same as the matrix representation $\Mmat$ of the spin model given in Eq.~\eqref{eq:exFig:GT:M} (we labeled the operators in the same way in Eq.~\eqref{eq:ex:fM} and~\eqref{eq:exFig:GT:M}).

\begin{figure}[h]
\centering
    \centering 
    \includegraphics[width=1.0\textwidth]{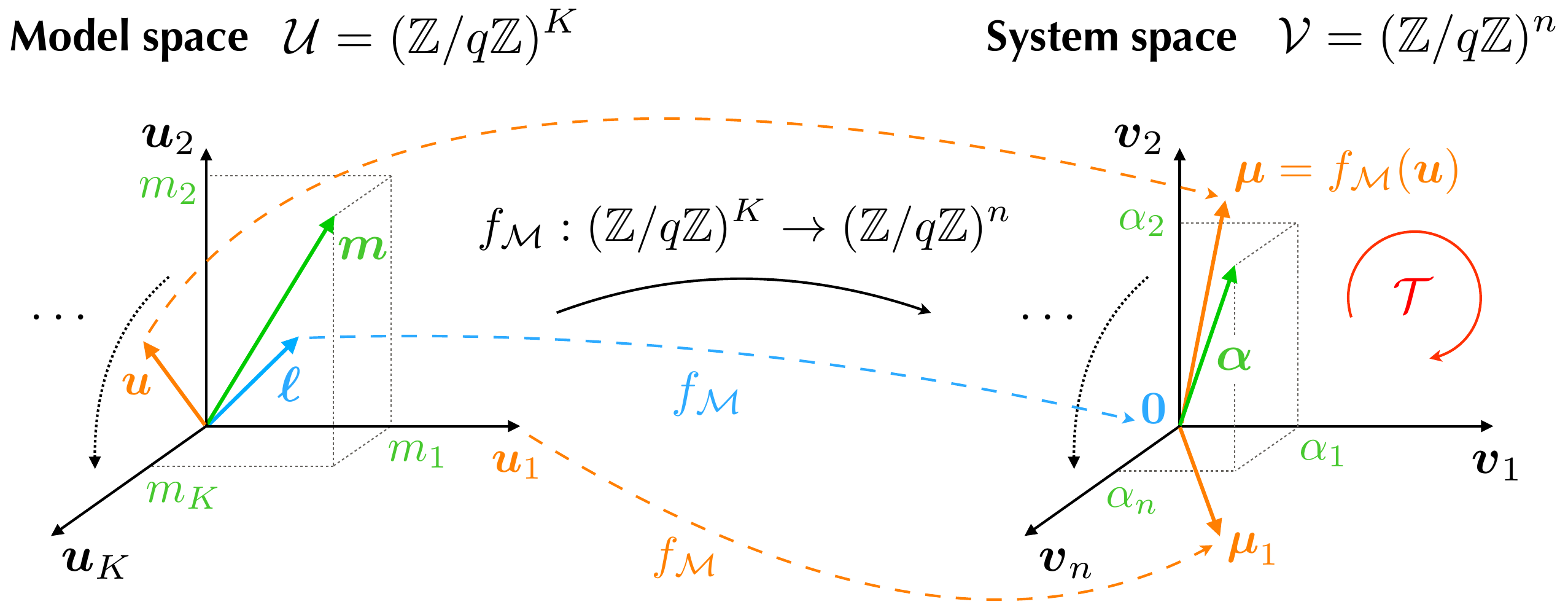} 
\caption{
    \label{fig:LinearMap}
    {\bf Illustration: Representation of $q$-state spin models as linear maps.}
    We consider a system with $n$ $q$-state spin variables and a spin model $\M$ with $K$ operators. 
    The space $\mathcal{V} =\Field^n$ (on the right) is the system space, in which each of the $n$~basis elements is associated with %[the color of] 
    one of the spin variables.
    %in which the $n$~basis elements are associated with the $n$~variables of the spin system. 
    %\textcolor{red}{In $\mathcal{V}$,}
    The state of the system %the $n$~spins
    is then represented by the random vector $\state=(\alpha_1,\dots,\alpha_n)\in\mathcal{V}$, where $\alpha_j$ is the color of the $j$-th spin, and a spin operator is represented as a vector $\vecmu\in\mathcal{V}$, which gives the decomposition of the operator in terms of the variables of the system.
    The space $\mathcal{U}=\Field^K$ (on the left) is the model space, in which each of the $K$~basis elements is associated with one of the operators of the model (which are the variables of the models).
    The state of the model %the $K$~spin operators of the model 
    is then represented by the random vector $\boldsymbol{m}=(m_1,\dots,m_K)\in\mathcal{U}$, where $m_k$ is the color of the $k$-th operator, $\phi^{\boldsymbol{u}_k}(\boldsymbol{m}) = z_q^{m_k}$, and each spin operator that can be decomposed over the operators of the model is represented by a vector $\boldsymbol{u}\in\mathcal{U}$.
    A model~$\M$ is represented by a linear map $f_{\M}$ that associates %maps [to]
    each operator represented by~$\boldsymbol{u}\in\mathcal{U}$ in model space to a representation~$\vecmu=f_{\M}(\boldsymbol{u})\in\mathcal{V}$ in system space. % (in terms of the variables of the system).
    The map $f_{\M}$ is defined in Eq.~\eqref{eq:def:f_M} by specifying the image $\vecmu_k=f_{\M}(\boldsymbol{u}_k)\in\mathcal{V}$ of each basis vector $\boldsymbol{u}_k\in\mathcal{U}$. 
    Gauge transformations~$\T$ are automorphisms of $\mathcal{V}$. Loops of $\M$ correspond to the vectors~$\vecell\in\mathcal{U}$ whose image is the null vector $f_{\M}(\vecell)=\veczero\in\mathcal{V}$.
}
\end{figure}

Figure~\ref{fig:LinearMap} gives an illustration of the linear map representation of $q$-state spin models. 
%\textcolor{red}{In the system space (right-hand side), the state of the spin system is represented by the random vector $\state=\sum_{j=1}^{n} \alpha_j \boldsymbol{v_j}$, corresponding to the coordinate vector $\state=(\alpha_1,\cdots,\alpha_n)\in\mathcal{V}$ (where the $\alpha_i$ are the $n$ discrete random variables).]}
%
The state of the system is characterized by the state of the $n$ discrete random variables~$\alpha_j$ (colors of the spins) and is represented in the system space~$\mathcal{V}$ by the random vector $\state=\sum_{j=1}^{n} \alpha_j \boldsymbol{v}_j$, corresponding to the coordinate vector $\state=(\alpha_1,\cdots,\alpha_n)$.
The state of the model is characterized by the state of the $K$ operators of the model. %which we characterize by the color variables $m_k$
We denote by $m_k\in\Field$ the color of the $k$-th spin operator %(such that $\phi^{\vecmu_k}(\state)=z_q^{m_k}$).}
(the state of this operator is thus~$z_q^{m_k}$).
With this notation, %This way, 
the state of the model is then described by %the state of 
the $K$ discrete random variables~$m_k$ %(colors of the operators of the model) 
and is represented in the model space~$\mathcal{U}$ by the random vector $\boldsymbol{m}=\sum_{k=1}^{K} m_k \boldsymbol{u}_k=(m_1,\cdots,m_K)$.
Using the linear map~$f_{\M}$, each operator $\boldsymbol{u}_k\in\mathcal{U}$ of the model~$\M$ is given a representation $\vecmu_k\in\mathcal{V}$ in the system space. 
The state $m_k$ of the $k$-th operator when the system is in a state $\state$ is then given by the dot product $m_k=\state\cdot\vecmu_k$ in~$\mathcal{V}$ (because $\phi^{\vecmu_k}(\state)=z_q^{\state\cdot\vecmu_k}=z_q^{m_k}$).
We can thus obtain the state of the model $\boldsymbol{m}\in\mathcal{U}$ when the system is in the state $\state\in\mathcal{V}$ by the vector-matrix product $\boldsymbol{m}=\state\,\Mmat$ over $\Field$, where $\state$ is considered as a row vector and $\Mmat$ is the matrix representation of $f_{\M}$.
%%%%%%%%%%%%%%%% %%%%%%%%%%%%%%%% %%%%%%%%%%%%%%%%
%Using this framework, we can also represent the state of all the operators of the model when the system is in a given state $\state\in\mathcal{V}$ by the coordinate vector $\boldsymbol{m}=(m_1, \dots,m_K)\in\mathcal{U}$, where the state $m_k$ of the $k$-th operator is given by the dot product $m_k=\state\cdot\vecmu_k$ in~$\mathcal{V}$. The vector $\boldsymbol{m}$ can thus be obtained by the vector-matrix product $\boldsymbol{m}=\state\,\Mmat$ over $\Field$, where $\state$ is considered as a row vector and $\Mmat$ is the matrix representation of $f_{\M}$.
%%%%%%%%%%%%%%%% %%%%%%%%%%%%%%%% %%%%%%%%%%%%%%%%

In an $n$-spin system, all the statistical properties of the system are captured by the ensemble averages of each the $(q^n-1)$ spin operators (i.e., all the high-order moments). 
In this framework, each spin operator whose statistics is constrained by the model has two representations: one in the system space~$\mathcal{V}$ and one in the model space~$\mathcal{U}$.
%%%%%%%%%%%%%%%%%%%%% System space:
{\bf In the system space}, any spin operator can be written as a product of spin variables (monomials). This product corresponds to a linear combination of the color variables $\alpha_i$, which is represented by a vector $\vecmu\in\mathcal{V}$. The random variable~$\state\cdot\vecmu\in\Field$, obtained from the dot product of $\vecmu$ with the random state vector~$\state$, then characterizes the state of the operator $\phi^{\vecmu}(\state)=z_q^{\state\cdot\vecmu}$ when the system is in the state~$\state$.
%%%%%%%%%%%%%%%%%%%%% Model space:
%Vectors in the model space represent operators that can be written as products over the operators of the models.
{\bf In the model space}, any spin operator that can be written as a product of operators of the model (monomials over the operators) is represented as a vector $\boldsymbol{u}=\sum_{k}u_k\boldsymbol{u}_k=(u_1,\dots,u_K)\in\mathcal{U}$.
For example, the vector $\boldsymbol{u}=\boldsymbol{u}_1+2\boldsymbol{u}_2 \in\mathcal{U}$ represents the spin operator $\phi^{\boldsymbol{u}}(\boldsymbol{m}) =\phi^{\boldsymbol{u_1}}(\boldsymbol{m})[\phi^{\boldsymbol{u_2}}(\boldsymbol{m})]^2$, where the operators $\phi^{\boldsymbol{u_k}}(\boldsymbol{m})$ are the operators of the model. %(represented in terms of a product of spin operators $\phi^{\boldsymbol{u_k}}(\boldsymbol{m})$ of the model).
The random variable~$\boldsymbol{m}\cdot\boldsymbol{u}\in\Field$, obtained from the dot product of $\boldsymbol{u}$ with the random state vector~$\boldsymbol{m}$, then characterizes the state of the operator $\phi^{\boldsymbol{u}}(\boldsymbol{m})=z_q^{\boldsymbol{u}\cdot\boldsymbol{m}}$ when the model is in the state~$\boldsymbol{m}$.
%
%%%%%%%%%%%%%%%%%%%%% Applying f_M:
{\bf For a chosen model~$\M$}, the linear map $f_{\M}$ associates each operator represented by $\boldsymbol{u}=\sum_{k=1}^K u_k\boldsymbol{u}_k\in\mathcal{U}$ in model space to a representation $\vecmu=f_{\M}(\boldsymbol{u}) = \sum_{k=1}^K u_k \vecmu_k\in\mathcal{V}$ in the system space. 
%we obtained the expression of the operator in the system space in terms of the variables $\state$.
The state of the spin operator can then be equivalently written in model space as $\boldsymbol{m}\cdot\boldsymbol{u}=\sum_{k}m_k u_k$ and in system space as $\state\cdot\vecmu=\sum_{k}u_k \state\cdot\vecmu_k=\sum_{k}u_k m_k$ (i.e., $\phi^{\boldsymbol{u}}(\boldsymbol{m})=\phi^{f_{\M}(\boldsymbol{u})}(\state)$).

%%%%%%%%%%%%%%%%%%%%%
%For example, the vector $\boldsymbol{u}=\boldsymbol{u}_1+2\boldsymbol{u}_2 \in\mathcal{U}$ represents the spin operator $\phi^{f_{\M}(\boldsymbol{u})}(\state)$, corresponding to the vector $f_{\M}(\boldsymbol{u})\in\mathcal{V}$. Applying $f_{\M}$, one obtains the expression of this operator in terms of the [state] variables~$\state$, i.e., $\phi^{f_{\M}(\boldsymbol{u})}(\state)=\phi^{\vecmu_1+2\vecmu_2}(\state)$.
%

%%%%%%%%%%%%%%%%%
%All the operators whose statistics are controlled by the model are represented by a vector in the operator space. 
%Operator space can be seen as the space of all the operator statistics that can controlled by the model.
%%%%%%%%%%%%%%%%%
If the rank of the model is smaller than $n$, then not all spin operators in~$\mathcal{V}$ have a representation in the model space~$\mathcal{U}$. The operators that can be decomposed in $\mathcal{U}$ are those whose statistical properties are controlled by the statistics of the operators of the model (namely the relevant statistics). %observables
%These are the operators whose statistics (statistical properties) in the model probability distribution are controlled by the model (i.e they depend on the statistics of the operators of the models. 
The other operators are given a uniform distribution in the model% 
\footnote{For example, for the model $\M=\{\phi^{\vecmu_1}(\spin)=s_1,\, \phi^{\vecmu_2}(\spin)=s_1s_2s_3\}$ with $q=2$, the spin operator $\phi^{\vecmu}(\spin)=s_2$ cannot be obtained %written
as a product of operators of the model, but the operator $\phi^{\vecmu}(\spin)=s_2s_3$ can. 
}.
Different operators $\boldsymbol{u}\in\mathcal{U}$ can be mapped by $f_{\M}$ to the same operator $\vecmu\in\mathcal{V}$, which means that the same spin operator (expressed in terms of the system's variables~$\state$) can be decomposed in different ways over the operators of the model\footnote{For example, for the model $\M=\{\phi^{\vecmu_1}(\spin)=s_1,\, \phi^{\vecmu_2}(\spin)=s_2,\, \phi^{\vecmu_3}(\spin)=s_3,\, \phi^{\vecmu_4}(\spin)=s_1s_2s_3\}$ with $q=2$, the operator $\phi^{\vecmu}(\spin)=s_1s_2$ can be decomposed as $\phi^{\vecmu}(\spin)=\phi^{\vecmu_1}(\spin)\phi^{\vecmu_2}(\spin)$ or as $\phi^{\vecmu}(\spin)=\phi^{\vecmu_3}(\spin)\phi^{\vecmu_4}(\spin)$.}.
Finally, the operators $\boldsymbol{u}\in\mathcal{U}$ that are mapped to $\veczero\in\mathcal{V}$ correspond to the loops of the model (because they identify products of operators of the model that are equal to identity in the system space).

Gauge transformations are changes of basis of the state space of the system, which, in this framework, correspond to automorphisms of the codomain $\mathcal{V}=\Field^n$ (see Fig.~\ref{fig:LinearMap}). These automorphisms can be represented by invertible $n\times n$-matrices over $\Field$ and the GT of a model is then obtained by the matrix product previously described in property~\ref{eq:model:matrix-representation}.
%precisely by the matrix product previously described in Eq.~

\begin{Definition}
    {\bf Gauge transformations.} GTs are automorphisms of the codomain of the linear map $f_{\M}$ representing the spin model $\M$.
\end{Definition}

%\begin{Definition}
%Using the ``state'' vector $\state=(\alpha_1,\dots,\alpha_n)\in\Field$, this definition can also be rewritten as the following scalar product in $\Field^n$: $f_{\M}(u_j) = \vecmu_j\cdot\state$.
%\end{Definition}

Using this framework, we can identify several properties of linear maps that are useful for characterizing spin models, such as the rank of the map and the properties of its kernel.
The notion of rank of a spin model $\M$ was already introduced in Definition~\ref{def:M:rank} %Sec.~\ref{sec:IM:strong-indep} 
as the cardinality of the largest %/maximally}
linearly independent set of operators in $\M$. This is the same as the rank of the matrix representation of $\M$, and %which therefore \textcolor{red}{corresponds to/coincides with} 
therefore it %this notion of rank 
coincides with the rank of the linear map associated with $\M$.
The kernel of a linear map is the set of all the vectors in the domain that are mapped to %/whose image is}
the zero vector in the codomain (as illustrated by the vector~$\boldsymbol{\ell}$ in Fig.~\ref{fig:LinearMap}). 
For example, 
%for the map above, the vector $\boldsymbol{u}_1+\boldsymbol{u}_3+2\boldsymbol{u}_5$ is in the kernel of $f_{\M}$ as its image is the zero vector: 
the vector $\boldsymbol{u}_1+\boldsymbol{u}_3+2\boldsymbol{u}_5$ is in the kernel of the map $f_{\M}$ defined in Eq.~\eqref{eq:ex:fM}, as its image is the zero vector: 
\begin{align}\label{eq:loop:ex1}
    f_{\M}(\boldsymbol{u}_1+\boldsymbol{u}_3+2\boldsymbol{u}_5)
        &=\vecmu_1+\vecmu_3+2\vecmu_5 = \boldsymbol{v}_1+\boldsymbol{v}_2+2(\boldsymbol{v}_1+\boldsymbol{v}_2)=\veczero\,.
\end{align}
Here, we can recognize one of the loops of $\M$, which is the first loop represented in Fig.~\ref{fig:GT_and_Loops}.b (top left loop). We recall that a loop of a model with operators $\{\vecmu_k\}_{1\leq k\leq K}$ is a vector $\vecell\in\Field^K$ that verifies that $\sum_{k=1}^{K}\ell_k \vecmu_k=\veczero$, where $\ell_k$ denotes the $k$-th coefficient of $\vecell$ (see Eq.~\eqref{eq:loop:V2}). Labeling the operators in the same order as in Eq.~\eqref{eq:ex:fM}, the loop above can be represented as %corresponds to 
the coordinate vector $\vecell=\boldsymbol{u}_1+\boldsymbol{u}_3+2\boldsymbol{u}_5=(1,0,1,0,2,0,0,0,0,0)$ %\in\mathcal{U}$. 
in~$\mathcal{U}$.
More generally, the loops of a model $\M$ are precisely the elements of the kernel of $f_{\M}$; they identify dependence relations in $\mathcal{V}$ between the operators $\vecmu_k$ of $\M$, i.e., between the images of the basis elements $\boldsymbol{u}_k$. Let's provide a more formal proof of this last point.
%The loop structure of a spin model in Eq.~\eqref{eq:loop:V2} corresponds to the \textcolor{red}{dependence relations} between its operators, which can be recognized as the kernel of the linear map associated with the model.
%For instance, Fig.~\ref{fig:GT_and_Loops}.b 

\begin{proof} %[Proof: {\rm of the last point above}:] 
Consider a spin model $\M=\{\vecmu_1,\dots,\vecmu_K\}\subseteq[\Field^n]^*$ with $K$ operators. We recall that the set of loops of $\M$ %, denoted \mathcal{L}_q(\M),
is defined in Eq.~\eqref{eq:loop:V2} as: %in Eq.~\eqref{eq:loop:V2}.
\begin{align}\label{eq:loop:V2:bis}
    \mathcal{L}_q(\M) = \bigg\{\vecell\in (\mathbb{Z}/q\mathbb{Z})^K\;\Big|\; \sum_{k=1}^K \ell_k\,\vecmu_k = \veczero\bigg\}\,,
\end{align}
where $\ell_k$ denotes the $k$-th coefficient of $\vecell$.
Using the matrix representation of the model $\Mmat = (\vecmu_1,\dots,\vecmu_K)$, in which the vectors $\vecmu_k$ are the columns of the matrix (see Definition~\ref{def:model:matrix-representation}),
and viewing $\vecell$ as a column vector,
%Let us define the matrix $M\in {\rm mat}_{n\times K}(\mathbb{Z}/q\mathbb{Z})$ whose columns are the elements of $\M$ (i.e. the $i$-th column of $M$ is $\vecmu_i$). 
we can rewrite $\mathcal{L}_q(\M)$ in the following form: %in the following way
\begin{align}\label{eq:proof:loop:kernel}
    \mathcal{L}_q(\M) 
        = \Big\{\vecell\in (\mathbb{Z}/q\mathbb{Z})^K\;\Big|\; \Mmat \,\vecell = \veczero\Big\}
        %= \bigg\{\vecell\in (\mathbb{Z}/q\mathbb{Z})^K\;\Big|\; \Mmat^{T}\vecell = \veczero\bigg\} 
        = \ker(\Mmat)\;.
\end{align}
%in which $\vecell$ is a column vector.
%This kernel is invariant under any bijection of the spin variables that corresponds to multiplying the matrix $M$ by an invertible matrix in $\Field$. These transformations correspond to the automorphisms of $\Field^n$. 
Because $\Mmat$ is also the matrix representation of the linear map $f_{\M}$ associated with the model~$\M$, Eq.~\eqref{eq:proof:loop:kernel} above means that the set of loops of $\M$ corresponds to % is [equal to]
the kernel of~$f_{\M}$. % corresponding to
\end{proof} 
%
%\noindent In summary:
\begin{Property}\label{ppty:linear-map}
    %{\bf Spin models inherit properties of the corresponding linear map:} % of its linear map representation
    %the rank of a model is the rank of the map and the set of loops (called loop structure) of a model corresponds to the kernel of the map. 
    Spin models $\M$ inherit properties of their linear map representation $f_{\M}$ in Def.~\ref{def:linearmap}:
    \begin{itemize}
        \item[--] the matrix representation of a model in Definition~\ref{def:model:matrix-representation} %$\M$ corresponds to 
        is the matrix representation of the map; % $f_{\M}$;
        \item[--] the rank of a model in Definition~\ref{def:M:rank} is the rank of the map, $\rank(\M)=\rank(f_{\M})$;
        \item[--] the set of loops of a model defined in Eq.~\eqref{eq:loop:V2} is %corresponds to 
        the kernel of the map, $\mathcal{L}_q(\M)=\ker(f_{\M})$. 
    \end{itemize}
\end{Property}

Figure~\ref{fig:GT_and_Loops}.b shows the loops of the spin model $\M$ taken as an example above. In this model, operators are included by conjugate pairs, just like for real-valued spin models (see Definition~\ref{eq:def:q:spin-model}). For clarity in Fig.~\ref{fig:GT_and_Loops}, we only displayed the elements of the reduced set of loops $\mathcal{L}^{cc}_q(\M)$ defined for real-valued spin models in Eq.~\eqref{eq:loop:V3}; in short, we didn't represent trivial loops resulting from complex conjugate operators (of the type $\vecmu_i+\vecmu_j=\veczero$) nor their combinations with other loops. 
For example, the first loop in the top left of the figure corresponds to $\vecmu_1+\vecmu_3+2\vecmu_5 = \boldsymbol{v}_1+\boldsymbol{v}_2+2(\boldsymbol{v}_1+\boldsymbol{v}_2)=\veczero$ in $(\mathbb{Z}/3\mathbb{Z})^n$. 
Considering all linear combinations of this loop with the trivial loop $\vecmu_1+\vecmu_2=\veczero$ from the c.c. operators $\vecmu_1$ and $\vecmu_2$, we also get the following loops:
$2\vecmu_1+\vecmu_2+\vecmu_3+2\vecmu_5=\veczero$, $2\vecmu_2+\vecmu_3+2\vecmu_5=\veczero$, and $\vecmu_2+2\vecmu_3+\vecmu_5=\veczero$, which are not represented %/shown 
in Fig.~\ref{fig:GT_and_Loops}.b.
Note that the choice of the loops in $\mathcal{L}^{cc}_q(\M)$ is not unique, but it is fully determined by the choice of the operators in the reduced model representation $\M_r$. 
%%%%%%% maybe, should be kept:
%The loop expansion in~\eqref{eq:Z:loop-expansion:V3} is independent of the choice of $\M_r$.
%%%%%%%%%%%%%%%%%%%%%%%%%%

%\textcolor{red}{In fact,} 
The reduced loop structure $\mathcal{L}_q^{cc}(\M)$ of a real-valued spin model can be also related to the kernel of the model's reduced representation (see Definition~\ref{def:reducedrep}).
More precisely, for real-valued spin models that can be written under the form\footnote{This is always the case when $q$ is prime and larger than $2$.} $\M=\M_1\cup\M_1^*$ with the decomposition of Definition~\ref{def:M:M0-M1}
(i.e., for which $\M_0=\emptyset$ -- such as the model taken as an example above), the reduced set of loops of $\M$, excluding trivial loops due to c.c.~operators, is equal to the kernel of its reduced matrix representation $\Mmat_1$: $\mathcal{L}_q^{cc}(\M) = \mathcal{L}_q(\M_1) =\ker(\Mmat_1)$, where $\Mmat_1$ is a matrix whose columns are the operators in $\M_1$ (see proof in App.~\ref{App:M_r}). 

%\begin{Property}
%    For real-valued spin models with $q>2$ and prime, the reduced loop structure $\mathcal{L}_q^{cc}$ of a spin model $\M$ is equal to the kernel of its reduced matrix $\Mmat_1$ defined in Definition~\ref{def:reducedrep}: $\mathcal{L}_q^{cc}(\M) = \ker(\Mmat_1)$. 
%\end{Property}

Finally, because $\mathcal{L}_q(\M)=\ker(f_{\M})$, the loop structure of a spin model inherits the properties of kernels of linear maps. 
%In particular, the kernel of $f_{\M}$ is a linear subspace of $\mathcal{U}=\Field^K$. [This subspace is a vector subspace if $q$ is prime, a sub-module otherwise.] If $q$ is prime, it is a vector subspace of dimension $\dim(\ker(f_{\M}))=K-\rank(f_{\M})$, where $K=|\M|$; we denote this dimension $\lambda$. 
In particular if $q$ is prime, the kernel of $f_{\M}$ is a vector subspace of $\mathcal{U}=\Field^K$ of dimension $\dim(\ker(f_{\M}))=K-\rank(f_{\M})$, where $K=|\M|$; we denote this dimension $\lambda$. 
Consequently, all the elements of $\ker(f_{\M})$ can be generated by a set of $\lambda$ linearly independent vectors in $\ker(f_{\M})$, and the cardinality of $\ker(f_{\M})$ is then~$q^{\lambda}$.
%The same results apply to the set of loops $\mathcal{L}_q(\M)$, because $\mathcal{L}_q(\M)=\ker(f_{\M})$. As a consequence, the number of loops in $\mathcal{L}_q(\M)$ is equal to: $|\mathcal{L}_q(\M)|=q^{\lambda}$.
This generalizes to $q>2$ and prime the result found in the binary case~\cite{beretta2018stochastic} that the set of loops of a binary spin model has the structure of an Abelian group generated by $\lambda$ independent loops, where $\lambda=K-\rank(\M)$ with $K=|\M|$.

\begin{Property}
    For $q$ prime, the set of loops $\mathcal{L}_q(\M)$ of a spin model $\M$ with $K=|\M|$ operators is a vector subspace %linear subspace 
    of $\Field^K$ of dimension $\lambda=K-\rank(\M)$. It is generated by $\lambda$ independent loops, and its cardinality is $|\mathcal{L}_q(\M)|=q^{\lambda}$.
\end{Property}

For real-valued spin models with $q$ prime, the model can be written under the form $\M=\M_1\cup\M_1^*$ ($\M_0=\emptyset$ for $q$ prime), where $\M_1$ is the reduce representation~$\M$.
%of the form $\M=\M_1\cup\M_1^*$ (which is always the case for $q$ prime),
Similarly, the reduced loop structure $\mathcal{L}^{cc}_q(\M)$  can be written as the kernel of the reduced model $\Mmat_1$, which is also a vector %linear 
subspace of $\mathcal{U}$ of dimension $\lambda_1=\dim(\ker(\Mmat_1))=|\M_1|-\rank(\M_1)$.
All the loops of $\mathcal{L}^{cc}_q(\M)$ can be generated by $\lambda_1$ independent loops and the number of loops in $\mathcal{L}^{cc}_q(\M)$ is thus $|\mathcal{L}^{cc}_q(\M_1)|=q^{\lambda_1}$.
For example, the model $\M$ in Fig.~\ref{fig:GT_and_Loops} has nine loops in the reduced set $\mathcal{L}^{cc}_3(\M)$ (including the empty loop and excluding trivial loops due to c.c.~operators). They are all generated by two independent loops, because $\lambda_1=|\M_1|-\rank(\M_1)=2$. 
Indexing the operators of $\M_1$ in the same order as in Eq.~\eqref{eq:exFig:GT:M1}, the two top left loops in the first row of Fig.~\ref{fig:GT_and_Loops}.b correspond to the two vectors
$\vecell_a=(1,1,2,0,0)$ and $\vecell_b=(0,0,1,2,2)$
in $\mathcal{U}$ reduced to the $|\M_1|=5$ elements of $\M_1$
(i.e., the $j$-th entry of $\vecell$ corresponds to the coefficient in front of the $j$-th operator of $\M_1$ in the definition of $\mathcal{L}_q^{cc}$ in Eq.~\eqref{eq:loop:V3}). All the other loops of $\M_1$ can be obtained by linear combinations of these two loops, which are identified %represented/indicated
by the two-dimensional vectors in $\Field^{\lambda_1}$ shown below each loop diagram in the figure. For example, in the first row, the third loop corresponds to $\vecell=\vecell_a+\vecell_b=(1,1,0,2,2)$ and the fourth to $\vecell=\vecell_a+2\vecell_b=(1,1,1,1,1)$.

%%%%%%%%%%%%%%%%%%% QUESTION
%- Introduce a matrix representation of the Kernel?
%%%%%%%%%%%%%%%%%%%
%\begin{itemize}
    %\item Highlight complex conjugate loops, which didn't exist in the binary case. For real $q$-state models (as oppose to complex models), each operator as its c.c. also in the model (could be the operators itself, or another operator) and each loop has a c.c. loop. Note that this set contains loops that are c.c. to each other. 
%\end{itemize}

%%%%%%%%%%%%%%%%%%%
%{\bf From probabilistic models to linear algebra.} 
%{\bf From algebraic to statistical properties of spin models.}
%{\bf From probabilistic to algebraic properties of spin models.}
%%%%%%%%%%%%%%%%%%%%%%%%%%%%%%%%%%%%%%

In summary, the algebraic representation of $q$-state spin models in terms of linear maps not only makes it easier to manipulate spin models (via vector-matrix operations), it also allows for a better characterization of their properties.
We saw in Sec.~\ref{sec:2a_LoopExpansion} that the loop expansion of the partition function $Z_{\M}$ of spin models makes a connection between the statistical properties of a model (captured by $Z_{\M}$) and its algebraic properties corresponding to linear dependencies between its operators (captured by the loop structure). The framework introduced here formalizes the latter as the kernel of the linear map representing the model, thus providing a formal algebraic characterization of the statistical properties of  $q$-state spin models.
%[The statistical properties of a $q$-state spin model are fully characterized by the properties of the kernel of the linear map associated to the models.]

\subsection{Equivalence classes of spin models, and their statistical and algebraic properties} %invariants
\label{sec:Ppty:EquivalentClasses}

In the binary case ($q=2$), Ref.~\cite{beretta2018stochastic} introduces the notion of equivalence classes of spin models, as classes of models related by GTs. 
They found that these models have the same partition function %up to permutation of their parameters 
and thus define the same statistical structure, only observed in different bases. In other words, models of the same class can be seen as different representations of the same abstract statistical model. In particular, they share common statistical and algebraic properties: they have the same loop structure, partition function, and model information-theoretic complexity. 
Ref.~\cite{beretta2018stochastic} calls these classes ``complexity classes''. Let us discuss these results in the more general context of $q$-state spin models with $q\geq 2$.

Consider a real-valued $q$-state spin model $\M$. A gauge transformation $\mathcal{T}_{\basis}$ of $\M$ is a bijection of the state space that rearranges the energy landscape of the model, assigning the energy of each state $\spin$ in $\M$ to another (unique) state $\spin'=\basis(\spin)$ in the transformed model $\M'=\mathcal{T}_{\basis}[\M]$, as described in Definition~\ref{def:GT_M}. As a result, the probability\footnote{We recall that for real-valued spin models, the probability to observe a state $\spin$ is proportional to exponential minus the energy $H(\spin)$ of the state: $p(\spin)\propto \exp(-H(\spin))$.} 
of finding the system in the state~$\spin$ in the model $\M$ is equal to the probability of finding the system in the transformed state $\spin^\prime = \basis(\spin)$ in the transformed model $\M^\prime = \mathcal{T}_{\basis}[\M]$:
\begin{align}
\label{GT:def_on_model}
p(\spin\,|\,\mathcal{M},\vecg)
    =p(\spin^{\prime}\,|\,\mathcal{M}^{\prime},\vecg^{\prime})\,.
    %=p\left(\basis(\spin)\,|\,\mathcal{T}_{\basis}[\vecg], \mathcal{T}_{\basis}[\M]\right)\,,
\end{align}
Here, the vector of parameters $\vecg^\prime=\mathcal{T}_{\basis}[\vecg]$ is a permutation of the parameters $\vecg$ of $\M$ such that each transformed operator $\phi^{\vecmu'}= \mathcal{T}_{\basis}[\phi^{\vecmu}]$ in $\M'$ is parameterized by the same parameter as the operator $\phi^{\vecmu}$ in $\M$ (see definition~\ref{def:GT_M}).  
In equation~\eqref{GT:def_on_model}, the reassignment in state space created by the GT shuffles the state probabilities in the model, but preserves the overall shape of the distribution.
In other words, Eq.~\eqref{GT:def_on_model} expresses that the probability distributions of model $\M$ and of its gauge transformation $\M'$ are the same up to permutation of the states.

More precisely, one can verify that the transformed spin model $\M'$ has the same statistical properties as the original model $\M$ by comparing their partition functions. 
%%%%%%%%%%%% %%%%%%%%%%%% %%%%%%%%%%%% %%%%%%%%%%%% 
%[The partition function of a spin model indeed carries all the information about its statistical properties.]
%%%%%%%%%%%% %%%%%%%%%%%% %%%%%%%%%%%% %%%%%%%%%%%% 
%It is computed by summing the exponential of minus the energy of the state over all possible states of the system.
%
The partition function~\eqref{eq:Z:Ising:def} of a spin model is obtained by summing, over all possible states $\spin$ of the spin system, a term that depends only on the energy of the state (exponential of minus the energy of the state).
This sum thus remains unchanged under the one-to-one state reordering of the energy landscape created by a GT (see proof in App.~\ref{App:GT}, which is also valid for spin models with complex-valued Hamiltonian and partition function).
In other words, just like in the binary case~\cite{beretta2018stochastic}, $q$-state spin models related by GTs have the same partition function. Because the partition function encapsulates %captures
all the statistical properties of a spin model, this means that models related by GTs describe %have
the same statistical structure, only represented in different bases.

From a statistical modeling perspective, the Fisher information matrix (when it is defined) is in general the mathematical object that captures all the statistical properties of a %\textcolor{red}{[general]} 
parametric statistical model %in general 
(not just models that have a partition function, such as %like 
spin models). In particular, in the context of information geometry, it can be used as a metric for the model manifold to study the statistical properties of the model~\cite{rao1945information, amari2000methods}. %amari2016information
For spin models, the Fisher information matrix is given by the Hessian of the log-partition function: 
\begin{align}\label{def:FIM}
    [\,\FIM_{\M}(\vecg)\,]_{\vecmu,\vecnu} = -\left\langle \partial_{g_{\vecmu}}\partial_{g_{\vecnu}} \log p(\spin\,|\,\M,\,\vecg) \right\rangle_{p(\spin|\vecg,\M)}=\partial_{g_{\vecmu}}\partial_{g_{\vecnu}} \log Z_{\M}(\vecg)\,,
\end{align}
%in which the partial derivations are over the parameters.
and is therefore identical for models related by GTs (assuming the parameters are labeled consistently between %in the same order
the models, as explained for Eq.~\eqref{GT:def_on_model}). 
%Spin models related by GTs thus correspond to the same abstract statistical model, observed in different basis.
This means that spin models related by GTs correspond to the same statistical manifold (up to permutation of the parameters).
Therefore, also from this geometrical perspective, models related by GTs are just different representations of the same abstract statistical model.

\begin{Property}\label{ppty:class:invariants-statistical}
    {\bf Statistical properties of spin models invariant under GTs.}
    $q$-state spin models related by gauge transformations have the same partition function (for a model~$\M$ and GT~$\T$, $Z_{\M}(\vecg)=Z_{\T[\M]}(\vecg)$\,).
    For real-valued spin models, which can be used as probabilistic models in the context of statistical inference, this implies that they have the same Fisher information matrix. %(up to permutation of the indices).
    In other words, they describe the same statistical structure represented in different spin bases. 
%    $Z_{\M}(\vecg)=Z_{\T[\M]}(\vecg)$, where $\vecg'=\T[\vecg]$ is only a permutation of the parameters in $\vecg$
\end{Property}

More generally, the invariance of the partition function under gauge transformations is valid for all $q$-state spin models (not just real-valued models). 
This is straightforward to see 
using the loop expansion of the partition function in Sec.~\ref{sec:2a_LoopExpansion} and the linear algebra framework introduced previously, in which a spin model $\M$ is associated to a linear map~$f_{\M}$.
Based on the loop expansion~\eqref{eq:Z:loop-expansion:V2}, 
the partition of a spin model is fully specified by its number of operators (each parameterized by a different parameter) and its loop structure. 
The number of operators %$K=|\M|$ 
of a model is invariant under GT (by definition~\ref{def:GT_M}), 
while the loop structure has been identified as %corresponds to
the kernel of the linear map, $\mathcal{L}_q(\M) = \ker(f_{\M})$. 
Because a GT is an automorphism of the co-domain of $f_{\M}$ (change of basis of the space of state variables -- see Fig.~\ref{fig:LinearMap}), it has no effect on the kernel of $f_{\M}$. %In other words, a spin model and its gauge-transformed model have the same loop structure. 
This can be written down explicitly using the matrix representations $\Mmat$ of $f_{\M}$ (and of the model~$\M$) %of $f_{\M}$ %(which is also the matrix representation of $\M$) 
and $\Tmat$ of a given GT (which is an $n\times n$ invertible matrix in $\Field$). The transformed model is represented by the matrix $\Mmat' =\Tmat^{-1}\Mmat$ and its loop structure is given by:
\begin{align}
    \mathcal{L}_q(\M') 
        = \Big\{\vecell\in (\mathbb{Z}/q\mathbb{Z})^K\;\Big|\; \Tmat^{-1}\Mmat \,\vecell = \veczero\Big\}
        = \Big\{\vecell\in (\mathbb{Z}/q\mathbb{Z})^K\;\Big|\; \Mmat \,\vecell = \Tmat\,\veczero\Big\} 
    =\mathcal{L}_q(\M)\;,
        \nonumber
\end{align}
where we multiplied by the matrix $\Tmat$ on both sides of the equation defining the set.
% where we multiplied the equation $\Tmat^{-1}\Mmat \,\vecell = \veczero$ by $\Tmat$ on both sides of the equality.
%
%This corresponds to multiplying the model matrix $\Mmat$ by an invertible $n\times n$-matrix in $\Field$ (representing an element of $Aut(\Field^n)$), as described in Ppty~\ref{eq:model:matrix-representation}. 
%Such operation produces the new model matrix $\Mmat'$, which has the same number of columns (i.e., the same number of operators), the same rank, and the same structure of the kernel (i.e., the loops are defined by the same equations).
%
In other words, a spin model and its gauge transformation have the same loop structure, and therefore we recover that they have the same partition function. %We thus obtain again that models related by a GT have the same partition function. %, up to permutation of their parameters.
In doing so, we also identify %found 
an algebraic property of spin models that is invariant under GT, namely their loop structure.
Moreover, the rank and the dimension of a spin model, which are respectively the cardinality of the maximal set of independent operators within the model, and the cardinality of the minimal independent set of operators that can generate all the operators of the model,
are also invariant under GT. 
%\textcolor{orange}{, as shown previously (see Sec.~\ref{Sec:GTSpinModel}).}
%\textcolor{orange}{NOT SURE THIS WAS DISCUSSED BEFORE...}

%\textcolor{orange}{\noindent Moreover, independent operators remain independent under GTs, which implies that the rank and the dimension of a spin models are invariant under GT.}

\begin{Property}\label{ppty:class:invariants-algebraic}
    {\bf Algebraic properties of spin models invariant under GTs.}
    The number of operators, the rank, the dimension, and the loop structure $\mathcal{L}_q$ of a $q$-state spin model are invariant under gauge transformations. For any $q$-state spin model~$\M$ and gauge transformation $\T$, $|\T[\M]|=|\M|$, $\rank(\T[\M])=\rank(\M)$, and $\mathcal{L}_q(\T[\M])=\mathcal{L}_q (\M)$.
\end{Property}

For example, take the top left loop of model $\M$ in Fig.~\ref{fig:GT_and_Loops}.b;
this loop corresponds to the vector $\vecell=\boldsymbol{u}_1+\boldsymbol{u}_3+2\boldsymbol{u}_5$, whose image by $f_{\M}$ is:
\begin{align}
    f_{\M}(\boldsymbol{u}_1+\boldsymbol{u}_3+2\boldsymbol{u}_5)
        &=\vecmu_1+\vecmu_3+2\vecmu_5 = \boldsymbol{v}_1+\boldsymbol{v}_2+2(\boldsymbol{v}_1+\boldsymbol{v}_2)=\veczero\,,
\end{align}
where the sums on the right-hand side are taken in $(\mathbb{Z}/3\mathbb{Z})^n$.
For the transformed model $\M'=\T[\M]$, the image of the same vector $\vecell$ by the linear map $f_{\M'}$ associated with $\M'$ is also the null vector:
\begin{align}
    f_{\M'}(\boldsymbol{u}_1+\boldsymbol{u}_3+2\boldsymbol{u}_5)
        &=\vecmu_1'+\vecmu_3'+2\vecmu_5'
        =(\boldsymbol{v}_2'+\boldsymbol{v}_3')+(\boldsymbol{v}_1'+2\boldsymbol{v}_2'+2\boldsymbol{v}_3')+2\boldsymbol{v}_1'=\veczero\,,
\end{align}
where the vectors $\vecmu_i'=\T[\vecmu_i]$ are given in Eq.~\eqref{eq:exFig:GT:M}.
The representation of the operators has changed from $\vecmu_j$ to $\vecmu_j'$, but not the dependence relations between them. In particular, $\vecell$, which was a loop of $\M$, remains a loop of $\M'$.

Like in the binary case~\cite{beretta2018stochastic}, GTs define an equivalence relation between $q$-state spin models (one can easily check for reflexivity, symmetry, transitivity\footnote{
Reflexivity: a spin model is a GT of itself, where the GT is the identity. Symmetry: consider a model $\M'$ obtained by the GT $\T$ of another model $\M$, then $\M$ is also the gauge transformation of $\M'$ using $\T^{-1}$. Transitivity: consider two GTs $\T_1$ and $\T_2$, and the transformed model $\M'=\T_1[\M]$ and $\M''=\T_2[\M']$, then $\M''=\T[\M]$ where $\T=\T_2[\T_1]$ is also a GT resulting from the composition of $\T_1$ and $\T_2$.}).
This equivalence relation divides the (superexponential) space of the $2^{q^n-1}$ spin models into equivalence classes of models that share common statistical and algebraic properties.
In section~\ref{sec:3_MCM}, we will focus on the subset of classes corresponding to minimally complex models~\cite{demulatier2024MCM}.

%\begin{Definition}
%    {\bf Equivalence relation between $q$-state spin models.} 
%    Spin models related by gauge transformations are equivalent, in the sense that they are described by the same partition function.
%    \textcolor{blue}{to define equivalent models as models related by GTs.}
%    \textcolor{blue}{We define an equivalence relation between spin models related by GTs.}
%\end{Definition}
%

\begin{Property}
    {\bf Equivalence classes of $q$-state spin models and their properties.} % characteristics/invariants.
    Gauge transformations define an equivalence relation between $q$-state spin models, dividing the space of models into equivalence classes. % \textcolor{red}{[of models related by GTs]}. 
    Models of the same class are different representations of the same abstract statistical model. % and are characterized by 
    Each class is thus characterized by the statistical and algebraic invariants given in Property~\ref{ppty:class:invariants-statistical} and~\ref{ppty:class:invariants-algebraic}. 
%In particular, they have:
%\begin{itemize} 
%    \item[--] the same number of operators;
%    \item[--] the same rank, which is the cardinality of the maximal set of independent operators in the model;
%    \item[--] the same dimension, which is the cardinality of the minimal independent set of operators that can generate all the operators of the model;
%    \item[--] the same loop structure $\mathcal{L}(\M)$;
%    \item[--] the same partition function.
%\end{itemize}
\end{Property}

\subsection
%{\textcolor{red}{[Generalized]} symmetries of discrete data and \textcolor{red}{[of]} spin models under gauge transformations} 
{Discussion: equivariant modeling and learning symmetries of data under gauge transformations}
%Equivariant model selection under GT.
%{Symmetry under gauge \textcolor{red}{transformation[s]}} % Models and data symmetries
% Symmetries of models and of data
%{Gauge invariances of the model or the data as symmetries of the spin system}
% Discussion: learning symmetries of the data under gauge transformations

%%%%%%%%%%%%%%%%% %%%%%%%%%%%%%%%%% %%%%%%%%%%%%%%%%%
%For stationary systems, symmetries of the data correspond to invariance of the empirical probability distribution under [certain] bijective transformations.
%%%%%%%%%%%%%%%%% %%%%%%%%%%%%%%%%% %%%%%%%%%%%%%%%%%

%%%%%%%%%%%%%%%%% %%%%%%%%%%%%%%%%% %%%%%%%%%%%%%%%%%
%%%%%%%%%%%%%%%%%\textbf{Discussion: equivariant model selection under GT.}
%%%%%%%%%%%%%%%%% %%%%%%%%%%%%%%%%% %%%%%%%%%%%%%%%%%
%\textcolor{purple}{This allows one to approach model selection from a representation invariant perspective, and ultimately design model selection procedures able to capture patterns of the data (such as symmetries, co-dependencies) in a way that is independent of the basis in which the data is initially observed. }
%%%%%%%%%%%%%%%%% %%%%%%%%%%%%%%%%% %%%%%%%%%%%%%%%%%

The gauge transformation of a discrete dataset simply re-writes the data in a different basis, without any loss or added information or noise. Ideally, a perfect model selection procedure would be able to find models that are consistent %if performed on 
for the original data and its transformed version, as illustrated in Figure~\ref{fig:Bob_and_Alice} in the introduction. 
In other words, this ideal model selection would be equivariant under GTs. Such approach is theoretically possible if we select among all $q$-state spin models, because this family of models is invariant under GTs (for any given~$q$). 
However, brute force selection of the best model among all is impossible in practice, due to the huge set of models (super-exponential in the number of spins). In Section~\ref{sec:3_MCM}, we extend to discrete data the work of Ref.~\cite{demulatier2024MCM} on minimally complex models, which are a sub-family of spin models within which one can perform equivariant model selection in practice.

%%%%%%%%%%%%%% %%%%%%%%%%%%%% %%%%%%%%%%%%%%
%More generally, one idea to approach model selection among all $q$-state spin models is to focus the selection on learning patterns of the data that are invariant under GTs. As a way to reduce the set of models that one can choose from.
%%%%%%%%%%%%%% %%%%%%%%%%%%%% %%%%%%%%%%%%%%

More generally, if one wants to tackle model selection among all $q$-state spin models, it is important to find principled ways to reduce the number of parameters and %restrict the set of possible models. 
the set of possible models for a given dataset.
One approach is to focus on modeling symmetries of the data, instead of specific correlation patterns. 
If a dataset is invariant (up to noise) under certain bijective transformations, then a good model should encode that invariance, i.e., it should be invariant under the same transformation.
%%%%%%%%%%
%One would expect data generated from a given models to be invariant under the same GTs as the model (up to noise level). 
%%%%%%%%%%
This reduces the set of candidate models. In other words, we simplify the problem of identifying relevant correlation patterns of the data by searching first for relevant symmetries of the data.
%%%%%%%%%%%%%%
%In other words, we move the problem of identifying relevant correlation patterns of the data to the problem of finding [first] relevant symmetries of the data.
%%%%%%%%%%%%%%
%In practice, identifying the set of most relevant (high-order) correlation patterns in the data is difficult. 
Indeed, it is in general %In general, it is 
difficult to identify in a robust manner the set of the most relevant (high-order) correlation patterns of a dataset, because these patterns are typically not independent from each other and, as a result, 
%many possible combinations of relevant patterns
many combinations of possible relevant patterns
can explain the data similarly\footnote{This problem is in fact equivalent to identifying the best spin model for the data, by definition of maximum entropy models~\cite{jaynes1957information}.}.
In contrast, one can treat symmetries of the data independently from each other, and each newly uncovered symmetry further reduces the pool of possible candidate models.
In the following, we formalize this idea based on the notion of invariance under GTs. %by using invariance under GTs as symmetries.

%%%%%%%%%%% %%%%%%%%%%% %%%%%%%%%%% %%%%%%%%%%% %%%%%%%%%%%
%%%%%%%%%%%{\bf Symmetries of discrete data under GTs.} 
%%%%%%%%%%% %%%%%%%%%%% %%%%%%%%%%% %%%%%%%%%%% %%%%%%%%%%%
As discussed in Sec.~\ref{sec:GT:data}, applying a GT to a dataset permutes the probability %number
of occurrence of the states in the empirical %\textcolor{red}{probability} 
distribution. % \textcolor{red}{of the data}.
However, certain GTs can leave the empirical distribution fully invariant, by only permuting states that have the same probability. Such invariance can be interpreted as a symmetry of the data under GT. 
The definition of GTs thus allows us to formalize the notion of generalized symmetries in discrete data, as the invariance of the empirical probability distribution under GT. These symmetries go beyond ordinary symmetries, such as translational and reflection symmetries for spatially organized variables, or symmetries under permutations of spin variables.
\begin{Definition}\label{def:symmetry:data}
    {\bf Generalized symmetry of (stationary) discrete data.}
    We call symmetry of a discrete dataset the invariance of its %\textcolor{red}{the/} 
    empirical distribution %\textcolor{red}{[of the data]} 
    under a gauge transformation.
\end{Definition}

%%%%%%%%%%% %%%%%%%%%%% %%%%%%%%%%% %%%%%%%%%%% %%%%%%%%%%%
%%%%%%%%%%%{\bf Symmetries of $q$-state spin models under GTs.} 
%%%%%%%%%%% %%%%%%%%%%% %%%%%%%%%%% %%%%%%%%%%% %%%%%%%%%%% 
For $q$-state spin models, gauge transformations are changes of representation of the models.
%Gauge transformations are changes of representation of $q$-state spin models. 
They transform a model's architecture (i.e., which operators are in the model), while preserving its statistical and algebraic properties (e.g., partition function and loop structure).
However, certain GTs can leave a given model architecture fully invariant, % (i.e., $\M$ and $\T[\M]$ have the same set of operators)
by transforming each operator of the model into a spin operator that is already in the model. 
In other words, a model $\M$ and its transformation $\T[\M]$ have the same set of operators and thus correspond to the same parametric model. 
For example, take Bob's model in the introduction Figure~\ref{fig:Bob_and_Alice}. The permutation of the spins $s_1$ and $s_2$ is a GT that leaves the model's architecture invariant. Equivalently, for Alice's model, the GT $(\sigma_1', \sigma_2',\sigma_3')=(\sigma_1,\,\sigma_1\sigma_2\sigma_3,\,\sigma_3)$ leaves the model invariant (as it permutes the field operator~$\sigma_2$ with the three-body operator~$\sigma_1\sigma_2\sigma_3$).
We call such invariance of the model architecture under GT a symmetry of the spin model.

Contrary to the definition~\ref{def:symmetry:data} of symmetry of data, the symmetry of a model under GT doesn't necessarily leave the model probability distribution invariant, because the parameters associated with the permuted operators are then also permuted. For the model probability distribution to remain invariant for any value of the parameters, the parameters of the permuted operators must be equal (degenerate parameters). 
%\textcolor{blue}{[Let us call such invariance a strong symmetry of the statistical model.]}

\begin{Definition}
    {\bf Generalized symmetry of $q$-state spin models.}
    We call symmetry of a spin model the invariance of its %the model's 
    architecture (which operators are in the model) under a gauge transformation.
    We call ``strong'' symmetry of a spin model the invariance of the model probability distribution under a GT. 
    Such invariance requires both an invariance of the model architecture and the degeneracy of the model parameters associated with the operators permuted by the GT. % (see example above).
\end{Definition}

%For the model probability distribution to be invariant, one would also need that the parameters of the GT models are the same than the one of the initial model.
%If we further consider that the model has degenerate parameters $g^{\vecmu_1}=g^{\vecmu_1}=g$.
%If the two interactions are parametrized by the same parameters ($g^{\vecmu_1}=g^{\vecmu_1}=g$; degenerate models). 
%In this case, the two models correspond to the same parametric probability distribution; the two models' parametric probability distributions are identical.
%%%%%%%%%%%%%%%%%%%%%%%%%
%In this case, the parametric probability distributions corresponding to these two models are [exactly] identical, at any values of the parameters.
%%%%%%%%%%%%%%%%%%%%%%%%%

%%%%%%%%%%%%%%%%%%% %%%%%%%%%%%%%%%%%%% %%%%%%%%%%%%%%%%%%%
%{\bf Relevant symmetries of data imply symmetries of the model.}
%{\bf Learning symmetries in/of data.} 
%%%%%%%%%%%%%%%%%%% %%%%%%%%%%%%%%%%%%% %%%%%%%%%%%%%%%%%%%
%In this context, one could learning ``relevant symmetries'' of the data and performing the model selection only among models that have this symmetry.
%
%\textcolor{red}{The idea is to} % One could exploit %that one can 
We can thus exploit %the 
generalized symmetries of a discrete dataset to reduce the pool of candidate spin models to those that have the same symmetries.
An invariance of the data under a GT should be reflected in the candidate models as a strong symmetry,
%A symmetry of the data under a GT should be reflected in the candidate models by a strong symmetry of the model under the same GT,
%%%%%%%%%%%%
%\textcolor{red}{Each symmetry of the data thus imposes a degeneracy of the parameters of the spin model.} %are reflected in the model by a degeneracy of the parameter.
%%%%%%%%%%%
%This
which imposes a degeneracy of the parameters associated with the operators permuted by the GT\footnote{
For example, take a binary dataset with $3$ spins that is invariant under the GT characterized by the inverse transformation $\Tmat^{-1}=\begin{pmatrix}%{ccc} %pmatrix/pNiceArray
    1 & 0 & 0 \\ 
    1 & 1 & 0 \\
    0 & 1 & 1 
\end{pmatrix}$. Then the operators $(100),(110), (101), (111)$ should be given the same parameters as well as the operators $(011)$ and $(010)$ for the model to be strongly symmetric under the same transformation (this can be checked by multiplying the matrix $\Tmat^{-1}$ with each of these operators as in Eq.~\eqref{def:eq:GT:Op:2}).
}: they must either all be absent from the model or all have the same parameter. The model selection procedure should then decide if this degenerate parameter is relevant or if these operators should be all be removed from the model%
\footnote{Note that a symmetry under GT will not reduce the set of models to a specific subset of equivalence classes, but instead it will invalidate candidate models %reduce the candidate models 
from all classes. A symmetry under a given GT is indeed representation-dependent.
%More generally, 
For each transformation under which a model $\Mmat$ is invariant, there is a corresponding transformation under which another model of the same class is invariant. Consider a transformation $\Tmat_1$ under which $\Mmat_1$ is invariant: $\Tmat_1\Mmat_1=\Mmat_1$. Consider another model $\Mmat_2$ of the same class: there exists a GT $\Tmat$ such that  $\Mmat_2=\Tmat\Mmat_1$. The model $\Mmat_2$ is then invariant under the GT $\Tmat_2=\Tmat\Tmat_1\Tmat^{-1}$, as it satisfies that $\Tmat_2\Mmat_2=\Mmat_2$. Furthermore, %in cases where 
if $\Tmat$ and $\Tmat_1$ commute, then %we have that 
$\Tmat_2=\Tmat_1$, which means that the models $\Mmat_1$ and $\Mmat_2$ are invariant under the same GT $\Tmat_1$.
}.

%%%%%%%%%%%%%%%%%% SHORTER VERSION:
%For example, consider a dataset from a two-spin system. Assuming sufficiently large data samples, we observe that the system is invariant under the permutation of the two spins. This means that the fields (if any) on each of the two spins have the same parameters. 
%This implies in particular that the architecture of the model should be invariant under this permutation (either the two fields are both there, or they are both absent).
%%%%%%%%%%%%%%%%%%
%This means that the architecture of the model should be invariant under this GT, and that the parameters of the operators permuted by the GT in the model should be identical.
%%%%%%%%%%%%%%%%%%

For real data, the probabilities of occurrence of the states are known up to some uncertainty that depends on the size of the data: the larger the number of datapoints, the smaller the uncertainty. 
To find all the symmetries of a finite dataset, one should take into account the level of uncertainty (even more so in undersampling regimes).
This can be done by reducing the noise in the empirical probability distribution, for instance, by assigning the same probability to states that have similar frequencies 
%at an appropriate level of precision, %(at the level of precision accessible), 
at the level of precision warranted by the data (e.g., following the procedure of~\cite{haimovici2015criticality}).
%which could be reduced using a similar approach to the one described in Sec.~1.
A smaller number of different relevant probabilities
%reduction in the relevant probability 
in the empirical distribution increases the number of symmetries observed in the data, and therefore reduces the number of parameters in the model.
%The smaller the number of different relevant probabilities, the larger the number of symmetries.
This is a consequence of the duality between mixture models and models of the exponential family~\cite{amari2002information}, 
%described by Amari~\cite{} 
which was exploited for instance for model reduction and parameter fitting for small systems in~\cite{Gresele2017}. %\textcolor{orange}{and which we exploited in Sec.~1}.
%which we exploited for instance in Sec.~1.
%
Such a procedure can significantly reduce the number of possible model parameters. A model selection procedure would then have fewer parameters to select from to identify if they are relevant or not. These parameters would also be associated with larger ``macro''-patterns (composed of multiple operators).
Such an approach has the potential to greatly simplify the model selection procedure and lead to more robust model discovery.

%\textcolor{purple}{{\bf Question:} Is the dimension of teh Kernel well defined when $q$ is not prime? I guess not. What about the formula $\dim(\ker(f_{\M}))=K-\rank(f_{\M})$. Does it work with the definition of dimension we give in Sec~2?}

%%%%%%%%%%%%%%%%%%%%%%%%%%%%%%%%%%%%%%%%%%
\section{Limiting the search to Minimally Complex Models (MCMs)}
\label{sec:3_MCM}

%\textcolor{orange}{For the general INTRODUCTION OF THE PAPER:\\ Introduce first MCMs as a factorization of the multi-variate probability distribution. Then say that using the structure of $q$-state spin models, this can be extended to all possible gauge-transformed basis, which opens the idea of performing a model selection that is independent of the original representation of the data.}

Despite the existence of equivalent models, the space of models is still extremely large, making it difficult to perform model selection. Ref.~\cite{demulatier2024MCM} proposes to perform model selection among a sub-family of spin models that they call Minimally Complex Models  (MCMs), for which all the quantities of interest in the context of statistical inference are computationally cheap to calculate, making model selection feasible in practice among these models.

In particular, Ref.~\cite{demulatier2024MCM} introduced MCMs as an alternative to pairwise models that respects the structure of gauge transformations (GTs), because the family of MCMs is invariant under GT (the GT of an MCM is also an MCM). This is not the case for families of spin models restricted to have interactions only up to a certain order, such as pairwise models. 
Working with MCMs thus brings a new perspective to statistical modeling, by making it possible to tackle data modeling problems in a way that is independent of the basis in which the data was originally represented.
%making it possible to develop model selection procedures that do not depend on the basis in which the data was originally represented.
%
This is interesting especially for systems where it is not clear if the current representation of the data is the most meaningful, or when one is interested in tackling a modeling problem from the most general perspective, in the absence of any information about the system.

\begin{figure}[!ht]
\centering
\includegraphics[width=0.80\linewidth]{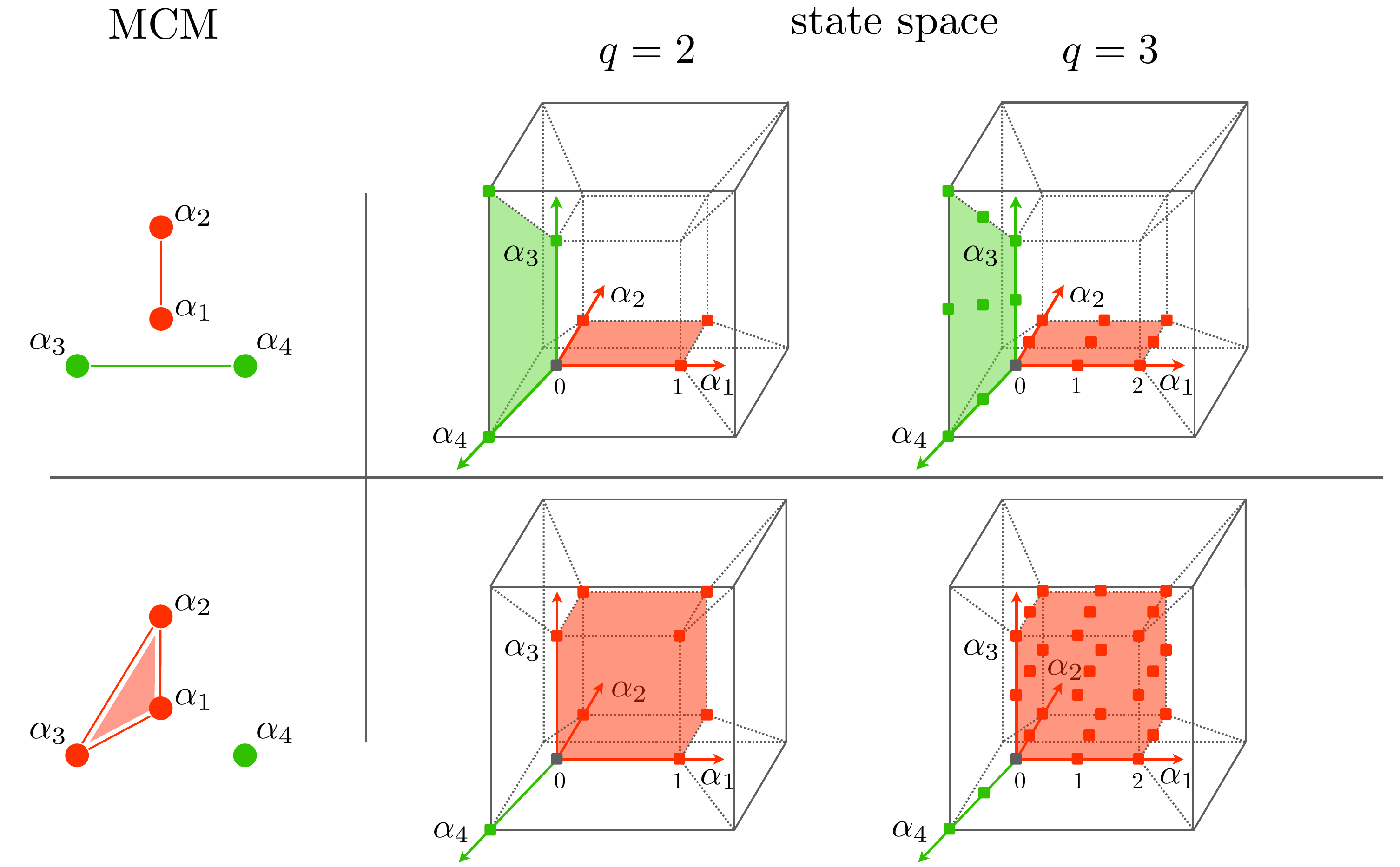}
\caption{(colors online)
    %{\bf .}
    Visualization of MCMs as factorization in state space.
    {\bf Left.} In each row, the left panel %gives a representation of 
    shows an MCM %represented in a preferred basis. 
    based on four discrete variables $(\alpha_1, \alpha_2, \alpha_3, \alpha_4)$ with two ICCs 
    %in which each ICC is 
    represented in different colors. %by different colors.
    The MCM is represented in a preferred basis, which means that each ICC contains all possible operators %can be read as complete 
    over the shown subset of $\alpha$-variables.
    The value of $q$ for the discrete variables is not yet specified and the diagrams can be read for any chosen value of $q$.
    %must be read with a chosen value of $q$ in mind. % considering that each ICC is complete over the subset of variables.
    For example, for the top row: for $q=2$, the red ICC has the $q^2-1=3$ operators $(\mu_1,\mu_2)\in\{(1,0), (0,1), (1,1)\}$; for $q=3$, the red ICC has the $q^2-1=8$ operators $(\mu_1,\mu_2)\in\{(1,0), (2,0), (0,1), (0,2), (1,1), (2,2), (1,2), (2,1)\}$.
    {\bf Right.} %\textcolor{red}{Factorization in the (real) space of the discrete variables~$\state$ (left) and in state space (right). The number of different possible factorizations of the $n$ basis variables doesn't depends on $q$.}
    The panel represents the factorization in state space (for $q=2$ and $q=3$) that results from the MCM partition of the basis variables~$\alpha_i$ %in the real space, %created by the MCM 
    shown in the left panel.
    %corresponds to the MCM partition of the basis variables shown in the left panel.
    The model probability distribution can be written as the product of joint distributions over lower dimensional subspaces shown in colors (line, surface, or volume). The color (green or red) %indicates/
    refers to the corresponding ICC in the left panel and each marker %dot
    represents a state of the %sub-system/
    system reduced to %/in 
    these subspaces (e.g., for the top panel and $q=2$, the pair of variables $(\alpha_1,\alpha_2)$ can take the values $\{(0,0), (1,0), (0,1), (1,1)\}$, hence the four markers in the red area).
    %Each dot represents a state of the system. The probability of the states represented in green is independent from the.\\
    The probability of the states that are not represented by a colored marker is obtained by products of the probabilities of the sub-states that are represented by a marker.
    When written in a preferred basis, the factorization of the state space resulting from % corresponding to
    the same MCM partition of the basis variables into ICC (as in the left panel) doesn't depend on the value %of $q$
    of the discretization $q$ of the variables. 
    %The only difference is that the number of states in each subspace increases with $q$.
    The only difference is the %increasing 
    increase of the number of states in each subspace with $q$.
    %%%%%%%%%%%%%%%%%%%%
    %The factorization in state space corresponding to the same MCM partition of the basis variables into ICC (in a preferred basis, as in the left panel) doesn't depend on the value of $q$. The only difference is that the number of states in each subspace increases with $q$.
    %%%%%%%%%%%%%%%%%%%%%%%%%%%%%%%%%%%%%%%%%%%%%%%%%%%%%%%%%%%%
    %\textcolor{red}{For a given system of discrete variables, MCMs for different values of $q$ corresponding to the same partition of the basis variables into ICCs also correspond to the same factorization of the state space; the only difference is the number of states in each subspace.}
    %%%%%%%%%%%%%%%%%%%%%%%%%%%%%%%%%%%%%%%%%%%%%%%%%%%%%%%%%%%%
    }
\label{fig:MCM:factorization-completness}
\end{figure}

Model selection with MCMs is also efficient thanks to the peculiar interaction structure of MCMs, in which interactions are grouped into independent sub-systems. 
%%%%%%%%%%%%%%%%%%%%%%
%In some representations, an MCM can be mapped to a partition of the basis variables, where variables are fully connected inside the parts, with all possible (high-order) interactions, and not connected at all between the parts.
%The MCM thus fully models all possible statistical patterns of the data inside the parts, but no pattern at all between the parts, as if they were independent.  
%%%%%%%%%%%%%%%%%%%%%% 
More precisely, in some representations, an MCM corresponds to a factorization of the model multivariate probability distribution over subsets of basis variables; 
i.e., the model assumes that these subsets are statistically independent from each other (see Fig.~\ref{fig:MCM:factorization-completness}). 
%For each MCM, there exists a basis in which the model probability distribution factorizes.
Performing model selection among MCMs thus %\textcolor{red}{
aims to % allows to 
uncover the factorization of the model probability distribution that best matches the structure of the data. 
%%%%%%%%%%%%%%%%%%%%%%%%%%%%%
%[Finding] the MCM that best models the data thus provides a factorization of the model (multivariate) probability distribution that best matches the structure of the data.
%%%%%%%%%%%%%%%%%%%%%%%%%%%%%
%The best MCM then provides an approximation of the empirical probability distribution in terms of a product of (lower-dimensional) [marginal] empirical distributions.
%%%%%%%%%%%%%%%%
%Modeling data with the best MCM corresponds to finding the best approximation of the empirical probability distribution in terms of a product of (lower dimensional) marginal distributions.
%%%%%%%%%%%%%%%%
% provides an optimal factorization of the empirical multivariate probability distribution.
%%%%%%%%%%%%%%%%
Such factorization reduces the dimensionality of the statistical inference problem, while preserving most of the relevant information within the data. 
%\textcolor{red}{
MCMs can thus be used to identify groups of variables that carry most of the multivariate information in the data, or to facilitate the modeling problem for large systems by breaking it down into lower-dimensional ones.
%
%For large systems, % sizes, 
%this approach allows identifying groups of variables that carry most of the multivariate %collective 
%information of the data. 
%MCMs can thus be used to identify emergent collective patterns in the data, or to facilitate the modeling problem by breaking it down into lower-dimensional ones.
Ref.~\cite{AgrimiCaputo2026} likens this approach to performing community detection for data (by analogy to community detection in network analysis), giving a simple interpretation to MCMs in terms of community structure.
%%%%%%%%%%%%%%%%%%%%%%%
% Variables within the parts share more (multivariate) information (are more correlated) than variables between the parts. This is similar to a community detection on networks, but for data.
%%%%%%%%%%%%%%%%%%%%%%%
%Allow to find an optimal factorization of the data, which significantly reduces the dimensionality of the problem.
%%%%%%%%%%%%%%%%%%%%

In this section, we focus on a subset of $q$-state spin models that generalize the framework of minimally complex models~\cite{demulatier2024MCM} to discrete systems. \\

\subsection{Definitions}
%\textcolor{orange}{After separating this paper: we will have to recall the definition of basis, rank, and GT at this point.}

Minimally complex models (MCMs) are defined in the context of binary data as models constructed from the union of independent complete components (ICCs)~\cite{demulatier2024MCM}. 
Here, we generalize to the discrete case the definitions of complete components (CCs), independent components (ICs), 
and MCMs. We call a ``component'' of a spin model $\M$ any subset of operators of $\M$.
%%%%%%%
%The notion of minimally complex $q$-state spin models can be defined for any integer values of $q$ (including non-prime values).
%%%%%%%
%The following definition is a direct generalization of the notion of complete components from the binary case to discrete data, but it is, however, only valid for $q$ prime.
%%%%%%%

For $q$ prime, we can directly extend to $q>2$ the definition of complete components from Ref.~\cite{demulatier2024MCM}.
%%%%%%%%%%%%%%%%%
%The definition of complete components in Ref.~\cite{demulatier2024MCM} can be [directly] extended to $q>2$, in the form of Definition~\ref{def:CC} below. However, for $q$ non-prime, this approach defines model components that we won't consider as ``complete'' in this paper. This definition is thus only valid when $q$ is prime, and we will give a more general definition below.
%%%%%%%%%%%%%%%%%
%For prime values of $q$, the definition of complete components in Ref.~\cite{demulatier2024MCM} can be directly extended to $q>2$, in the form of Definition~\ref{def:CC} below. For $q$ non-prime, this approach defines model components that we won't consider as ``complete'' in this paper. We will give a more general definition below.
%%%%%%%%%%%%%%%%%
%
\begin{Definition}\label{def:CC:q-prime}
    {\bf Complete Component (CC), for $q$ prime.}
    A complete component is a set of operators $\M_a\subseteq[\Field^n]^*$ that verifies that, for any two operators $(\vecmu,\vecnu)\in\mathcal{M}_a^{\,2}$, 
    all the operators obtained by linear combinations of $\vecmu$ and $\vecnu$ in $\Field^n$ are in~$\mathcal{M}_a\cup\veczero$\,, i.e.:
    \begin{align}\label{eq:CC}
        \forall (\vecmu,\vecnu)\in\M_a^2\,,\qquad
        \forall (c,d)\in\Field^2, \qquad c\,\vecmu + d\,\vecnu \in\M_a\cup\veczero\,.
    \end{align}
    \noindent In other words, $\M_a\cup\veczero$ is a vector subspace of $\Field^{n}$ of dimension $r_a$, %/isomorphic to $\Field^{r_a}$,
    where $r_a$ is the cardinality of the largest linearly independent set in~$\M_a$ (namely the rank of $\M_a$ -- see Definition~\ref{def:M:rank}). 
\end{Definition}
\noindent 
In this definition, the set~$\M_a$ of operators can be a subset of operators of a model %, which is why we use the term ``component'' and not ``model''.}
(hence the use of the term ``component'' instead of ``model'').
%We use the term ``component'' and not ``model'' in this definition because the set~$\M_a$ of operators can be a subset of a model.
We recall that the interaction vector $c\,\vecmu + d\,\vecnu$ is associated with the spin operator $\phi^{c\,\vecmu + d\,\vecnu}(\state)\doteq[\phi^{\vecmu}(\state)]^c\,[\phi^{\vecnu}(\state)]^d$. 
For $q>2$, because of conjugate operators, the vector $c\,\vecmu + d\,\vecnu$ can be equal to $\veczero$ even for $(c,d)\neq(0,0)$, which is why we added the element $\veczero$ to $\M_a$ in Eq.~\eqref{eq:CC}.  
This doesn't happen in the binary case~\cite{demulatier2024MCM}.

Extending Definition~\ref{def:CC:q-prime} to non-prime values of $q$ would define as being ``complete'' model components that we don't want %would prefer not to 
to consider complete in the context of MCMs. 
For example, take the component $\M_a=\{(2,2)\}$ with a single pairwise interaction for $n=2$ spins and $q=4$. This component satisfies Eq.~\eqref{eq:CC}, because the only multiples of the operator $(2,2)$ in $(\mathbb{Z}/4\mathbb{Z})^2$ are $(0,0)$ and itself, which are both in $\M_a\cup\veczero$.
%%%%%%%%%%%%%%
%The elements in this model are self-dependent. 
%The model only has one operator and not $(4^1-1)$ as we would expect based on the results above.
%This is model spans $r_a=1$ dimensions, but instead of spanning a sub-vector space $\Field^{r_a}$, it is instead over a space $(\mathbb{Z}/2\mathbb{Z})^{r_a}$. 
%%%%%%%%%%%%%%
The component~$\M_a$ is in some sense complete, but it is isomorphic to $\mathbb{Z}/2\mathbb{Z}$ and not to $\mathbb{Z}/q\mathbb{Z}$ (here $q=4$).
Considering such component as complete is problematic, %makes it more difficult to work with MCMs,
because an MCM would then correspond to a factorization over ``CC'' defined for variables of different types (i.e., with different values of $q$). 
We would also lose the re-parametrization property of~CC (see Property~\ref{Ppty:ICC:re-parametrization} below), which is necessary for the efficient computation of many quantities of interest for modeling data with MCMs (see Sec.~\ref{sec:MCM:inference}). 
%To preserve the re-parametrization property.
%%%%%%%%%%%%%
% In order to preserve the re-parametrization property of CC in terms of the state probabilities.
%%%%%%%%%%%%%
%We exclude this possibility and give the following general definition of complete components, valid for any integer value of $q$.
To exclude this possibility, we give the following general definition of complete components, valid for any integer value of $q$.

\begin{Definition}\label{def:CC}
    {\bf (for all $q$)\, Complete Component (CC).\;}
    A set of operators is said to be complete if it is equal to the set of all the operators generated by an independent set of operators (see Definition~\ref{def:independent:operators} of independent operators).
    Formally, consider a set of operators $\M_a\subseteq[\Field^n]^*$ and let $I=\{\vecmu_1,\dots,\vecmu_{r_a}\}$ be a maximal independent set of operators in $\M_a$. 
    % We call the cardinality $r_a$ of such set the rank of the component and denote it $\rank(\M_a)$.
    The model component $\M_a$ is said complete if its elements are all the operators generated by $I$, %i.e. if
    \begin{align}
        \M_a=\langle I\rangle_{gen}^*
        =\bigg\{\,\sum_{j=1}^{r_a} c_j\,\vecmu_j \;\Big|\; 
        \,(c_1,\dots,c_{r_a})\in[\Field^{r_a}]^*\bigg\}
        \,,
    \end{align} 
    where we used the notation $*$ to indicate that we exclude the null vector (which doesn't correspond to an operator).
    In other words, $\M_a\cup\veczero$ is a submodule of $\Field^{n}$ isomorphic to $\Field^{r_a}$, where $r_a$ is the cardinality of $I$ (namely, the rank of $\M_a$, denoted $\rank(\M_a)$).
\end{Definition}
\noindent From this definition, it follows that the number of operators in a complete component of rank $r_a$ is equal to the number of nonzero vectors in $\Field^{r_a}$, which is $q^{r_a}-1$. 
\begin{Property} \label{ppty:CC:nb-operators}
{\bf Number of operators in a CC.} The number of operators in a complete component~$\M_a$ is $|\M_a|=q^{r_a}-1$, where $r_a$ denotes the rank of $\M_a$.
\end{Property}

We can then directly extend the definition of independence between complete components from the binary case to the more general discrete case.
\begin{Definition}\label{def:CC:indep}
    {\bf Independence between CC and Minimally Complex Models (MCMs).\;}
    A set of complete components $\{\M_a\}_{a\in\mathcal{A}}$ is said to be {\it independent} if none of the operators of a given component $\M_a$ can be obtained by linear combinations of operators of the other components (i.e., generated by the set of operators $\cup_{a'\in\mathcal{A}\backslash a}\M_{a'}$).
    %(i.e. by linear combinations of operators of the set $(\cup_{a'}\M_{a'})\backslash\M_a$).
    This then leads to the definition of Minimally Complex Models (MCMs) as models formed by the union of independent complete components (ICCs), $\M=\cup_{a\in\mathcal{A}}\M_a$.
\end{Definition}

In the following, we give a more general definition of independence between components, for components that are not necessarily complete. 
%Our goal is to build a general formalism for modeling data in a basis independent way, which could be used for any $q$-state spin models (not just MCMs).
Our %\textcolor{red}{[long-term]} 
goal is to build a general formalism that can be used for any $q$-state spin model (not just MCMs) and which allows
%%for approaching the data modeling problem in a basis independent way.
%%for equivariant modeling of data,
for modeling data in a basis independent way.
%
%In this context of general spin models,
For general spin models,
the notion of independence between components %for general spin models 
is directly linked to a factorization property of the model probability distribution (see Property~\ref{Ppty:Factorization} below); this factorization can be %which can be 
in the original representation %basis
of the data or in another gauge-transformed basis. %of $q$-state spin models, 
In this %Within this 
framework, MCMs % which can identify such factorizations
are a natural stepping stone %could be used as a first step 
towards modeling the data with more complex spin models.
To construct this formalism, we first introduce the definition of {\it basis of a component}, which we will then use to define independent components.

\begin{Definition}{\bf Basis and dimension of a component.} \label{def:basis-component}
A basis of a component $\M_a$, denoted $B_{\M_a}$, is a minimal set of $d_a$ independent operators\footnote{The operators of a basis of a component $\M$ are not necessarily in %elements of $\M$
$\M$.} that can generate all the operators of~$\M_a$: 
\begin{align}\label{eq:def:basis:M}
    \Basis_{\M_a}=(\vecmu_1,\cdots,\vecmu_{d_a})\subseteq [\Field^n]^*\;\; \textrm{such that}\; \M_a\subseteq \langle \Basis_{\M_a} \rangle_{gen}\;.
\end{align}
We call the cardinality~$d_a$ of such basis the {\it dimension} of the component and denote it $\dim(\M_a)$.
By definition, in an $n$-spin system, $\dim(\M_a) \leq n$\, for any model component $\M_a$. 
In the following, we consider the operators of the basis to be ordered, which we indicated in Eq.~\eqref{eq:def:basis:M} by using parentheses instead of curly brackets in $(\vecmu_1,\cdots,\vecmu_{d_a})$ and a bold symbol for $\Basis_{\M_a}$.\\
This definition is also valid if the component coincides with the whole model, thus defining the basis and the dimension of a $q$-state spin model.
\end{Definition}

%%%%%%%%%%%%%%%%%%%%%%%%%%%%%%%%%%%%%%
%\begin{Definition}{\bf Dimension of a component.} The dimension of a model component $\M$, denoted $\dim(\M)$, is the cardinality of the minimal independent set of operators that can generate all the operators of~$\M$.
%\end{Definition}
%%%%%%%%%%%%%%%%%%%%%%%%%%%%%%%%%%%%%%%

\noindent The basis of a component is not necessarily unique, but its dimension is (by definition).
Note the difference between the rank and the dimension of a component:
the rank of a component is the cardinality of the largest independent set within the component, whereas its dimension is the cardinality of the smallest independent set containing the component.
If $q$ is prime, the dimension and the rank of a component are always equal, but this is not necessarily the case if $q$ is not prime, for which the dimension can be larger than the rank\footnote{For example, consider a two-spin system with $q=4$ and the model %with a single operator 
$\M=\{(2,0)\}$. This model has a dimension $\dim(\M)=1$ because it can be generated by the single independent operator $(1,0)$ (and not less), but it's rank is $\rank(\M)=0$ because $\vecmu=(2,0)$ is not an independent operator over $(\mathbb{Z}/4\mathbb{Z})^2$ (as $2\vecmu=\veczero$).}. 
However for CC, any maximal independent set of operators in the component can generate the whole component, by definition~\ref{def:CC} of completeness. This means that such a set is also a basis of the CC and therefore that the dimension of a CC is equal to its rank for all values of~$q$. %\textcolor{purple}{Add a comment that Spin models in general are difficult to work with when $q$ is not prime, but not MCMs.}
In general, %Generally,} %Importantly,} %we would like to point out that} %highlight that
$q$-state spin models are %generally 
difficult %non-trivial 
to work %to analyze
with if $q$ is not prime, due to $\Field$ not being a field,
but thanks to this last property (or more generally, to the completeness of the components), %\textcolor{red}{conveniently} %interestingly and conveniently, 
this is not an issue %not the case
for MCMs.
%(or more generally, thanks to the completeness of the components).}
%the completeness of its independent components. } 
% because CC are isomorph to $\Field^q$. 

\begin{Property} {\bf Basis, dimension, and rank of a CC.}  \label{ppty:CC:dim-rank}
Any maximal independent set of operators in a CC is also a basis of the component.
Consequently, the dimension and the rank of a CC are equal, for all integer values of $q$ (prime or not). 
In the following, we will denote by $r_a=\rank(\M_a)=\dim(\M_a)$ both the rank and the dimension of a CC $\M_a$.
\end{Property}

We have already introduced a definition of spin basis in Sec.~\ref{Sec:GT:MatrixRepresentation}, in the context of the definition~\ref{def:GT} of gauge transformations (GTs). We defined a spin basis for an $n$-spin system as an ordered set of $n$ independent operators, denoted $\basis=(\phi^{\vecmu_1},\cdots,\phi^{\vecmu_n})$, and a GT as a change of spin basis of the form $\mathcal{T}:\spin\to\spin'=\basis(\spin)$.
A basis of a component in Definition~\ref{def:basis-component} also corresponds to a spin basis, but it is reduced to the (minimal) subspace in which the component is defined.
%in which [the operators of] the component is defined.
%
Note that, in Definition~\ref{def:basis-component} and throughout the paper, we used the term ``operator'' for both the spin operator~$\phi^{\vecmu}$ (proper use of the term)
and its vector representation~$\vecmu$.
For clarity and coherence of the notations, we will denote a basis of a component $\M_a$ by $\basis_{\M_a}=(\phi^{\vecmu_1},\cdots,\phi^{\vecmu_{d_a}})$ when written in terms of the spin operators~$\phi^{\vecmu}$ and by $\Basis_{\M_a}=(\vecmu_1,\cdots,\vecmu_{d_a})$ when written in terms of their vector representation~$\vecmu$.

Using the definition of basis of a component, we now define independent components.
\begin{Definition}\label{def:ICs}
    {\bf Independent Components (ICs).} 
    Consider a set of operators $\M\subseteq[\Field^n]^*$ formed by the union of disjoint components~$\M_a$\,:\, $\M = \cup_{a\in\mathcal{A}} \M_a$, where $\mathcal{A}$ is a set of labels identifying the components. 
    Consider a basis $\Basis_{\M_a}$ %$\basis_a$
    for each component $\M_a$. The set of components $\{\M_a\}_{a\in\mathcal{A}}$ is said to be independent if, for each component $\M_a$, none of the operators generated by $\Basis_{\M_a}$ can be generated by the combined %the operators of the 
    bases of the other components (except for the null vector):
    \begin{align}\label{eq:def:IC}
        \forall \vecmu\in \langle \Basis_{\M_a}\rangle_{gen}^*\,,\qquad \vecmu\notin
    \bigg\langle \displaystyle\bigcup_{\substack{a'\in\mathcal{A}\\ a'\neq a}} \Basis_{\M_{a'}} 
    \bigg\rangle_{gen}\;,
    \end{align}
    where $\langle S\rangle_{gen}$ denotes the set of all the operators generated by a set $S$ of operators over $\Field^n$.
    In other words, the bases $\Basis_{\M_a}$ for all $a\in\mathcal{A}$ generate independent subspaces of $\Field^n$ (each isomorphic to $\Field^{d_a}$, where $d_a=\dim(\M_a)$), and the operators of each component~$\M_a$ belong only to the subspace defined by $\Basis_{\M_a}$.
\end{Definition}
\noindent For example, any choice of decomposition of an independent model (see Definition~\ref{def:independent:operators:bis}) into components gives a set of independent components.

As a direct consequence of the independence between the components $\M_{a,\,a\in\mathcal{A}}$, one can construct a basis~$\Basis_{\M}$ of a model $\M=\cup_{a\in\mathcal{A}} \M_a$ by combining the elements of the basis~$\Basis_{\M_a}$ of each component. 
The resulting set of operators, $\Basis_{\M}=\cup_{a\in\mathcal{A}} \Basis_{\M_a}$, is indeed still independent, thanks to Eq.~\eqref{eq:def:IC}.
This is generally not the case for random choices of decomposition of $\M$ in components $\M_a$, as the components can belong to overlapping (minimal) subspaces of $\Field^n$.
We recall that a basis was defined as an ordered set; here we assume that the elements in $\mathcal{A}$ have an order, and we create %define
the combined basis $\Basis_{\M}$ by ``gluing'' together the operators of the basis-component $\Basis_{\M_a}$ in that order (same for the notation $\basis_{\M}=\cup_{a\in\mathcal{A}}\basis_{\M_a}$ that will also be used below). %to create an ordered set of $d=\dim(\M)$ operators.} 
In practice, the specific choice of ordering of these basis components doesn't matter, but it must be consistent throughout the analysis.
%. The important point/aspect is to keep track of which is which and stay consistent throughout the analysis.}

\begin{Property}\label{ppty:MCM:basis-of-ICs}
    {\bf Basis and dimension of ICs.} Consider a model $\M$ formed by the union of components, %~$\M_a$, 
    $\M=\cup_{a\in\mathcal{A}}\M_a$, and consider a basis $\Basis_{\M_a}$ for each component. % $\M_a$. 
    If the components $\M_a$ are independent, then the set of operators $\Basis=\cup_{a\in\mathcal{A}}\Basis_{\M_a}$\,
    forms a basis of $\M$, and
    the dimension of $\M$ is thus equal to the sum of the dimensions of the ICs $\M_a$, $\dim(\M)=\sum_{a\in\mathcal{A}} \dim(\M_a)$.
\end{Property}

Finally, we recover the definition~\ref{def:CC:indep} of 
minimally complex models by combining the definitions of complete components and independent components\footnote{%Indeed, 
If all the independent components~$\M_a$ in Def.~\ref{def:ICs} are also complete, then, in Eq.~\eqref{eq:def:IC}, the set $\langle \Basis_{\M_a}\rangle^*$ of operators generated by a basis $\Basis_{\M_a}$ of a component is equal to the %precisely the whole CC~$\M_a$, 
component, $\langle \Basis_{\M_a}\rangle^*=\M_a$ (by definition of CC), and %we recover 
Def.~\ref{def:ICs} becomes %the definition
Def.~\ref{def:CC:indep} of independence between CC.}.

\begin{Definition}\label{def:MCM}
    {\bf Minimally Complex Models (MCMs).}
    A $q$-state spin model is a minimally complex model if its set of operators, $\M\subseteq[\Field^n]^*$, is formed by the union of \textbf{independent complete components (ICCs)}~$\mathcal{M}_a$:
    \begin{align}\label{eq:defMCM}
    \mathcal{M}=\bigcup_{a\in\mathcal{A}}\mathcal{M}_a\,,
    \end{align} 
    where $\mathcal{A}$ is a set of labels identifying the ICCs. 
    In other words, $\{\M_a\}_{a\in\mathcal{A}}$ is a set of independent components and, for all $a\in\mathcal{A}$\,, the component $\M_a$ is complete.
    Each ICC thus completely models an independent subspace of $\Field^n$. 
    %%%%%%%%%
    %The ICCs thus completely model independent subspaces of $\Field^n$.
\end{Definition}
\noindent Note that contrary to the binary case, an independent model, as defined in Definition~\ref{def:strongindependent:model}, is not an MCM. 
This is because each independent subspace of dimension $d_a=1$ requires $(q-1)$ operators to be fully modeled by a $q$-state spin model, while an independent model only has one operator per dimension (which is strictly less than $(q-1)$ for $q>2$). 
However, we can define a notion of independent MCM in the discrete case, which reduces to that of independent model in the binary case.

\begin{Definition}\label{def:IMCM}
    {\bf Independent Minimally Complex Models (IMCMs).}
    MCMs for which each ICC has rank 1. %is based on a single variable.
\end{Definition}
\noindent These models will be useful %in the development
for the search algorithms for %to find 
the best MCM proposed in Sec.~\ref{Sec:MCM:Search-algo}.

Finally, MCMs inherit all the properties of CCs and ICs mentioned above. 
\begin{Property}
    {\bf Number of operators, basis, dimension, and rank of an MCM.}
    Consider an MCM~$\M=\cup_{a\in\mathcal{A}}\M_a$ with ICCs $\{\M_a\}_{a\in\mathcal{A}}$.
    As a consequence of the completeness of the components, the dimension and the rank of each component~$\M_a$ are equal (see Property~\ref{ppty:CC:dim-rank}), denoted $r_a=\rank(\M_a)=\dim(\M_a)$, 
    and the number of operators in $\M$ is given by (see Property~\ref{ppty:CC:nb-operators}):
    \begin{align}
        |\M| = \sum_{a\in\mathcal{A}} (q^{r_a}-1)\,.
    \end{align}
    Thanks to the independence between the components~$\M_a$, one can obtain a basis of $\M$ %of the MCM $\M=\cup_{a\in\mathcal{A}}\M_a$ 
    by combining the bases $B_{\M_a}$ of its components (see Property~\ref{ppty:MCM:basis-of-ICs}): 
    \begin{align}
        B_{\M}=\cup_{a\in\mathcal{A}} B_{\M_a}\,.
    \end{align}
    The dimension of $\M$ is thus equal to the sum of the dimensions of its ICCs~$\M_a$, and so is its rank: $\rank(\M)=\dim(\M)=\sum_{a\in\mathcal{A}}r_a$.
    In the following, we will denote by $r$ both the rank and the dimension of an MCM $\M$.
\end{Property}

\begin{figure}[h]
    \centering
    \includegraphics[width=0.93\textwidth]{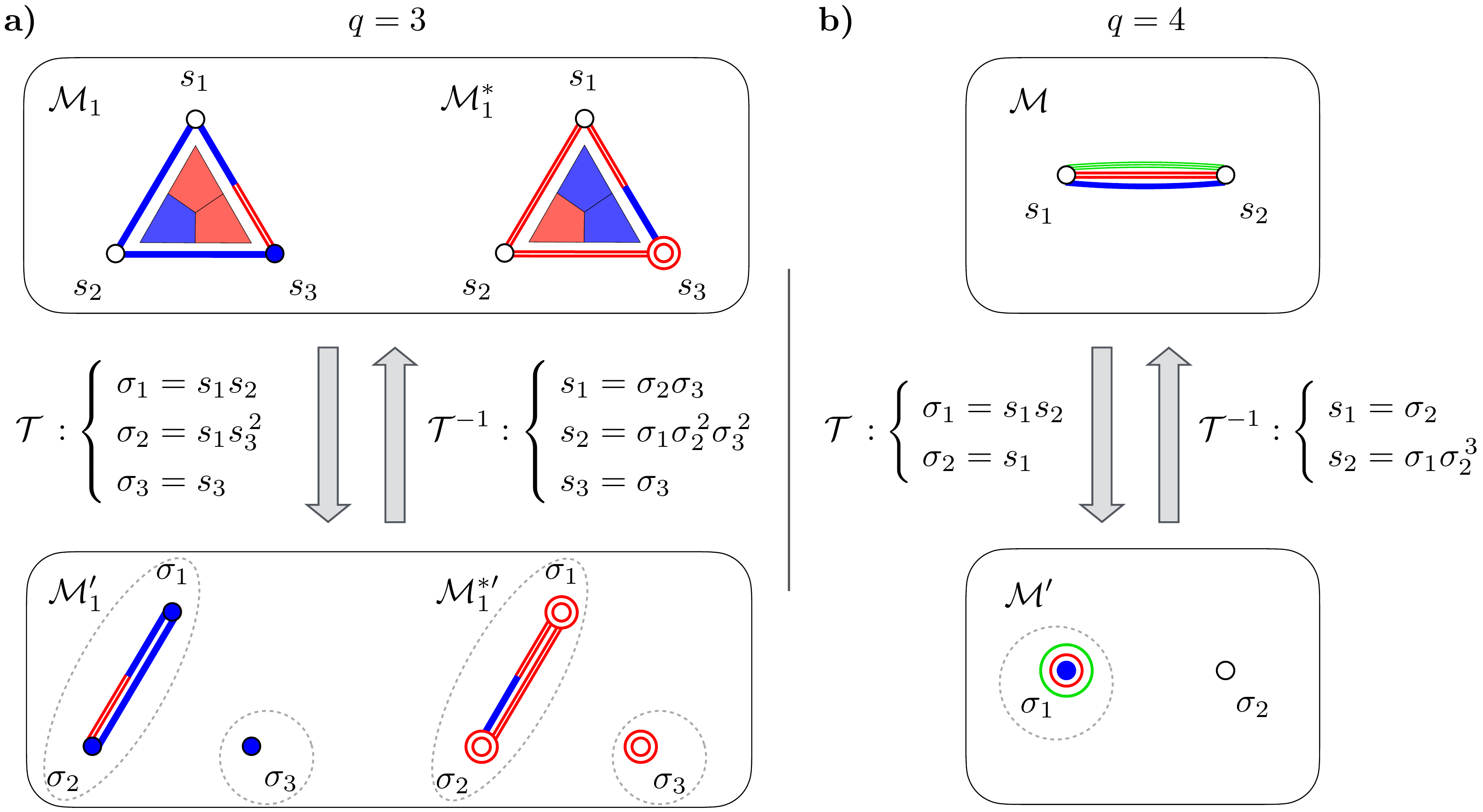}
\caption{{\bf Examples of MCMs and their GTs to a preferred basis.} 
Each box contains an example of an MCM. 
Operators are represented using the same convention as in Fig.~\ref{fig:GT_and_Loops}. In panel b), the green color is also used to indicate that a spin is raised to the third power in the corresponding operator (e.g., in model $\M$, the green pairwise link represents the operator $s_1^{\,3}s_2^{\,3}$). For each example, we use the GT $\T$ to transform the top model into the bottom one, and the inverse GT $\T^{-1}$ to go back to the top model.  
The bottom model $\M'$ is in a preferred basis, in which each ICC appears in a compact complete form over a subset of spins (circled %contoured
with gray dashed lines). The matrix representations of these models and of their GTs are available in App.~\ref{app:GT:MCM:ex}.
{\bf a)} Example of two equivalent MCMs with $q=3$ for $n=3$ spins; these models have two ICCs with rank~$2$ and~$1$ respectively. For clarity we represented the models using the decomposition $\M=\M_1\cup\M_1^*$, where the operators in $\M_1^*$ are the c.c. of the operators in $\M_1$. The transformed model is obtained using $\M'=\M_1'\cup\M_1^{*\prime}$, where $\M_1'=\T[\M_1]$ (see Sec.~\ref{Sec:GT:RealM}).
{\bf b)} Example of two equivalent MCMs, $\M$ and $\M'$, with $q=4$ for $n=2$ spins; these models have a single ICC of rank~$1$. 
}
\label{fig:MCM:examples}
\end{figure}

Figure~\ref{fig:MCM:examples} gives four examples of MCMs. Observe that in some representations, the complete components $\M_a$ of an MCMs can spread over more than $r_a=\rank(\M_a)$ variables. 
For example, the two models in panel (b) both have a single ICC of rank $r_a=1$. In the top model, this ICC spreads over two variables (with pairwise operators), whereas in the bottom model, the ICC is compact, based solely on one variable, and we can immediately see that the component is complete and of rank 1.  
Similarly, the two models in panel (a) both have two ICCs of rank $r_a=1$ and $2$ respectively.
It is not easy %straightforward 
to see that the top model is an MCM, because the ICC of rank 2 is spread over three variables. In contrast in the bottom model, one can clearly identify the two ICCs, with all possible operators based on $\sigma_1$ and $\sigma_2$ for one ICC, and all the operators based on $\sigma_3$ for the other ICC.
In these examples, the top and bottom models of each panel are related by gauge transformations, which means that they are different representations of the same abstract MCM (see sec.~\ref{Sec:GT:MatrixRepresentation}). 
In the following section, we discuss GTs of MCMs.

\subsection{Gauge transformations of MCMs, equivalence classes, and counting}\label{sec:MCM:ppties}

Similarly to the binary case, the family of MCMs for $q>2$ is invariant under gauge transformations (GTs), meaning that the GT of an MCM is also an MCM (see Def.~\ref{def:GT} for the definition of GTs).
Indeed, %More generally, 
under a GT, a complete component remains complete and with the same rank, and a set of independent components remains independent (see App.~\ref{app:proof:GT:MCM}).
%%%%%%%%%%%%
%(i.e., applying the same GT to all the components of the set results in a set of ICs).
%(i.e., the GT of each component in a set of independent components are still independent). 
%(i.e., applying the same GT to all the components of a set of ICs results in a set of ICs).}
%%%%%%%%%%%%
This is an interesting property of MCMs if one wants to tackle model selection of discrete data from a more general perspective, independently of the basis in which the data is originally represented.
In the following, we lay out the foundations to work with MCMs in a basis-independent way.  % enabling model selection with MCMs in a representation independent manner.

%Let us start with the invariance of the family of MCMs under GT (see Definition~\ref{def:GT} of GTs). 
%We show in \textcolor{purple}{App.~\ref{app:proof:GT:MCM}} that, under a gauge transformation, a complete component remains complete and with the same rank, and independent components remain independent 
%(i.e., the GTs of each component in a set of independent components are still independent). 
%(i.e., applying the same GT to all the components of a set of ICs results in a set of ICs).
%Combining these properties gives that:
%%%%%%%%%%%%%%%%%%%%%%%%%%%%%%%%%%%%%%%%%%%
\begin{Property}\label{Th:GT:MCM}
    {\bf Gauge transformation of an MCM.}
    The gauge transformation of an MCM is also an MCM and has ICCs of the same ranks as the original model.  %and has the same number of ICCs, with the same ranks as in the original model. 
\end{Property}
Figure~\ref{fig:MCM:examples} shows examples of GTs of MCMs.  
In particular, it highlights %illustrates
the transformation of MCMs into a {\it preferred basis}, a useful notion introduced in the binary case~\cite{demulatier2024MCM}.
As noted previously, in some representations, a complete component $\M_a$ can spread over more than $r_a=\rank(\M_a)$ basis variables. 
For example, the top model for $q=4$ in Fig.~\ref{fig:MCM:examples}.b has a single CC $\M=\{(1,1),(2,2), (3,3)\}$ with operators of order~two despite having only a rank $r=1$.
Using the GT $\sigma_1=s_1s_2$ and $\sigma_2=s_1$, %(equivalently, $\alpha_1'=\alpha_1+\alpha_2$ and $\alpha_2'=\alpha_1$), 
$\M$ is transformed into the model $\M'=\{(1,0),(2,0), (3,0)\}$ (see App.~\ref{app:GT:MCM:ex}), in which all the operators are based solely on $r=1$ variable (see Fig.~\ref{fig:MCM:examples}.b bottom).
Ref.~\cite{demulatier2024MCM} introduced the notion of {\it preferred bases} of an MCM, which are spin bases in which each ICC $\M_a$ of the MCM appears in a compact form, based solely on $r_a=\rank(\M_a)$ basis elements just like for $\M'$ %in the example 
above. %or for the model in Fig.~\ref{fig:MCM:examples}.a bottom.
The model in Fig.~\ref{fig:MCM:examples}.a bottom is another example of an MCM represented in a preferred basis.
%%%%%%%%%%%%%%%%%%
%In such basis, an MCM corresponds to a partition of a subset of the basis variables, in which each part identifies an ICC (see dashed outlines in the figure)}: the variables are fully connected within each part, with all operators of all orders, and not connected between the parts.

\begin{Definition}
    {\bf Preferred basis of an MCM.} 
    A preferred basis of an MCM is a spin basis, $\basis^*=(\phi^{\vecmu_1}, \cdots, \phi^{\vecmu_n})$, in which each ICC~$\M_a$ is based solely on $r_a$ basis elements, where $r_a$ is the rank (and dimension) of $\M_a$. In such basis, each ICC appears in a compact form and the MCM corresponds to a partition of a subset of the basis variables in which each part identifies an ICC (see dashed outlines in Fig.~\ref{fig:MCM:examples} bottom -- the variables are fully connected within each part, with %all 
    operators of all orders, and not connected between the parts).
    %
    %The preferred basis of an MCM is not necessarily unique.
\end{Definition}
\noindent In other words, a preferred spin basis $\basis^*$ for an MCM defines a GT $\T:\spin\to\spin'=\basis^*(\spin)$ (see Definition~\ref{def:GT}) that transforms the MCM into its compact representation, with each ICC based on a minimal number of basis variables.
More generally, one can define a preferred basis for any set of ICs (for which the components are not necessarily complete) as a basis in which each component $\M_a$ is only expressed over $d_a=\dim(\M_a)$ basis variables.

%%%%%%%%%%%%%%%%%%%%%%%%%%%%
%Below we recall useful definitions and properties introduced in the binary case ($q=2$)~\cite{demulatier2024MCM} and that still hold for the more general case of $q$ larger than $2$.
%For a more formal definition of preferred basis of an MCM, we start by recalling the definition of basis of a spin model.
%%%%%%%%% DEF BASIS RECALL:
%We recall that a {\it basis} $\basis$ of a model component $\M_a$ is a minimal set of $d_a$~independent operators, $\basis_a=(\phi_1,\cdots,\phi_{d_a})$, that can generate all the operators of $\M_a$. 
%%%%%%%%%%%%%%%%%%%%%%%%%%%%

%%%% %%%%%%%%%%%%%%%%%%%%%%%%%%%% %%%%%%%%%%%%%%%%%%%%%%%%%%%%
%Thanks to the independence between ICCs, one can construct a basis of an MCM by taking the union of a basis for each ICC (see Property~\ref{ppty:MCM:basis-of-ICs}).
%%%%% %%%%%%%%%%%%%%%%%%%%%%%%%%%% %%%%%%%%%%%%%%%%%%%%%%%%%%%% 

It is always possible to construct a preferred basis for an MCM $\M=\cup_{a\in\mathcal{A}} \M_a$ (and more generally for a set of ICs), simply by combining the elements of a basis $\basis_{\M_a}$ for each IC~$\M_a$, thanks to the independence between the ICCs (see Property~\ref{ppty:MCM:basis-of-ICs}). 
We obtain a spin basis of the form $\basis^*=\cup_{a\in\mathcal{A}}\,\basis_{\M_a}\cup \overline{\basis_{\M}}$\,, where $\overline{\basis_{\M}}$ is a set of $(n-\dim(\M))$ independent operators chosen to complete the spin basis. 
In such a basis, the operators of each component $\M_a$ would be based solely on their respective basis variables $\basis_{\M_a}$ (by definition~\ref{def:ICs} of ICs), which is precisely what we want for a preferred basis.
%For each ICC~$\M_a$, any choice of a set of $r_a=\rank(\M_a)$ independent operators in $\M_a$ can be used to define its basis\,~$\basis_{\M_a}=(\phi^{\vecmu^a_{1}}, \cdots,\phi^{\vecmu^a_{r_a}})$.
Moreover, for MCMs, because the component~$\M_a$ are complete, we can choose any set of $r_a=\rank(\M_a)$ independent operators in $\M_a$ to define its basis\,~$\basis_{\M_a}=(\phi^{\vecmu^a_{1}}, \cdots,\phi^{\vecmu^a_{r_a}})$ (see Property~\ref{ppty:CC:dim-rank}).

This is precisely the construction we followed in each panel of Fig.~\ref{fig:MCM:examples} to define the GT~$\T$ (from top to bottom) of the model~$\M$ into a preferred basis.
In panel b), the top model $\M$ has a single CC, which can be fully generated by the operator $\vecmu_1=(1,1)$. We used this operator to create a preferred basis for $\M$, by defining the first basis element of the GT $\T$ as $\sigma_1=\phi^{\vecmu_1}(\spin)=s_1s_2$ and choosing the second basis element to be independent from $\sigma_1$.
In panel a), the model $\M$ has two ICCs; we constructed a preferred basis by combining the elements of a basis~$\basis_a$ for each ICC~$\M_a$, for which any choice of $r_a$ independent operators in $\M_a$ can be used.
The operators $\vecmu_1=(1,1,0)$ and $\vecmu_2=(1,0,2)$ defining the first two variables of the preferred basis, %first two basis variables
$\sigma_1=\phi^{\vecmu_1}(\spin)=s_1s_2$ and $\sigma_2=\phi^{\vecmu_2}(\spin)=s_1s_3^2$\,,
are two independent operators of the rank-$2$ ICC of~$\M$.
The operator $\vecmu_3=(0,0,1)$, which defines the third basis element $\sigma_3=\phi^{\vecmu_3}(\spin)=s_3$, is an independent operator of the other ICC of $\M$ (which is already in a compact form).
%%%%%%%%%%%%
Upon the GT $\T$, the model~$\M$ is transformed into the bottom model~$\M'$, which appears in a compact form with the rank-2 ICC based solely on the first two variables.

\begin{Property}\label{ppty:MCM:preferred-basis}
    {\bf Construction of a preferred basis of an MCM.} 
    A preferred basis $\basis^*$ of an MCM $\M=\cup_{a\in\mathcal{A}}\M_a$ is %can be
    constructed by combining the elements of a basis $\basis_{\M_a}$ for each ICC~$\M_a$: 
    \begin{align}
        \basis^*=\cup_{a\in\mathcal{A}}\,\basis_{\M_a}\cup \overline{\basis_{\M}}\,,
    \end{align}
    where $\overline{\basis_{\M}}$ is a set of $(n-r)$ independent operators that complete the spin basis, with $r=\rank(\M)$. 
    For each component $\M_a$, any set of $r_a=\rank(\M_a)$ independent operators in $\M_a$ can be chosen to define the basis $\basis_{\M_a}=(\phi^{\vecmu^a_{1}}, \cdots,\phi^{\vecmu^a_{r_a}})$. 
    The preferred basis of an MCM is therefore not necessarily unique.
\end{Property}
%
%%%%%%%%%%%%%%%%%
%The preferred basis of an MCM is not [necessarily] unique, as any maximal independent set of operators in each ICC can be used for its definition.
%%%%%%%%%%%%%%%%%
%\noindent We recall that the preferred basis is a spin basis for an $n$~spin system, which means that it is composed of $n$~independent operators and that it can be used to defined a GT of the form $\spin'=\basis^*(\spin)$ (see definition~\ref{def:GT}).
%%%%%%%%%%%%%%%%%

Using the construction above, any MCM can be represented in a preferred basis, in which the MCM corresponds to a partition of the basis variables, where each part of $r_a$~variables identifies an ICC of rank~$r_a$.
%From this, we deduce that 
This implies that MCMs with ICCs of the same ranks are related by a GT, because they can be mapped to the same partition in a preferred basis, and therefore they belong to the same equivalence class. %\textcolor{orange}{(see proof in App.)}.
This gives the following property, characterizing equivalence classes of MCMs:

\begin{Property}\label{Th:GT:MCM:inverse}
    {\bf Equivalence classes of MCMs.}
    Consider an MCM with $m$ ICCs and denote by $m_{r_a}$ the number of ICCs of rank~$r_a$ in the model, such that $m = \sum_{r_a=1}^n m_{r_a}$. For a given value of $q$ and number of spins $n$, all the MCMs with the same sequence of multiplicities $\{m_{r_a}\}_{1\leq r_a\leq n}$ of their ICC ranks are equivalent under gauge transformation, i.e. for any two such model, there exists a GT that can transform one into the other. They all belong to the same equivalence class of MCMs, which is identified by the sequence of multiplicities $\{m_{r_a}\}_{1\leq r_a\leq n}$.
\end{Property}

\begin{proof}
We give a sketch of the proof. 
Consider two MCMs $\M_x$ and $\M_y$ that have the same sequence of multiplicities $\{m_{r_a}\}_{1\leq r_a\leq n}$ (i.e., the same rank list $\{r_a\}_{a\in\mathcal{A}}$). 
We specify two GTs, $\T_x$ for $\M_x$ and $\T_y$ for $\M_y$, that bring each model into a preferred basis (in which each MCM corresponds to a partition of the basis elements).
We design the two GTs such that they map ICCs of the same rank in each model to the same basis elements (for example, an ICC of rank 2 could be mapped to the variables $\sigma_3$ and $\sigma_4$ for both MCMs).
This is possible because the two models have ICCs of the same ranks.
As a result, the transformed models $\T_x[\M_x]$ and $\T_y[\M_y]$ are identical MCMs (they correspond to the same partition of the basis elements), which means that $\M_x$ and $\M_y$ are equivalent. Composing the GTs, we can also write that $\M_y=\T_y^{-1}[\T_x[\M_x]]$, where $\T_y^{-1}\circ\T_x$ is a GT.
%\textcolor{orange}{[To put in App.]}
\end{proof}

As a consequence, the number of different classes of MCMs with the same rank $r$ is obtained by enumerating all possible (unordered) rank list $\{r_a\}_a$ satisfying the constraint that $\sum_{a\in\mathcal{A}}r_a=r$. This corresponds to all possible integer partitions of $r$.
\begin{Property}
    {\bf Number of classes of MCMs of rank $r$.} In a $q$-state spin system with $n$ spins, the number of equivalence classes of MCMs of rank $r$ (where $r\leq n$) is given by the number of integer partitions of $r$, independently of the value of $q$.
\end{Property}
\noindent This number is the same as in the binary case~\cite{demulatier2024MCM},
because MCMs of different classes only differ by the rank of their ICCs, independently of the value of~$q$ (see %as illustrated in 
Fig.~\ref{fig:MCM:factorization-completness} and~\ref{fig:MCM_complexity}). 
For example, for any integer value of $q$, there are two equivalence classes of MCMs with rank $r=2$ (the class of MCMs with two ICCs of rank $1$ each, and the class of MCMs with a single ICC of rank $2$), three equivalence classes for $r=3$, and five for $r=4$ (see complexity landscape in Fig.~\ref{fig:MCM_complexity}).
The value of $q$ instead changes the number of MCMs in each equivalence class.

\begin{Property}\label{ppty:NbMCM:inClass}
    {\bf Number of MCMs in a class.} 
    For a $q$-state spin system with $n$ spins, the number of models in the class of MCMs defined by the sequence of multiplicities $\{m_{r_a}\}_{1\leq r_a\leq n}$ (where $m_{r_a}$ denotes the number of ICCs of rank $r_a$ in the models of the class) is given by (see App.~\ref{app:ppty:NbMCM:inClass}):
    \begin{align}\label{eq:Nb_MCM:inClass}
    \mathcal{N}_{\rm MCM}(n,\{m_{r_a}\}_{1\leq r_a\leq n}) 
    = \displaystyle\frac{\mathcal{N}_q(n, r)}{\displaystyle\prod_{r_a=1}^n (m_{r_a}!)\;\, \mathcal{N}_q(r_a, r_a)^{m_{r_a}}}
    \;.
    \end{align}
    Here, %the factors $1/m_{r_a}!$ account for [the possible] permutations among ICCs of the same size, and 
    $\mathcal{N}_q(n, r)$ is the number of different ways to choose $r$ independent operators in a $q$-state spin system with $n$ spins, which is: 
    \begin{align}
    \mathcal{N}_q(n, r) = 
    \begin{cases}
        \;\;\displaystyle\prod_{i=0}^{r-1} (q^n-q^i)\,,
            &\qquad{\rm for }\;q\;\textrm{prime},\\[5pt]
        \;\;\displaystyle\prod_{k=1}^{K} p_k^{\,(m_k-1)n^2}\,
        \prod_{i=0}^{r-1} (p_{k}^{\,n}-p_{k}^{\,i})\,,
            &\qquad{\rm for }\;q\;\textrm{not prime}.
    \end{cases}
    \end{align}
    The result for $q$ prime was obtained using the same reasoning as in the binary case~\cite{demulatier2024MCM}.
\end{Property}
\noindent 
For example, for $n=3$ and $q=3$, there are $117$ MCMs with two ICCs of respective rank $r_1=1$ and $r_2=2$; Fig.~\ref{fig:MCM:examples}.a shows two of them.
The total number of spin models that are MCMs in an $n$-spin system can be obtained by summing the value in Eq.~\eqref{eq:Nb_MCM:inClass} over all possible integer partitions of~$r$, for all~$r$ from~$1$ to~$n$.

The invariance of the family of MCMs under gauge transformations allows one to approach model selection from a representation invariant perspective, and ultimately design model selection procedures able to capture patterns of the data (such as symmetries, co-dependencies) in a way that is independent of the basis in which the data is initially observed. 
This is not possible for instance if one restricts the selection to 
%%%%% to models with operators up to a fix order, such as pairwise models
pairwise models, because the family of pairwise spin models is not invariant under gauge transformation. %, \textcolor{red}{as shown in the example of Figure~\ref{fig:GT_and_Loops}.a, where a pairwise model is transformed into a model with operators of higher order (in this case, there is a $3$-spin operator).}
Figure~\ref{fig:GT_and_Loops}.a gives an example of a GT that transforms a pairwise model into a model with operators of higher order (in this case, there is a $3$-spin operator).

\subsection{Statistical properties of MCMs}
\label{sec:MCM:ppties}

\begin{figure}[!ht]
\centering
\includegraphics[width=0.93\linewidth]{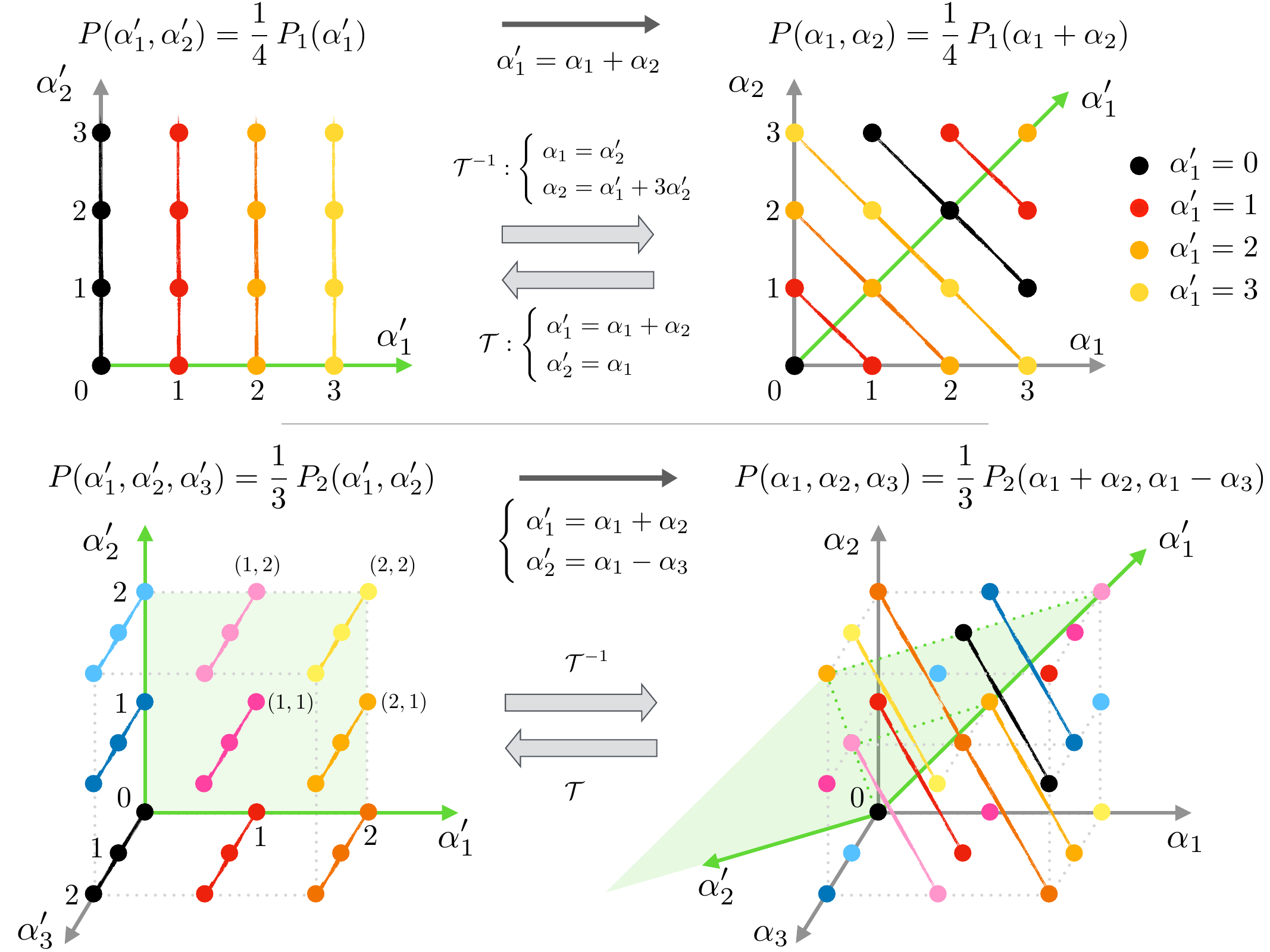}
\caption{(colors online)
    Representation of the model discrete probability distribution in state space for two ICCs of Fig.~\ref{fig:MCM:examples}. Each dot represents a state of the spin system and the color represents the probability of the state. States with the same color have the same probability.
    {\bf Top panel.} Representation of the probability distribution of the MCM with a single ICC in Fig.~\ref{fig:MCM:examples}.b. In the best basis, defined as the color variables $(\alpha_1',\alpha_2')$ in the bottom panel of Fig.~\ref{fig:MCM:examples}.b, one can see directly that the model probability distribution factorizes and is uniform along the direction of $\alpha_2'$, which gives the state space representation in the top left panel. After the GT $(\alpha_1,\alpha_2)=\T^{-1}[(\alpha_1',\alpha_2')]$ (top MCM in Fig.~\ref{fig:MCM:examples}.b), the probability distribution remains uniform in the direction of $\alpha_2'$, which is perpendicular to the direction of  $\alpha_1'=\alpha_1+\alpha_2$. This is a non-trivial factorization when looking at the system of variables $(\alpha_1,\alpha_2)$.
    {\bf Bottom panel.} Representation of the probability distribution of the single ICC based on the spin variables $(\sigma_1,\sigma_2)$ in the bottom panel of Fig.~\ref{fig:MCM:examples}.a. We denote by $(\alpha_1',\alpha_2',\alpha_3')$ the corresponding color variables of this basis, which is a preferred basis for the ICC, and by $(\alpha_1,\alpha_2,\alpha_3)$ the basis variables for the top panel of Fig.~\ref{fig:MCM:examples}.a. Independently of the basis representation, the probability distribution for this ICC is uniform in the direction of $\alpha_3'$ (indicated here by the color remaining identical in this direction), which is perpendicular to the plane defined by $\alpha_1'$ and $\alpha_2'$. 
    The independencies in the model probability distribution 
    (here indicated by the uniform direction)
    are non-trivial in the basis of the variables $(\alpha_1,\alpha_2,\alpha_3)$ (bottom right panel),
    but much easier to see in the basis of the variables $(\alpha_1',\alpha_2',\alpha_3')$ (bottom left panel).
    }
\label{fig:MCM:StateSpace-rotation}
\end{figure}

Just like in the binary case~\cite{demulatier2024MCM}, when represented in a preferred basis, an MCM corresponds to a partition of a subset of the basis variables in which each part identifies an ICC: the variables are fully connected inside the parts, with all operators of all orders, and not connected between the parts (see the models in Fig.~\ref{fig:MCM:examples} bottom).
The MCM thus completely models all the statistical dependencies within each part (each ICC), but no statistical dependencies at all between the parts, as if they were independent (see Fig.~\ref{fig:MCM:factorization-completness}, in the right panel, subspaces of different colors are considered statistically independent by the MCM in the left panel).
This peculiar structure of MCMs endows them with properties that make modeling data with MCMs relatively efficient, in comparison to other common modeling approaches.
In particular, the independence between the ICCs allows writing the model probability distribution as a product of marginal distributions over each ICC, when expressed in a preferred basis (see Fig.~\ref{fig:MCM:factorization-completness} and~\ref{fig:MCM:StateSpace-rotation}).
Moreover, the completeness of the ICCs enables a re-parametrization of the components, with which most of the quantities of interest in the context of statistical inference can be easily computed (see next section~\ref{sec:MCM:inference}).
%%%%%%%%%%%%%%%%%%%%%%%%%%%%%%%%%%%%%%%%%
%%%%%%%%%%%%%%%%%%%%%%%%%%%%%%%%%%%%%%%%%
% Property 1: Factorization of the probability distribution of MCMs over their ICCs in a preferred basis, 
% Property 2: Equivalent parametrization of ICCs in terms of the state probabilities.
%%%%%%%%%%%%%%%%%%%%%%%%%%%%%%%%%%%%%%%%%
%%%%%%%%%%%%%%%%%%%%%%%%%%%%%%%%%%%%%%%%%
Below, we formalize these two properties.

When written in a preferred basis, there are no dependencies between the basis variables corresponding to different ICCs of an MCM. This means that the multivariate probability distribution of the MCM factorizes over the basis elements of its ICCs. 
We recall that a preferred basis $\basis^*$ is a spin basis for the $n$~spin system, which means that it can be used to define a GT~$\T$ of the form $\spin'=\basis^*(\spin)$ (see definition~\ref{def:GT}). 
%, or equivalently $\state'=\Tmat_{\basis}\, \state$, where $\Tmat$ is the matrix representation of the GT.

\begin{Property}\label{Ppty:Factorization}
    {\bf Factorization of MCMs over their ICCs.}
    As a consequence of the independence between ICCs, the probability distribution of an MCM~$\M=\cup_{a\in\mathcal{A}}\,\M_a$ factorizes over the probability distributions of its ICCs $\M_a$, when expressed
    % in a preferred basis, i.e. in terms of [...] 
    in terms of the operators 
    $\basis^*(\spin)=\cup_{a\in\mathcal{A}}\,\basis_a(\spin)\cup\overline{\basis_{\M}}(\spin)$ of a preferred basis of $\M$ (where each $\basis_a$ denotes a basis of the ICC~$\M_a$ -- see property~\ref{ppty:MCM:preferred-basis}): 
\begin{equation}\label{eq:factorization}
    p(\spin\,|\,\vecg,\mathcal{M})=\frac{1}{q^{(n-r)}}\,\prod_{a\in\mathcal{A}}
\,p_a\big(\basis_a(\spin)\,|\,\vecg_a^\prime,\mathcal{M}_a^\prime\big)\,.
\end{equation}
    Each $\M_a^\prime$ denotes the ICC $\M_a$ represented in terms of the basis variables $\basis_a(\spin)$, and each $p_a$ is a probability distribution over these $r_a=\rank(\M_a)$ spin variables only. 
    The vector $\vecg_a^\prime$ of parameters of the ICC $\M_a'$ corresponds to a permutation of the parameters of the vector $\vecg_a$ that results from the gauge transformation $\spin'=\basis^*(\spin)$ (see Def.~\ref{def:GT_M}).
    The prefactor corresponds to a uniform distribution over the $(n-r)$ $q$-state variables not modeled by $\M$.
\end{Property}
\noindent In equation~\eqref{eq:factorization}, we used that for a GT, the probability distribution of the model $\M'=\T[\M]$ in the new basis $\spin'=\basis^*(\spin)$ is related to that of the model $\M$ in the original spin basis~$\spin$ by
$p(\spin\,|\,\vecg,\M)=p(\spin'\,|\,\vecg',\M')$ (see Eq.~\eqref{GT:def_on_model}).
For each ICC~$\M_a$, the vector~$\basis_a(\spin)$ corresponds to a subset of variables of the new basis~$\spin'=\basis^*(\spin)$; it can be interpreted as the transformed state~$\spin'$ reduced to the subspace modeled by~$\M_a$.

%\textcolor{orange}{The GT $\spin'=\basis^*(\spin)$ can be equivalently written in terms of the corresponding color variables as $\state'=\Basis^*(\state)$, where $\Basis^*(\state)=(\state\cdot\vecmu_1, \dots,\, \state\cdot\vecmu_n)$ is the vector with $n$ elements resulting from the gauge transformation of $\state$ into the basis $\Basis^*=(\vecmu_1,\cdots,\vecmu_n)$ (see Property~\ref{def:GT:matrix}). Here $\Basis^*$ is a preferred basis of $\M$, which takes the form $\Basis^*=\cup_{a\in\mathcal{A}} \Basis_{\M_a}\cup \overline{\Basis_{\M}}$.}

The GT $\spin'=\basis^*(\spin)$ can be written equivalently for the corresponding color variables as the vector-matrix product $\state'=\state\,\Tmat_{\basis^*}$,
where $\Tmat_{\basis^*}=(\vecmu_1,\,\dots,\,\vecmu_n)$ is an $n\times n$-matrix whose columns are the $n$ operators of the preferred basis $\Basis^*=\cup_{a\in\mathcal{A}} \Basis_{\M_a}\cup \overline{\Basis_{\M}}$ (see Property~\ref{def:GT:matrix}). 
To use a notation analogous to that used for the spin variables,
we denote by $\Basis^*(\state)=\state\,\Tmat_{\basis^*}$ the vector resulting from the gauge transformation of $\state$ into the basis $\Basis^*=(\vecmu_1,\cdots,\vecmu_n)$, i.e. $\Basis^*(\state)=(\state\cdot\vecmu_1, \dots,\, \state\cdot\vecmu_n)$.
The gauge transformation can thus be written as $\state'=\Basis^*(\state)$. Similarly, the vector~$\Basis_a(\state)=(\state\cdot\vecmu^{a}_1, \dots,\, \state\cdot\vecmu^a_{r_a} )$ denotes the transformed state $\state'$ reduced to the subspace modeled by the ICC $\M_a$.
Equation~\eqref{eq:factorization} can then be written in terms of the color variables~$\state$ as:
\begin{equation}\label{eq:factorization:2}
    p(\state\,|\,\vecg,\mathcal{M})=\frac{1}{q^{(n-r)}}\,\prod_{a\in\mathcal{A}}
\,p_a\big(\Basis_a(\state)\,|\,\vecg_a^\prime,\mathcal{M}_a^\prime\big)\,.
\end{equation}

When written in a preferred basis, each ICC~$\M_a$ of the MCM is a complete model (see Sec.~\ref{sec:def:HO:vectorPottsModel}) for the sub-system of $r_a$~variables $\state_a'=\Basis_a(\state)$, where $\Basis_a$ is a basis of~$\M_a$.
This means that $\M_a$ can model any possible statistical patterns of the discrete variables~$\state_{a}'$.
The component has $(q^{r_a}-1)$ parameters $g_{\vecmu}$ that fully parametrize the $r_a$-dimensional sub-system and is thus equivalently specified by the $(q^{r_a}-1)$ state probabilities $p(\state_a')$ for all states $\state_a'\in[\Field^{r_a}]^*$ of the sub-system (the last probability being fixed by normalization, $p(\veczero)=1-\sum_{\state_a'\neq\veczero} p(\state_a')$).
The ICC~$\M_a'$ can thus be re-written as the following parametric family of probability distributions: 
\begin{align}\label{eq:ICC_PDF}
    p_a(\state_a'\,|\,\,\boldsymbol{\eta}_a,\M_{a}') = \sum_{\vecnu\in[\Field^{r_a}]^*}\, \eta_{\vecnu} \, \delta(\state_a'-\vecnu)+\bigg(1-\sum_{\vecnu\in[\Field^{r_a}]^*} \eta_{\vecnu}\bigg)\,\delta(\state_a'-\boldsymbol{0})\,,
\end{align}
where each $\eta_{\vecnu}$ is a parameter that represents the probability that the system is in the state $\vecnu=\state_a'$, and $\boldsymbol{\eta}_a$ is a vector that contains all the $(q^{r_a}-1)$ parameters $\eta_{\vecnu}$.

Within each ICC, there is therefore a bijection between the $(q^{r_a}-1)$ parameters $g_{\vecmu}$'s and $\eta_{\vecnu}$'s (see Sec.~\ref{sec:def:HO:vectorPottsModel}). 
For any choice of the parameters $g_{\vecmu}$'s of the ICC, the parameters $\eta_{\vecnu}$'s 
correspond to the state probabilities of the system given by the definition of the complete spin model in Eq.~\eqref{eq:expansion:Fourier_basis}, $\eta_{\vecnu}=p(\vecnu\,|\,\vecg_a',\M_a')$ for each state~$\state_a=\vecnu$ of the sub-system (which corresponds to an inverse discrete Fourier transform). Reciprocally, the parameters $g_{\vecmu}$ can be obtained from the parameters $\eta_{\vecnu}$ by the discrete Fourier transform of the log-probabilities, as in Eq.~\eqref{eq:pdf:FT:1}.

\begin{Property}\label{Ppty:ICC:re-parametrization}
{\bf Equivalent parametrization of ICCs in terms of the state probabilities.}
As a consequence of the completeness of the ICCs, the probability distribution of an ICC~$\M_a$ can be equivalently parametrized by the $(q^{r_a}-1)$ state probabilities
    $\eta_{\state_{a}^{\prime}} = p_a\big(\state_{a}^{\prime}\,|\,\vecg_a^{\prime},\,\mathcal{M}_a^\prime\big)$ for all state $\state_a'\in[\Field^{r_a}]^*$ 
(one of the state probabilities is fixed by normalization), where $\state_{a}^{\prime} = \Basis_a(\state)$ is the transformed states reduced to the $r_a$-dimensional sub-system modeled by~$\M_a'$.
%%%%%%%%%%%%%%%
%can be equivalently parametrized by the $(q^{r_a}-1)$ state probabilities $\eta_{\vecnu} = p_a\big(\vecnu\,|\,\vecg_a^{\prime},\,\mathcal{M}_a^\prime\big)$ for all state $\vecnu\in[\Field^{r_a}]^*$ of the variables $\state_a'= \Basis_a(\state)$.
\end{Property}

The factorization and completeness properties of MCMs mean that, at best fit, an MCM corresponds to an approximation of the data that neglects all statistical dependencies between the variables of different ICCs and preserves all the other statistical patterns.
In the context of model selection, the best MCM for a given dataset therefore corresponds to a factorization of the model probability distribution that best matches the structure of the data, optimally balancing goodness-of-fit of the data against model complexity (methods for this selection are discussed in the next section). This optimal factorization could be in the original basis of the data or in any other gauge-transformed basis (see Fig.~\ref{fig:MCM:StateSpace-rotation}).
%
%At best fit, this MCM thus provides an optimal factorization of the empirical distribution.
At best fit, this MCM thus provides a factorization of the empirical distribution that optimally approximates the data, by neglecting statistical dependencies between variables where they are the least relevant (variables assigned to different ICCs) and preserving them where they are the most relevant (within ICCs).

In a given basis, the number of possible factorizations over $r$ basis elements corresponds to the number of partitions of these $r$ variables into distinct parts, which is equal to the Bell number of $r$, denoted $B_r$. The number of different MCMs that factorize in the same preferred basis (denoted $MCM^*$) is therefore obtained by summing this number over all possible model ranks, from $r=0$ to $n$, and accounting for the possible choices of the $r$ basis operators on which the model is based:
% We denote $MCM^*$ the set of MCMs that factorizes in the same preferred basis.
\begin{align}
    \mathcal{N}_{MCM^*}(n)=\sum_{r=0}^n {n \choose r} B_r\,.
\end{align}
This number doesn't depend on the value of $q$ and is thus identical to the binary case (see Fig.~\ref{fig:MCM:factorization-completness}). %\textcolor{orange}{To move elsewhere?}

%\begin{Lemma}
%    {\bf Number of MCMs with a common preferred basis.} 
%The number of MCMs [that have the same/in a given] preferred basis % that share the same preferred basis
%is given by the number of partitions of the $n$ variables of the basis in distinct parts, which is equal to the Bell number of $n$, independently of the value of $q$. \textcolor{purple}{Summed for all $r$ between 0 and $n$.}
%\end{Lemma}

\subsection{Statistical inference and model selection with MCMs}
\label{sec:MCM:inference}

A significant advantage of Minimally Complex Models over other parametric models for discrete data is that most useful quantities for statistical modeling have a closed-form analytical expression and are thus cheap to compute numerically.
Similarly to the binary case~\cite{demulatier2024MCM}, the factorization property~\ref{Ppty:Factorization} of the model probability distribution implies that the likelihood and the evidence~\eqref{eq:evidence} of an MCM also factorize over the ICCs when written in a preferred basis, and the equivalent parametrization %of ICCs 
property~\ref{Ppty:ICC:re-parametrization} makes these quantities easy to compute analytically for each ICC.
Below, we give closed-form expressions for the maximum likelihood, the geometric complexity appearing in the MDL principle Eq.~\eqref{eq:MDL},  
and the %marginal likelihood 
model evidence~\eqref{eq:evidence}.
These results (see Appendix~\ref{app:ICC} for proofs) are direct extensions of the formulae obtained in the binary case~\cite{demulatier2024MCM} to the more general discrete case.

Consider a $q$-state spin system with $n$ spins represented by the discrete variables $\state = (\alpha_1, \cdots, \alpha_n) \in \Field^n$ and a dataset composed of $N$ observations of the system denoted $\data = (\state^{(1)}, \cdots, \,\state^{(N)})$.
We assume that the system is stationary and that the observed states are randomly sampled from the same underlying distribution, $p(\state\,|\,\vecg,\,\M)$, defined by a parametric model $\M$.
Under this assumption, the probability that the model $\M$ with parameters~$\vecg$ assigns to the data, namely the likelihood of the model, is $P(\data\,|\,\vecg,\M)=\prod_{i=1}^{N}p(\state^{(i)}\,|\,\vecg,\M)$.
In the following, we take the model~$\M$ to be an MCM %we model the data with an MCM 
$\M=\cup_{a\in\mathcal{A}}\M_a$. We consider a preferred basis of~$\M$ of the form $\Basis^*=\cup_{a\in\mathcal{A}}\,\Basis_a\cup\overline{\Basis_{\M}}$, where each $\Basis_{a}$ is a basis of the ICC $\M_a$, and denote by $\M'=\cup_{a\in\mathcal{A}}\M_a'$ the representation of $\M$ in this basis (i.e., $\M'=\T_{\Basis^*}[\M]$ and $\M_a'=\T_{\Basis^*}[\M_a]$ for all $a\in\mathcal{A}$).
We denote by $r$ the rank of $\M$ and by $r_a$ the rank of each ICC $\M_a$.

Using the factorization Property~\ref{Ppty:Factorization}, the likelihood of an MCM can be written as the product of the likelihood of each ICC, when expressed in a preferred basis: 
\begin{equation} \label{eq:likelihoodMCM}
    P(\data\,|\,\vecg,\,\M) 
    \,= P(\Basis^*(\data)\,|\,\vecg',\,\M')
    =\frac{1}{q^{N(n-r)}}\prod_{a\in\mathcal{A}} P_a\big(\data_a^{\prime}\,|\,\vecg_a^{\prime},\,\mathcal{M}_a^\prime\big)\,,
\end{equation}
where, for each ICC~$\M_a$, $\data_a^{\prime} = \Basis_a(\data)$ denotes the dataset $\data$ transformed in the basis $\Basis^*$ (preferred basis of $\M$) and reduced to the basis variables $\state_a'=\Basis_a(\state)$ (i.e., reduced to the subspace modeled by the ICC).
Each term $P_a\big(\data_a^{\prime} \,|\,\vecg_a^{\prime},\,\mathcal{M}_a^\prime\big)=\prod_{i=1}^N p_a(\state_a'^{(i)}\,|\,\vecg_a^{\prime},\,\mathcal{M}_a^\prime)$ is the likelihood of the ICC $\M_a'$ reduced to the variables $\state_a'$\,; it is a function of the $(q^{r_a}-1)$ parameters in the vector $\vecg_a'$.
Using Property~\ref{Ppty:ICC:re-parametrization}, this likelihood is equivalently parametrized by the $(q^{r_a}-1)$ state probabilities $\eta_{\vecnu} = p_a\big(\vecnu\,|\,\vecg_a^{\prime},\,\mathcal{M}_a^\prime\big)$,
where $\vecnu\in[\Field^{r_a}]^*$ denotes the state of the basis variables $\state_{a}^{\prime}$. %reduced to the ICC.
The likelihood of an ICC expressed in terms of the parameters~$\eta_{\vecnu}$ is then:
\begin{align}\label{MCM:likelihood:ICC}
    P_a(\data_a'\,|\,\boldsymbol{\eta}_a,\M_a')
     & =\prod_{\vecnu\in\Field^{r_a}} \,\eta_{\vecnu}^{\;\,k_{\vecnu}(\data_a')}\,,
\end{align}
where $k_{\vecnu}(\data_{a}')$ is the number of times the state $\state_a'=\vecnu$ occurs in the dataset $\data_a'$, and $\boldsymbol{\eta}_a$ denotes the vector of all the parameters $\eta_{\vecnu}$ of $\M_a'$ (excluding $\eta_{\veczero}$ which is fixed by normalization).
Consequently, the values of the parameters $\hat{\boldsymbol{\eta}}_a={\rm argmax}_{\boldsymbol{\eta}_a}P(\data_a'\,|\,\boldsymbol{\eta}_a,\M_a')$ that best fit the data within each ICC are directly equal to the empirical probabilities: 
\begin{Property}
    \textbf{Maximum likelihood estimates.} 
    For each ICC~$\M_a'$, %at best fit, the parameters $\eta_{\state_{a}^{\prime}}$ are equal to/given by the [corresponding] empirical probabilities: 
    the best fit 
    %$\boldsymbol{\hat{\eta}}_a
    %=\big(\hat{\eta}_{\vecnu}\big)_{\vecnu}
    of the parameters $\boldsymbol{\eta}_a=\big(\eta_{\vecnu}\big)_{\vecnu}$ is given by the empirical probabilities:
    \begin{equation}\label{eq:max-likelihood:parameters}
        \hat{\eta}_{\vecnu}=\frac{k_{\vecnu}(\data_a^{\prime})}{N}\,,
        \qquad \textrm{for all}\; \vecnu\in[\Field^{r_a}]^*\,,
    \end{equation}
    where $k_{\vecnu}(\data_a^{\prime})$ is the number of times the state $\state_a^{\prime}=\vecnu$ occurs in the %transformed and 
    reduced dataset $\data_a^{\prime}=\Basis_a(\data)$.
\end{Property}
\noindent The bijection between the parameters $\vecg$ and the $\boldsymbol{\eta}$ greatly simplifies the inference step, allowing one to compute the maximum likelihood parameters without having to resort to any parameter fitting (e.g., through gradient descent), which are often computational demanding for high-dimensional models.
Replacing these parameter values in the likelihood function~\eqref{MCM:likelihood:ICC}, we obtain a closed-form expression for the maximum likelihood of an MCM: 
\begin{Property}
\label{ppty:max-log-L}
\textbf{Maximum likelihood of MCMs.} The maximum likelihood of an MCM $\M=\cup_{a\in\mathcal{A}}\M_a$ for a dataset $\data$ is:
    \begin{equation}\label{eq:max-likelihoodMCM}
        P(\data\,|\,\vecghat,\mathcal{M})=\frac{1}{q^{N(n-r)}}\,\prod_{a\in\mathcal{A}}
    \left[
        \prod_{\vecnu\in\data_a^{\prime}}\,
        \left(\frac{k_{\vecnu}(\data_a^{\prime})}{N}\right)^{\vspace{-5mm}k_{\vecnu}(\data_a^{\prime})}
    \right]\,,
        %P(\data\,|\,\vecghat,\mathcal{M})=\frac{1}{q^{N(n-r)}}\,\prod_{a\in\mathcal{A}}
    %\left[
    %\prod_{\state_a^{\prime}\in\data_a^{\prime}}\,
    %\left(\frac{k_{\state_a^{\prime}}}{N}\right)^{\vspace{-4mm}k_{\state_a^{\prime}}}\right]\,,
    \end{equation}
    where the second product is over %all
    the distinct %unique
    states $\state_a^{\prime}=\vecnu$ occurring in the reduced dataset $\data_a^{\prime} = \Basis_a(\data)$ (written in the basis $\Basis_a$ of the ICC $\M_a$)
    %(written in the preferred basis $\Basis^*=\cup_a\Basis_a$ of $\M$ and reduced to the subset of variables $\data_a^{\prime} = \Basis_a(\data)$),
    and $k_{\vecnu}(\data_a^{\prime})$ is 
    %the number of these occurrences.
    the number of times they occur.
    The maximum log-likelihood can then be written as: % in the form:
    \begin{equation}\label{eq:max-log-likelihoodMCM}
    \log P(\data\,|\,\vecghat,\mathcal{M})
       =-N(n-r)\log(q)
       -N\,\sum_{a\in\mathcal{A}}
        S(\data^{\prime}_a)\,,
    \end{equation}
    where
    \begin{equation}\label{eq:Entropy:ICC}
    %S(\data^{\prime}_a)=-\sum_{\state_a^{\prime}\in\data_a^{\prime}}\,
        %\frac{k_{\state_a^{\prime}}}{N}\,\log
        %\left(\frac{k_{\state_a^{\prime}}}{N}\right)
    S(\data^{\prime}_a)=-\sum_{\vecnu\in\data_a^{\prime}}\,
        \frac{k_{\vecnu}(\data_a^{\prime})}{N}\,\log
        \left(\frac{k_{\vecnu}(\data_a^{\prime})}{N}\right)
    \end{equation}
    is the entropy of the dataset $\data_a^{\prime}=\Basis_a(\data)$ 
    %which corresponds to the original data $\data$ reduced to the subspace modeled by the ICC $\M_a$.
    (which is a reduction of the original data $\data$ to the subspace modeled by the ICC $\M_a$).
    %$S(\data^{\prime}_a)$ is the entropy of the data reduced to the subspace modeled by the ICC $\M_a$.
\end{Property}
%
%The maximum log-likelihood in Eq.~\eqref{eq:max-log-likelihoodMCM} could have been obtained more directly, by using that the maximum log-likelihood of a model is equal to minus $N$ times the sum of the entropy of the data and of the Kullback-Leibler divergence between the model at best fit and the data. 
%The log-likelihood of an MCM is the sum of the log-likelihood of its ICCs \textcolor{red}{[for the data $\data_a'=\basis_a(\data)$ reduced to their respective sub-systems,]} and we can then use the previous identity to compute the maximum log-likelihood of an ICC $\M_a$ for the dataset $\data_a'=\basis_a(\data)$ reduced to the sub-systems of variables $\state_a'$. 
%The entropy of the reduced data is $S(\data^{\prime}_a)$ and the Kullback-Leibler divergence between the ICC and the data is zero at best fit, because the ICC is complete over the subspace based on the variables $\state_a'$. 
%This leads back to Eq.~\eqref{eq:max-log-likelihoodMCM}.
%
\noindent The maximum log-likelihood in Eq.~\eqref{eq:max-log-likelihoodMCM} can also be obtained more directly by expressing the %(maximum)
log-likelihood of each ICC $\M_a$ for the system reduced to the variables $\state_a'=\Basis_a(\state)$, as minus $N$ times the sum of the entropy of the reduced data $S(\data^{\prime}_a)$ and of the Kullback-Leibler divergence between the ICC and the data. This latter is zero at best fit, because the ICC is complete over the subspace based on the variables $\state_a'$.

A useful quantity for model selection is the model evidence (or marginal likelihood) defined in Eq.~\eqref{eq:evidence}.
Selecting the model with the largest evidence, instead of the largest maximum likelihood, allows basing the selection of the model not only on the goodness of its fit to the data, but also on its complexity.
Within a family of models, the model with the largest evidence achieves an optimal balance between goodness-of-fit and simplicity.
Here we compute the evidence assuming Jeffreys' prior~\cite{jeffreys1946invariant} over the parameters (see App.~\ref{app:ICC:JeffreysPrior}). Using the factorization property~\ref{Ppty:Factorization} of MCM, Jeffreys' prior and the model evidence factorize over the ICCs, and, using the completeness of the ICCs, we obtain a closed-form expression for the model evidence (see App.~\ref{app:ICC:Evidence}):
\begin{Property}\label{ppty:log-E:MCM}
    \textbf{Evidence (or marginal likelihood) of MCMs.}
    The evidence of an MCM $\M=\cup_{a\in\mathcal{A}}\M_a$ %with Jeffreys' prior
    is the product of the evidence of each ICC when expressed in a preferred basis %in terms of the operators
    $\Basis^*(\state)=\cup_{a\in\mathcal{A}}\,\Basis_a(\state)\cup\overline{\Basis_{\M}}$ %of a preferred basis
    of $\M$: 
    \begin{equation}\label{eq:evidenceMCM}
        P(\data\,|\,\mathcal{M}) 
        =\frac{1}{q^{N(n-r)}}\prod_{a\in\mathcal{A}} 
        P_a\big(\Basis_a(\data)\,|\,\mathcal{M}_a^\prime\big)\,.
    \end{equation}
    Using Jeffreys' prior, the evidence of each ICC $\M_a'$ for the reduced dataset $\dataprime_{a} = \Basis_a(\data)$ is given by (see proof in App.~\ref{app:ICC:Evidence}):
    \begin{align}\label{eq:evidenceICC}
    P_a(\dataprime_{a}\,|\,\M_{a}') 
    \;=\;
        \frac{\displaystyle\Gamma\left(\frac{q^{r_a}}{2}\right)}{\displaystyle\Gamma\left(N+\frac{q^{r_a}}{2}\right)}
        \;\prod_{\vecnu\in\data_a^{\prime}}\left(\frac{\displaystyle\,\Gamma\left(k_{\vecnu}(\data_{a}^{\prime})+\frac{1}{2}\right)}
        {\displaystyle\sqrt{\pi}}\right)\;,
    \end{align}
where $\Gamma$ denotes the gamma function. The product is over all the distinct states $\state_a^{\prime}=\vecnu$ observed in the reduced dataset $\data_{a}^{\prime}$ and $k_{\vecnu}(\data_{a}^{\prime})$ is the number of times they %that state occurs 
occur. % in the dataset.}
%is the number of occurrences of the state $\vecnu$ in the data $\data_{a}^{\prime}$. 
For $q=2$, we recover the result derived by Ref.~\cite{demulatier2024MCM} in the binary case.
\end{Property}
\noindent The evidence of an ICC in Eq.~\eqref{eq:evidenceICC} only depends on its rank $r_a$, on the empirical distribution of the reduced dataset $\dataprime_{a} = \Basis_a(\data)$, and on the number $N$ of datapoints. The evidence of a statistical model is usually expensive to compute from Eq.~\eqref{eq:evidence}, due to the high-dimensional integral. Thanks to the closed-form expression~\eqref{eq:evidenceICC}, the evidence of any MCM can be computed efficiently with minimal computational resources, making it feasible in practice to perform model selection among MCMs on systems of reasonably large sizes (with hundreds of variables). In Sec.~\ref{Sec:Applications}, we apply this approach to real data. 

%\textcolor{red}{Another common approach to compare [the performance of] statistical models is the minimum description length (MDL) principle.}
Models can also be selected following the minimum description length (MDL) principle~\cite{rissanen1986stochastic, rissanen1996fisher, grunwald2007minimum}, for which the best model is the one providing the most compressed representation of the data. Equation~\eqref{eq:MDL} gives the description length of a spin model for a sufficiently large number $N$ of samples (expanded to order $O(1)$ in $N$). The last term $c_{\M}$ in the expansion is the geometric complexity of the model. It is defined as a high-dimensional integral~\eqref{app:def:cM}, which can be computed analytically %we can compute 
for MCMs (see App.~\ref{app:ICC:Complexity}).

\begin{Property}
    \textbf{Geometric complexity of MCMs.} Thanks to the independence between ICCs, the geometric complexity of an MCM is given by the sum of the complexity of its ICCs:
    \begin{align}\label{eq:MCM:Complexity}
    &c_{\M} =\sum_{a\in\mathcal{A}}  \,c_{\M_{a}}\,,%\\
    %&[{\rm where}\;\; c_{\M_{a}} = \log \int d \vecg_a \sqrt{\det \FIM_{\M_a}(\vecg_a)}.] \nonumber
\end{align}
    where the complexity of an ICC only depends on its rank~$r_a$ and is given by (see proof in App.~\ref{app:ICC:Complexity}):
\begin{align}\label{eq:c_M}
    c_{\M_a} = \frac{q^{r_a}}{2} \log \pi - \log \Gamma\left(\frac{q^{r_a}}{2}\right)\,,
\end{align}
where $\Gamma$ is the gamma function. For $q=2$, one recovers the result obtained by~\cite{beretta2018stochastic} in the binary case. 
\end{Property}
\noindent Figure~\ref{fig:MCM_complexity} shows the complexity landscape of MCMs for different values of $q$. 
%Fig.~\ref{fig:MCM_complexity} shows the shift in the complexity landscape as one increases the value of $q$.
Observe a shift of the entire landscape towards larger complexities for increasing values of $q$, reflecting %which reflects
the increase in the number of parameters with $q$. 
Note the two limit lines, sub-complete models (MCMs with a single ICC) are the simplest among all MCMs with the same number of parameters and rank, while IMCMs are the most complex.
For binary systems, Ref.~\cite{demulatier2024MCM} has hypothesized that classes of MCMs are the classes of lowest complexity among models with the same number of parameters and the same rank (hence their name). Although they verified that this is the case for %models of 
systems with $n=4$ spins or less, this hypothesis has not been proved in the more general case.
However, it is interesting to point out that MCMs likely span %the whole 
a large part of the complexity landscape of all $q$-state spin models, because they cover a large set of combinations of parameter number and rank. Indeed, at fixed number of parameters, it is expected that the higher the rank of the model (number of degrees of freedom of the model), the higher its complexity~\cite{beretta2018stochastic}.
In the binary case, Ref.~\cite{beretta2018stochastic} showed that the sub-complete models are the simplest among all models with the same number of parameters and rank, and the independent models (which are MCMs when $q=2$) are the most complex.
%
%This is another reason why it is useful/interesting to use them as a first processing step in model selection.

%\textcolor{orange}{Comment: all speculation after the ``However''.}

\begin{figure}[!ht]
\centering
\caption{(colors online)
    {\bf Complexity $c_{\mathcal{M}}$ of all MCMs with $n=4-5$ spins for varying values of $q$.}
    For a given value of $q$, all the models with the same rank $r$ and rank sequence $\{r_a\}_{a\in\mathcal{A}}$ of their ICCs are equivalent. As a consequence, the number of classes of MCMs in an $n$-spin system %is independent of the value of $q$.
    is the same for all values of $q$. Here, one observes the shift in complexity of these classes as one increases the value of~$q$.
    %complexity landscape or table for all MCMs for $n=4$ (or more). 
    %\textcolor{purple}{Counting + complexity landscape of the MCM only (complexity of other models would take too much time to compute -- avoid).} 
    The two dashed lines indicate the complexity of IMCMs (red) and the complexity of sub-complete models (black).
    }
\label{fig:MCM_complexity}
\end{figure}

For a large number $N$ of samples, the expansion of the negative log-evidence assuming Jeffreys' prior matches the expansion of the description length~$L(\data\,|\,\M)$ in Eq.~\eqref{eq:MDL}~\cite{balasubramanian1997statistical, myung2000counting}. In App.~\ref{app:log-Evidence:expansion}, we verify that this is indeed the case for the closed-form expression~\eqref{eq:evidenceICC} of the log-evidence of MCMs and give the next order term of the expansion.
%and found that the next order term in the expansion is of the form:.
%
The advantage of using the closed-form expression~\eqref{eq:evidenceICC} of the model evidence over the description length expansion~\eqref{eq:MDL} is that it is exact for any value of $N$ and it is cheap to compute in practice. 
In particular, it provides an accurate measure for comparing models when the number of datapoints is small or, more precisely, in undersampling cases. 
%
%\textcolor{orange}{One could also compute the exact value of the MDL for the family of MCMs, but, in principle, it would require a computationally demanding sum over possible datasets.}
%\textcolor{orange}{Maybe move the expansion of log-E here? [Add here description of each term. In particular comment on larger penalty for larger values of $q$. OR this can be done later, in the example where artificially increasing $q$ in binary data.]}

\subsection{Measuring data complexity with MCMs}
\label{sec:complexity}
%Stochastic complexity of data
%Estimating the stochastic complexity of data with MCMs

The minimum description length (MDL) principle was introduced by Rissanen in the context of measuring the complexity of a noisy dataset. %motivated by the problem of quantifying the complexity of noisy data.
Without compression, a dataset with $n$ $q$-state discrete variables and $N$ datapoints requires $N\times n$ qits to be encoded, i.e., one qit per variable per datapoint, where one qit corresponds to a unit of information of $(\log q)$ nats (we use the term %call this unit
``qit'' for this unit by analogy to bit and trit).
Using a statistical model, redundant patterns in the data can be encoded more concisely (encoding the patterns one time instead of multiple times), which yields a compression of the data. 
The resulting number of qits required to encode a dataset~$\data$ using a statistical model $\M$ is given by the description length $L(\data\,|\,\M)$ (taking the logarithm in base $q$).
%
%\textcolor{red}{In Ref.~\cite{rissanen1986stochastic}, Rissanen interprets the [...].}
The minimum description length achieved (i.e., maximal compression) within a family of statistical models is interpreted by Rissanen~\cite{rissanen1986stochastic} as a measure of stochastic complexity of the data, % (for that family of models),
by analogy with algorithmic definitions of complexity (such as Kolmogorov complexity).
Extending the parallel between the description length and the negative log-evidence to finite values of~$N$, we propose to use the negative log-evidence of the best MCM as a measure of stochastic complexity of the data within the family of MCMs. %, [as also suggested in~[17]]. % Ref.~\cite{AgrimiCaputo2026} also shows that this measure do indeed reflect a property of the data. is more of a property of the data than a property of the model.
In Sec.~\ref{Sec:Applications}, we give an example of the application of this quantity as a measure of data complexity. 
\begin{Definition}{\bf Complexity of the data.}
    We propose to use the negative log-evidence of the best MCM (which is the MCM with the largest evidence) %with maximal evidence.
    as a measure of stochastic complexity of the data within the family of MCMs:
    \begin{align}
        {\rm COMP}_{MCM}(\data)=- \max_{\M\in {\rm MCM}}\log P(\data\,|\,\M)\,,
    \end{align}
    where the evidence $P(\data\,|\,\M)$ of an MCM is given in Property~\ref{ppty:log-E:MCM}.
\end{Definition}

%\textcolor{orange}{TO USE FOR COMMENTING ON RESULTS OF THE MEASURE COMPLEXITIES.\\
%The stochastic complexity measured by the MDL is a consequence of, on one side, the complexity of the data patterns captured by the model (which contribute to the MDL as the complexity of the model), and on the other side, the entropic part of the data (pure noise, which can't be compressed). }

%%%%%%%% IMPORTANT: TO KEEP MAYBE FOR FINAL VERSION  %%%%%%%%%%%%
%%%%%%%%%%%%%%%%%%%%%%%%%%%%%%%%%%%%%%%%%%%%%%%%%%%%%%%%%%%%%%%%%
%Given that the entire family of $q$-state spin models is a complete family of statistical models for discrete data (i.e., they can model any [type of] discrete data), %we can consider that 
%the minimum description length obtained %among all of them 
%within that family of models (after comparing all the models) gives a value of the ``true'' stochastic complexity of the data~\cite{rissanen1986stochastic}. %
%%%%%%%%%%%%%%%%%%%%%%%%%%%%%%%%%%%%%%%%%%%%%%%%%%%%%%%%%%%%%%%%%
%Note that other classes of models, [corresponding to other choices of the decomposition of the log-probability over a family of basis functions], such as mentioned in Sec.~1, could lead to lower values of complexity. 
%However, the family of MCMs is universal in the sense that, [being a factorization model], it doesn't dependent on the choice of the decomposition of the log-probability over a family of basis functions, [and it will exist in all these classes of models].
%%%%%%%%%%%%%%%%%%%%%%%%%%%%%%%%%%%%%%%%%%%%%%%%%%%%%%%%%%%%%%%%%
%%%%%%%%%%%%%%%%%%%%%%%%%%%%%%%%%%%%%%%%%%%%%%%%%%%%%%%%%%%%%%%%%

\section{Applications to real data} \label{Sec:Applications}

\subsection{Search algorithms: optimization of the log-Evidence of MCMs}\label{Sec:MCM:Search-algo}

Despite the advantage provided by the closed-form expression for the evidence in Eq.~\eqref{eq:evidenceICC}, the number of MCMs is still large and in practice searching exhaustively for the best MCM among all remains computationally untractable for systems with more than ten variables. 
Fortunately however, finding the best MCM (or near-optimal MCM) among those sharing a common {\it preferred basis} is computationally accessible, thanks to a direct mapping between these models and partitions of the basis variables.
This one-to-one correspondence with partitions allows for the development of efficient search algorithms beyond exhaustive enumeration.
%and the number of possible partitions grows much slower than the total number of MCMs (as $(n/log(n))^n$ -- see Bell number of $n$).} 
%%%%%
In a given preferred basis, finding the best MCM corresponds to finding %an optimal
the factorization of the model probability distribution over the basis variables that best matches the structure of the data.
Ref.~\cite{AgrimiCaputo2026} likens finding such MCM factorization in the original basis of the data to a high-order community detection, and thus is interesting on its own.
To simplify the model selection procedure in the binary case, Ref.~\cite{demulatier2024MCM} also hypothesized that the best MCM overall would be obtained in the basis formed by the best independent model (i.e., the one with the largest evidence) and thus proposed to divide the selection procedure in two steps: first find the best independent model and use it as a new basis for the data, and then find an optimal factorization in this basis.
We follow a similar approach for the discrete case, focusing on two separate steps: searching for an optimal spin basis and searching for the best MCM in a chosen preferred basis.
%using this basis as a preferred basis.

\subsubsection{Search for the optimal MCM in a given preferred basis}
In a given {\em preferred} basis, each MCM maps to a partition of the spin variables into independent parts, independently of the value of $q$. This makes it possible to use the same search algorithms as in the binary case, with the only difference being that the model evidence is now computed using the more general formula~\eqref{eq:evidenceICC}.  
Following Ref.~\cite{demulatier2024MCM}, we used two different algorithms: 1) an exhaustive search, going through all possible partitions of the $n$~variables using a Gray code~\cite{ehrlich1973loopless, knuth2011art}, which can be used for small systems with up to about $15$ variables; and 2) a greedy search, that starts from the independent model and successively merges the two ICCs that lead to the largest increase in evidence until the evidence reaches a maximum, which can be used for larger systems (our current implementation~\cite{DeClerq_2024_Python} can be used with up to 128 variables). 
%\textcolor{blue}{``Finally, we also implemented an approach based on the idea of Divide and Conquer, also available in Ref.~\cite{DeClerq_2024}: we start with all the variables in a single community and search for the partition in 2 parts that maximizes the log-evidence, with then iterate within each part until the log-evidence cannot be increased. [plus add issue with log-E not necessarily increasing.]''} 
Ref~\cite{DeClerq_2024_Python} also implements two other search algorithms not used in this paper, a simulated annealing search and a greedy search using divisive hierarchical clustering.

\subsubsection{Search for an optimal basis}

In the binary case, Ref.~\cite{demulatier2024MCM} proposes to use the best independent model with $n$~operators as a preferred basis for the MCM search. The $n$ independent operators of this model correspond to the spin basis in which the empirical probability distribution is the most likely to factorize over each variable as if there were independent.
%\textcolor{red}{over which the empirical probability distribution is the most likely to be expressed as independent factors over each variable.}
%%%%%%%%%%%%%%%%%%%%%%%%%%%%%%%%%%%%%%
We extend this idea to the more general discrete case, by searching for a spin basis in which the empirical probability distribution is most likely to factorize. %is best expressed as factorizing
%over each variable as if they were independent.
%\textcolor{blue}{in which the model assuming that the basis variables are independent is the closest to the data.}
In other words, we search for the best IMCM for the data and any basis for that IMCM equally satisfies the requirement above.
%and can then be used to define the {\em preferred} basis for the MCM search described in the previous paragraph 
Using any of these bases as the preferred basis for the MCM search (in the previous section) will give the same set of MCMs to choose from (because ICCs always include all possible multiples of the basis operators). In the binary case, an IMCM is (also) an independent model, and we recover the approach proposed in~\cite{demulatier2024MCM}.

%%%%%%%%%%%%%%%% %%%%%%%%%%%%%%%% %%%%%%%%%%%%%%%% %%%%%%%%%%%%%%%%
%Gauge transformations are not easily defined when $q$ is not prime. In the applications of Sec.~\ref{Sec:Applications} for $q=4$, we select gauge transformations defined by bases of invertible operators.
%%%%%%%%%%%%%%%% %%%%%%%%%%%%%%%% %%%%%%%%%%%%%%%% %%%%%%%%%%%%%%%%
The best IMCM with $n$~ICCs (each of rank~$1$) is defined as the one with the largest evidence.
%To find an optimal basis for the MCM, we search for the best IMCM with $n$~ICCs (each of rank~$1$), which is the one with the largest evidence.
Because all IMCMs with $n$~ICCs %have the same complexity (they are in the same equivalence class), 
are in the same equivalence class, the model with the largest evidence among them is also the one that has the largest maximum likelihood. Using Eq.~\eqref{eq:max-log-likelihoodMCM}, the maximum log-likelihood of an IMCM with~$n$ ICCs based on the spin basis $\Basis=(\vecmu_1,\,\dots,\,\vecmu_n)$ (where each %one-dimensional [vector]
$\Basis_a=(\vecmu_a)$ is a basis of one of the ICC) 
is given by: 
\begin{align}\label{eq:max-likelihood:IMCM}
    \log P(\data\,|\,\vecghat,\mathcal{M}_{\rm imcm})=-N\,
    \sum_{a=1}^n
    %S(\data_a')\,,
    S_{\data}[\alpha^{\prime}_a]\,, 
    \;\;\;
    {\rm where}\;
    S_{\data}[\alpha^{\prime}_a] 
    =-
    \sum_{\alpha_a^{\prime}=0}^{q-1}\,
    \frac{k_{\alpha_a^{\prime}}(\data)}{N}
    \log\left(\frac{k_{\alpha_a^{\prime}}(\data)}{N}\right)
\end{align}
%%%%%%%%%% SHORTER VERSION:   %%%%%%%%%%%%%%%%%%%%%%%%
%\begin{align}\nonumber
%    \log P(\data\,|\,\vecghat,\mathcal{M}_{\rm imcm})=-N\,
%    \sum_{a=1}^n S(\data^{\prime}_a)\,,
%    \;\;\; {\rm where}\;
%    S(\data^{\prime}_a) =S_{\data}[\alpha^{\prime}_a]
%    =-\sum_{\alpha_a^{\prime}=0}^{q-1}\,\frac{k_{\alpha_a^{\prime}}(\data)}{N} \log\left(\frac{k_{\alpha_a^{\prime}}(\data)}{N}\right)
%\end{align}
%can be recognized as Shannon entropy of the discrete variable $\alpha_a^{\prime} = \vecmu_a\cdot\state$ in the dataset~$\data$ (because the vector $\state_a'=\Basis_a(\state)=(\alpha_a')$ has only one variable).
%%%%%%%%%%%%%%%%%%%%%%%%%%%%%%%%%%%%%%%%%%%%%%%%%%%%%%%
is the Shannon entropy of the discrete variable $\alpha_a^{\prime} = \vecmu_a\cdot\state$ in the dataset~$\data$.
Here, the entropy of the reduced dataset $S(\data_a')$ appearing in Eq.~\eqref{eq:Entropy:ICC} was re-written as $S(\data_a')=S_{\data}[\alpha^{\prime}_a]$ because the vector $\state_a'=\Basis_a(\state)=(\alpha_a')$ has only one variable.
We used the $n$ independent operators $\{\vecmu_1,\dots,\vecmu_n\}$ to define the new basis variables $\alpha_{a}^{\prime} = \vecmu_a\cdot\state$ for all $a\in\{1,\,\dots,\, n\}$ (see Definition~\ref{def:GT}), and each term $k_{\alpha_a^{\prime}}(\data)$ denotes the number of times $\alpha_a'$ %the variable $\alpha_a'= \vecmu_a\cdot\state$ 
takes a given value $\alpha_a^{\prime}\in\{0,\,\dots,\,q-1\}$ in the dataset~$\data$ (or equivalently the number of times the basis variables $\state_a'=\Basis(\state)=(\alpha_a')$ takes a given value in the transformed and reduced dataset $\data_a'=\Basis_a(\data)$). 
Because each state of $\alpha_a'$ corresponds to a state of the operator $\phi^{\vecmu_a}(\state)$, %the term 
$S_{\data}[\alpha^{\prime}_a]$ is also the Shannon entropy of the operator $\vecmu_a$ in the data: $S_{\data}[\alpha^{\prime}_a]=S_{\data}[\phi^{\vecmu_a}(\state)]$.
Equation~\eqref{eq:max-likelihood:IMCM} therefore implies that the most likely IMCM is given by the set of the $n$ least entropic independent operators in the data\footnote{
%%%%%%%%%%%%%%%% footnote:
Note that for $q>2$, finding the most likely IMCM is different from finding the most likely independent model. An independent model with $n$ operators can be written as a set of $n$ independent components with one operator in each component, but these components are not complete. This means that, for an independent model, the Kullback-Leibler divergence between the single operator and the data within each independent component is not zero~(contrary to IMCMs). This term would then have to be taken into account in the computation of the maximum likelihood of the independent model.}. %[if we were to try to find the most likely independent model].
%%%%%%%%%%%%%%%%%%%%%%%%%%%%
Exactly as in the binary case, finding the most likely IMCM corresponds to finding the set of $n$ independent operators with the lowest entropy. The only difference is that now there are multiple bases of operators that correspond to the same IMCM and thus work equivalently as a preferred basis (they identify the same set of MCMs on that basis).

To find these operators for small systems (with $n\lesssim 15$), we extend the exhaustive algorithm of Ref.~\cite{demulatier2024MCM}. We compute the entropy of all the operators of the system (excluding certain operators, listed below) and create an $(n\times K)$-matrix containing in columns the discrete representations $\vecmu_j$ of the $K$ least entropic operators, ordered from low to high entropy. The elements of this matrix are in $\Field$. Identifying the $n$ least entropic independent operators then consists of finding the independent columns in this matrix starting from the left-hand side. This can be done using Gaussian elimination in $\Field$: the columns with leading coefficients identify the set of least entropic independent operators.
For $q$ non-prime, some interactions are ``not invertible'' and are directly excluded from this procedure; these are the interactions $\vecmu$ for which there exists a coefficient $c\in\Field$ such that $c\vecmu=\veczero$ in $\Field^n$ (see Sec.~\ref{sec:GT:def}). 
%Note that for $q>2$, any operator~$\vecmu$ that can be used to define a spin basis has a non-identical c.c.~operator $-\vecmu$ ($\vecmu$ cannot be self-dependent).
Moreover for $q>2$, all the considered (invertible) operators $\vecmu$ have a non-identical complex conjugate $-\vecmu$; conjugate operators have the same empirical entropy and for the preferred basis it doesn't matter if one uses $\vecmu$ or $-\vecmu$ as a basis element because they identify the same preferred basis\footnote{It is reassuring that c.c.~operators have the same entropy, because they identify the same IMCM. Different entropies would give different values of maximum likelihood~\eqref{eq:max-likelihood:IMCM} to the same IMCM depending on which operator is used in the basis, which would be problematic.}
(any partition of the basis elements into ICCs will give the same MCM independently of using $\vecmu$ or $-\vecmu$ as a basis element). 
%So this choice has no impact on the final selected MCM.
Therefore, only one element from each conjugate pair is included in the procedure (in our implementation~\cite{DeClerq_2024_Python}, this is done by keeping only the interaction vectors $\vecmu$ whose first entry is smaller or equal to $q/2$).
Similarly, we can also exclude from the selection all multiples of the already selected operators,
%\textcolor{red}{operator multiples} 
because they identify %corresponds to 
different bases of the same IMCMs\footnote{Any partition of the basis elements into ICCs will give the same MCM independently of which operator multiple is used, %This is the case 
because ICCs always include all possible multiples of the basis operators.}. This wasn't implemented, but would %lead to \textcolor{red}{additional/further} speed up of the code. % can improve the code
improve further the speed of the code.

With this exhaustive approach, the number of operators to consider grows exponentially as~$q^n$, making this algorithm unusable in practice for systems of more than $25$ variables. 
%\footnote{To give a rough idea, for the $n=15$ first variables of the big five data analyzed in Sec.~\ref{Sec:MCM:Big5}, which has around $6.10^5$ datapoints, the exhaustive search algorithm for the best basis takes of the order of \textcolor{green}{xx minutes for $q=2$, XX minutes of $q=3$ and xx minutes for $q=5$ on a laptop.}}.
For many real systems, it seems reasonable to assume that the least entropic operators will be of low order (i.e., of order 1, 2, 3, or 4), %\textcolor{red}{and a similar algorithm could be used by exhaustively searching for the best basis among operators up to a fixed order $k$ only.}
and one could reduce the exhaustive search to consider operators up to a fixed order $k$ only.
In the binary case, Ref.~\cite{demulatier2024MCM} also proposes an iterative algorithm that can consider operators of any order while keeping the computational time similar to fixing~$k$. This algorithm can be easily extended to larger values of $q$; it %the code
is available in~\cite{DeClerq_2024_Python} but not used in this paper.

%\textcolor{orange}{Why not using the log-E instead of log-L (as it is easy to compute anyway)? Hopefully they should give the same basis, right? I.e., they give the same value, but with a constant shift due to model complexity. But not sure, one has the log-gamma functions instead of logarithms...}

\begin{figure}[p]
    \centering
    \includegraphics[width=\textwidth]{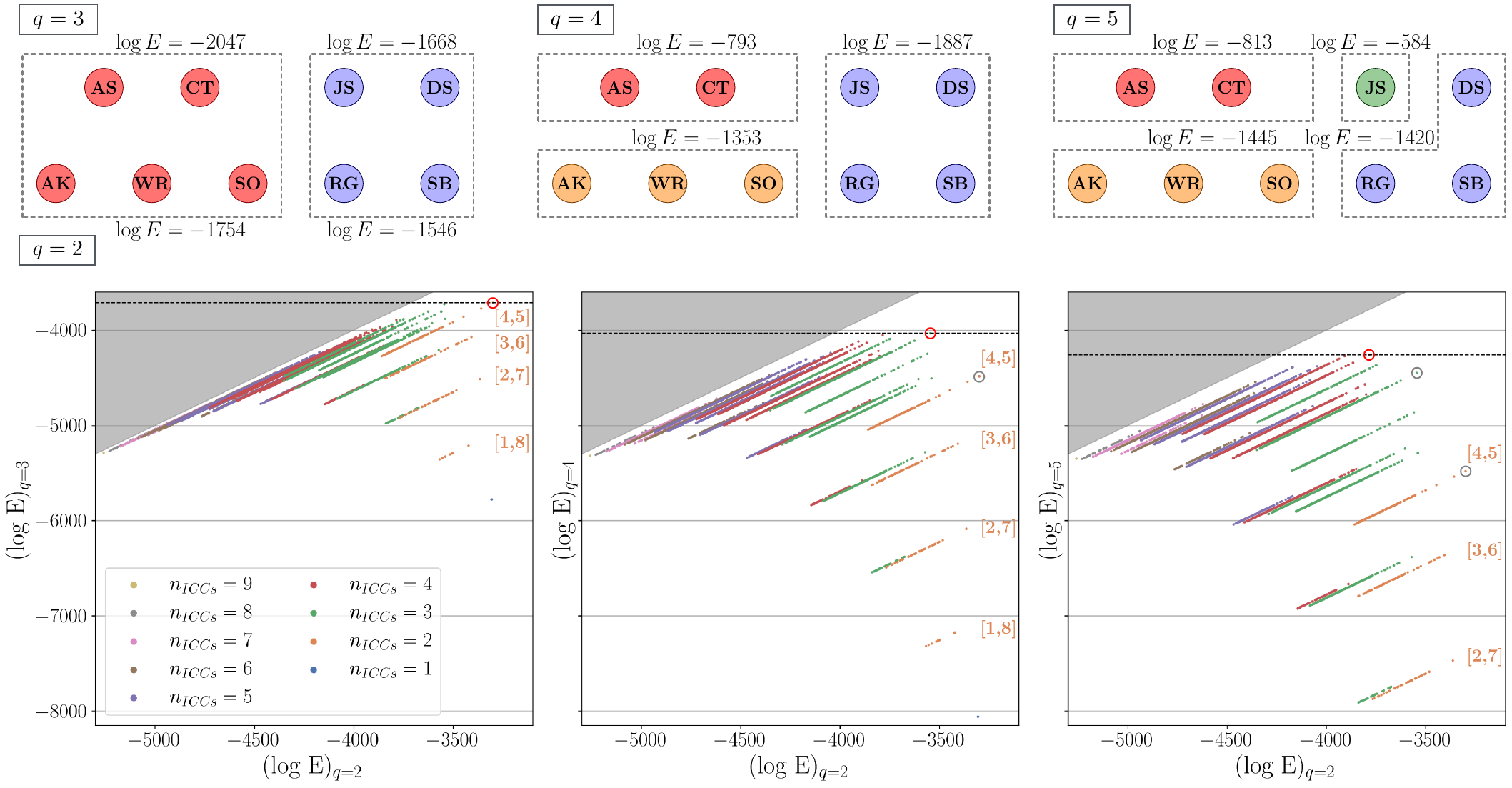}
    \caption{{\bf MCM analysis of the US Supreme Court data in its original basis for different embeddings.} 
    The original dataset is binary and is embedded in the larger $q$-state space for this analysis. % for different values of $q$.}
    %after artificially varying the value of $q$.} 
    %\textcolor{red}{[We must better combine the figure here: add $q=5$; move partition models as inset in the lower plots.]}
    {\bf Top.} %Representation of the best MCM found in the original basis of the data for different values of $q$. 
    Optimal MCM found by exhaustive search in the original basis of the data for different values of $q$. 
    The circles represent the 9 justices labeled by their initials: Ruth Bader Ginsburg (RG), John P. Stevens (JS), David Souter (DS), Stephen Breyer (SB), Sandra Day O'Connor (SO), William Rehnquist (WR), Anthony Kennedy (AK), Clarence Thomas (CT), and Antonin Scalia (AS).
    MCMs are represented as partitions of the basis variables, with components identified by different colors and delimited by dashed lines. The best MCMs for $q=2$ and $q=3$ correspond to the same partition. The value of the log-evidence is indicated for each component. 
    %The value of the log-evidence [for each component]  is indicated near each component.
    The log-evidence of the MCM (reported in Table~\ref{tab1}) is the sum of the values of its components.
    {\bf Bottom.} 
    Scatter plot of the log-evidence of MCMs for all possible partitions of the basis variables, for $q=2$ (along the horizontal axis) against, respectively, $q=3$, $4$ and $5$ (along the vertical axis).
    Each dot corresponds to one partition, whose color indicates the number of parts, i.e., the number $n_{ICCs}$ of ICCs of the corresponding MCMs. % denoted $n_{ICCs}$.
    In each plot, the partition corresponding to the MCM with the largest log-evidence %for $q=3,\,4,\,5$ respectively 
    is identified by a red circle for each $q$; they are the best MCMs shown in the top panel. The grey circles indicate the partitions corresponding to the best MCMs for smaller values of $q$.
    The log-evidence of MCMs corresponding to the same partition decreases with $q$, due to the increase in complexity penalty.
    %%%%%%%%%% For the same partition, the log-evidence strictly decreases as one increases $q$. 
    In particular, no partition lies above the $(x=y)$-line (indicated by the shaded area) and the vertical distance between this line and a dot corresponds to the increase in penalty in the log-evidence when changing $q$ from $2$ to, respectively, $3$, $4$, or $5$ for the corresponding partition.
    MCMs with ICCs of the same sizes are penalized by the same amount and corresponds to aligned dots parallel to the $(x=y)$-line. For the MCMs with $n_{ICCs}=2$ (orange dots), we indicated the size of the ICCs near the alignment: %aligned dots: 
    for example ``$[4,5]$'' denotes MCMs with one ICC of size 4 and one of size 5.
    %%%%%%%%%%%%%%%%%
    %The smaller the number of ICCs, the larger their average size, and thus the larger is the complexity of the MCM.
    %%%%%%%%%%%%%%%%%
    Models with fewer ICCs have a larger average ICC size and thus get a stronger penalty with increasing $q$ (see orange dots moving faster down than other dots from one panel to the next).
    %which leads to structural changes in the models with the largest log-evidence.\\
    }
    \label{fig:SCOTUS_BestMCM}
\end{figure}

\subsection{Code availability and reproducibility}
The codes for all three algorithms can be found in our Python package in Ref.~\cite{DeClerq_2024_Python}
%\footnote{
%The repository also has two other search algorithms for finding the best MCM, one uses simulated annealing and the other is a greedy search \textcolor{green}{using a divisive hierarchical clustering}. %\textcolor{green}{divisive hierarchical greedy approach.}
%}
(C++ codes with Python bindings).
%\textcolor{blue}{``Finally, we also implemented an approach based on the idea of Divide and Conquer, also available in Ref.~\cite{DeClerq_2024}: we start with all the variables in a single community and search for the partition in 2 parts that maximizes the log-evidence, with then iterate within each part until the log-evidence cannot be increased. [plus add issue with log-E not necessarily increasing.]''} . 
A notable difference with respect to the codes for binary MCMs provided by Ref.~\cite{demulatier2024MCM} is that datapoints and operators cannot be easily encoded by binary integers anymore. This is because the state vectors~$\state$ and the interaction vectors~$\vecmu$ are not necessarily binary, but more generally defined in $\Field^n$.
To avoid losing efficiency, we encoded these vectors using several 128-bit integers to encode 128~$q$-state variables (i.e., each $q$-state variable is encoded by $\log_2(q)$ bits). This is to compensate for the lack of built-in data structure to encode integers in base $q$ in our chosen language (C++) but also in most common programming languages. 

As a proof of concept, we revisit two datasets previously analyzed with MCMs in the binary setting~\cite{demulatier2024MCM}.
We study the impact of the choices made for how to discretize the data and of the resolution $q$ of discretization on the structure of the resulting MCM.
%The data analyses presented in this section are provided \textcolor{green}{as Jupyter notebooks in~\cite{DeClerq_2024_notebook}}.
%\textcolor{red}{The repository also contains Python notebooks with the three data analyses performed in this section.}

%\paragraph{\bf Questions:}
%\begin{itemize}
%    \item Prove that conjugate operators have the same entropy in App.
    %\item Search algorithms for finding the best basis: Are you using an exhaustive search up to fix order $k$? Bottleneck: computing the entropy. \textcolor{red}{Is the non-exhaustive search used in the following analysis?}
    %\item Should we focus only on the cases where $q$ is prime? I am not sure what happens when $q$ is not prime. Can we even map an element of $\M_0$ to an element of $\M_1$, seems not possible (because of complex conjugate), but not impossible either?! If it is possible, can it be that then the two conjugate operators both map to the same element in $\M_0$?
%\end{itemize} 

\subsection{Example 1: binary dataset with artificially increased values of $q$, embedded in a larger space}
\label{Sec:Ex:SCOTUS}
For illustration purposes, we first analyze a binary dataset using MCMs with varying values of $q$, as if the (binary) variables were taking values in $[0,q-1]$. The dataset comes from the US Supreme Court database~\cite{USSCdatabase} and consists of the votes of the $n=9$ justices on $N=895$ cases debated during the second Rehnquist Court.  
For $q=2$, voting outcomes were labeled as in Ref.~\cite{lee2015statistical} using the political inclination (if known) of each debated case, such that a judge casts a vote $\alpha_j=1$ if the decision is conservative-oriented, or $\alpha_j=0$ if it is liberal-oriented. Although the datapoints are binary, one can embed them into the larger state space $\Field^n$, effectively increasing the number of possible configurations while keeping the observed states unchanged. Embedding the data into a larger space is common in data analysis. For instance, when using principal component analysis on discrete data, the data is analyzed as if the variables were taking real values even though they are discrete.

\subsubsection{Embedding choice}
For $q>2$, to reflect the opposite political valence between conservative and liberal oriented votes, %we gave opposite values of $\alpha_j$}
%For $q>2$, to reflect \textcolor{red}{the symmetry} between conservative and liberal oriented votes, %opposite political orientations/inclinations of the votes
we gave opposite values of $\alpha_j$ to opposite votes, assigning $\alpha_j=1$ to liberal-oriented votes and $\alpha_j=-1\, (\bmod{q})=(q-1)$ to conservative-oriented votes. This choice confers conjugate spin values $s_j=\exp(2 \I \pi\alpha_j/q)$ to these two states. All the other states of $\alpha_j$ are not observed in the data.
The specific choice of mapping of $\alpha_j$ from binary to discrete values will not impact the outcome of the best MCM in the original basis of the data (compared to any other permutations of the values of $\alpha_j$).
%The outcome of the best MCM in the original basis of the data is independent of the choice of mapping of $\alpha_j$ from binary to discrete values (compared to any other permutations of the values of $\alpha_j$).
%The choice of mapping of $\alpha_j$ from binary to discrete values doesn't affect the outcome of the best MCM in the original basis of the data with respect to selecting any other permutations of the values of $\alpha_j$ (because it doesn't change the dataset per se). 
However, it can impact the interactions selected in the best basis (and the interactions of the best spin model overall) and change the interpretation of what a given interaction means (see Sec.~\ref{sec:Potts:interpretation} on interpreting 
the interactions).
%\textcolor{blue}{However, it will change the interpretation of what a given interaction means. In particular, it can impact the interactions selected for the best basis.}
%A priori
Intuitively, we expect that labeling the discrete data in a way that best reflects the meaning associated with the values taken by the variables in the studied system would make it easier to interpret the uncovered interactions.
%\textcolor{blue}{would help interpret the uncovered interactions.}
%In this context, it is best to construct the discrete data in a way that best respect the meaning associated to the values that can be taken by the variables (in particular, trying to respect any possible symmetry in the interpretation of the values), leading to a much easier interpretation of the uncovered interactions.
On the other hand, assigning the values at random can make it more difficult to interpret the model.
Overall, the final result from the modeling procedure (i.e., the interpreted model) %should not depend on the choice of the mapping; it 
should be consistent across mapping choices. In other words, the uncovered models for different choices should be related mathematically by the (bijective) transformation of variables associated with the different mapping choices). %\textcolor{green}{For instance, see Example 8 in Sec.1, if a system as important anti-symmetric-like patterns, between specific states, then shuffling the values may make it harder to interpret that there's antim-symetric.}} 

In the following, we first analyze the impact of the embedding %of increasing $q$ 
on the optimal MCM factorization in the original basis of the data, and then on the overall best MCM.

\subsubsection{Optimal MCM factorization in the original basis}
%[Let us first discuss the impact of increasing values of $q$ on the optimal factorization found by the MCM analysis in a given preferred basis.]
Figure~\ref{fig:SCOTUS_BestMCM} Top shows the best MCMs found by the exhaustive search in the original basis of the data for varying values of $q$. For $q=3$, 
we found the same factorization as in the binary case~\cite{demulatier2024MCM},  with the five conservative-oriented justices grouped in one ICC and the four more liberal-oriented justices grouped in a second ICC. 
Further increasing $q$ breaks these components into smaller ICCs,
%results in the division/\textcolor{red}{split} the components into smaller ICCs,
first splitting the conservative group at $q=4$ and then the liberal group at $q=5$.
%[indicating that larger values of $q$ favor smaller components].
The reduction of the ICC sizes with larger $q$ can be understood intuitively. For larger values of $q$, ICCs based on the same number~$r_a$ of variables require more parameters (precisely, $q^{r_a}-1$ parameters) and are thus more penalized. %\textcolor{blue}{[resulting in smaller ICCs being favored.]}
% which results in smaller ICCs
%as MCMs corresponding to the same partition require more parameters for larger values of $q$ and are thus more penalized, resulting in smaller parts.
Because the observed states remain unchanged, 
we therefore expect the ICCs of the best MCM to get smaller %and smaller 
as we artificially increase $q$. 

Formally, %precisely, 
consider a dataset with variables taking values %in
$\state\in(\mathbb{Z}/q_1\mathbb{Z})^n$, 
and consider two MCMs, $\M^{q_1}$ and $\M^{q_2}$, corresponding to the same partition of the $n$ basis variables but for two different values of $q$, respectively $q_1$ and $q_2$ with $q_2 > q_1$.
%More precisely, consider two MCMs, $\M_{q_1}$ and $\M_{q_2}$, corresponding to the same partition of the $n$ basis variables but for two different values of $q$, respectively $q_1$ and $q_2$. Consider a dataset with variables taking values in $\state\in(\mathbb{Z}/q_1\mathbb{Z})^n$. 
For these models, %(with rank $r=n$),
the log-evidence takes the form (using $r=n$ in Eq.~\eqref{eq:evidenceMCM} and \eqref{eq:evidenceICC}):
\begin{align}\label{eq:log:evidence:MCM}
    \log P(\data\,|\,\M) = \sum_{a\in\mathcal{A}} \left[
    \log \frac{\displaystyle\Gamma\left(\frac{q^{r_a}}{2}\right)}{\displaystyle\Gamma\left(N+\frac{q^{r_a}}{2}\right)}
    + 
    \sum_{\state_a\in\data_{a}} 
    \log\left(\frac{\displaystyle\,\Gamma\left(k_{\state_a}(\data_{a})+\frac{1}{2}\right)}
        {\displaystyle\sqrt{\pi}}\right)
    \right] 
    \;.
\end{align}
Here, $\mathcal{A}$ denotes the common partition identifying both $\M^{q_1}$ and $\M^{q_2}$, such that each element $a\in\mathcal{A}$ identifies an ICC %$\M^{q}_{a}$
in each model based on the same groups of variables~$\state_a$.
%where $\mathcal{A}$ is the set of ICCs of $\M$.
%where $\mathcal{A}$ denotes the set of parts of the common partition identifying both $\M^{q_1}$ and $\M^{q_2}$, such that each element $a\in\mathcal{A}$ %corresponds to a [distinct] ICC in each model. %to an ICC in both models. % [respectively] in each model
%identifies an ICC $\M^{q}_{a}$ in \textcolor{blue}{each} model based on the same groups of variables~$\state_a$.
%We label the elements $a\in\mathcal{A}$ such that the same $a$ corresponds to the ICC in each model identified by the same part.
%%%%%%%%%%%%
%For these two MCMs, the sum over ICCs in Eq.~\eqref{eq:log:evidence:MCM} corresponds to a sum over the same parts: for a given $a\in\mathcal{A}$, the rank $r_a$ (which is the size of the part) is the same in both models and the variables~ $\state_a$ modeled by the ICCs are the same in both models.  %the ICCs have the same respective rank
%%%%%%%%%%%
Because the dataset itself remains unchanged (with states $\state\in(\mathbb{Z}/q_1\mathbb{Z})^n$) and because the two MCMs correspond to the same partition of the basis variables, the second sum in Eq.~\eqref{eq:log:evidence:MCM} takes the same value for both models. Only the complexity part (first term within the first sum) changes with %as one varies
$q$, becoming more negative for larger~$q$ at fixed value of~$r_a$. 
%\textcolor{red}{[i.e., Models get more and more penalized as one increases $q$.]}
%becoming more and more negative as one increases~$q$. %[(for $r_a$ fixed)].
%%%%%%%%
%Because the dataset itself remains unchanged, only the complexity part (first term) in Eq.~\eqref{eq:log:evidence:MCM} changes as one varies $q$, becoming more and more negative as one increases $q$.
%%%%%%%%
%
Figure~\ref{fig:SCOTUS_BestMCM} Bottom illustrates how this increase in penalty with $q$ impacts different models.
MCMs with ICCs of the same sizes $\{r_a\}$ (which belong to the same complexity class)
%(i.e.  with the same values of $r_a$'s) 
%MCMs that belong to the same complexity class,
are penalized by the same amount as one increases $q$ (see the alignment of the dots %dot alignement; alignment of the dots
parallel to the $(x=y)$-line)
and the penalty increases faster with $q$ for MCMs with larger average ICC size (i.e. with a smaller number of ICCs). % (and therefore a larger average ICC size). 
% the largest the size $r_a$ of the components, the stronger the effect
These effects can also be deduced from the dependence in $\{r_a\}$ of the complexity term in Eq.~\eqref{eq:log:evidence:MCM}.
As a result, the best MCM can structurally change when increasing~$q$, favoring models with smaller ICCs (see changing location of the red circle towards MCMs with smaller ICCs).
%The division of ICCs into smaller ICCs thus only result from the difference in complexity, which is illustrated in Figure~\ref{fig:SCOTUS_BestMCM} Bottom.

\subsubsection{Best MCM factorization overall}

\begin{figure}[h]
    \centering
    \includegraphics[width=\textwidth]{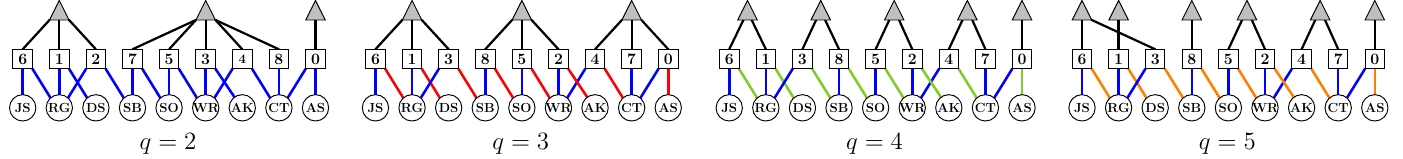}
    \caption{{\bf Factor graph representation of the best MCM found by exhaustive search for the US Supreme Court dataset for different values of $q$.} 
    %ranging from 2 to 5.} 
    %These model were found by exhaustive search
    The circles represent the original variables, which correspond to the 9 justices labeled by their initials as in Fig.~\ref{fig:SCOTUS_BestMCM}. 
    The squares represent the operators of the best basis found by exhaustive search, numbered from low to high entropy.
For each basis operators (squares), the colored lines indicate which ``justice'' variables (circles) are included in the operator and the color specifies the power to which each variable is raised: one in blue, two in red, three in green, and four in orange. For example, for $q=3$, the basis operator of lowest entropy (square with number ``$0$'') is the operator $\phi_0(\spin)=s^{\,}_{\tiny CT}\,s^2_{\tiny AS}$.
    %The colored lines between operators (squares) and justice variables (circles) indicate which variables are included in each operator; the color indicates the power to which each variable is raised: one in blue, two in red, three in green and four in orange.
    The triangles represent the ICCs of the best MCM, a line from an operator to a component %indicates
    identify the presence of that operator in the component.
    }
    \label{fig:SCOTUS_BestBasis}
\end{figure}

Figure~\ref{fig:SCOTUS_BestBasis} shows the best MCM found overall for different values of $q$, by performing first an exhaustive search for the optimal basis, and then an exhaustive search for the optimal MCM on that basis.

The nine independent operators of the best basis are represented by squares in the figure. Except for one single spin operator on AS, all these operators are pairwise and identify interactions between the same pairs of justices for all values of~$q$. % (see the lines between the variables and the operators).
In the binary case, \cite{demulatier2024MCM} reported that these $9$ interactions alone account for $86\%$ of the information extracted by a fully connected pairwise model with $45$ interactions. %[This likely remains true for larger $q$, because the data is truly binary.]
A notable difference with the binary case, is that these pairwise operators are of the form $\vecmu_{ij}=(1,\,q-1)$ between two justices $i$ and $j$, and not of the form $\vecmu_{ij}=(1,\,1)$ as one could have naively extrapolated from the binary case (where $\vecmu_{ij}$ denotes the interaction vector $\vecmu$ reduced to the $i$-th and $j$-th entries).
This is because the most frequent pattern between these pairs of justices is the symmetric voting pattern $\alpha_i=\alpha_j$, which is a pattern that is best captured %modeled
by this type of pairwise interaction (see the explanations in Example~\ref{Ex:Pairwise_interactions} of Sec.~\ref{Sec:HOI:interpretation}).
Indeed, the justices SO and SB, whose interaction is the most entropic pairwise interaction in the basis, vote identically in $78\%$ of the cases, and AS and CT, whose pairwise interaction is the least entropic interaction among all, vote identically in $93\%$ of the cases.
Note that the choice of embedding doesn't impact the best basis in this case, because the symmetric patterns $\alpha_i=\alpha_j$ identified by the best basis remain unchanged under permutation of the state values.
In Fig.~\ref{fig:SCOTUS_BestBasis}, the basis operators are numbered from the least to the most entropic. Observe that this ordering changes slightly from $q=2$ to $q=3$ (see the operators ranked 7 and 8, and 2 and 3). 
%%%%%%%%%%%%
%\textcolor{blue}{This is because for $q\geq 3$ the relative occurrence of the non-symmetric states enters in the computation of the entropy, while for $q=2$ the entropy of an operator only distinguishes the probability of the operator to be even (symmetric state $\alpha_i=\alpha_j$) or odd (antisymmetric states $\alpha_i=-\alpha_j$).}
%
%\textcolor{red}{This is because for $q\geq 3$ the entropy of a pairwise operator depends on the relative occurrence of the non-symmetric states, while for $q=2$ it only depends on the probability of the operator to be even (symmetric state $\alpha_i=\alpha_j$) or odd (antisymmetric states $\alpha_i=-\alpha_j$).}
%%%%%%%%%%%%
This is because for $q=2$ the entropy of an operator only depends on the probability of the operator to be even (symmetric state $\alpha_i=\alpha_j$) or odd (antisymmetric states $\alpha_i=-\alpha_j$), while for $q\geq 3$ the entropy of a pairwise operator also depends on the relative occurrence of the non-symmetric states. % \textcolor{green}{of the non-symmetric states  (examples? what other states are there?) [with respect to the non-observed states?]}.

The ICCs of the best MCMs found on the optimal basis for each $q$  are identified by the triangles in Fig.~\ref{fig:SCOTUS_BestBasis}. 
%First, observe that the best MCMs found in the best basis 
These MCMs have ICCs of smaller sizes (on average) %average size
than those found in the original basis for the same value of $q$ (shown in Fig.~\ref{fig:SCOTUS_BestMCM}), which means that they correspond to less complex models (with fewer interactions). The log-evidence of these models (see table~\ref{tab1}) is also larger than that of the MCMs based on the original basis. This means that they are better models for the data, which is what we expected from the two-step procedure.
Similarly to the analysis in the original basis, ICCs are %\textcolor{red}{getting/becoming} 
smaller for larger $q$, which can be explained by the increase in penalty with $q$ while the data patterns remain mostly unchanged. % (\textcolor{green}{the difference is only due to the increase in [the fraction of] non-observed states VS observed states}).
%Similarly to what happens in the original basis, one can also observe that the ICCs are getting smaller for larger $q$, and 
ICCs for larger $q$ also seem to lie mostly within the ICCs for smaller $q$, although this doesn't hold for all ICCs. 
%\textcolor{green}{In fact, 
%it is unclear what happens here, as the fact that many states aren't observed (due to embedding) can bias the measured high-order correlations.
%it is unclear what the impact of the many unobserved states (due to the embedding at larger $q$) is on the high-order correlation structure of the data (``shrinking the relevant patterns in comparison to non-relevant ones''). These many unobserved states could bias the structure of the data. It could in fact even be thought of as surprising that one can still observe a similar structure at this point.}

%\textcolor{blue}{After artificially increasing $q$, we found that the best basis is mostly formed of pairwise interactions of the form $\phi^{1(q-1)}$ that encode symmetric pairwise voting patterns between the judges.}

%\textcolor{red}{As an example, the most prominent voting patterns in the US Supreme Court dataset analyzed in Sec.~\ref{Sec:Ex:SCOTUS} are symmetric votes between pairs of justices, which explains why the best basis uncovered in that case (after artificially increasing $q$) has mostly interactions of the type $\vecmu=(1,\,q-1)$ (and not $\vecmu=(1,\,1)$ as one could naively expect).}

\subsection{Example 2: 
%\textcolor{red}{the Big Five Personality Test:}
%Discussion on the choice of discretization -- 
Choice of discretization %\textcolor{red}{scheme} 
and standardization of discrete data.}
%Importance of the choice of discretization.
\label{Sec:MCM:Big5}
In this second example, we analyze answers to the Big Five Personality Test \cite{goldberg1992development}. In particular, we discuss possible choices of discretization schemes and their impact on the best MCM factorization in the original basis.

The test consists of $n=50$ questions designed to probe the personality of the respondents along five dimensions generally considered as describing the main traits of personality: extraversion, emotional stability, agreeableness, conscientiousness, and openness to experience.
Each trait is measured based on the answers to ten statements on a scale of $x=1$~(disagree) to $5$~(agree). For example, extraversion is probed by statements such as {\it ``I am the life of the party''} or {\it ``I am quiet around strangers''}.
We obtained a dataset from the Open-Source Psychometrics Project~\cite{Big5data} with answers from $N=602\,591$ individuals (after cleaning for errors %non-valid answers
and identical IP addresses).
The dataset was then converted from the original 1-to-5 scale to the desired discrete formats, with respectively $q=2$, $3$, or $5$, using two different discretization schemes.

In the following, we only perform the MCM analysis in the original basis of the data. This is because we are mainly interested in studying the impact of different choices of discretization of the data %Moreover, the Big Five test was originally designed with the variables as being meaningful for the tested traits.
and the original variables were purposely chosen %have an intrinsic meaning 
in the design of the Big Five test.
As a side note, in the binarized version, \cite{demulatier2024MCM} found that the best MCM in the best basis happened to be is identical to the best MCM found in the original basis.

%the original variables are meaningfull for the tested traits.
%we are interested, which is why we stick with these variables
%Given the nice division with the result, it is likely that the test was designed using PCA.

\subsubsection{Discretization Scheme 1: direct mapping}
For the first scheme, we re-assign the values of the variables $x_j\in\{1,\,2,\,3,\,4,\,5\}$ in the original data (where $x_j$ denotes the answer to the $j$-th question) to new discrete values~$\alpha_j\in\Field$ for $q=2,\,3$ and $5$, as described in Table~\ref{tab:Big5} Scheme~1.
To respect the opposite valence between positive and negative answers, we associated the ``neutral'' answer to $\alpha_j=0$ and gave opposite values\footnote{corresponding to conjugate values of $s_j=\exp(2 i \pi\alpha_j/q)$} 
to opposite answers (e.g., $\alpha_j=1$ to ``agree'' and $\alpha_j=-1\,({\rm mod}\,q)=(q-1)$ to ``disagree'').
%%%%%%%%%%%%
Choosing this assignment of~$\alpha_j$ doesn't affect the outcome of the best MCM with respect to selecting any other permutations of the values of $\alpha_j$ (because permuting the values of $\alpha_j$ doesn't change the numbers $k_{\state_a}$ of occurrence of the states used in Eq.~\eqref{eq:log:evidence:MCM}).
%\textcolor{red}{However, as mentioned previously, this choice can be useful/can give models that are easier to interpret/can make it easier to interpret the uncovered model, if the final goal is to obtain a more precise high-order spin model for the data, beyond MCMs.}
However, as mentioned previously, if the final goal is to obtain a more precise high-order spin model for the data (e.g., beyond MCMs), then different choices can lead to different optimal models, and a choice informed by domain knowledge %an informed choice 
can make it easier to interpret the uncovered model.
%%%%%%%%%%%%%%%%%%%%%%%%%%%%%%%%%%%%%%%%%%%%%%
%\textcolor{red}{However, it can change the interactions selected in the best basis and in the best model overall (when going beyond minimally complex models). And, as mentioned in the previous section/previously, it can be useful to do this if the final goal is to get a more precise high-order spin model for the data.}
%%%%%%%%%%%%%%%%%%%%%%%%%%%%%%%%%%%%%%%%%%%%%%

Figure~\ref{fig:log_ev_Big5:q-bits} shows the best MCMs found in the original basis of the data for Scheme 1.
From $q=2$ to $q=3$, the sizes of the ICCs remain large with 9 to 10 variables: the dataset is sufficiently rich at $q=3$ for the balance between fitting and complexity to require ICCs of a similar sizes than at $q=2$ (contrary to Example 1). In other words, the increase in size of the state space is compensated by an increase in richness of the data.
From $q=3$ to $q=5$, the ICCs break down into smaller components. 
Although questions are mostly grouped into ICCs corresponding to the same traits, we also find mismatches (especially for larger $q$). 
Additionally, we observe that some statements are more biased than others towards agreeing or disagreeing (e.g., ``I insult others''), which may give rise to spurious correlations. %then may appear as correlated. 
The next scheme will attempt to reduce this effect by standardizing the data.

\subsubsection{Discretization Scheme 2: standardization}
In the second scheme, the value of each variable $x_j$ was re-centered around its empirical mean $\bar{x_j}$ following Scheme~2 in Table~\ref{tab:Big5}. 
Besides, for $q=3$ and $5$, we defined small intervals on the values of $x_j$ of size $2\sigma_j\,\epsilon$ within which $\alpha_j$ takes a constant value, where $\sigma_j$ denotes the standard deviation of $x_j$.
The value of~$\epsilon$ was then adjusted to maximize the %lowest  
entropy of the variables in the data (see Fig.~\ref{SI:fig:Big5-Entropy} in Appendix).
Here as well, we reflected the opposite valence of the data between positive and negative answers by giving opposite values of $\alpha_j$ to opposite intervals in~$x_j$.
The introduction of Scheme~2 is motivated by the observation that some statements are significantly more biased than others. %, \textcolor{red}{[resulting in some answers being much less entropic than others]}.
For example, the answer to the statement ``{\it I insult others}'' is more often negative,
with $62\%$ of the respondents who disagree and only $21\%$ who agree ($\bar{x}_j\simeq 2.27$).
%\textcolor{red}{However, such patterns are more informative about the design of the question than about %[the respondents'] 
%[characterizing] personality.}
However, such patterns are inherent to the choice of the question,
and are not directly informative about personality.
%and are not interesting to model [per se].  
%\textcolor{purple}{Instead, one may assess personality by understanding how respondents answer in comparison %/with respect/compared 
%to the average behavior in the population.}
Instead, one may %assess/
understand personality from the way respondents answer with respect to the average behavior in the population.
%personality traits as being defined by how different a person tends to behave in comparison to the average behavior of the population.}
In this context, it would be more interesting to model patterns between answers based on their deviation from the population average instead of raw dependencies.
%%%%%%%%%%%%%%%%%
We achieved this in Scheme 2 by standardizing each variable and maximizing their entropy before modeling them, so that individual variables carry as little information as possible.
%%%%
%The questionnaire was likely designed to measure correlation between the answers to the questions, i.e. correlation between how much your answers deviate from the mean of the overall population.
%%%%%%% 
For example, staying ``neutral'' for the statement ``{\it I insult others}'' ($17\%$ of the respondents) tends to be more negative than the majority of people, and recentering the data will then move ``neutral'' answers to the negative side.
Similarly, the statement ``{\it I spend time reflecting on things}'' has the lowest entropy (see Fig.~\ref{SI:fig:Big5-Entropy} in Appendix), with $48\%$ of the respondents ``strongly agreeing`` and $34\%$ ``agreeing`` ($\bar{x}_j\simeq 4.22$); recentering the data will then give ``neutral`` answers to respondents ``agreeing`` in their response. 

\begin{figure}[p!]
    \centering
    \includegraphics[width=\textwidth]{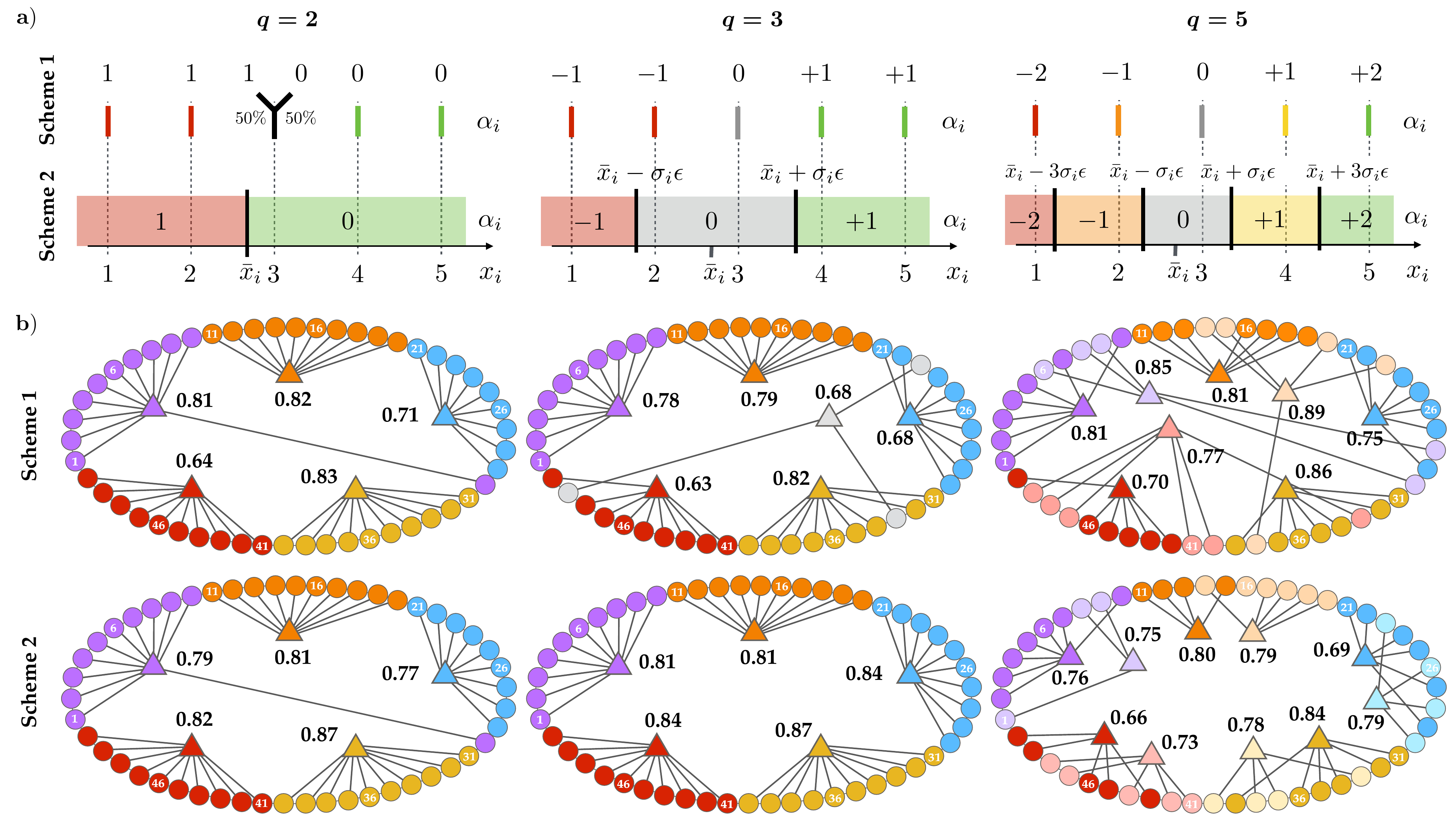}
    \caption{
    {\bf MCM analysis of the Big Five Personality Test data~\cite{Big5data} for varying values of $q$.} 
    {\bf a)}~Illustration of the two discretization schemes described in Table~\ref{tab:Big5}.
    {\bf b)~Best MCMs found in the original basis of the data
    for each discretization scheme}
    for $q=2$, $3$, and $5$.
    The dots (some numbered) correspond to the 50 variables $\alpha_i$ associated with the answers to the 50 questions ordered (clockwise) as in the original dataset~\cite{Big5data}: questions 1-10 assess Extraversion (mostly in purple), questions 11-20 Emotional stability (mostly in orange), questions 21-30 Agreeableness (mostly in blue), questions 31-40 Conscientiousness (mostly in yellow), and questions 41-50 Openness to experience (mostly in red).
    Each colored triangle represents an ICC. The absolute value of the log-evidence for each ICC is indicated near each triangle in qits per variable per datapoint. This value is between $0$ and $1$, and the smaller the value the stronger the compression. % within the ICCs. 
    %of the data based on the variables of the ICC.
    }
    \label{fig:log_ev_Big5:q-bits}
\end{figure}

Figure~\ref{fig:log_ev_Big5:q-bits} shows the best MCMs found in the original basis of the data for Scheme~2. They reveal a clearer separation of the five traits than in Scheme~1, in particular for $q=3$ and~$5$.
%%%%%%
%This guarantees a better focus on the true correlation between the variables.
%By removing/reducing the bias of each variable, one is able to better focus on the correlation structure
In this example, reducing the bias of the variables allows us to better focus the model on the (high-order) correlation structure of the data, 
following a principle similar to using %standardized 
$z$-scored variables in Principal Component Analysis and revealing in this case a clearer separation between the traits.
%%%%%%%%%
%Likely closer to the way the test was originally designed.
%The questionnaire was likely designed to measure correlations between the answers to the questions, i.e. correlation between how much your answers deviate from the mean of the overall population.
%
We note that the standardization of the data is not a requirement of the technique, 
but is a specific choice that can be made depending on what type of dependencies one is interested in modeling (i.e., row dependencies or correlations). %dependencies on deviation from the mean/correlations).
%Recentering the data allows one to study the correlation structure between the answers with respect to their mean [in the population].

%%%%%%%%%%%%%%%%%%%%%%%%%%%%%%%%%%%%%%%%%%%
\begin{table}[h] 
\caption{\label{tab:Big5}
{\bf Discretization schemes for the Big Five Personality Test data.} In the original dataset, the answers $x_i$ to the $i$-th statement can take five different values: 1~(disagree), 2~(slightly disagree), 3~(neutral/no opinion), 4~(slightly agree), and 5~(agree).
The table gives the two mapping schemes we used 
between the values of the original variable $x_i$ and the new %discrete 
variable $\alpha_i$ for each value of~$q$, % in the two schemes used.
where $\bar{x}_i$ and $\sigma_i$ are respectively the empirical average and the standard deviation of $x_i$ and $\epsilon$ is a parameter. % \textcolor{red}{that we adjusted/optimized for}.
}

\begin{tabularx}{\textwidth}{
    *{1}{>{\centering\arraybackslash}p{1.2cm}}|
    *{1}{>{\centering\arraybackslash}p{4cm}}|
    >{\centering\arraybackslash}X 
    >{\centering\arraybackslash}X}
\toprule
&$\alpha_i$ 
& \makecell{$x_i$ in \textbf{Scheme 1}}
    & \makecell{$x_i$ in \textbf{Scheme 2}}\\
\toprule
\multirow{3}{*}{$q=2$}  & $1$ & 1-2:~(slightly) disagree & $x_i<\bar{x}_i$\\
    & $0$ ($50\%$)\; or\; $1$ (50\%)  %$0$ or $1$ (50/50) 
    & $3$: neutral & \\
    & $0$ & $4$-$5$: (slightly) agree & $\bar{x}_i < x_i$\\
\midrule
\multirow{3}{*}{$q=3$}  & $-1$ & $1$-$2$:~(slightly) disagree & $x_i < \bar{x}_i - \sigma_i\,\epsilon$\\
    & $0$ & $3$: neutral & $\bar{x}_i -  \sigma_i\,\epsilon \leq x_i \leq \bar{x}_i + \sigma_i\,\epsilon$\\
    & $+1$ & $4$-$5$: (slightly) agree & $\bar{x}_i + \sigma_i\,\epsilon < x_i$\\
\midrule
\multirow{5}{*}{$q=5$} & $-2$ & $1$: disagree & $x_i < \bar{x}_i - 3\,\sigma_i\,\epsilon$ \\
 & $-1$ & $2$: slightly disagree & $\bar{x}_i - 3\,\sigma_i\,\epsilon \leq x_i < \bar{x}_i - \sigma_i\,\epsilon$\\
 & $0$ & $3$: neutral    & $\bar{x}_i - \sigma_i\,\epsilon \leq x_i \leq \bar{x}_i + \sigma_i\,\epsilon$\\
 & $+1$ & $4$: slightly agree & $\bar{x}_i + \sigma_i\,\epsilon < x_i \leq \bar{x}_i + 3\, \sigma_i\,\epsilon$\\
 & $+2$ & $5$: agree & $\bar{x}_i + 3\,\sigma_i\,\epsilon < x_i$\\
\bottomrule
\end{tabularx}
\end{table}

\clearpage
\clearpage

\subsection{Discussion about compressibility and stochastic complexity}

\newcolumntype{C}{>{\centering\arraybackslash}X}

\begin{table}[h] 
\center
\caption{
{\bf Log-evidence values of MCMs for the US Supreme Court voting data.}
Raw values (in nats) of the log-evidence of the MCMs found for the US Supreme Court voting data at different values of $q$ and the corresponding absolute values expressed in qits per datapoint.
These latter provide an estimate of the complexity of the data (see Sec.~\ref{sec:complexity}), which can be interpreted as the average number of qits per datapoint needed to encode the data using MCMs.
\label{tab1}}
\begin{tabularx}{\textwidth}{CCCCC}
%{0.85\textwidth}{
%    *{1}{>{\centering\arraybackslash}p{1.5cm}}
%    >{\centering\arraybackslash}X
%    >{\centering\arraybackslash}X 
%    >{\centering\arraybackslash}X 
%    >{\centering\arraybackslash}X}
\toprule	
    & \multicolumn{2}{c}{\makecell{\textbf{log-evidence}\\ (in nats)}}
    &\multicolumn{2}{c}{\makecell{\textbf{-log-evidence}\\ (in qits per datapoint)}}\\[2mm]
\midrule
\textbf{q}	
    & \textbf{Original basis}
    & \textbf{New basis}
    & \textbf{Original basis}
    & \textbf{New basis}\\
\midrule
2		& -3300		& -3154  & 5.32 & 5.08 \\
3		& -3715		& -3588  & 3.78 & 3.65 \\
4		& -4033		& -3763  & 3.25 & 3.03 \\
5		& -4262		& -3848  & 2.96 & 2.67  \\
\bottomrule
\end{tabularx}
%\noindent{\footnotesize{\textsuperscript{1} Tables may have a footer.}}
\end{table}
%For $q=2$, the best model has $\log E = -3300 = 5.32$ bits/datapoint; for $q=3$, the best model has $\log E = -3715 = 5.99 $ bits/datapoint $= 3.78$ trits/datapoint; for $q=4$, the best model has $\log E = -4033 = 6.50 $ bits/datapoint $= 3.25$ \,$4$-its/datapoint; for $q=5$, the best model has $\log E = -4262 = 6.87$ bits/datapoint $= 2.96$\, $5$-its/datapoint.\\

\begin{table}[h] 
\caption{%Case where scheme 2 is with $\pm \sigma\epsilon$.
{\bf Log-evidence values of MCMs for the Big Five Personality Test data} (with $N=602\,591$ datapoints).
Raw value of the log-evidence in nats (first column) and absolute values of the log-evidence in qits per datapoints (second column). This latter value gives the average number of qits per datapoint needed to express the dataset using the best MCM (this is a value between $0$ and $n=50$), which can be thought of as giving an estimate of the complexity of the data. %providing an estimate of the complexity of the data. 
%\textcolor{orange}{Number of qits needed to express the Big Five Personality Test data (with $N=602\,591$ datapoints) using MCMs.}
\label{tab:logE:Big5}}
%\newcolumntype{C}{>{\centering\arraybackslash}X}
\begin{tabularx}{\textwidth}{CCCCC}
\toprule
    & \multicolumn{2}{c}{\makecell{\textbf{log-evidence}\\ (in nats)}}
    &\multicolumn{2}{c}{\makecell{\textbf{-log-evidence}\\ (in qits per datapoint)}}\\[2mm]
\midrule
\textbf{q}  & {\bf Scheme 1} & {\bf Scheme 2} & {\bf Scheme 1} & {\bf Scheme 2}\\
\midrule
$2$	& $-1590\times10^4$ & $-1699\times10^4$ & $38.06$ & $40.68$ \\
$3$	& $-2437\times10^4$ & $-2758\times10^4$ & $36.81$ & $41.66$ \\
$5$	& $-3892\times10^4$ & $-3672\times10^4$ & $40.13$ &  $37.86$\\
\bottomrule
\end{tabularx}
\end{table}

%%%%%%%%%%%%%%%%%%%%%%%%%%%%%%%%%%%%%%%%%%%%%%%%%
\section{Discussion}
\label{sec:discussion}
%%%%%%%%%%%%%%%%%%%%%%%%%%%%%%%%%%%%%%%%%%
%\section{Discussion}

In this work, we introduced a complete family of maximum entropy models designed %to study higher-order correlation patterns in discrete data. 
to model any high-order patterns of discrete data.
This comprehensive framework %acts as a ``bulldozer'' approach, 
is capable of capturing the full complexity of multivariate discrete systems. %state spaces. 
%Crucially, 
In contrast to the version of the Potts models typically used in statistical inference (see Eq.~\eqref{Eq:Ex:Potts:stat_inference}), our models avoid over-parametrization and are able to capture high-order patterns. Focusing strictly on the pairwise case, for example, the maximum number of parameters used corresponds exactly to the number of pairwise patterns that can be fitted to the data. Furthermore, mapping these maximum entropy models to discrete Fourier analysis not only confirms that they represent a complete model family for discrete data, but also allows us to straightforwardly invert the parameters from empirical frequencies via a discrete Fourier transform.

A key analytical result of our framework is the cluster expansion of the partition function of $q$-state spin models, which we express in terms of the loops of the models. In practice, these loops can be extracted easily using the matrix representation of spin models. %This expansion is highly versatile and proves useful across numerous theoretical and computational contexts, such as sampling from the free energy \cite{papers}. 
Furthermore, the cluster expansion of the partition function %we have 
establishes a direct connection between the statistical properties of the models and the algebraic properties of their matrix representation. This transition to a linear algebraic perspective facilitates the %much easier} 
manipulation of spin models and allows for a rigorous characterization of statistical constraints, model complexity, and equivalence classes. In particular, we extended the framework of equivalence classes of spin models by generalizing the notion of gauge transformations (GTs) to the discrete setting. We showed that GTs provide a more general definition of symmetries in discrete data, extending well beyond standard spatial reflections, rotations, or permutations. More importantly, GTs enable the development of model selection procedures that are entirely independent of the original basis representation of the data. This opens the way for equivariant modeling, allowing systems to be analyzed based on their inherent symmetries rather than arbitrary representational choices. 

%\subsection{Practical Applicability and Minimally Complex Models}
While the %complete 
family of $q$-state spin models %high-order models 
has a super-exponential number of models, we make the framework %\textcolor{red}{practically usable}/
more widely usable in practice by focusing on a sub-family of models, the minimally complex models (MCMs). MCMs operate essentially as factorization models: they allow us to find a factorization of the multivariate probability distribution that optimally matches the underlying structure of the data by carefully balancing goodness-of-fit against complexity. The benefit of working with MCMs is that all quantities of interest %\textcolor{red}{are analytically computable/tractable} / 
have a closed-form expression allowing for efficient model selection in reasonably large systems. Moreover, the family of MCMs is invariant under GTs enabling %allowing for 
equivariant model selection within this family. %the class of MCMs.
We demonstrated the efficacy of this approach through real-world applications. % on US Supreme Court data and Big Five Personality Test data. 
In particular, we discussed the importance of the choice of the discretization scheme and illustrated how they affect the resulting models in our examples.
%\textcolor{blue}{In particular, we emphasize the importance of the choice of the discretization scheme by illustrating how they affect the resulting models in our examples.}
%we discuss how different discretization scheme affect the resulting model.
%Because many real-world applications require raw data to be discretized prior to analysis, the choice of discretization can profoundly impact the inferred structures. We discuss how different discretization schemes may affect the resulting model.
%
To ensure broad accessibility, we have also released an %easy-to-use, 
open-source package~\cite{DeClerq_2024_Python} capable of analyzing systems with up to $128$ discrete variables. 

%\subsection{Future Directions}
Looking forward, interesting research directions %our primary objective 
are the %to 
development of %more highly optimized 
more efficient search algorithms within the space of MCMs for scaling the analysis to larger systems, %(to reach several hundreds to a thousand of variables)}. 
%Expanding these algorithms to efficiently search for both the best MCM and the optimal basis will be critical for scaling analyzes to even larger systems without sacrificing computational efficiency. 
as well as extending our general modeling framework to continuous and non-stationary data. In particular, extending the MCM approach will broaden the scope of optimal factorization techniques to an even wider array of complex systems.
%Furthermore, we plan to explore how the concept of data-driven factorization can be generalized to non-stationary data and continuous variables, broadening the scope of optimal factorization techniques to an even wider array of complex systems.
Another interesting idea is the use of generalized symmetries of discrete data and spin models (thanks to the definition of gauge transformations) for %model selection, 
%to improve model selection.
%in the context 
the development of equivariant model selection. % techniques.

%%%%%%%%%%%%%%%%%%%%%%%%%%%%%%%%%%%%%%%%%%%%%%%%%
\newpage
%\appendixtitles{yes} % Leave argument "no" if all appendix headings stay EMPTY (then no dot is printed after "Appendix A"). If the appendix sections contain a heading then change the argument to "yes".
%\appendixstart
%\appendix
\addcontentsline{toc}{section}{Appendices}
\section*{\LARGE Appendices}

\appendixtitleon
\appendixtitletocon

\renewcommand{\theequation}{A\arabic{equation}}
\setcounter{equation}{0}

\renewcommand{\thefigure}{A\arabic{figure}}
\setcounter{figure}{0}

\begin{appendices}
%\section[\appendixname~\thesection]{High-order vector Potts models}
\section{$q$-state spin models}
\label{App:Potts:def}
\subsection{From complex to real Hamiltonian.} \label{App:H:CtoR}
\begin{proof}[Proof of Eq.~\eqref{eq:g:constraint}] 
    Let us denote $g_{\vecmu}=A_{\vecmu}+ i B_{\vecmu}$. We recall that $\phi^{-\vecmu}(\state) = [\phi^{\vecmu}(\state)]^{*}$. For $p(\state\,|\,\vecg)$ in Eq.~\eqref{eq:def:CompleteM} to be real, one needs that, for all $\state\in\Field^n$:
    %$A_{\vecmu}$ and $B_{\vecmu}$ respectively the real and imaginary part of $g_{\vecmu}$: 
    \begin{align}
        {\rm Im}\left[g_{\vecmu}\,\phi^{\vecmu}(\state)+g_{-\vecmu}\,\phi^{-\vecmu}(\state)\right]&=0 \\
        \Leftrightarrow\qquad
        (B_{\vecmu}+B_{-\vecmu})\cos\bigg(\frac{2\pi}{q} \state\cdot\vecmu\bigg)
        +(A_{\vecmu}-A_{-\vecmu})\sin\bigg(\frac{2\pi}{q} \state\cdot\vecmu\bigg)
        &=0\,.
    \end{align}
    Because %$\varphi^{\vecmu}(\state)=$
    $\cos\left(\frac{2\pi}{q}\state\cdot\vecmu\right)$ and %$\psi^{\vecmu}(\state)=$
    $\sin\left(\frac{2\pi}{q}\state\cdot\vecmu\right)$ are orthogonal vectors of $\R[\Field^n]$, this latter equation is equivalent to $A_{-\vecmu}=A_{\vecmu}$ and $B_{-\vecmu}=-B_{\vecmu}$, which gives %and thus to 
    $g_{-\vecmu}=g_{\vecmu}^*$.
\end{proof}

\begin{proof}[Proof: re-writting Eq.~\eqref{eq:def:CompleteM} under the form of \eqref{eq:def:CompleteM:real}]
    Let us denote $g_{\vecmu}=A_{\vecmu}+ i B_{\vecmu}=g_{-\vecmu}^*$. We have that $\phi^{\vecmu}(\state)=\varphi^{\vecmu}(\state)+i\psi^{\vecmu}(\state)$, where $\varphi^{\vecmu}(\state)=\cos(2\pi\state\cdot\vecmu/q)$ and $\psi^{\vecmu}(\state)=\sin(2\pi\state\cdot\vecmu/q)$, %are given by Eq.~\eqref{}
    and that $\phi^{-\vecmu}(\state) = [\phi^{\vecmu}(\state)]^{*}$. Grouping conjugate terms together for all $\vecmu$ such that $-\vecmu\neq\vecmu$, the Hamiltonian in Eq.~\eqref{eq:def:CompleteM} takes the form:
    \begin{align}
        H(\state) = \sum_{\vecmu=-\vecmu} g_{\vecmu}\varphi^{\vecmu}(\state)
            + \sum_{\vecmu<-\vecmu} \left(g_{\vecmu}\phi^{\vecmu}(\state) + g^*_{\vecmu}\phi^{-\vecmu}(\state)\right)\,,
    \end{align}
    where we used the notation $\vecmu<-\vecmu$ to ensure that the second sum includes each conjugate pair only once. For $-\vecmu=\vecmu$, $\phi^{\vecmu}(\state)=\varphi^{\vecmu}(\state)$ and, to match Eq.~\eqref{eq:def:CompleteM} to %with the expression in
    Eq.~\eqref{eq:def:CompleteM:real}, we define the parameter %we directly get that 
    $a_{\vecmu}=2 g_{\vecmu}$ . 
    For~$-\vecmu\neq\vecmu$, we compute the local Hamiltonian:
    \begin{align}\label{app:eq:H:CtoR}
        H^{\vecmu}(\state) = g_{\vecmu}\phi^{\vecmu}(\state) + g^*_{\vecmu}\phi^{-\vecmu}(\state) = 2 A_{\vecmu} \varphi^{\vecmu}(\state) -  2 B_{\vecmu} \psi^{\vecmu}(\state)\,,
    \end{align}
    from which we define $a_{\vecmu}=2 A_{\vecmu}$ and $b_{\vecmu}=-2 B_{\vecmu}$ to match Eq.~\eqref{eq:def:CompleteM:real}, i.e. $g_{\vecmu}=\frac{a_{\vecmu}-i b_{\vecmu}}{2}$.
\end{proof}

\subsection{Orthogonality of complex spin operators.}
\label{app:FourierBasis:C}
\begin{proof}[Proof of Eq.~\eqref{eq:linearly_indep_monimials}:]
for all $(\vecmu, \vecnu)\in[\Field^n]^2$:
    \begin{align}
    %\label{eq:proof:linearly_indep_monimials:1}
    \langle \phi^{\vecnu}\,|\, \phi^{\vecmu}\rangle
        &= \frac{1}{q^n}\,\sum_{\state} \phi^{\vecmu}(\state)\,[\phi^{\vecnu}(\state)]^*
        = \frac{1}{q^n}\,\sum_{\state} \phi^{\vecmu}(\state)\,\phi^{-\vecnu}(\state)
        \,,\\[1pt]
        &\overset{(a)}{=} \frac{1}{q^n}\,\sum_{\state} \phi^{\vecmu-\vecnu}(\state) 
        \overset{(b)}{=} \frac{1}{q^n}\,\times\,
            \begin{cases}
                \;\;q^n \qquad&{\rm if}\;\;\vecmu-\vecnu = \veczero\,,\\
                \;\;0 \qquad&{\rm otherwise}
            \end{cases}\nonumber\\[1pt]
    \langle \phi^{\vecnu}\,|\, \phi^{\vecmu}\rangle
        &= \delta_{\vecmu, \vecnu}\,.\nonumber        
\end{align}
where $\phi^{-\vecnu}(\state) =[\phi^{\vecnu}(\state)]^*$ is the inverse of $\phi^{\vecnu}(\state)$ in $\Omega$.
For (a), we used the closure property of $\Omega$ %expressed 
in Eq.~\eqref{eq:monomials:group_structure}, and for (b) 
we used the property Eq.~\eqref{eq:ppty:monomials}.
\end{proof}

\subsection{Orthogonality of real spin operators in $\Omega_{\rm R}$.}
%{Basis of real spin operators $\Omega_{\rm R}$.}
%{Orthogonality of real spin operators in $\Omega_{\rm R}$ and basis.}
%{Fourier basis for $\R[\Field^n]$.}
\label{app:FourierBasis:R}
%
%Replacing the expression of $\phi^{\vecmu}(\alpha)$ in Eq.~\eqref{eq:ppty:monomials} and  identifying respectively the real and the imaginary parts on both sides of the equality, the equi-repartition property~\eqref{ppty:sum:phi} can also be written as: $\forall\vecmu \in\Field^n$:
%\begin{align}\label{eq:ppty:monomials:cos-sin} 
%    \sum_{\alpha\in\Field^n} \cos\left(\frac{2\pi}{q}\state\cdot\vecmu\right)= q^n\,\delta_{\vecmu, \veczero}
%    \qquad{\rm and}\qquad
%    \sum_{\alpha\in\Field^n} \sin\left(\frac{2\pi}{q}\state\cdot\vecmu\right)=0\,.
%\end{align}
We denote by $\R[\Field^n]$ the vector space of real functions on $\Field^n$ %from $\Field^n$ to $\R$
endowed with the following inner product:
for two real functions $f(\state)$ and $h(\state)$,
\begin{align}\label{eq:def:HermitianInnerProd:R}
    \langle \,f\,|\, h\,\rangle
        &= \frac{1}{q^n}\sum_{\state\in\Field^n} f(\state)\,h(\state)\,,
\end{align}
Applying the scalar products for any two operators of the forms $\varphi^{\vecmu}(\state)=\cos\left(\frac{2\pi}{q}\state\cdot\vecmu\right)$ and $\psi^{\vecmu}(\state)=\sin\left(\frac{2\pi}{q}\state\cdot\vecmu\right)$, one gets that:\;\;
for all $(\vecmu, \vecnu)\in[\Field^n]^2$:
%\begin{align}\label{eq:linearly_indep_monimials:cos-sin}
%\begin{cases}
%    \left\langle \cos\left(\frac{2\pi}{q}\state\cdot\vecmu\right)\,\big|\,\sin\left(\frac{2\pi}{q}\state\cdot\vecmu\right) \right\rangle&=0\\
%    \left\langle \cos\left(\frac{2\pi}{q}\state\cdot\vecmu\right)\,\big|\,\cos\left(\frac{2\pi}{q}\state\cdot\vecmu\right) \right\rangle&= \delta_{\mu,\nu}\\
%    \left\langle \sin\left(\frac{2\pi}{q}\state\cdot\vecmu\right)\,\big|\,\sin\left(\frac{2\pi}{q}\state\cdot\vecmu\right) \right\rangle&= \delta_{\mu,\nu}
%\end{cases}
%\end{align}
%
%\begin{align}\label{eq:linearly_indep_monimials:cos-sin}
%\begin{cases}
%    \left\langle \varphi^{\vecmu}\,\big|\,\psi^{\vecnu}\right\rangle&=0\\
%    \left\langle \varphi^{\vecmu}\,\big|\,\varphi^{\vecnu}\right\rangle&=\delta_{\mu,\nu}\\
%    \left\langle \psi^{\vecmu}\,\big|\,\psi^{\vecnu}\right\rangle&=\delta_{\mu,\nu}
%\end{cases}
%\end{align}
\begin{align}\label{eq:linearly_indep_monimials:cos-sin}
    \left\langle \varphi^{\vecmu}\,\big|\,\psi^{\vecnu}\right\rangle
        =0\;,\qquad
%    \qquad {\rm and} \qquad
    \left\langle\varphi^{\vecmu}\,\big|\,\varphi^{\vecnu}\right\rangle
        =\frac{\delta_{\vecmu,\vecnu}+\delta_{\vecmu,-\vecnu}}{2}
    \qquad {\rm and} \qquad
    \left\langle \psi^{\vecmu}\,\big|\,\psi^{\vecnu}\right\rangle
        =\frac{\delta_{\vecmu,\vecnu}-\delta_{\vecmu,-\vecnu}}{2}
\end{align}

The set $\Omega_R$ of real spin operators~\eqref{eq:Omega:real} is a set of $q^n$ orthogonal elements of $\R[\Field^n]$, which is a vector space of dimension $q^n$. The set $\Omega_R$ thus forms a basis of $\R[\Field^n]$.

\begin{proof}{Eq.~\eqref{eq:linearly_indep_monimials:cos-sin}} can be proven easily using trigonometric formulae and Eq.~\eqref{eq:ppty:monomials:cos-sin}:
\begin{align}
    \left\langle \varphi^{\vecmu}\,\big|\,\psi^{\vecnu}\right\rangle
        &= \frac{1}{q^n}\sum_{\state\in\Field^n} 
        \cos\left(\frac{2\pi}{q}\state\cdot\vecmu\right)\,\sin\left(\frac{2\pi}{q}\state\cdot\vecnu\right)\,,\\
        &=\frac{1}{2}\,\frac{1}{q^n}\bigg[
            \sum_{\state\in\Field^n} \sin\left(\frac{2\pi}{q}\state\cdot(\vecmu+\vecnu)\right)
            -\sum_{\state\in\Field^n} \sin\left(\frac{2\pi}{q}\state\cdot(\vecmu-\vecnu)\right)
            \bigg]\\
        &=0\,.
\end{align}
\begin{align}
    \left\langle \varphi^{\vecmu}\,\big|\,\varphi^{\vecnu}\right\rangle
        &= \frac{1}{q^n}\sum_{\state\in\Field^n} 
        \cos\left(\frac{2\pi}{q}\state\cdot\vecmu\right)\,\cos\left(\frac{2\pi}{q}\state\cdot\vecnu\right)\,,\\
        &=\frac{1}{2}\,\frac{1}{q^n}\,\bigg[
            \sum_{\state\in\Field^n} \cos\left(\frac{2\pi}{q}\state\cdot(\vecmu+\vecnu)\right)
            +\sum_{\state\in\Field^n} \cos\left(\frac{2\pi}{q}\state\cdot(\vecmu-\vecnu)\right)
            \bigg]\nonumber\\
        &=\frac{1}{2}\left[\delta_{\vecmu,-\vecnu} + \delta_{\vecmu,\vecnu}\right]\,.
\end{align}
The proof can be done the same way for the last identity,
 using that $\sin a\sin b=(\cos(a-b)-\cos(a+b))/2$.
\end{proof}

%Besides, \textcolor{blue}{These operators satisfy that $\varphi^{-\vecmu}(\state)=\varphi^{\vecmu}(\state)$ and $\psi^{-\vecmu}(\state)=-\psi^{\vecmu}(\state)$ and thus there are only two distinct operators (and not four) for each pair of vectors $(\vecmu, -\vecmu)$, which can also be written as the constraint, $a_{-\vecmu}=a_{\vecmu}$ for the coefficient associated to $\varphi^{\vecmu}$ and $b_{-\vecmu}=-b_{\vecmu}$ for the coefficient associated to $\psi^{\vecmu}$ (this is a direct consequence of the constrained $g_{-\vecmu}=g_{\vecmu}^*$).
%\textcolor{red}{Each pair of conjugate operators is represented by a single operator $\varphi$ and $\psi$.}
%Let us denote $Z_0 = \{\vecmu\in\Field \mid -\vecmu=\vecmu\}$ and $Z_1\cup \bar{Z}_1 = \{\vecmu\in\Field  \mid -\vecmu\neq\vecmu\}$ such that $\forall\vecmu\in Z_1$,\; $-\vecmu\in\bar{Z}_1$.}

%\textcolor{red}{Using the relations in Eq.~\eqref{eq:ppty:monomials:cos-sin}, one can easily show that the set of all real spin operators $\Omega_R=\{\varphi^{\pm\vecmu}(\state),\,\psi^{\pm\vecmu}(\state) \mid \vecmu\in  \Field^n\}$ \textcolor{red}{form[s]} a set of $q^n$ basis functions for the vector space of real functions over $\Field^n$ (see App.~\ref{app:FourierBasis:R}).}

\subsection{Fourier coefficients and inverse Fourier transform.}\label{app:Fouriercoeff}
\begin{proof}[{\bf Proof of Eq.~\eqref{eq:pdf:FT}:}] Starting from Eq.~\eqref{eq:expansion:Fourier_basis} for the log-probability:
    \begin{align}
        \log P(\state) = \sum_{\vecmu\in\Field^n} g_{\vecmu}\, \phi^{\vecmu}(\state)\;,
        %- \log Z\,,
    \end{align}
we take the inner product with the operator $\phi^{\vecnu}(\state)$ for a given $\vecnu\in\Field^n$:
\begin{align}
    %\,\frac{1}{q^n} \sum_{\state\in\Field^n} \, \log P(\state) \, \left[\phi^{\vecnu}(\state)\right]^* &=
    \left\langle \phi^{\vecnu}\mid\log P\right\rangle %&
        = \sum_{\vecmu\in\Field^n} g_{\vecmu}\, \left\langle \phi^{\vecnu}\mid\phi^{\vecmu}\right\rangle 
        =\sum_{\vecmu\in\Field^n} g_{\vecmu} \,\delta_{\vecmu, \vecnu} %\nonumber\\
        %&
        =g_{\vecnu} \,,
\end{align}
where we used that spin operators are orthogonal~\eqref{eq:linearly_indep_monimials}.
\end{proof}

\begin{proof}[{\bf Proof of Eq.~\eqref{eq:pdf:FT:a_b}:}] We start from Eq.~\eqref{eq:def:CompleteM:real} for the log-probability, written %as a function of the 
in terms of the variables~$\state$ (with $a_{\veczero}=-\log Z(\vecg)$):
    \begin{align}\label{app:logP:real}
        \log P(\state) = \sum_{\vecmu\leq -\vecmu} 
        \left(a_{\vecmu} \varphi^{\vecmu}(\state)
        + b_{\vecmu} \psi^{\vecmu}(\state) \right)\,, % - \log Z\,,
    \end{align}
where the notation $\vecmu\leq -\vecmu$ prevents including twice the same conjugate pair in the sum, and where $b_{\vecmu}=0$ for all $\vecmu$ such that $-\vecmu=\vecmu$. % and $a_{\veczero}=-\log Z(\vecg)$
%we multiply both sides by $\varphi^{\vecnu}(\state)$ for $\vecnu\neq\veczero$ and sum over the states of the system:
We take the inner product with the operator $\varphi^{\vecnu}(\state)$ for a given $\vecnu\in\Field^n$: % and sum over the states of the system:
\begin{align}
    \left\langle \varphi^{\vecnu}\mid\log P\right\rangle 
    %\frac{1}{q^n}\sum_{\alpha\in\Field^n}\log P(\state) \,\varphi^{\vecnu}(\state) 
        &= \sum_{\vecmu\leq -\vecmu} \bigg[a_{\vecmu} \,\langle\varphi^{\vecnu}\,|\,\varphi^{\vecmu} \rangle
        + b_{\vecmu} \,\big\langle\varphi^{\vecnu}\,|\,\psi^{\vecmu} \big\rangle \bigg]
        %- \log Z\sum_{\alpha\in\Field^n}\varphi^{\vecnu}(\state)\,,
        %\nonumber\\
        =\sum_{\vecmu\leq -\vecmu} a_{\vecmu} \,\frac{\delta_{\vecnu,\vecmu}+\delta_{\vecnu,-\vecmu}}{2}
        %= \frac{a_{\vecnu}+a_{-\vecnu}}{2}
        \nonumber\\
        %&=a_{\vecmu}\,,\nonumber
        &=
        \begin{cases}
            \, a_{\vecnu}\qquad\quad&{\rm for\quad} \vecnu=-\vecnu\\[2pt]
            \, a_{\vecnu}/2 %\displaystyle\frac{a_{\vecnu}}{2}
            \qquad\quad&{\rm for\quad} \vecnu\neq-\vecnu\\
        \end{cases}
\end{align}
where we used the orthogonality between real spin operators~\eqref{eq:linearly_indep_monimials:cos-sin}.
%that the real spin operators are orthogonal~\eqref{eq:linearly_indep_monimials:cos-sin}.
Similarly, the relation for $b_{\vecmu}$ in Eq.~\eqref{eq:pdf:FT:a_b} can be obtained by taking the inner product of $\log P(\state)$ with $\psi^{\vecnu}(\state)$.
\end{proof}

\begin{proof}[Proof of Eq.~\eqref{eq:pdf:IFT}] There's nothing to prove, as this equation directly comes the initial decomposition of $\log P$ over the Fourier basis in Eq.~\eqref{eq:expansion:Fourier_basis}. However, it can be interesting to show how to go back from the Fourier coefficients~\eqref{eq:pdf:FT} to $\log P$ using inner products. Starting from Eq.~\eqref{eq:pdf:FT} for the coefficients:
    \begin{align}
        G(\vecmu) = g_{\vecmu} = \frac{1}{q^n} \sum_{\state'\in\Field^n} \log P(\state')\, [\phi_{\state'}(\vecmu)]^*\;,
    \end{align}
where we introduced the functions $\phi_{\state}(\vecmu)\doteq\phi^{\vecmu}(\state)$ for all $\vecmu$ and $\state$ in $\Field^n$. % (see App.~\ref{app:Fouriercoeff}).
Let's take the inner product of $G$ with the operator $\phi_{\state}$ %(\vecnu)
for a given state $\state\in\Field^n$:
\begin{align}
    \left\langle G^* \mid \phi_{\state}\right\rangle
        = \frac{1}{q^n}\sum_{\state'\in\Field^n} \log P(\state')\, \left\langle \phi_{\state'}\mid\phi_{\state}\right\rangle 
        =\frac{1}{q^n}\sum_{\state'\in\Field^n} \log P(\state')\,\delta_{\state', \state}
        =\frac{1}{q^n} \log P(\state)\,,\nonumber
\end{align}
where we used the completeness property~\eqref{ppty:op:complete} of spin operators.
%%%%% Proof without using the inner product notation: %%%%%
%\begin{align}
%    \left\langle G^* \mid \phi_{\state}\right\rangle
%        &= \frac{1}{q^n}\sum_{\vecmu\in\Field^n} G(\vecmu)\,\phi_{\state}(\vecmu)
%        = \frac{1}{q^n}\sum_{\vecmu\in\Field^n} \frac{1}{q^n} \sum_{\state'\in\Field^n} \log P(\state')\, [\phi_{\state'}(\vecmu)]^*\,\phi_{\state}(\vecmu)\nonumber\\
%        &= \frac{1}{q^n} \sum_{\state'\in\Field^n} \log P(\state')\, \left\langle \phi_{\state'}\mid\phi_{\state}\right\rangle
%        =\frac{1}{q^n}\sum_{\state'\in\Field^n} \log P(\state')\,\delta_{\state', \state}
%        =\frac{1}{q^n} \log P(\state)\,,\nonumber
%\end{align}
%%%%% %%%%% %%%%% %%%%% %%%%% %%%%% %%%%% 
\end{proof}

\subsection{Entropy of models with a single interaction}\label{app:entropy:single-mu}

Take %a system of $q$-state variables and a 
a $q$-state spin model for $n$ variables with a single interaction $\vecmu$ encoded by the operator $\phi^{\vecmu}$ (and its c.c. if $\vecmu\neq -\vecmu$). For all states $\state$, the operator $\phi^{\vecmu}(\state)=z_q^{\vecmu\cdot\state}$ only takes values in the $q$-th roots of unity (corresponding to $q$ %evenly spaced 
locations on the unit circle). Besides, if $q$ is prime or if for all integer $c\in[0,q-1]$ the vector $\vecmu$ verifies that $c\vecmu\neq \veczero$ in $\Field^n$, then there are the same number of states  at each root (i.e., $q^{n-1}$ states).
For a given $j\in\Field$, all the states $\state$ for which $\vecmu\cdot\state=j\,$ have the same interaction energy~\eqref{eq:energy_projection}, $H^{\vecmu}(\state)=-2\,\vec{g_{\vecmu}^*}\cdot\vec{z_q}^j$, where $\vec{z_q}^j$ is the direction of the root $z_q^j$ in the complex plane, and the same probability:
\begin{align}
    P(\state\,|\,\{\vecmu\},\, g_{\vecmu}) = \frac{p_j}{q^{n-1}}\,,
    \qquad{\rm where}\qquad p_j=\frac{e^{\,2\,\vec{g_{\vecmu}^*}\cdot\vec{z_q}^j}}{Z_q}\qquad{\rm and}\qquad Z_q=\sum_{j=0}^{q-1} e^{\,2\,\vec{g_{\vecmu}^*}\cdot\vec{z_q}^j}
\end{align}
is a normalization factor. 
%Indeed, to compute the entropy of an operator, one must compute 
The probability that the operator $\phi^{\vecmu}(\state)$ takes the value $z_q^j$ is then:
%must compute the probability distribution over the values $z_q^j$ that can be taken by the operator:
\begin{align}
    P[\phi^{\vecmu}(\state)=z_q^j] 
        &= \sum_{\state:\;\vecmu\cdot\state=j} P(\state\,|\,\{\vecmu\},\,g_{\vecmu})
        =q^{n-1} \frac{p_j}{q^{n-1}}
        =p_j\,,
\end{align}
because there are exactly $q^{n-1}$ states satisfying $\vecmu\cdot\state=j$. Hence, the entropy of the operator $\phi^{\vecmu}$ in the model with the single-interaction~$\vecmu$ is
\begin{align}
    S[\phi^{\vecmu}] = -\sum_{j=0}^{q-1} p_j\,\log p_j\,,
\end{align}
and the entropy of the model probability distribution is
\begin{align}
    S[P(\state\,|\,\{\vecmu\},\, g_{\vecmu})] 
        &= -\sum_{\state} %\in\Field^n
        P(\state\,|\,\{\vecmu\},\, g_{\vecmu})\log P(\state\,|\,\{\vecmu\},\, g_{\vecmu})
        =-q^{n-1}\sum_{j=0}^{q-1} \frac{p_j}{q^{n-1}}\log \frac{p_j}{q^{n-1}}\,,
        \nonumber\\
        &=-\sum_{j=0}^{q-1} p_j \log p_j + (n-1)\log q\,.
\end{align}
The first part this last equation corresponds to the entropy of the operator $\phi^{\vecmu}$, while the second corresponds to the entropy of $(n-1)$ purely random $q$-state variables. 

\begin{figure}[h]
    %\centering
    \includegraphics[width=\textwidth]{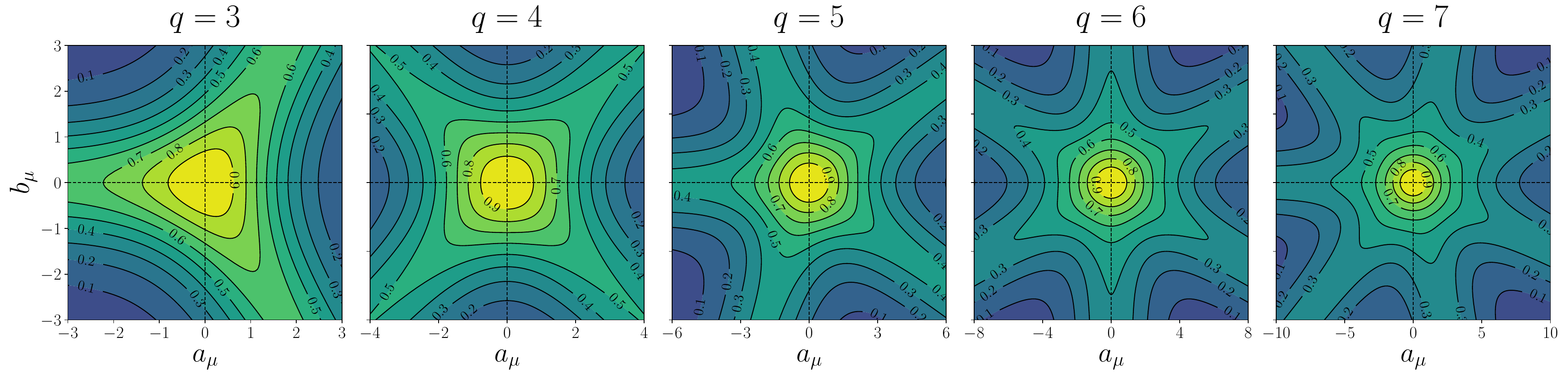}
\caption{
    Contour plot of the entropy (in quits -- i.e., we divided the value by $\log q$) of the operator $\phi^{1}(\alpha_1)=\exp(\frac{2\pi i}{q}\alpha_1)$ as a function of the parameters $a_1$ and $b_1$ for different values of~$q$. 
    %\textcolor{red}{Dependency of the value of the entropy as a function of the parameter.}
    For $q$ prime, all the other operators have the same entropy plot as this operator. 
    %(the only thing that changes from this operator is that the orientations of the chosen operator for a given state $\state$ will be permuted). 
    %(the difference is in the direction of the vector $\vec{\phi}^{\vecmu}(\state)$ for different values of $\state$).
    For $q$ non-prime, the entropy plot can become the one of a smaller $q$ if $\vecmu$ is divisible. For instance, the operator $\phi^{2}(\alpha_1)=\exp(\frac{2\pi i}{4}\,2\alpha_1)=z_q^{\,2\alpha_1}$ for $q=4$ is equal to the operator $\phi^{1}(\alpha_1)=\exp(\frac{2\pi i}{2}\alpha_1)=z_q^{\,\alpha_1}$ for $q=2$, and the entropy plot of this operator $\phi^{2}$ becomes %like the plot for an operator when $q=2$.
    the one of a typical operator at $q=2$ (instead of $q=4$).
    }
\label{app:fig:Entropy}
\end{figure}

\section%[\appendixname~\thesection]
{Expansions of the partition function}\label{app:Z:loops}
%{Loop expansion of the partition function}

%\subsection[\appendixname~\thesubsection]{Trick: generalize Euler formula}
\subsection{Generalized Euler formula} %Trick:
\label{app:EulerFormula}
Similarly to the binary case, we re-write the exponential under the sum of $q$ hyperbolic functions. In the binary case ($q=2$), Ref.~\cite{beretta2018stochastic} uses that:
\begin{align}\label{eq:Euler}
e^{g_{\vecmu}\phi^{\vecmu}(\spin)} 
    = \cosh(g_{\vecmu}) + \phi^{\vecmu}(\spin)\sinh(g_{\vecmu})\,,
    \qquad {\rm for} \; \phi^{\vecmu}(\spin)\in\{-1;+1\}\,.
\end{align}
For larger values of $q$, %In the more general case
the monomials $\phi^{\vecmu}(\state)$ are the $q$-th roots of unity:
\begin{align}
\phi^{\vecmu}(\state) 
    &= e^{\frac{2i\pi}{q}\state\cdot\vecmu}\,\qquad \forall \vecmu\in(\mathbb{Z}/q\mathbb{Z})^n\,,\;\;\forall \state\in(\mathbb{Z}/q\mathbb{Z})^n \\
\phi^{\vecmu}(\state) 
    &\in \{e^{\frac{2i\pi}{q}k}\,|\,k\in[\![0,\,q-1]\!]\}=\sqrt[q]{1}\,.
    %\lceil N\rfloor
\end{align}
One can use the generalized version of Euler's formula~\cite{muldoon1996beyond} (see proof below):
\begin{align}\label{eq:Euler:general}
\boxed{
\qquad
    e^{g_{\vecmu}\phi^{\vecmu}(\state)} 
    = 
    \sum_{r=0}^{q-1} c_{r}(g_{\vecmu}) \,
    \left[\phi^{\vecmu}(\state)\right]^{r}\,,
    \qquad {\rm where}\;\;
    c_r(x) 
    =\sum_{k\geq 0}\frac{x^{kq+r}}{(kq+r)!}\,.\qquad
}
\end{align}
The functions $c_r$ 
%\begin{align}
%c_i(x) 
%    = F_{q,i}^{1}(x) 
%    =\sum_{k\geq 0}\frac{x^{kq+i}}{(kq+i)!} 
%\end{align}
are the generalized hyperbolic functions of order $q$: $c_r(x) = F_{q,r}^{1}(x)$~\cite{muldoon1996beyond}. %(see also \href{https://mathworld.wolfram.com/GeneralizedHyperbolicFunctions.html}{here}). 
For $q=2$, one has respectively for $r=0$ and $r=1$:
\begin{align}
    c_0(x)=F_{2,0}^{1}(x) = \cosh(x)
    \qquad {\rm and}\qquad
    c_1(x)=F_{2,1}^{1}(x) = \sinh(x)\,,
\end{align}
which gives back Eq.~\eqref{eq:Euler}.
For $q=3$, one has:
\begin{align}
e^{g_{\vecmu}\phi^{\vecmu}(\state)} 
    = 
    F_{3,0}^{1}(g_{\vecmu})
    + F_{3,1}^{1}(g_{\vecmu})\,
    \phi^{\vecmu}(\state)
    +F_{3,2}^{1}(g_{\vecmu})
    \left[\phi^{\vecmu}(\state)\right]^{2}\,,
\end{align}
where
\begin{align}\label{eq:Euler:coeff:q=3}
&c_0(x)=F_{3,0}^{1}(x) 
    =\frac{1}{3}
        \left[e^x+2\,e^{-x/2}\cos\left(\frac{\sqrt{3}}{2}\,x\right)\right]\nonumber\\
&c_1(x)=F_{3,1}^{1}(x)
    =\frac{1}{3}
        \left[e^x+2\,e^{-x/2}\cos\left(\frac{\sqrt{3}}{2}\,x+\frac{1}{3}\pi\right)\right]\\
&c_2(x)=F_{3,2}^{1}(x)
    =\frac{1}{3}
        \left[e^x+2\,e^{-x/2}\cos\left(\frac{\sqrt{3}}{2}\,x-\frac{1}{3}\pi\right)\right]\nonumber
\end{align}

\begin{proof}[Proof of Eq.~\eqref{eq:Euler:general}]
%Using 
We start from the power series expansion of the exponential function:
\begin{align}
e^{g_{\vecmu}\phi^{\vecmu}(\state)} 
    = \sum_{j\geq 0} \frac{\left[g_{\vecmu}\phi^{\vecmu}(\state)\right]^j}{j!}
    = \sum_{j\geq 0} \,\frac{g_{\vecmu}^{\,j}}{j!}\,\left[\phi^{\vecmu}(\state)\right]^{j\,{\rm mod}\,q}\,,
\end{align}
where we used that $[\phi^{\vecmu}(\state)]^q=1$ ($j\,{\rm mod}\,q$ denotes $j$ modulo $q$). 
%Factorizing 
We then group all the terms that have the same power $r$ %$r = j\,{\rm mod}\,q$
of $\phi^{\vecmu}$ (for which $j$ is of the form $j = kq+r$ for all $k\geq 0$): %, one gets: %grouping all the terms of the form:
\begin{align}
e^{g_{\vecmu}\phi^{\vecmu}(\state)} 
    &= \sum_{r=0}^{q-1}\; \sum_{k\geq 0} \;\frac{g_{\vecmu}^{\,kq+r}}{(kq+r)!}\,\left[\phi^{\vecmu}(\state)\right]^{r}\,,
\end{align}
which gives Eq.~\eqref{eq:Euler:general}.
\end{proof}

\subsection{Euler formula, special case $\vecmu\in\M_0$}
\label{app:Euler:M0}
\begin{proof}[Proof of Eq.~\eqref{eq:Euler:general:M0:maintext}]
From the definition~\eqref{eq:M0:def}, all the elements $\vecmu\in\M_0$ verify that $2\,\vecmu=\veczero$ in $\Field^n$ and, as a consequence, that $\left[\phi^{\vecmu}(\state)\right]^2=1$ for all state $\state\in\Field^n$. % (because each entree $\mu_i$ satisfies $\mu_i=(-\mu_i\bmod{q})$, which means that $2\mu_i=(0\bmod{q})$).
This means that for $\vecmu\in\M_0$ one can write %the term 
$\exp(g_{\vecmu}\phi^{\vecmu}(\state))$ as a sum over two terms only after grouping, respectively, odd and even powers of $\phi^{\vecmu}$ in the generalized Euler formula~\eqref{eq:Euler:general}. One then simply recovers the original Euler formula~\eqref{eq:Euler}:
\begin{align}
    \label{eq:Euler:general:M0}
    &\forall \vecmu\in\M_0, \qquad
        e^{g_{\vecmu}\phi^{\vecmu}(\state)}
        = A_{0}(g_{\vecmu}) + A_{1}(g_{\vecmu})\; \phi^{\vecmu}(\state)\,,\\
    \label{eq:Euler:general:M0:A}
    &{\rm where}\;\;\; 
        %A_{0}(g_{\vecmu}) = \sum_{k=0}^{\frac{q}{2}-1} c_{2k} (g_{\vecmu})
        A_{0}(x) = \sum_{k=0}^{\frac{q}{2}-1} c_{2k} (x) = \cosh(x)
        \;\;\, {\rm and}\;\;\, 
        %A_{1}(g_{\vecmu}) = \sum_{k=0}^{\frac{q}{2}-1} c_{2k+1} (g_{\vecmu})\,.
        A_{1}(x) = \sum_{k=0}^{\frac{q}{2}-1} c_{2k+1} (x) = \sinh(x)\,.
\end{align}
We recall that $q$ is even (and thus $q/2$ is an integer), because $\M_0=\emptyset$ for odd values of $q$.
This result can also be obtained directly by observing that for all $\vecmu\in\M_0$, $\phi^{\vecmu}(\state)=\pm 1$, which immediately gives Eq.~\eqref{eq:Euler}. 
It is however interesting to recover it from the general formula by grouping odd and even powers of $\phi^{\vecmu}$, because a similar reasoning can also be used for other prime factors of $q$ when $q$ is not prime (as in Sec.~\ref{sec:cluster:q-non-prime}). 
\end{proof}

\subsection{Euler-like formula for complex-conjugate terms $(\vecmu,-\vecmu)$} 
%\textcolor{red}{Contributions to $Z$ from complex-conjugate terms}
\label{app:Euler:M1}
\begin{proof}[Proof of Eq.~\eqref{eq:Euler:general:M1}]
One can expand each exponential in the following equation using the generalized Euler formula~\eqref{eq:Euler:general}:
\begin{adjustwidth}{-1cm}{-1cm}
\begin{align}
    e^{\;g_{\vecmu}\phi^{\vecmu}(\state)}\,e^{g_{\vecmu}^*\phi^{-\vecmu}(\state)}
    &=\sum_{m=0}^{q-1} \sum_{k=0}^{q-1} 
        c_{m}(g_{\vecmu})c_{k}(g_{\vecmu}^*)\left[\phi^{\vecmu}(\state)\right]^{m-k}\nonumber\\
    &= \underbrace{\sum_{m=0}^{q-1} c_{m}(g_{\vecmu})c_{m}(g_{\vecmu}^*)}_{k=m}
    + \underbrace{\sum_{m=0}^{q-1}\sum_{k=0}^{m-1} c_{m}(g_{\vecmu})c_{k}(g_{\vecmu}^*)\left[\phi^{\vecmu}(\state)\right]^{m-k}}_{k<m}
    + \underbrace{\sum_{m=0}^{q-1}\sum_{k=m+1}^{q-1} c_{m}(g_{\vecmu})c_{k}(g_{\vecmu}^*)\left[\phi^{\vecmu}(\state)\right]^{q+m-k}}_{k>m}\nonumber
\end{align}
\end{adjustwidth}
in which, for the case $k>m$, we used that $[\phi^{\vecmu}(\state)]^q=1$ for all $\state,\vecmu$. For the double sum with $k<m$, %in the case $k<m$, 
the exponent $(m-k)$ is in the interval $[1,q-1]$ for all the terms in the sum. In the following, we make the change of variables $r=(m-k)$ replacing $k$ by $r$. 
For the double sum with $k>m$, %in the case $k>m$, 
the exponent $(m-k)$ is in the interval $[-(q-1), -1]$ for all the terms in the sum, and, %with the addition of $q$, 
by adding $q$, we get that the exponent $(q+m-k)\in[1,q-1]$. % we recover an exponent between $1$ and $q-1$.
In the following, we make the change of variables $r=(q+m-k)$ replacing $k$ by $r$. 
\begin{adjustwidth}{-1cm}{-1cm}
\begin{align}
    e^{\;g_{\vecmu}\phi^{\vecmu}(\state)+g_{\vecmu}^*\phi^{-\vecmu}(\state)}
    = \underbrace{\sum_{m=0}^{q-1} c_{m}(g_{\vecmu})c_{m}(g_{\vecmu}^*)}_{k=m}
    + \underbrace{\sum_{m=0}^{q-1}\sum_{r=1}^{m} c_{m}(g_{\vecmu})c_{m-r}(g_{\vecmu}^*)\left[\phi^{\vecmu}(\state)\right]^{r}}_{k<m}
    + \underbrace{\sum_{m=0}^{q-1}\sum_{r=m+1}^{q-1} c_{m}(g_{\vecmu})c_{q+m-r}(g_{\vecmu}^*)\left[\phi^{\vecmu}(\state)\right]^{r}}_{k>m}\nonumber
\end{align}
\end{adjustwidth}
Note that the coefficient in front of $[\phi^{\vecmu}]^r$ in each term can be written under the form: $c_{m}(g_{\vecmu})c_{(m-r)\bmod{q}}(g_{\vecmu}^*)$.
Indeed, in the first double sum, the values taken by $m$ and $r$ are such that $(m-r)\in[0,q-1]$, and thus $(m-r)\bmod{q}=(m-r)$. In the second double sum, the values taken by $m$ and $r$ are such that $(m-r)\in[-(q-1),-1]$, and thus $(m-r)\bmod{q}=q+(m-r)$. This gives:
\begin{align}
    e^{\;g_{\vecmu}\phi^{\vecmu}(\state)+g_{\vecmu}^*\phi^{-\vecmu}(\state)}
    =
    \sum_{m=0}^{q-1}\sum_{r=0}^{q-1} c_{m}(g_{\vecmu})c_{(m-r)\bmod{q}}(g_{\vecmu}^*)\left[\phi^{\vecmu}(\state)\right]^{r}\,,
\end{align}
which finally gives Eq.~\eqref{eq:Euler:general:M1} after switching the order of the two sums.
\end{proof}

\subsection{Loop expansion of the partition function} %{Expansion over loops of the model}
\label{app:Z:loops:V1}
\begin{proof}[Proof of Property~\ref{ppty:Z:loop-expansion}]
Consider a model $\M\subseteq (\mathbb{Z}/q\mathbb{Z})^n$. We denote by $K=|\M|$ the number of operators in the model and arbitrarily index the operators from $1$ to $K$: $\M = \{\vecmu_1,\,\cdots,\,\vecmu_K\}$.
Using the generalized Euler formula~\eqref{eq:Euler:general} in the expression of the partition function~\eqref{eq:Z:def}:
\begin{align}%\label{eq:Z:def}
    Z_{\M}(\vecg) 
    &= \sum_{\state} \prod_{\vecmu\in\M} 
    \left(
    \sum_{r=0}^{q-1} c_{r}(g_{\vecmu}) \,
    \left[\phi^{\vecmu}(\state)\right]^{r}
    \right)\,,
    \qquad {\rm where}\;\;
    c_r(x) 
    =\sum_{k\geq 0}\frac{x^{kq+r}}{(kq+r)!}\,,\\
    &=\sum_{\state} \sum_{\vecr\in(\mathbb{Z}/q\mathbb{Z})^K}
    \prod_{k=1}^K \big[c_{r_k}(g_{\vecmu_k}) \,
    \left[\phi^{\vecmu_k}(\state)\right]^{r_k}\big]\,,
    %\,,
\end{align}
in which $r_k$ denotes the $k$-th element of the vector $\vecr$. Using that $[\phi^{\vecmu}]^r(\state) = \phi^{r\vecmu}(\state)$ and that $\phi^{\vecmu_i}(\state)\phi^{\vecmu_j}(\state) = \phi^{\vecmu_i + \vecmu_j}(\state)$, where the sum and product in the exponents are taken over $(\mathbb{Z}/q\mathbb{Z})^n$, one gets:
\begin{align}%\label{eq:Z:def}
    Z_{\M}(\vecg) 
    &=\sum_{\vecr\in(\mathbb{Z}/q\mathbb{Z})^K}
    \left(\prod_{k=1}^K c_{r_k}(g_{\vecmu_k}) \right)\,
    \sum_{\state} 
    \phi^{\sum_{k=1}^K r_k\vecmu_k}(\state)\,,
    %\,,
\end{align}
Observing that, due to Property~\eqref{ppty:sum:phi}, %\textcolor{blue}{(see property)}:
\begin{align}
    \sum_{\state} \phi^{\sum_{k=1}^{K} r_k\vecmu_k}(\state)
    = 
    \begin{cases}
        \;\; q^n\,,\qquad &{\rm if } \;\;\sum_{k=1}^{K} r_k\vecmu_k = \veczero\qquad\textrm{in }\Field^n\\
        \;\; 0\,, \qquad &{\rm else}\,,
    \end{cases}
\end{align}
we define the loops of $\M$ as the vectors $\vecr=(r_1,\,\dots,\,r_K)$ for which this sum is non-zero. The set of loops of a model $\M$ is then: %as the set of vectors $\vecr$ for which this sum is non-zero:
\begin{align}\label{eq:loop:V2:app}
\boxed{
    \qquad
    \mathcal{L}_q(\M) = \bigg\{\vecell\in (\mathbb{Z}/q\mathbb{Z})^K\;\Big|\; \sum_{k=1}^K \ell_k\,\vecmu_k = \veczero\bigg\}\;,
    \qquad}
\end{align}
and, the partition function can be computed as a sum over the loops of the model as:
\begin{align}\label{eq:Z:loop-expansion:V2:app}
    \boxed{
    \qquad
Z_{\M}(\vecg) 
    =q^n\,
    \sum_{\vecell\in\mathcal{L}_q(\mathcal{M})}
    \prod_{k=1}^K c_{\ell_k}(g_{\vecmu_k})\,,
    \qquad {\rm where}\;\;
    c_{\ell}(x) 
    =\sum_{k\geq 0}\frac{x^{kq+\ell}}{(kq+\ell)!}\,.
    \qquad}
\end{align}
\end{proof}

\subsection{Loop expansion without trivial loops from c.c.~operators}
%{Cluster expansion of the partition function}
\label{app:ClusterExpansion}
\begin{proof}[Proof of Property~\ref{ppty:LoopExpansion:R-Model}]
Consider a spin model $\M\subseteq (\mathbb{Z}/q\mathbb{Z})^n$, which we write under the form  $\M=\M_0\cup\M_1\cup\M_1^*$ as described in Definition~\ref{def:M:M0-M1}. We denote $K_0 = |\M_0|$ and $K_1 = |\M_1|$ the number of operators respectively in $\M_0$ and in $\M_1$ and arbitrarily index the operators: % of the two sets: 
\begin{align}
    \M_0 = \{\vecmu_{0, 1}, \cdots,\,\vecmu_{0, K_0}\}
    \qquad {\rm and} \qquad
    \M_1 = \{\vecmu_{1, 1}, \cdots,\,\vecmu_{1, K_1}\}\,.
\end{align} 
Let's replace Eq.~\eqref{eq:Euler:general:M1} and Eq.~\eqref{eq:Euler:general:M0:maintext} %\eqref{eq:Euler:general:M0} 
in the expression of $Z_{\M}$ from Eq.~\eqref{eq:Z:def:2}:
%\begin{adjustwidth}{-1cm}{0cm}
\begin{align}
Z_{\M}(\vecg) \label{app:eq:Z:separation-M0-M1-Euler}
    &= \sum_{\state} 
    \prod_{\vecmu_0\in\M_0} \Big[A_{0}(g_{\vecmu_0}) + A_{1}(g_{\vecmu_0})\; \phi^{\vecmu_0}(\state)\Big]
    \prod_{\vecmu_1\in\M_1} 
    \left[\,\sum_{r=0}^{q-1} B_{r}(g_{\vecmu_1}) \left[\phi^{\vecmu_1}(\state)\right]^r \,\right]\\
    &= \sum_{\state} 
    \Bigg[
    \sum_{\vecr_0\in (\mathbb{Z}/2\mathbb{Z})^{K_0}}
    \prod_{i=1}^{K_0} 
    \left(
    A_{r_{0,i}}(g_{\vecmu_{0,i}}) \,[\phi^{\vecmu_{0,i}}(\state)]^{r_{0,i}}
    \right)
    \Bigg]
    \Bigg[
    \sum_{\vecr_1\in (\mathbb{Z}/q\mathbb{Z})^{K_1}}
    \prod_{j=1}^{K_1} 
    \left(
    B_{r_{1,j}}(g_{\vecmu_{1,j}}) \,[\phi^{\vecmu_{1,j}}(\state)]^{r_{1,j}}
    \right)
    \Bigg]\,,
\end{align}
%\end{adjustwidth}
where $r_{0,i}$ (resp. $r_{1,i}$) denotes the $i$-th element of the vector $\vecr_0$ (resp. $\vecr_1$). %Here, we used the fact that the products of a sum of elements is equal to the sum of sub-products between the terms that 
We use that $[\phi^{\vecmu}]^r(\state) = \phi^{r\vecmu}(\state)$ and that $\phi^{\vecmu_i}(\state)\phi^{\vecmu_j}(\state) = \phi^{\vecmu_i + \vecmu_j}(\state)$, where the operations in the exponents are taken over $(\mathbb{Z}/2\mathbb{Z})^n$ if $\vecmu\in\M_0$ and over $(\mathbb{Z}/q\mathbb{Z})^n$ if $\vecmu\in\M_1$. This leads to:
%\begin{adjustwidth}{-1cm}{0cm}
\begin{align}
Z_{\M}(\vecg) 
    &=  
    \sum_{\vecr_0\in (\mathbb{Z}/2\mathbb{Z})^{K_0}}
    \sum_{\vecr_1\in (\mathbb{Z}/q\mathbb{Z})^{K_1}}
    \left[
    \prod_{i=1}^{K_0} A_{r_{0,i}}(g_{\vecmu_{0,i}}) 
    \prod_{j=1}^{K_1} B_{r_{1,j}}(g_{\vecmu_{1,j}})
    \right]
    \sum_{\state}
    \phi^{
    \sum_{i=0}^{K_0}r_{0,i}\vecmu_{0,i}+
    \sum_{j=0}^{K_1}r_{1,j}\vecmu_{1,j}
    }(\state)\,.
\end{align}
%\end{adjustwidth}
Using that (see Property~\ref{ppty:sum:phi}):
%\begin{adjustwidth}{-\extralength}{0cm}
\begin{align}
    \sum_{\state} \phi^{\sum_{i=0}^{K_0}r_{0,i}\vecmu_{0,i}+
    \sum_{j=0}^{K_1}r_{1,j}\vecmu_{1,j}}(\state)
    = 
    \begin{cases}
        \;\; q^n\,,\qquad &{\rm if } \;\;
        \sum_{i=0}^{K_0}r_{0,i}\vecmu_{0,i}+
    \sum_{j=0}^{K_1}r_{1,j}\vecmu_{1,j} = \veczero\qquad\textrm{in }\Field^n\\
        \;\; 0\,, \qquad &{\rm else.}
    \end{cases}
\end{align}
%\end{adjustwidth}
we defined the set of (non-trivial) loops of a model $\M=\M_0\cup\M_1\cup\M_1^*$ as:
%\begin{adjustwidth}{-\extralength}{0cm}
\begin{align}\label{eq:loop:V3:app}
\boxed{
    \qquad
    \mathcal{L}_q^{cc}(\M_0,\M_1) = \bigg\{(\vecr_0,\vecr_1)\in (\mathbb{Z}/2\mathbb{Z})^{K_0}\times(\mathbb{Z}/q\mathbb{Z})^{K_1}\;\Big|\; \sum_{i=1}^{K_0} r_{0,i}\,\vecmu_{0,i} + \sum_{j=1}^{K_1} r_{1,j}\,\vecmu_{1,j} = \veczero\bigg\}\;.
    \qquad}
\end{align}
%\end{adjustwidth}
Finally, the partition can be computed as a sum over these loops, % non-CC loops of the model, 
as:
\begin{align}\label{eq:Z:loop-expansion:V3:app}
    &\boxed{
    \qquad
Z_{\M}(\vecg) 
    =q^n\,
    \sum_{(\vecr_0, \vecr_1)\in\mathcal{L}_q^{cc}(\mathcal{M}_0,\mathcal{M}_1)}
    \prod_{i=1}^{K_0} A_{r_{0,i}}(g_{\vecmu_{0,i}}) 
    \prod_{j=1}^{K_1} B_{r_{1,j}}(g_{\vecmu_{1,j}})\,,
    \qquad}\\
&{\rm where}\qquad 
        A_{0}(g_{\vecmu}) = \sum_{k=0}^{\frac{q}{2}-1} c_{2k} (g_{\vecmu})
        \qquad {\rm and}\qquad 
        A_{1}(g_{\vecmu}) = \sum_{k=0}^{\frac{q}{2}-1} c_{2k+1} (g_{\vecmu})\,,\\
&%{\rm and}\qquad 
    B_{r}(g_{\vecmu}) = \sum_{m=0}^{q-1} c_m(g_{\vecmu})\,c_{(m-r)\bmod{q}}(g^*_{\vecmu})\,,
\qquad{\rm and}\qquad
    c_r(x) 
    =\sum_{k\geq 0}\frac{x^{kq+r}}{(kq+r)!}\,.
\end{align}
\end{proof}
For systems with $q$ prime or $q = 2\,p$ where $p$ is prime, the contribution from the independent part of a model $\M$ (strong and weak) can be directly identified from this last expression, as the term in the sum for which %$\ell_0=\veczero$ and $\ell_1=\veczero$:
$(\vecell_0, \vecell_1)=(\veczero, \veczero)$:
\begin{align}
%\;{\rm For}\;q\;{\rm prime},\;{\rm or}&\;q=2p\;{\rm with}\;p\;{\rm prime}:\nonumber\\[2pt]
Z_{\M_{ind}}(\vecg) \label{app:eq:Z:loop-expansion:V3:ind}
    &=q^n\,
    \prod_{\vecmu_0\in\M_0} A_{0}(g_{\vecmu_{0}}) \,
    \prod_{\vecmu_1\in\M_1} B_{0}(g_{\vecmu_{1}})\,,\\
&{\rm where}\; A_0(x) = \cosh(x)
\qquad{\rm and}\qquad
B_0(x) = \sum_{m=0}^{q-1} c_{m}(x)\,c_{m}(x^*)\,.
\end{align}
Note that $c_{m}(x^*) = [c_{m}(x)]^*$ and thus $c_{m}(x)\,c_{m}(x^*) = \lVert c_{m}(x) \rVert^2$.

%    \qquad {\rm where}\;\;
%    c_r(x) 
%    =\sum_{k\geq 0}\frac{x^{kq+r}}{(kq+r)!}\,.

\section{Gauge transformations}
\subsection{Counting gauge transformations}
\label{app:GT:counting}
We give the main lines of the proof of Eq.~\eqref{eq:nb_GT:q-non-prime}.
The group of gauge transformations $\mathcal{G}_q(n)$ is given by the general linear group $GL(\Field^n)$. We can compute the number of gauge transformations using the Chinese Remainder Theorem. Let the prime factorization of $q$ be $q = p_1^{a_1} p_2^{a_2} \cdots p_k^{a_k}$. The ring $\mathbb{Z}/q\mathbb{Z}$ is then isomorphic to the product of rings $\prod_{i=1}^k \mathbb{Z}/p_i^{a_i}\mathbb{Z}$. This induces a group isomorphism:
\begin{align}
    GL(\Field^n) \cong \prod_{i=1}^k GL((\mathbb{Z}/p_i^{a_i}\mathbb{Z})^n)\,.
\end{align}
It follows that
\begin{align}
    |GL(\Field^n)| = \prod_{i=1}^k |GL((\mathbb{Z}/p_i^{a_i}\mathbb{Z})^n)|\,.
\end{align}
It remains to determine the order of $GL((\mathbb{Z}/p^a\mathbb{Z})^n)$ for a prime $p$ and $a \in \mathbb{N}$.

Consider the canonical ring homomorphism $\rho: GL((\mathbb{Z}/p^a\mathbb{Z})^n) \to GL((\mathbb{Z}/p\mathbb{Z})^n)$ which reduces each matrix entry modulo $p$. This map is a surjective group homomorphism. By the first isomorphism theorem for groups, we have:
\begin{align}
    |GL((\mathbb{Z}/p^a\mathbb{Z})^n)| = |\ker(\rho)| \cdot |GL((\mathbb{Z}/p\mathbb{Z})^n)|\,.
\end{align}
The kernel, $\ker(\rho)$, consists of matrices $M \in GL((\mathbb{Z}/p^a\mathbb{Z})^n)$ such that $M \equiv I \pmod p$, where $I$ is the identity matrix. Any such matrix can be written in the form $M = I + pA$, where $A$ is an $n \times n$ matrix with entries in $\mathbb{Z}/p^{a-1}\mathbb{Z}$. There are $(p^{a-1})^{n^2}$ possible matrices for $A$. Therefore, the order of the kernel is:
\begin{align}
    |\ker(\rho)| = (p^{a-1})^{n^2} = p^{n^2(a-1)}\,.
\end{align}
The remaining term, $|GL_n(\mathbb{Z}/p\mathbb{Z})|$, is the order of the general linear group over the finite field $\mathbb{F}_p$. This is a standard result obtained by counting the ways to choose $n$ linearly independent column vectors from $\mathbb{F}_p^n$. The formula is:
\begin{align}
    |GL(\mathbb{F}_p^n)| = (p^n - 1)(p^n - p)\cdots(p^n - p^{n-1}) = \prod_{k=0}^{n-1} (p^n - p^k)\,.
\end{align}
Combining these results, the order for the prime power case is:
\begin{align}
    |GL_n(\mathbb{Z}/p^a\mathbb{Z})| = p^{n^2(a-1)} \prod_{k=0}^{n-1} (p^n - p^k)\,.
\end{align}
The total number of gauge transformations for an arbitrary $q$ is found by substituting this into the product over the prime factors of $q$.

\subsection{Gauge transformation of a spin model}
\label{App:GT}
Consider a spin model $\M$ with parameters $\vecg$.
Under a gauge transformation $\mathcal{T}_{\basis}$, the energy of a state $\spin$ in the model can be rewritten in terms of the new basis variables~$\spin'$, by using the inverse GT $\spin=\basis^{-1}(\spin')$:
%To transform the spin model $\M$ Transforming a spin model with a gauge transformation $\mathcal{T}_{\basis}$ consists in re-writing the Hamiltonian (energy landscape) of the model in the new basis $\spin'=\basis(\spin)$, 
\begin{align}\label{app:eq:GT:model}
        H(\spin\mid \vecg,\,\M) 
        &= \sum_{\vecmu\in\M} g_{\vecmu}\,\phi^{\vecmu}(\spin)
        = \sum_{\vecmu\in\M} g_{\vecmu}\,\phi^{\vecmu}(\basis^{-1}(\spin')) %\,,\\&
        = \sum_{\vecmu\in\M} g_{\vecmu}\,\mathcal{T}_{\basis}[\phi^{\vecmu}](\spin')\,,
\end{align}
where we used the definition of gauge transformed operators~\eqref{def:eq:GT:Op} in the last equality.
Because the transformation of a spin operator $\mathcal{T}_{\basis}[\phi^{\vecmu}]$ is still a spin operator (see Definition~\ref{def:GT:Op}), the last expression in Eq.~\eqref{app:eq:GT:model} corresponds to the Hamiltonian of a spin model with operators $\mathcal{T}_{\basis}[\phi^{\vecmu}](\spin')=\phi^{\vecmu'}(\spin')$ where $\vecmu' = \Tmat^{-1}\vecmu$.
The energy of a state $\spin$ under a model $\M$ is thus mapped by a GT to the energy of a new state $\spin'=\basis(\spin)$ of a new spin model $\M'$: 
\begin{align}
        H(\spin\mid \vecg,\,\M) 
        &= \sum_{\vecmu'\in\M'} g_{\vecmu'}\,\phi^{\vecmu'}(\spin')
        =H(\spin'\mid \vecg',\,\M')\,,
\end{align}
where $\M'=\mathcal{T}_b[\M]=\{\vecmu'=\Tmat^{-1}\vecmu\mid\vecmu\in\M\}$, and % the vector of parameters
$\vecg'=\mathcal{T}_b[\vecg]$ is a permutation of the parameters in $\vecg$ that is such that $g_{\vecmu'}=g_{\vecmu}$ for $\vecmu'=\Tmat^{-1}\vecmu$.

\begin{proof}[Proof: models related by GTs have the same partition function (for Sec.~\ref{sec:Ppty:EquivalentClasses}).]
    Consider a $q$-state spin model $\M=\{\phi^{\vecmu_1},\dots,\phi^{\vecmu_K}\}$ and a GT of this model, $\M'=\T_{\basis}[\M]=\{\phi^{\vecmu'_1},\dots,\phi^{\vecmu'_K}\}$, with $\phi^{\vecmu'_k}=\T_{\basis}[\phi^{\vecmu_k}]$ for all $k\in[1,K]$. 
    We labeled the operators in the same order in both models such that 
    %We labeled the operators in both models such that $\phi^{\vecmu'_k}=\T_{\basis}[\phi^{\vecmu_k}]$. Using this labeling, 
    the $k$-th parameter of $\M'$ is equal to the $k$-th parameter of $\M$: $g_{\vecmu'_k}=g_{\vecmu_k}$\,, where $\vecmu_k'=T_{\basis}^{-1}\vecmu_k$ are related by the GT.\\
    The partition function of $\M'$ is given by:
    \begin{align}
        Z_{\M'}(\vecg')=\sum_{\spin'\in\Omega} 
            \exp\left({\;\displaystyle \sum_{k=1}^K g_{\vecmu_k'} \phi^{\vecmu_k'}(\spin')}\right)\,,
    \end{align}
    where $\Omega$ denotes the set of all possible $q^n$ spin states and 
    $\vecg'=(g_{\vecmu'_1},\dots,g_{\vecmu'_K})$ is the vector of parameters of $\M'$. %Let's use the definition of gauge transform operator
    The gauge transformation $\T_{\basis}$ corresponds to the change of spin basis $\spin'=\basis(\spin)$. The gauge transform operators are then related to the original operators by %is defined as~\eqref{}: 
    $\phi^{\vecmu'}(\spin')=\T_{\basis}[\phi^{\vecmu}](\spin')=\phi^{\vecmu}[\basis^{-1}(\spin')]$ (see Def.~\ref{def:GT:Op}), which gives:
    \begin{align}
        Z_{\M'}(\vecg')=\sum_{\spin'\in\Omega} 
            \exp\left({\;\displaystyle \sum_{k=1}^K g_{\vecmu_k'} \phi^{\vecmu_k}[\basis^{-1}(\spin')]}\right)\,.
    \end{align}
%    Here we use the bijection $\spin=\basis^{-1}(\spin')$
%    Using the GT $\spin=\basis^{-1}(\spin')$, we can rewrite the sum in terms of the variables $\spin$
    %We use the GT $\spin=\basis^{-1}(\spin')$ to re-write the first sum as:
    We use the fact that a GT %$\spin=\basis^{-1}(\spin')$
    is an automorphism of the state space $\Omega$ (bijection between states in $\Omega$) % is a bijective transformation
    to perform the change of variable $\spin=\basis^{-1}(\spin')$ in the first sum and re-write it as:
    \begin{align}
        Z_{\M'}(\vecg')=\sum_{\spin\in\Omega} 
            \exp\left({\;\displaystyle \sum_{k=1}^K g_{\vecmu_k'} \phi^{\vecmu_k}(\spin)}\right)
            =\sum_{\spin\in\Omega} 
            \exp\left({\;\displaystyle \sum_{k=1}^K g_{\vecmu_k} \phi^{\vecmu_k}(\spin)}\right) 
        = Z_{\M}(\vecg)\,.
    \end{align}
    Here, in the second equality, we used that $k$-th parameter of $\M$ is equal to the $k$-th parameter of $\M'$, $g_{\vecmu'_k}=g_{\vecmu_k}$\,, following the labeling choice made above.
%    Here we used that the $k$-th parameter of $\M$ is equal to the $k$-th parameter of $\M'$ in the labeling choice made above. 
%    Using that $g_{\vecmu_k}=g_{\vecmu_k}'$, as mentioned previously.
\end{proof}

\subsection{Real-valued spin models and reduced model}
\label{App:M_r}

\begin{proof}[Proof of Property~\ref{ppty:GT:cc-op}]
Consider two c.c.~operators $(\vecmu,-\vecmu)\in[\Field^n]^2$. We recall that the notation $-\vecmu$ was defined in Sec~\ref{Sec:1.1} as
$-\vecmu = (q-\mu_1, \dots, q-\mu_n)$. %Element-wise, the $j$-the elements of $\vecmu'$ and $\vecmu$ are related by: $\mu'_j = (q-\mu_j)$.
Consider a gauge transformation $\mathcal{T}$ of these operators with matrix representation~$\Tmat$. We denoted by $t_{ij}$ the elements of the inverse matrix $\Tmat^{-1}$. We then have that:
\begin{align}
    \mathcal{T}[-\vecmu] = \Tmat^{-1}(-\vecmu)
    =\bigg(\Big( \sum_{j=1}^n t_{ij} \,(q-\mu_j)\Big) {\rm mod}\,q\bigg)_i
    = -\Tmat^{-1}\vecmu= -\mathcal{T}[\vecmu]\,,\nonumber
\end{align}
where were we used that $(\sum_{j=1}^n t_{ij} \,(q-\mu_j)) ({\rm mod}\,q) = -(\sum_{j=1}^n t_{ij}\,\mu_j)({\rm mod}\,q)$.
\end{proof}

For the two proofs below, we used the decomposition of Definition~\ref{def:M:M0-M1} for both the spin model $\M$ and its transformed model $\M'=\mathcal{T}[\M]$ under a GT $\T$, i.e. we decompose the models as $\M=\M_0\cup\M_1\cup\M_1^*$ and $\M'=\M_0'\cup\M_1'\cup\M_1'^*$. We show that $\T[\M_0]=\M_0'$ and that $\T[\M_1\cup\M_1^*]=\M_1'\cup\M_1'^*$.

\begin{proof}[Proof that:] {\bf for all $\vecmu\in\M_0$, \;$\T[\vecmu]\in\M_0'$\;.}
%\noindent {\bf Proof (for App):} %We recall that under a GT $T[-\vecmu]=-T[\vecmu]$ (ppty). 
%{\bf Proof that for all $\vecmu\in\M_0$, $\T[\vecmu]\in\M_0'$.}
For all $\vecmu\in\M_0$, $-\vecmu=\vecmu$. This implies that $\T[-\vecmu]=\T[\vecmu]=\vecmu'$. Using that under a GT $\T[-\vecmu]=-\T[\vecmu]=-\vecmu'$ (see Property~\ref{ppty:GT:cc-op}), one gets that %$T[\vecmu]=-T[\vecmu]$.
the operator $\vecmu'$ verifies $\vecmu'=-\vecmu'$, and thus is an element of $\M_0'$.
\end{proof}

\begin{proof}[Proof by contradiction that:] {\bf for all $\vecmu_1\in\M_1\cup\M_1^*$,\; $\T[\vecmu_1]\in\M_1'\cup\M_1'^*$\,.} %by contradiction for $\M_1$. 
Take an operator $\vecmu\in\M_1\cup\M_1^*$, by definition $-\vecmu\neq\vecmu$ in $\Field^n$.
Suppose that its transformation $\vecmu'=\T[\vecmu]$ is an element of $\M_0'$. By definition of $\M_0'$, the operator then verifies that $-\vecmu'=\vecmu'$ in $\Field^n$. Using the inverse GT on both side of this equality we get that: $\T^{-1}[-\vecmu']=\T^{-1}[\vecmu']$, where $\T^{-1}[\vecmu']=\vecmu$ and $\T^{-1}[-\vecmu']=-\T^{-1}[\vecmu']$ as $\T^{-1}$ is a GT (see Property~\ref{ppty:GT:cc-op}). We finally get that $-\vecmu=\vecmu$, which is in contradiction with the fact that $\vecmu\in\M_1\cup\M_1^*$. This means that $\vecmu'=\T[\vecmu]$ is not an element of $\M_0'$, and is therefore in $\M_1'\cup\M_1'^*$.
\end{proof}

\begin{proof}[Proof of the reduced loop structure]
    Consider a real-valued spin model $\M$ that can be written under the form $\M=\M_1\cup\M_1^*$ with the decomposition of Definition~\ref{def:M:M0-M1} (i.e., for which $\M_0=\emptyset$).
    This is always the case for a real-valued models with $q$ prime and $q>2$ (for which all the operators are included by c.c.~pairs and verify that $-\vecmu\neq\vecmu$). 
    %For a real-valued model $\M$ with $q>2$ and prime, all the operators are included by c.c.~pairs and verify that $-\vecmu\neq\vecmu$. Thus, using the decomposition in Definition~\ref{def:M:M0-M1}, we can write $\M=\M_1\cup\M_1^*$ (i.e., $\M_0=\emptyset$). 
    The reduced set of loops of $\M$, excluding trivial loops due to c.c.~operators, are given by Eq.~\eqref{eq:loop:V3}:
    \begin{align}\label{eq:loop:V3:app:bis}
    \mathcal{L}^{cc}_{q}(\M) = \bigg\{ \vecell \in (\mathbb{Z}/q\mathbb{Z})^{K_1}\;\Big|\;\sum_{j=1}^{K_1} \ell_{j}\,\vecmu_{j} = \veczero\bigg\}\;,
   \end{align}
   where $K_1=|\M_1|$ is the number of operators in $\M_1$. Using the reduced matrix representation $\Mmat_1 = (\vecmu_1, \dots, \vecmu_{K_1})$ of $\M_1$ and viewing $\vecell$ as a column vector, we can rewrite $\mathcal{L}^{cc}_{q}(\M)$ as follows:
    \begin{align}
    \mathcal{L}^{cc}_{q}(\M) = \bigg\{ \vecell \in (\mathbb{Z}/q\mathbb{Z})^{K_1}\;\Big|\; \Mmat_1 \vecell = \veczero\bigg\} = \ker(\Mmat_1)\;.
   \end{align}
   One could have also directly identified in Eq.~\eqref{eq:loop:V3:app:bis} that $\mathcal{L}^{cc}_{q}(\M)=\mathcal{L}_{q}(\M_1)$, which gives the same result.
\end{proof}

\section{Properties of MCMs}
\label{app:ICC}

The proofs in this section are almost identical to the one derived for the binary case in Ref.~\cite{beretta2018stochastic} (Supplementary Materials) and Ref.~\cite{demulatier2024MCM}, just generalized to the case $q>2$.

\subsection{Matrix representations of the MCMs and GTs in Figure~\ref{fig:MCM:examples}} %Examples of GTs of MCMs
\label{app:GT:MCM:ex}

For the models in Fig.~\ref{fig:MCM:examples}.a, we have that $q=3$, which means that the matrix representations and operations are over $\mathbb{Z}/3\mathbb{Z}$. The matrix representations of $\M_1$ and $\M_1'$ are respectively: % (in $\mathbb{Z}/3\mathbb{Z}$):
\begin{align}
\begingroup
    \setlength{\arraycolsep}{1.4pt}
    \renewcommand{\arraystretch}{1} % Default value: 1
\Mmat_1 = 
\begin{pNiceArray}{ccccc}[first-row,last-col,nullify-dots]
\textcolor{red}{\mu_1} & \textcolor{red}{\mu_2} & \textcolor{red}{\mu_3} & \textcolor{red}{\mu_4} & \textcolor{blue}{\mu_5} &\hspace{-5mm}\\
    \textcolor{red}{1} & \textcolor{red}{1} & \textcolor{red}{2} & \textcolor{red}{0} & \textcolor{blue}{0} & s_1\\ 
    \textcolor{red}{1} & \textcolor{red}{0} & \textcolor{red}{1} & \textcolor{red}{1} & \textcolor{blue}{0} & s_2\\
    \textcolor{red}{0} & \textcolor{red}{2} & \textcolor{red}{2} & \textcolor{red}{1} & \textcolor{blue}{1} & s_3
\end{pNiceArray}
    \qquad{\rm and}\qquad
    %\;\;\;\;\;\;\;\;
\Mmat_1' = 
\begin{pNiceArray}{ccccc}[first-row,last-col,nullify-dots]
\textcolor{red}{\mu_1'} & \textcolor{red}{\mu_2'} & \textcolor{red}{\mu_3'} & \textcolor{red}{\mu_4'} & \textcolor{blue}{\mu_5'} &\\
    \textcolor{red}{1} & \textcolor{red}{0} & \textcolor{red}{1} & \textcolor{red}{1} & 0 & \sigma_1\\ 
    \textcolor{red}{0} & \textcolor{red}{1} & \textcolor{red}{1} & \textcolor{red}{2} & 0 & \sigma_2\\
    0 & 0 & 0 & 0 & \textcolor{blue}{1} & \sigma_3
\end{pNiceArray}
    \;\;.
\endgroup
\nonumber
\end{align}
We recall that the full model is obtained by combining the operators of $\M_1$ with their c.c., i.e. $\M=\M_1\cup\M_1^*$ (idem for $\M'$). %The models $\M$ and $\M'$ are MCMs, and they each have two ICCs. % \textcolor{red}{with respective rank $2$ and $1$}. 
The models $\M$ and $\M'$ are two MCMs that belong to the same equivalence class; they each have two ICCs of respective rank $2$ and $1$.
In the matrices above, we indicated with different colors the operators that belong to the two ICCs. % in $\Mmat$, and to their transformed operators.
We highlighted the fact that the second MCM appears in a preferred basis by coloring only the entries of $\Mmat_1'$ corresponding to the subset of basis variables $\sigma_i$ over which the ICC is based, highlighting the block diagonal form of the matrix.
%the entries of the matrix that correponds to the subset of $\sigma_i$ over which the corresponding ICC is based.
For the gauge transformation, we used the same GT and inverse GT as in Fig.~\ref{fig:GT_and_Loops}.a, which are represented respectively by the matrices $\Tmat$ and $\Tmat^{-1}$ in Eq.~\eqref{eq:exFig:GT:T}. One can easily check that $\Mmat_1'=\Tmat^{-1}\Mmat_1$ in $\mathbb{Z}/3\mathbb{Z}$.

For the models in Fig.~\ref{fig:MCM:examples}.b, we have that $q=4$, which means that the matrix representations and operations are over $\mathbb{Z}/4\mathbb{Z}$. The matrix representations of the models $\M$ and $\M'$ are respectively: % in Fig.~\ref{fig:MCM:examples}.b are: % (in $\mathbb{Z}/4\mathbb{Z}$):
\begin{align}
\begingroup
    \setlength{\arraycolsep}{1.4pt}
    \renewcommand{\arraystretch}{1} % Default value: 1
\Mmat = 
\begin{pNiceArray}{ccc}[first-row,last-col,nullify-dots]
\mu_1 & \mu_2 & \mu_3 &\hspace{-5mm}\\
    1 & 2 & 3 & s_1\\ 
    1 & 2 & 3 & s_2
\end{pNiceArray}
    \qquad{\rm and}\qquad
    %\;\;\;\;\;\;\;\;
\Mmat' = 
\begin{pNiceArray}{ccc}[first-row,last-col,nullify-dots]
\mu_1' & \mu_2' & \mu_3' &\\
    1 & 2 & 3 & \sigma_1\\ 
    0 & 0 & 0 & \sigma_2
\end{pNiceArray}
    \;\;.
\endgroup
\nonumber
\end{align}
The models $\M$ and $\M'$ are two MCMs that are belong to the same equivalence class; they each have a single ICC of rank $1$. One can see in the matrix $\Mmat'$ above that the model $\M'$ is in a preferred basis, as its operators are all based on a single variable (all the entries based on $\sigma_2$ are zeros).
The matrix representations of the GT $\T$ and inverse GT $\T^{-1}$ are respectively:
\begin{align}\label{app:eq:exFig:GT:T}
   \qquad 
    \Tmat = 
    \begin{blockarray}{ccc}
    \sigma_1 & \sigma_2 &\\
    \begin{block}{(cc)c}
    1 & 1 & s_1\\ 
    1 & 0 & s_2\\
    \end{block}
    \end{blockarray}
    \qquad\Leftrightarrow\qquad
  \Tmat^{-1}=
  \begin{blockarray}{ccc}
    s_1 & s_2 &\\
    \begin{block}{(cc)c}
    0 & 1 & \sigma_1\\ 
    1 & 3 & \sigma_2\\
    \end{block}
  \end{blockarray}
  \;\;.
\end{align}
One can easily check that $\Tmat\Tmat^{-1}=\Imat_2$ over $\mathbb{Z}/4\mathbb{Z}$, and that $\Mmat'=\Tmat^{-1}\Mmat$ or equivalently that $\Mmat=\Tmat\Mmat'$.
Formally, the matrices should be read in terms of the color variables $\alpha_i$ and $\alpha_i'$, instead of the spin variables (see Sec.~\ref{Sec:GT:MatrixRepresentation}). For example, the GT~$\T$ in Fig.~\ref{fig:MCM:examples}.b is equivalently written in terms of the color variables as %(equivalently, 
 $\alpha_1'=\alpha_1+\alpha_2$ and $\alpha_2'=\alpha_1$.
The matrix $\Tmat$ in Eq.~\eqref{app:eq:exFig:GT:T} is the matrix representation of this linear transformation.

\subsection{GTs of MCMs}
\label{app:proof:GT:MCM}

\subsection{Number of MCMs in a class of MCMs}
\label{app:ppty:NbMCM:inClass}
We give the main lines of the proof of %To prove 
the results in Lemma~\ref{ppty:NbMCM:inClass}.
Consider a system with $n$ $q$-state spin variables.
Let us denote $\mathcal{N}_q(n,r)$ the number of different ways to choose $r$ independent operators in this $n$-spin system. For the special case $r=n$, this number corresponds to the number of possible gauge transformations in the $n$ spin system, $\mathcal{N}_q(n,r)=|\mathcal{G}_q(n)|$, given in Eq.~\eqref{eq:nb_GT} and~\eqref{eq:nb_GT:q-non-prime}. Using the same approach as the one used to compute $|\mathcal{G}_q(n)|$, we obtain more generally for any $r\in\{1,\dots,n\}$ that:
\begin{align}
    \textrm{for}\;q\;\textrm{prime},\;\qquad
    &\mathcal{N}_q(n, r)  
    = \prod_{i=0}^{r-1} (q^n-q^i)\,;\\
    \textrm{for}\;q\;\textrm{not prime},\;\qquad
    &\mathcal{N}_q(n, r)
    = \prod_{k=1}^{K} p_k^{\,(m_k-1)n^2}\,
    \prod_{i=0}^{r-1} (p_{k}^{\,n}-p_{k}^{\,i})\,.
\end{align}

From this result, we can compute %, \textcolor{red}{for real-valued spin models over $n$ spins,} 
the number of (strongly) independent models with rank $r$, among all real-valued spin models of $n$ spins. This number is given by:
%%%%%
%Consider real-valued spin models over $n$ spins. From the previous result, we can compute the number of (strongly) independent models with rank $r$:
%%%%%
\begin{align}
    \mathcal{N}_{ind}(n,r) = 
    \begin{cases}
        &\displaystyle\frac{\mathcal{N}_q(n, r)}{r!}\,,\qquad{\rm for }\;q=2\,,\\[5pt]
        &\displaystyle\frac{\mathcal{N}_q(n, r)}{2^r\,r!}\,,\qquad{\rm for }\;q>2\,.
    \end{cases}
\end{align}
%where the $r!$ account for 
This corresponds to the number of ways of choosing $r$ independent operators in the set of all possible $q^n-1$ operators, divided by the number of possible permutations of these $r$ elements. For $q>2$, the division by $2^r$ accounts for the fact that using an operator or its c.c. in the independent set %defining an independent model with an operator or with its c.c.
defines the same independent model (because real-valued models contain both conjugate operators, and 
%because operators that are their own conjugate for $q>2$ cannot be part of an independent set of operators).
because, for $q>2$, operators that can be used in an independent set cannot be their own conjugate).

Take an MCM~$\M$ with a single complete component of rank $r$. 
The number of different MCMs in the same class as $\M$ (i.e., with a single C.C. of rank $r$) 
%The total number of possible such model %with a single complete component (CC) of rank $r$, 
is given by the total number of possible GTs of%the single component in
~$\M$, divided by the number of these GTs that leaves $\M$ invariant. This is equal to the number of ways to choose $r$ independent (basis) operators among all possible $(q^n-1)$ operators, divided by the number of ways to choose $r$ independent (basis) operators among the $(q^r-1)$ operators that are already in the $\M$, which is:
\begin{align}
    \mathcal{N}_{\rm CC}(n,r) 
    = \displaystyle\frac{\mathcal{N}_q(n, r)}{\mathcal{N}_q(r, r)}
    =
    \begin{cases}
        \;\;\displaystyle\prod_{i=0}^{r-1}\,\frac{q^n-q^i}{q^r-q^i}\,,&\qquad{\rm for }\;q\;{\rm prime},\\[4pt]
        \;\;\displaystyle\prod_{k=1}^{K}p_k^{(m_k-1)(n^2-r^2)}\prod_{i=0}^{r-1}\frac{p_k^n-p_k^i}{p_k^r-p_k^i}\,,&\qquad{\rm for }\;q\;\textrm{not prime}.
    \end{cases}
\end{align}

Finally, take an MCM~$\M$ with $m$ ICCs and denote by $m_{r_a}$ the number of ICCs of rank $r_a$ in $\M$. The number of different MCMs in the same class as $\M$ (i.e., with the same sequence of multiplicity $\{m_{r_a}\}_{1\leq r_a\leq n}$ of their ICCs) can be obtained by combining the two previous arguments:
\begin{align}
    \mathcal{N}_{\rm MCM}(n,\{m_{r_a}\}_{1\leq r_a\leq n}) 
    = \displaystyle\frac{\mathcal{N}_q(n, r)}{\displaystyle\prod_{r_a=1}^n (m_{r_a}!)\;\, \mathcal{N}_q(r_a, r_a)^{m_{r_a}}}
    \;,
    %=
    %\begin{cases}
    %    \;\;\displaystyle\prod_{i=0}^{r-1}\,\frac{q^n-q^i}{q^r-q^i}\,,&\qquad{\rm for }\;q\;{\rm prime},\\[4pt]
    %    \;\;\displaystyle\prod_{k=1}^{K}p_k^{(m_k-1)(n^2-r^2)}\prod_{i=0}^{r-1}\frac{p_k^n-p_k^i}{p_k^r-p_k^i}\,,&\qquad{\rm for }\;q\;\textrm{not prime}.
    %\end{cases}
\end{align}
where the factors $1/m_{r_a}!$\, account for permutations among ICCs of the same size.

\subsection{Notations %Set up
for the proofs in the sections below}
\label{app:ICC:notation}

Throughout the paper, we consider a system of $n$ discrete random variables $\alpha_i$ that can take integer values between $0$ and $(q-1)$ with $q\geq 2$. A state of the system is denoted $\state = (\alpha_1, \cdots, \alpha_n) \in \Field^n$, and
%and $\data = (\state^{(1)}, \cdots, \,\state^{(N)})$ denotes a dataset composed of $N$ observations of the system. 
a dataset is composed of $N$ observations %of the state 
of the system and is denoted $\data = (\state^{(1)}, \cdots, \,\state^{(N)})$. We assume the system to be stationary and that the observed states % its states
are randomly sampled from the same underlying distribution, $p(\state\,|\,\vecg,\M)$, defined by a parametric model~$\M$.

In the following, we consider that $\M$ is an MCM with a single Independent Complete Component (ICC) $\M_a$ of rank $r_a$ (with $r_a\leq n$). 
%%%%%%%%%%%%%%%
%For the sake of simplicity of the derivations, we reduce the system to the $r_a$ variables modeled by $\M_a$. This corresponds to reducing the space to the variables $\state_a'=\Basis_a(\state)$, where $\Basis_a$ is a basis of $\M_a$, and considering the model distribution $p_a(\state_a'\,|\,\vecg_a',\M_a')$ of the transformed model $\M_a'=\T_{\Basis_a}[\M_a]$, which is a probability distribution over $r_a$ variables only, instead of $n$. 
%%%%%%%%%%%%%%%
For the sake of simplicity of the derivations, we consider that $\M$ is already written in a preferred basis and we reduce the system to the $r_a$ variables modeled by $\M_a$. This corresponds to reducing the space to the basis variables $\state_a$ on which the ICC $\M_a$ is based, and considering the model distribution $p_a(\state_a\,|\,\vecg_a,\M_a)$, which is a distribution over $r_a$ variables only (instead of $n$).
To simplify the notation, we then drop the index ``$a$'' used to label the ICCs. We denote the model $\M_{icc}$ and its rank~$r$, and use the notation $\state = (\alpha_1, \cdots, \alpha_r) \in \Field^r$ for the state of the system reduced to the $r$ variables (strictly speaking, these variables can be any subset of the variables $\state$ introduced in the previous paragraph).

\subsection{Alternative parametrization of an Independent Complete Components (ICC).}\label{app:equivalent-param}

%Consider an Independent Complete Component (ICC) over a subset of $r$ variables, with $r\leq n$. To simplify the notations, lets denote these $r$ variables: $\state = (\alpha_1, \cdots, \alpha_r) \in \Field^r$ (strictly speaking these variables could be any subset of the variables $\state$ introduced in the previous paragraph). 

The ICC $\M_{icc}$ %over the $r$ variables 
is a complete model~(see Sec.~\eqref{sec:def:HO:vectorPottsModel}) for the sub-system of $r=\rank(\M_{icc})$ variables. The model has $(q^r-1)$ parameters $\vecg$ and can be (equivalently) defined as the following parametric family of probability distributions: %with $(q^r-1)$ parameters:
\begin{align}\label{app:def:ICC}
    p(\state\,|\,\boldsymbol{\eta},\M_{icc}) = \sum_{\vecmu\in\Field^r}\, \eta_{\vecmu} \, \delta(\state-\vecmu)\,,
\end{align}
where each $\eta_{\vecmu}$ is the probability that the system is in the state $\state = \vecmu$, here considered as a parameter.
%where %$\vecp = (p_{1}, \,\cdots,\,p_{q^r-1})$ 
We denote $\boldsymbol{\eta}$ %$ = (p_{\vecmu}\,|\,)$
the vector of the $(q^r-1)$ parameters $\eta_{\vecmu}$ 
%for $\vecmu\neq\veczero$.
for all non-zero $\vecmu$.
The remaining probability $\eta_{\veczero}=p(\veczero)$ is defined by the normalization: %the normalization of the probability distribution: 
\begin{align}\label{eq:normalisation}
    \eta_{\veczero} = 1 - \sum_{\vecmu=1}^{q^r-1} \eta_{\vecmu}\,.
\end{align}
In this equation, the sum is over all the non-zero vectors $\vecmu\in[\Field^r]^*$. %, but for clarity, 
We denoted each vector $\vecmu$ by its integer value in base $q$, %(i.e., going from $1$ to $(q^r-1)$). 
which allows ordering the vectors $\vecmu$ from $1$ to $(q^r-1)$.
For instance, the vector $\vecmu=1$ has its first entry equal to $1$, and all the other entries are equal to $0$. The vector $\vecmu=(q^r-1)$ has all its $r$ entries equal to $(q-1)$. 
Substituting the expression of $\eta_{\veczero}$ 
%in the previous equation, we get:
in Eq.~\eqref{app:def:ICC}, we get:
\begin{align}\label{app:eq:ICC_PDF}
    p(\state\,|\,\boldsymbol{\eta},\M_{icc}) = \sum_{\vecmu=1}^{q^r-1}\, \eta_{\vecmu} \, \delta(\state-\vecmu)+\left(1-\sum_{\vecmu=1}^{q^r-1} \eta_{\vecmu}\right)\,\delta(\state-\boldsymbol{0})\,,
\end{align}
where $\vecmu=\boldsymbol{0}$ is the state where all the variables have the value $0$.

\subsection{Fisher Information Matrix (FIM) for an ICC.}

For an MCM (with multiple ICCs), the Fisher Information Matrix (FIM) in Eq.~\eqref{def:FIM} is block diagonal, where each block corresponds to the FIM restricted to a single ICC. This is because the model probability distribution of an MCM factorizes over its ICCs, which also have non-overlapping sets of parameters. %(i.e., to second order derivative/Hessian of the log-partition function according to the parameters of one ICC only).} \\
Consequently, the determinant of the FIM of an MCM is the product of the determinant of the FIM for each ICC, 
i.e., for an MCM $\M=\cup_{a\in\mathcal{A}}\M_a$ with parameters $\vecg=\cup_{a\in\mathcal{A}}\vecg_a$: 
\begin{align}\label{app:eq:MCM:detFIM}
    \det \FIM(\vecg)=\prod_{a\in\mathcal{A}}\det \FIM_a(\vecg_a)\,.
\end{align}

Let us now compute the FIM $\FIM_a$ for a single ICC. For simplicity, we consider %reduce the system to a 
a system reduced to a %with 
single ICC with the notations specified %[at the beginning of] 
in Sec.~\ref{app:ICC:notation} (in particular, we drop the index ``$a$'' used to label ICCs).
For the ICC probability distribution with the new parametrization in Eq.~\eqref{app:eq:ICC_PDF}, 
the Fisher Information Matrix (FIM) 
%is the $(q^r-1)$ square matrix with coefficients defined as:
is the square matrix of order $(q^r-1)$ with elements:
%whose coefficients are defined as:
\begin{align}\label{app:def:FIM}
    \FIM_{\vecmu \vecnu}(\boldsymbol{\eta})
        &=-\left\langle\partial_{\eta_{\vecmu}}\partial
    _{\eta_{\vecnu}}\log p(\state\,|\,\boldsymbol{\eta},\M_{icc})\right\rangle\,,\qquad \forall \, \vecmu,\vecnu\in[\Field^r]^*\,,
\end{align}
where the average is taken over the model distribution $P(\state\,|\,\M_{icc}, \,\boldsymbol{\eta})$.
Substituting Eq.~\eqref{app:eq:ICC_PDF} and taking the two derivatives gives, 
$\forall \, \vecmu,\vecnu\in[\Field^r]^*$:
\begin{align}\label{app:result:FIM}
    \FIM_{\vecmu \vecnu}(\boldsymbol{\eta})
        &=\left\langle 
        \frac{(\delta(\state-\vecmu)-\delta(\state-\veczero))(\delta(\state-\vecnu)-\delta(\state-\veczero))}
        {\left[P(\state\,|\,\M_{icc}, \,\boldsymbol{\eta})\right]^2}
        %{\left[\sum_{\veckappa\in\Field^r}\, p_{\veckappa} \, \delta(\state-\veckappa)\right]^2} 
        \right\rangle\\
        &= \sum_{\state\in\Field^r}\,
        \frac{\delta(\state-\vecmu)\delta(\state-\vecnu)+\delta(\state-\veczero)}
        {P(\state\,|\,\M_{icc}, \,\boldsymbol{\eta})}\\
    \FIM_{\vecmu \vecnu}(\boldsymbol{\eta})
        &=\frac{\delta_{\vecmu,\vecnu}}{\eta_{\vecmu}}+\frac{1}{\eta_{\veczero}}\,,
\end{align}
in which $\eta_{\veczero}$ is given by Eq.~\eqref{eq:normalisation}. We recall that $\boldsymbol{\eta}$ is a vector of the $(q^r-1)$ parameters 
$\eta_{\vecmu}$ for all $\vecmu\in[\Field^r]^*$.

%\paragraph{Determinant of FIM.} 
For the following sections,
it is useful to compute the determinant of $\FIM$. 
To do so, we observe that $\FIM$ can be written under the form $\FIM=\mathbf{D} + \mathbf{u}\mathbf{v}^{\top}$, 
where $\mathbf{D}$ is a diagonal matrix with entries
$\mathbf{D}_{\vecmu\vecmu} = 1/\eta_{\vecmu}$ for $\forall \, \vecmu\in[\Field^r]^*$, 
%and $\mathbf{u}$ is a column vector of $1$'s and $\mathbf{v}$ a vector of $1/p_0$.
and where $\mathbf{u}$ and $\mathbf{v}$ are two $(q^r-1)$-dimensional column vectors with, respectively, all entries equal to $1$ and all entries equal to $1/\eta_0$.
%Using the matrix determinant lemma, the determinant of the FIM is then given by:
The matrix determinant lemma then 
%gives the determinant of the FIM:
allows to compute the determinant of the FIM:
\begin{align}
\det \FIM(\boldsymbol{\eta})
    &=(1 + \mathbf{v}^\top \mathbf{D}^{-1} \mathbf{u})\, \det \mathbf{D}
    = \left(1+\frac{1}{\eta_{\veczero}}
    %\sum_{\vecmu\in[\Field^r]^*}p_{\vecmu}\right)\\
    %\sum_{\vecmu\neq\veczero}p_{\vecmu}\right)
    %\prod_{\vecmu\neq\veczero}\frac{1}{p_{\vecmu}}\\
    \sum_{\vecmu=1}^{q^r-1}\eta_{\vecmu}\right)\,
    \prod_{\vecmu=1}^{q^r-1}\frac{1}{\eta_{\vecmu}}\\
\det \FIM(\boldsymbol{\eta})
    &=\prod_{\vecmu=0}^{q^r-1}\;\frac{1}{\eta_{\vecmu}}\,,
    \label{app:eq:detFIM}
\end{align}
in which $\eta_{\veczero}$ is given by Eq.~\eqref{eq:normalisation}. 

\subsection{Geometric complexity of an ICC.}\label{app:ICC:Complexity}
%\paragraph{MDL geometric complexity.} 

Assuming Jeffreys' prior over the model parameters, the geometric complexity of a spin model is given by~\cite{beretta2018stochastic}:
\begin{align}\label{app:def:cM}
    c_{\M} = \log \int\sqrt{\det \FIM(\vecg)}\,d\vecg\,,
\end{align}
where the integral is over all possible values of the parameters.
Using the factorization of $\det\FIM(\vecg)$ in Eq.~\eqref{app:eq:MCM:detFIM}, the geometric complexity of an MCM can be written as the sum of the geometric complexity of its ICCs, as in Eq.~\eqref{eq:MCM:Complexity}.
In the following, we compute the complexity of a single ICC of rank~$r$, which we denote $c_{icc}(r)$.
For simplicity, we consider %reduce the system to a 
a system reduced to a %with 
single ICC with the notations specified in %[at the beginning of]
Sec.~\ref{app:ICC:notation}.

\begin{proof}[Proof of the geometric complexity of an ICC in Eq.~\eqref{eq:c_M}.]
    %Proof of the geometric complexity of an ICC in Eq.~\eqref{eq:c_M}.
The geometric part of the MDL complexity~\cite{rissanen1996fisher, rissanen2001strong} is defined as $c_{icc}(r) = \log V_{icc}(r)$, where $V_{icc}(r)$ is the volume of the $K=(q^r-1)$-dimensional model manifold: % given by:
%\begin{align}
%    c_{icc} = \log V_{ICC}\,,
    %\qquad\qquad    {\rm where}\;V_{ICC} = 
%\end{align}
\begin{align}\label{app:eq:Vicc:def}
V_{icc}(r) = \int_{[0,1]^K}\sqrt{\det \FIM(\boldsymbol{\eta})}\,d\boldsymbol{\eta}\,.
\end{align}
Substituting Eq.~\eqref{app:eq:detFIM}, one gets:
\begin{align}\label{app:eq:Vicc:def}
V_{icc}(r)
    = \int_{[0,1]^K} \eta_{\veczero}^{-1/2}\,\prod_{\vecmu=1}^{q^r-1} \eta_{\vecmu}^{-1/2}\,d\boldsymbol{\eta}\,,
    \qquad
    {\rm in \;which}\;\; 
    \eta_{\veczero}=1-\sum_{\vecmu=1}^{q^r-1} \eta_{\vecmu}\,.
\end{align}
Here one can recognize the multivariate beta function:
\begin{align}\label{app:eq:Vicc:result}
    V_{icc}(r)
        &= B\underset{q^r\;{\rm terms}}{\underbrace{\left(\frac{1}{2},\,\cdots,\,\frac{1}{2}\right)}}
        =\frac{\displaystyle\prod_{i=0}^K \,\Gamma\left(\frac{1}{2}\right)}{\displaystyle\Gamma\left(\sum_{i=0}^K \frac{1}{2}\right)}
    %V_{icc}
        =\frac{\displaystyle\pi^{\frac{q^r}{2}}}{\displaystyle\Gamma\left(\frac{q^r}{2}\right)}\,,
\end{align}
which gives the complexity term:
\begin{align}\label{app:eq:cm:icc}
    c_{icc}(r) = \frac{q^{r}}{2} \log \pi - \log \Gamma\left(\frac{q^{r}}{2}\right)\,.
\end{align}
\end{proof}
\noindent For $q=2$, one recovers the result obtained in the binary case by Ref.~\cite{beretta2018stochastic}:
%For the binary case $q=2$, one recovers the result obtained in Ref.~\cite{beretta2018stochastic}:
\begin{align}
    c_{icc}(r) = 2^{r-1} \log \pi - \log \Gamma(2^{r-1})\,.
\end{align}

\subsection{Likelihood function for an ICC.}
Using the definition of the ICC in Eq.~\eqref{app:def:ICC}, we compute the likelihood of the ICC for the dataset $\data$:
\begin{align}
    P(\data\,|\,\boldsymbol{\eta},\M_{icc})
        &=\prod_{i=1}^N p(\state^{(i)}\,|\,\boldsymbol{\eta},\M_{icc}) \\
        &
        =\prod_{\vecmu=0}^{q^r-1} \;\eta_{\vecmu}^{\;k_{\vecmu}(\data)}
        \label{app:eq:likelihood}
\end{align}
in which $\eta_0$ satisfies Eq.~\eqref{eq:normalisation} and where $k_{\vecmu}(\data)$ is 
%the number of times $\state=\vecmu$ in the dataset
the number of times the state $\state=\vecmu$ occurs in the dataset $\data$. 

\subsection{Jeffreys' prior.}\label{app:ICC:JeffreysPrior}

Jeffreys' prior for a parametric model with parameters~$\vecg$ is defined as~\cite{jeffreys1946invariant}:
\begin{align}\label{app:def:JeffreysPrior:general}
    P_0(\vecg) 
    \propto %\frac{1}{V_{icc}} \,
    \sqrt{\det \FIM(\vecg)}\,,
\end{align}
where $\FIM(\vecg)$ is the FIM [of the model].
Using the factorization of $\det\FIM(\vecg)$ in Eq.~\eqref{app:eq:MCM:detFIM}, Jeffreys' prior for an MCM factorizes over its ICCs: $P_0(\vecg)=\prod_{a\in\mathcal{A}} P_0(\vecg_a)$, for an MCM $\M=\cup_{a\in\mathcal{A}}\M_a$ with parameters $\vecg=\cup_{a\in\mathcal{A}}\vecg_a$.
The factorization of the prior over the parameters is used for the factorization of the model evidence in Eq.~\eqref{eq:evidenceMCM}.

%%%%%%%%%%%%%%%% %%%%%%%%%%%%%%%% %%%%%%%%%%%%%%%% %%%%%%%%%%%%%%%%
%The general lines to show that Jeffreys' prior for MCMs factorizes over the ICCs is that the FIM is block diagonal, where each block corresponds the FIM restricted to a single ICC (i.e., to second order derivative/Hessian of the log-partition function according to the parameters of one ICC only).\\
%
%Indeed, for an MCM, the model probability distribution factorizes over the ICCs (which have distinct/non-intersecting sets of parameters), [which means that the log-partition function of the model is the sum of the log-partition function for each ICC.] As a result, the FIM of an MCM, which is given by the Hessian of the log-partition function~\eqref{}, is block diagonal, where each block corresponds the FIM restricted to a single ICC. Consequently, the determinant of the FIM is the product of the determinant of the FIM for each ICC and Jeffreys' prior~\eqref{} factorizes over the ICCs.
%%%%%%%%%%%%%%%% %%%%%%%%%%%%%%%% %%%%%%%%%%%%%%%% %%%%%%%%%%%%%%%%

In the following, we obtain a closed-form expression for Jeffreys' prior for a single ICC.
For simplicity, we consider %reduce the system to a 
a system reduced to a %with 
single ICC with the notations specified in %[at the beginning of] 
Sec.~\ref{app:ICC:notation}.
For the ICC distributions~\eqref{app:def:ICC} parametrized by %the parameters 
$\boldsymbol{\eta}$, Jeffreys' Prior is computed as:
\begin{align}\label{app:def:JeffreysPrior}
P_0(\boldsymbol{\eta}) 
    = \frac{1}{V_{icc}} \,\sqrt{\det \FIM(\boldsymbol{\eta})}\,,
\end{align}
where $\FIM(\boldsymbol{\eta})$ is the FIM and the normalization $V_{icc}$ is the volume in Eq.~\eqref{app:eq:Vicc:def}. Substituting the expression of $\det \FIM(\boldsymbol{\eta})$ from Eq.~\eqref{app:eq:detFIM}, one gets:
\begin{align}\label{app:eq:JeffreysPrior}
P_0(\boldsymbol{\eta}) 
    = \frac{1}{V_{icc}} \,\prod_{\vecmu=0}^{q^r-1} \,\eta_{\vecmu}^{-1/2}\,,
    \qquad
    {\rm in \;which}\;\; 
    \eta_{\veczero}=1-\sum_{\vecmu=1}^{q^r-1} \eta_{\vecmu}\,.
\end{align}
Here, one can recognize the symmetric Dirichlet distribution with hyperparameter %equal to 
$1/2$.

\subsection{Evidence of an ICC.}\label{app:ICC:Evidence}
\begin{proof}[Proof of Eq.~\eqref{eq:evidenceICC}]
Combining the expressions in Eq.~\eqref{app:eq:likelihood} for the likelihood function and in Eq.~\eqref{app:eq:JeffreysPrior} for Jeffreys' prior $P_0(\boldsymbol{\eta})$, 
the evidence (or marginal likelihood) of the ICC is finally given by:
\begin{align}
    P(\data\,|\,\M_{icc}) 
        &=\int_{[0,1]^K} P(\data\,|\,\boldsymbol{\eta},\M_{icc})\,P_0(\boldsymbol{\eta})\,d\boldsymbol{\eta} \\
        &=\frac{1}{V_{icc}}\,\int_{[0,1]^K}
        \,\prod_{\vecmu=0}^{q^r-1}
        \eta_{\vecmu}^{\,k_{\vecmu}(\data)}\, \eta_{\vecmu}^{-1/2}\,
        d\eta_{\vecmu}\,.
\end{align}
where $K=(q^r-1)$ is the number of parameters in the ICC.
%where we used the expression in Eq.~\eqref{app:eq:JeffreysPrior} for Jeffreys' prior $P_0(\vecp)$.
Here we recognize the (multivariate) Beta function of the $q^r$ variables $(k_{\vecmu}(\data)+1/2)$:
%\begin{align}
%    P(\data\,|\,\M_{icc}) 
%        &=\frac{\displaystyle B\left(k_0+\frac{1}{2},\,\cdots,\,k_{q^r-1}+\frac{1}{2}\right)}{\displaystyle B\left(\frac{1}{2},\,\cdots,\,\frac{1}{2}\right)}\\
%    \Aboxed{
%    P(\data\,|\,\M_{icc}) 
%        &=\frac{\displaystyle\Gamma\left(\frac{q^r}{2}\right)}{\displaystyle\pi^{\frac{q^r}{2}}}
%        \;\frac{\displaystyle\prod_{\vecmu=0}^{q^r-1}\,\Gamma\left(k_{\vecmu}+\frac{1}{2}\right)}
%        {\displaystyle\Gamma\left(N+\frac{q^r}{2}\right)}\,,
%    }
%\end{align}
\begin{align}
    P(\data\,|\,\M_{icc}) 
    \;=\;
        \frac{\displaystyle B\left(k_0+\frac{1}{2},\,\cdots,\,k_{q^r-1}+\frac{1}{2}\right)}{\displaystyle B\left(\frac{1}{2},\,\cdots,\,\frac{1}{2}\right)}
    \;=\;
        \frac{\displaystyle\Gamma\left(\frac{q^r}{2}\right)}{\displaystyle\pi^{\frac{q^r}{2}}}
        \;\frac{\displaystyle\prod_{\vecmu=0}^{q^r-1}\,\Gamma\left(k_{\vecmu}(\data)+\frac{1}{2}\right)}
        {\displaystyle\Gamma\left(N+\frac{q^r}{2}\right)}\;,
\end{align}
where $k_{\vecmu}$ is the number of times that the state $\vecmu$ occurs in the dataset $\data$.
For $q=2$, we recover the results obtained for the binary case in Ref.~\cite{demulatier2024MCM}.
%
%The evidence can be written under the form:
Rearranging the terms:
\begin{align}\label{SM:eq:CM:LogEvidence}
    P(\data\,|\,\M_{icc}) 
    \;=\;
        \frac{\displaystyle\Gamma\left(\frac{q^r}{2}\right)}{\displaystyle\Gamma\left(N+\frac{q^r}{2}\right)}
        \;\prod_{\vecmu=0}^{q^r-1}\left[\frac{\displaystyle\,\Gamma\left(k_{\vecmu}(\data)+\frac{1}{2}\right)}
        {\displaystyle\sqrt{\pi}}\right]\;,
\end{align}
and one can observe that, 
%Note that 
for all the states $\vecmu$ that don't occur in the dataset ($k_{\vecmu}(\data)=0$), one has $\Gamma(k_{\vecmu}(\data)+1/2)/\sqrt{\pi}=1$, and therefore the evidence can be re-written as:
\begin{align}
    P(\data\,|\,\M_{icc}) 
    \;=\;
        \frac{\displaystyle\Gamma\left(\frac{q^r}{2}\right)}{\displaystyle\Gamma\left(N+\frac{q^r}{2}\right)}
        \;\prod_{\vecmu\in\data}\left[\frac{\displaystyle\,\Gamma\left(k_{\vecmu}(\data)+\frac{1}{2}\right)}
        {\displaystyle\sqrt{\pi}}\right]\;,
\end{align}
where the product is now only over the states observed in the dataset~$\data$.
\end{proof}

\subsection{Expansion of the log-evidence for large $N$ and MDL principle.}
\label{app:log-Evidence:expansion}

Bayesian model selection and Minimum Description length principle are already known to be equivalent for large dataset, assuming Jeffreys' prior over the parameters. For completeness, we re-derive this result in the particular case of MCM. % and give the next order term.
Let us focus on the evidence of the complete model, %first,
as the extension to %expansion/result for 
MCM is straightforward from that result. %the results for complete models. %being straightforward from result .
Taking the logarithm of the evidence in Eq.~\eqref{SM:eq:CM:LogEvidence} yields:
\begin{align}
    \hspace{-2mm}
    \log P(\data\,|\,\M_{icc}) = 
        \sum_{\state\in\Field^r}\, \log \Gamma\left(k_{\state}(\data)\hspace{-1mm}+\frac{1}{2}\right)
        -\log \Gamma\left(N+\frac{q^r}{2}\right)
        -\log \frac{\pi^{\frac{q^r}{2}}}{\Gamma\left(\frac{q^r}{2}\right)}\,,
\end{align}
where the last term can be recognized as the geometrical complexity $c_{icc}(r)$ of an ICC of rank~$r$ in Eq.~\eqref{app:eq:cm:icc}. 
We assume that the dataset~$\data$ is sufficiently well-sampled, such that the empirical probability $p_{\state} = k_{\state}(\data)/N$ is non zero for all state (each state is observed at least a few times).
%Assuming that the dataset~$\data$ is \textcolor{red}{well-sampled/far from the undersampling regime} (i.e., it is sufficiently large and each state is observed at least a few number times, $k_{\spin} = N\,p_{\spin}$, 
%\textcolor{red}{where the probability $p_{\spin}$ is assumed non zero: NOT NEEDED: $p_s$ are just the empirical probabilities defined as $p_s = k_s/N$}),
%at least once
%then one can 
We then take the expansion of %$\log \Gamma(N+2^{n-1})$ and of $\log \Gamma\left(k_{\spin}\hspace{-1mm}+\frac{1}{2}\right)$ 
the two log-gamma function for large~$N$. %to the order of $1/N$, 
%then one can take the expansion of $\log \Gamma(N+2^{n-1})$ and of each term $\log \Gamma\left(k_{\spin}\hspace{-1mm}+\frac{1}{2}\right)$ respectively to the order of $1/N$ and $1/k_{\spin}$:
This gives the following expansion of the log-evidence to the order of $1/N$:
\begin{align}\label{SM:eq:CM:LogEvidence:expansion}
    \hspace{-2mm}
        \log P(\data\,|\,\M_{icc}) = 
        N\,\sum_{\state\in\Field^r} p_{\state}\,\log p_{\state} - \frac{K}{2} \log\left(\frac{N}{2\pi}\right) - c_{icc}(r) +O\left(\frac{1}{N}\right)\,,
\end{align}
%\begin{align}\label{SM:eq:CM:LogEvidence:expansion}
%    \hspace{-2mm}
%        \sum_{\spin}\, \log \Gamma\left(k_{\spin}\hspace{-1mm}+\frac{1}{2}\right)
%        -\log \Gamma(N+2^{n-1})
%        =
%        \sum_{\spin} k_{\spin}\,\log \frac{k_{\spin}}{N} - \frac{K}{2} \log\left(\frac{N}{2\pi}\right) +O\left(\frac{1}{N}\right)\,.
%\end{align}
where $K=q^r-1$ is the number of parameters in the ICC.
The first term in the expansion is the maximum log-likelihood of the ICC (see Property~\ref{ppty:max-log-L}), %$P(\spin\,|\,\vecp,\,\M^c) = - N \sum_{\spin} p_{\spin}\log p_{\spin}$, 
in which $p_{\state} = k_{\state}(\data)/N$ is the empirical probability of state $\state$ in the dataset $\data$.
The second term is the first complexity term in the MDL principle, penalizing models with a large number $K$ of parameters. 
Comparing Eq.~\eqref{SM:eq:CM:LogEvidence:expansion} with Eq.~\eqref{eq:MDL}, one can see that selecting the MCM with the largest log-evidence, $\log P(\data\,|\,\M)$, is equivalent at large $N$ to selecting the MCM with the smallest description length, $L(\data\,|\,\M)$.

%\section{Details on the algorithms (if needed)}
%\textcolor{red}{Figure for the algorithms. (if needed)} 
%In particular regarding the use of several (three) 128-bit integers to encode 128 $q$-state variables.

\section{Additional Figures for data analysis examples}
\subsection{Entropy of the variables in the data of the Big Five Personality Test.}

\begin{figure}[!ht]
    \centering
    \includegraphics[width=\linewidth]{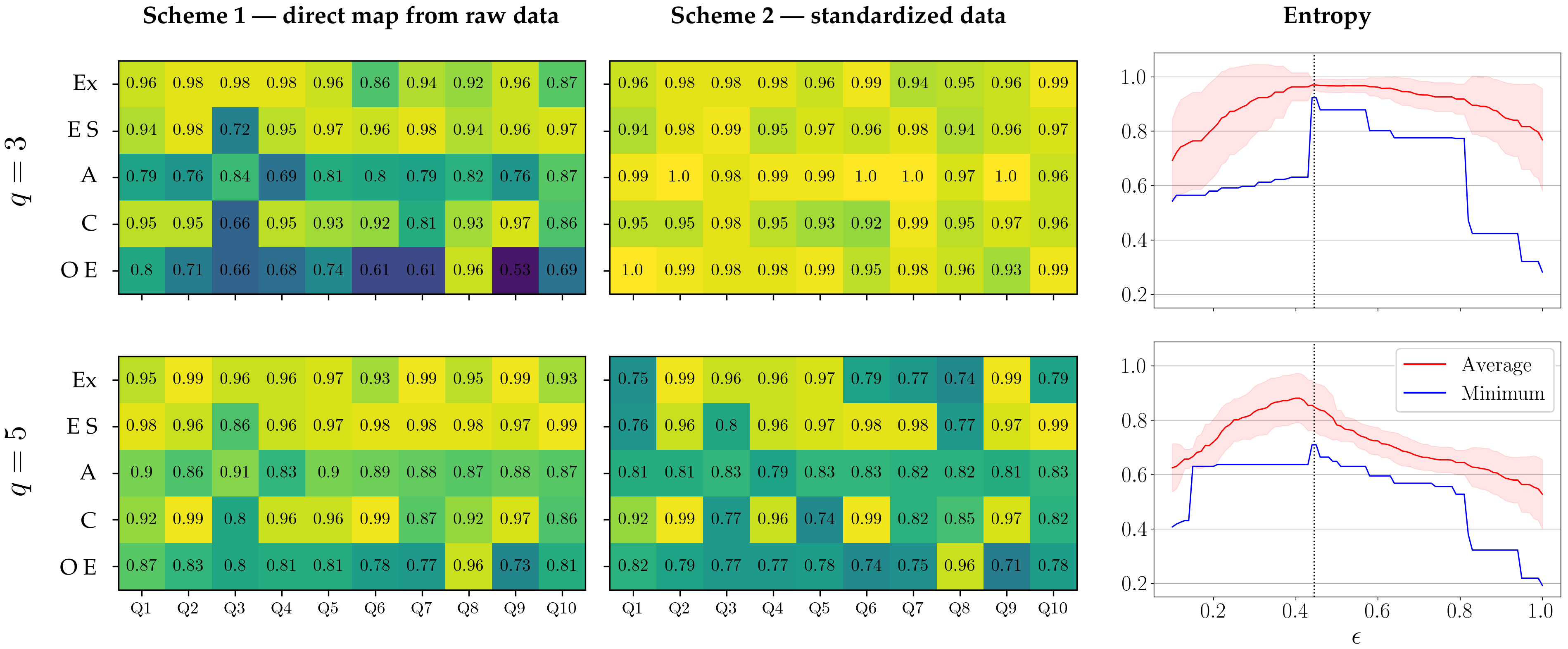}
\caption{
    \label{SI:fig:Big5-Entropy}
    {\bf Entropy (in base $q$) of the $50$ variables in the Big Five Personality Test data} for the two discretization schemes described in Sec.~\ref{Sec:MCM:Big5} with $q=3$ and $q=5$.
    %Scheme 1 = original discretization; Scheme 2 = re-centered discretization. 
    %Scheme 1: direct mapping; Scheme 2: standardization of the data.
    {\bf The two left panels} give the value of the entropy in base $q$ (or ``qits'') for each variable for each discretization choice. %is given in base $q$, or ``$q$-its'', (i.e., divided by $\log q$), and should thus vary between $0$ and $1$.
    The color indicates the magnitude of the entropy, which can vary between $0$ and $1$ qit.
    The variables are displayed in rows from left to right in the same order as in the original dataset~\cite{Big5data} and each row corresponds to one trait (``Ex'' for Extraversion, ``ES'' for Emotional stability, ``A'' for Agreeableness, ``C'' for Conscientiousness, and ``OE'' for Openness to experience).
%    The color intensity reflects the magnitude of the entropy. The value of the entropy is given in base $q$, or ``$q$-its'', (i.e., divided by $\log q$), and should thus vary between $0$ and $1$.
    For example, in the raw data (i.e., scheme~1 with $q=5$), the previous to last variable is the least entropic and corresponds to the question ``{\it I spend time reflecting on things}''. 
    {\bf The right panel} shows the average entropy (red curve) and minimal entropy (blue curve) obtained as one varies the value of $\epsilon$ in Scheme 2. For the analysis in Sec.~\ref{Sec:MCM:Big5}, the value of epsilon was chosen to maximize the entropy of the least entropic variable
    %the average entropy 
    (as indicated by the vertical dashed line). This is the same value of $\epsilon$ that is used in the middle panel.
    }
\end{figure}

\end{appendices}

%%%%%%%%%%%%%%%%%%%%%%%%%%%%%%%%%%%%%%%%%%%%%%%%%
\bibliographystyle{unsrt} %plain
\bibliography{biblio}

\end{document}